%% file: IRAM_CO_SF_bars.tex
\documentclass{aa}
\usepackage{graphicx}
\usepackage{txfonts}
\usepackage[utf8]{inputenc}
\usepackage{amssymb,amsmath}
\usepackage{lscape}
\usepackage{longtable}
\usepackage{float}
\usepackage{natbib}
\usepackage{multicol}
\usepackage[bookmarks=false,colorlinks,allcolors=blue]{hyperref}
\usepackage[para,multiple]{footmisc}
\begin{document} 
\title{Molecular gas and star formation \\ within 12 strong galactic bars observed with IRAM-30m
\thanks{
Tabulated CO(1–0) and CO(2–1) spectra are only available at the
CDS via anonymous ftp to cdsarc.u-strasbg.fr (130.79.128.5)
or via \href{http://cdsarc.u-strasbg.fr/viz-bin/cat/J/A+A/654/A135}{http://cdsarc.u-strasbg.fr/viz-bin/cat/J/A+A/654/A135}
}
}
\titlerunning{Molecular gas and star formation within 12 strong galactic bars observed with IRAM-30m}
  \author{S. D\'iaz-Garc\'ia\inst{1,2,3}
          \and        
          U. Lisenfeld \inst{4, 5}
          \and 
          I. P\'erez \inst{4, 5}
          \and 
          A. Zurita \inst{4, 5}
          \and 
          S. Verley \inst{4, 5}
          \and 
          F. Combes \inst{6}
          \and 
          D. Espada \inst{4, 7}
          \and
          S. Leon \inst{8}
          \and      \\
          V. Mart\'inez-Badenes \inst{9}
          \and
          J. Sabater \inst{10,11}
          \and      
          L. Verdes-Montenegro \inst{12}
          }
  \institute{Instituto de Astrof\'isica de Canarias, E-38205, La Laguna, Tenerife, Spain \\  
              \email{simondiazgar@gmail.com}
         \and
             Departamento de Astrof\'isica, Universidad de La Laguna, E-38205, La Laguna, Tenerife, Spain
         \and 
             Department for Physics, Engineering Physics and Astrophysics, Queen's University, Kingston, ON K7L 3N6, Canada
         \and
             Departamento de F\'isica Te\'orica y del Cosmos, Campus de Fuentenueva, Universidad de Granada, E18071-Granada, Spain
         \and 
            Instituto Carlos I de F\'\i sica Te\'orica y Computacional, Facultad de Ciencias, 18071 Granada, Spain
         \and 
            Observatoire de Paris, LERMA, Coll\`ege de France, CNRS, PSL University, Sorbonne University, 75014, Paris, France
         \and 
            SKA Organization, Lower Withington, Macclesfield, Cheshire SK11 9DL, UK
         \and 
            Joint ALMA Observatory, Alonso de Cordova 3107, Vitacura, Santiago 763-0355, Chile
         \and        
            Universidad Internacional de Valencia (VIU), C. del Pintor Sorolla 21, E-46002 Valencia, Spain
         \and 
            SUPA, Institute for Astronomy, Royal Observatory, Blackford Hill, Edinburgh, EH9 3HJ, UK
         \and
            UK Astronomy Technology Centre, Royal Observatory, Blackford Hill, Edinburgh, EH9 3HJ, UK
         \and 
            Instituto de Astrof\'isica de Andaluc\'ia (IAA-CSIC), Glorieta de la Astronom\'ia, 18008 Granada, Spain
             }
  \date{Received 26 February 2021; accepted 22 June 2021}
  \abstract
  {
\emph{Context.} 
While some galactic bars show recent massive star formation (SF) along them, some others do not. 
Whether bars with low level of SF are a consequence of low star formation efficiency, low gas inflow rate, or dynamical effects remains a matter of debate.
\\
\emph{Aims.}
In order to study the physical conditions that enable or prevent SF, we perform a multi-wavelength analysis of 12 strongly barred galaxies 
with total stellar masses ${\rm log}_{10}(M_{\star}/M_{\odot})\in [10.2,11]$, 
chosen to host different degrees of SF along the bar major axis without any prior condition on gas content. 
We observe the CO(1-0) and CO(2-1) emission within bars with the IRAM-30m telescope 
(beam sizes of 1.7-3.9 kpc and 0.9-2.0 kpc, respectively; 7-8 pointings per galaxy on average).
\\
\emph{Methods.} 
We estimated molecular gas masses ($M_{\rm  mol}$) from the CO(1-0) and CO(2-1) emissions. 
SF rates (SFRs) were calculated from \emph{GALEX} near-ultraviolet (UV) and \emph{WISE} 12~$\mu$m 
images within the beam-pointings, covering the full bar extent (SFRs were also derived from far-UV and 22~$\mu$m).\\
\emph{Results.} 
We detect molecular gas along the bars of all probed galaxies. 
Molecular gas and SFR surface densities span the ranges 
${\rm log}_{10}(\Sigma_{\rm mol}/[M_{\odot}$ pc$^{-2}]) \in [0.4,2.4]$ and 
${\rm log}_{10}(\Sigma_{\rm SFR}/[M_{\odot}$ pc$^{-1}$ kpc$^{-2}]) \in [-3.25,-0.75]$, respectively. 
The star formation efficiency (SFE; i.e. SFR/$M_{\rm  mol}$) in bars varies between galaxies by up to an order of magnitude 
(SFE $\in [0.1,1.8]\,$Gyr$^{-1}$). On average, SFEs are roughly constant along bars. 
SFEs are not significantly different from the mean value in spiral galaxies reported in the literature ($\sim 0.43$ Gyr$^{-1}$), 
regardless of whether we estimate $M_{\rm  mol}$ from CO(1-0) or CO(2-1). 
Interestingly, the higher the total stellar mass of the host galaxy, the lower the SFE within their bars. 
In particular, the two galaxies in our sample with the lowest SFE and $\Sigma_{\rm SFR}$ 
(NGC~4548 and NGC~5850, SFE $\lesssim 0.25$ Gyr$^{-1}$, $\Sigma_{\rm SFR} \lesssim 10^{-2.25} \, M_{\odot}$ yr$^{-1}$ kpc$^{-2}$, 
$M_{\star}\gtrsim 10^{10.7} M_{\odot}$) are also those hosting massive bulges and signs of past interactions with nearby companions.
\\
\emph{Conclusions.}
We present a statistical analysis of the SFE in bars for a sample of 12 galaxies. 
The SFE in strong bars is not systematically inhibited (either in the central, middle, or end parts of the bar).
Both environmental and internal quenching are likely responsible for the lowest SFEs reported in this work.\\
}
\maketitle
%
%
%
\section{Introduction}
%
%
Stellar bars are frequent stellar structures 
\citep[e.g.][]{1963ApJS....8...31D,2000ApJ...529...93K, 2000AJ....119..536E,2009A&A...495..491A,2012ApJ...761L...6M,2016A&A...587A.160D}. 
The bar potential creates lanes of molecular gas and dust in  disk galaxies 
\citep[e.g.][]{1963AJ.....68..278D,1990A&A...233...82C,1992MNRAS.259..345A,1993A&A...268...65F,1993RPPh...56..173S,
1995ApJ...454..623K,2002MNRAS.337..808K,2004ARA&A..42..603K,2004A&A...424..799P,2009ApJ...706L.256C,2015MNRAS.450.2670S}. 
These lanes are the loci of shocks in the gas flow \citep[][]{1962dmim.conf..217P} that can induce star formation (SF). 

Different distributions of SF in bars have been found based on the detection of ionised gas, 
which is typically traced by H$\alpha$ or ultraviolet (UV) emission \citep[e.g.][and references therein]{2020MNRAS.495.4158F,2020A&A...644A..38D}. 
H{\sc\,ii} regions are detected all along the bar in some galaxies, 
whereas SF is only detected in the nuclear or circumnuclear regions in others 
\citep[e.g.][]{1997A&A...326..449M,2002AJ....124.2581S,2007A&A...472..121V}. 
With the aid of stacking techniques applied to far-UV images, \citet[][]{2020A&A...644A..38D} showed that massive barred galaxies 
(total stellar masses $M_{\ast} \gtrsim 10^{10} \, M_{\odot}$) 
are typically characterised by a dip in the radial distribution of SF that is not seen in non-barred systems \citep[see also][]{2009A&A...501..207J}. 
These `SF deserts' indeed exist in barred galaxies \citep[][]{2016MNRAS.457..917J,2017MNRAS.465.3729S,2018MNRAS.474.3101J,2019MNRAS.489.4992D}.

Bar-driven central starbursts have also been proposed as the mechanism that  depletes the gas, and thus induces SF quenching 
\citep[e.g.][]{2013ApJ...779..162C,2015A&A...580A.116G,2020MNRAS.492.4697N}. 
Specifically, simulations by \citet[][]{2018A&A...609A..60K} of gas-rich disk isolated galaxies show 
that stellar bars reduce the star formation rates (SFRs) by a factor of ten in less than 1 Gyr, right after the bar strength reaches its saturation level. 
This supports the important role of the Galactic bar on the star formation history of the Milky Way \citep[e.g.][]{2016A&A...589A..66H}.

What drives SF in bars? 
In this work our aim is to compare the molecular gas mass ($M_{\rm mol}$) with the  SFR  along galactic bars, 
quantifying star formation efficiencies (SFE; i.e. SFR/$M_{\rm mol}$). 
This comparison allows us to distinguish whether the lack of SF in some bars 
is due to a small amount of gas (most likely caused by a low inflow rate) or rather due to a low star formation efficiency.

Here we estimated SFRs  based on both Galaxy Evolution Explorer (GALEX) UV 
imaging and Wide-field Infrared Survey Explorer (WISE) data. 
The UV radiation from young massive stars is partially absorbed by interstellar dust, which gets heated and emits infrared (IR) photons 
\citep[e.g.][]{2005ARA&A..43..727L}. The mid-IR emission in galaxies mainly comes from dust continuum associated with very small grains (VSGs) 
and polycyclic aromatic hydrocarbons (PAHs) \citep[e.g.][]{1984A&A...137L...5L,1989ARA&A..27..161P}, 
and can be used as a robust SFR indicator \citep[e.g.][]{2008ApJ...686..155Z}. 
There is a strong correlation between the 8 $\mu$m and 24 $\mu$m emission of H{\sc\,ii} regions and 
their H$\alpha$ and UV counterparts \citep[e.g.][]{2004AAS...205.6003C}. 
Not only can UV-based SFRs   be underestimated, but also IR-based SFRs, since not all the starlight 
is re-processed by dust \citep[][]{2009ApJ...703.1672K}; a hybrid combination of different tracers is advocated \citep[e.g.][]{2019ApJS..244...24L}. 

Giant molecular clouds (GMCs), from which stars are formed, are mainly composed of H$_{2}$, dust, and other molecules such as carbon monoxide 
\citep[CO;][]{1987ApJ...319..730S,2009ApJ...699.1092H,2011MNRAS.418..664N}. GMCs can be probed from the CO emission \citep[][]{1986ApJ...309..326D}. 
Using the IRAM-30m radiotelescope (Sierra Nevada, Spain), we  observed the $J=1-0$ and $J=2-1$ rotational transitions of CO 
along the strong galactic bars of 12 nearby massive spiral galaxies ($M_{\ast} \gtrsim 10^{10.2} \, M_{\odot}$). 
We  chose bars with different degrees of SF, hosting and lacking H{\sc\,ii} regions, 
as traced from ancillary H$\alpha$ images, and with no prior information about the molecular gas content.

Since the pioneering work of \citet[][]{1959ApJ...129..243S}, it has been a highly important endeavour to find a universal power law that relates 
the SFR surface density ($\Sigma_{\rm SFR}$) and the molecular gas surface density 
\citep[$\Sigma_{\rm mol}$, e.g.][]{1996AJ....112.1903Y,1998ApJ...498..541K,2007ApJ...671..333K,2010A&A...510A..64V}. 
The molecular gas mass within GMCs is indeed correlated with the SF activity in galaxies \citep[e.g.][]{2008AJ....136.2782L}. 
\citet[][]{2011ApJ...730L..13B} analysed the $\Sigma_{\rm mol}-\Sigma_{\rm SFR}$ relation by exploiting CO data from 
the HERA CO Line Extragalactic Survey \citep[HERACLES;][]{2009AJ....137.4670L}, 
GALEX far-UV, and \emph{Spitzer} 24 $\mu$m imaging, over thousands of positions in 
30 nearby disks at a resolution of 1 kpc. The authors imply a median molecular gas depletion time 
($\tau=1/$SFE=$M_{\rm mol}$/SFR; including helium) of $\sim 2.35$ Gyr ($\sigma=0.24$ dex), 
that is, a median SFE of 0.43 Gyr$^{-1}$ \citep[see also][]{2008AJ....136.2846B}; these values are used as a benchmark in this work. 
Our  aim is to assess the relationship between $\Sigma_{\rm SFR}$ and $\Sigma_{\rm mol}$, 
known as the Kennicutt-Schmidt (KS) law, within stellar bars. 
The analysis of the KS law in the central regions of gas-rich simulated galaxies by \citet[][]{2018A&A...609A..60K} 
reveals a flattening of its slope in the bar region.

In general, SFEs in bars have been only studied in a few objects 
\citep[e.g.][]{1996MNRAS.283..251K,2016PASJ...68...89M,2019PASJ...71S..13Y}. 
It has been shown that SF can be suppressed even in the presence of molecular gas \citep[][and references therein]{2020MNRAS.495.3840M}. 
The SFE in the spiral arms has been shown to be higher than 
in the bars \citep[e.g.][and references therein]{1991IAUS..146..156H,2010ApJ...721..383M,2014PASJ...66...46H}. 
In this work we search for possible gradients in SFEs along bars.

Some studies using CO indicate that the onset of SF along bars depends on the strength of shocks and shear stress 
\citep[e.g.][]{1998A&A...337..671R,2004A&A...413...73Z}. 
The formation of GMCs can also be inhibited by diverging streamlines in bars \citep[e.g.][]{1997AJ....114..965R,2002AJ....124.2581S}. 
However, observations of H{\sc\,ii} regions along bars have not always confirmed these hypotheses 
\citep[e.g.][]{1997A&A...326..449M,2008A&A...485....5Z}. 
Thus, several physical processes might come into play to explain the occurrence of SF in bars; 
for instance, \citet[][]{2000ApJ...532..221S} suggest that the disruption of GMCs in bars can be neutralised by self-gravity. 
Disentangling the physical conditions that enable or prevent SF in bars remains an unsolved astrophysical problem, 
despite all the observational and theoretical work referred to above.
%
%
\input{sample_table.dat}
%
%
\begin{figure*}
\centering
\includegraphics[width=0.15\textwidth,trim={0 0 4cm 0},clip]{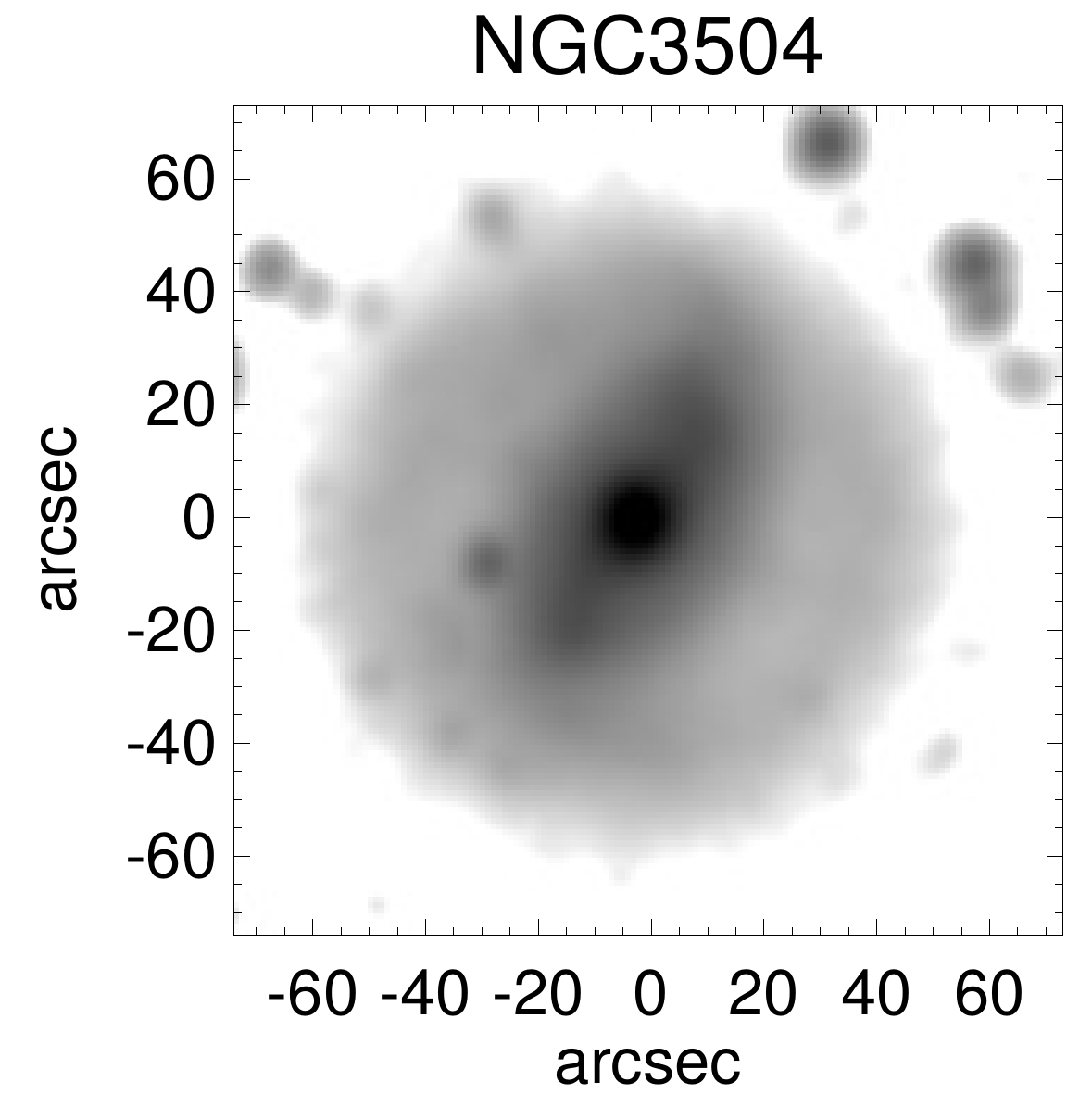}
\includegraphics[width=0.15\textwidth,trim={0 0 4cm 0},clip]{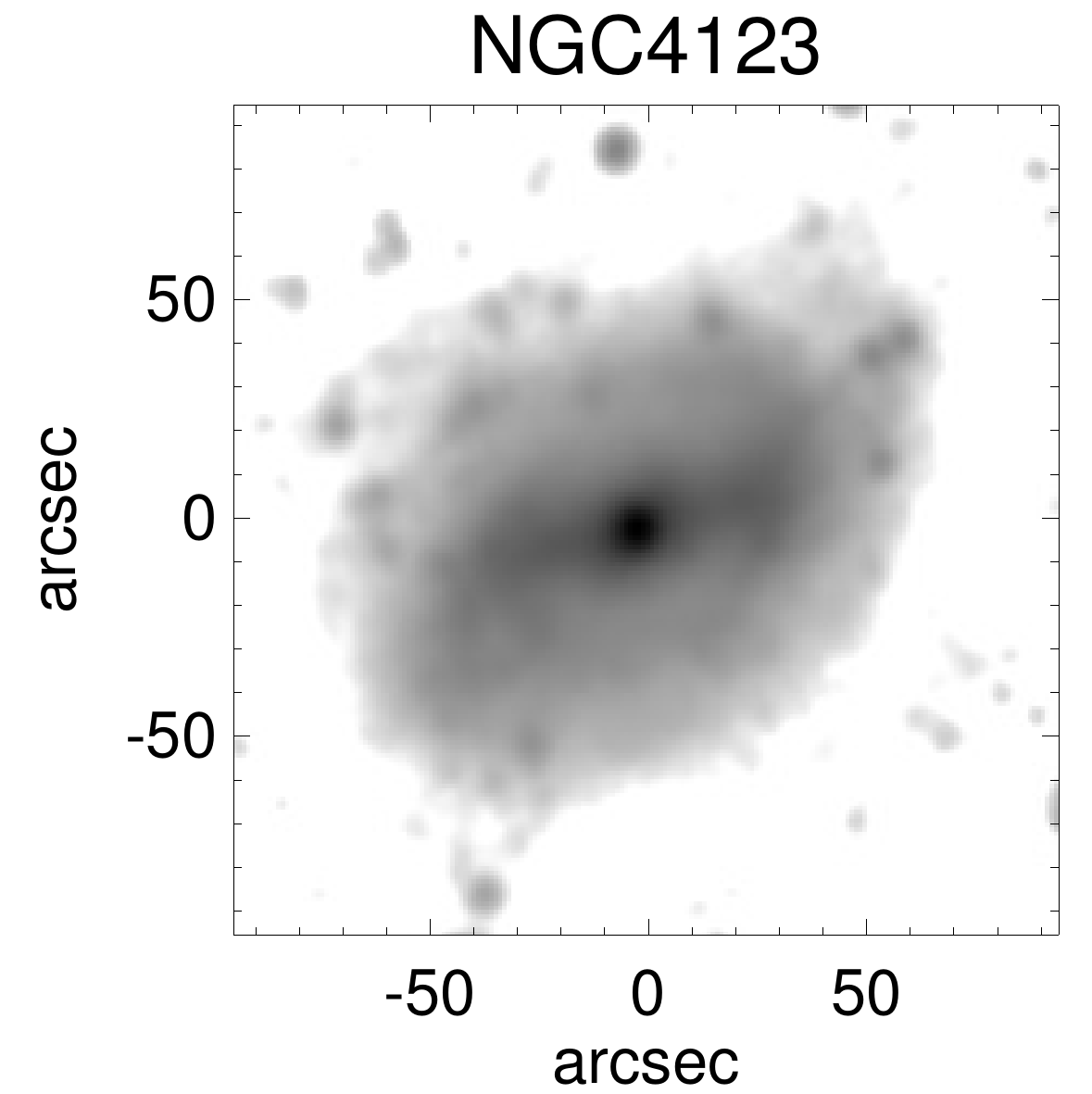}
\includegraphics[width=0.15\textwidth,trim={0 0 4cm 0},clip]{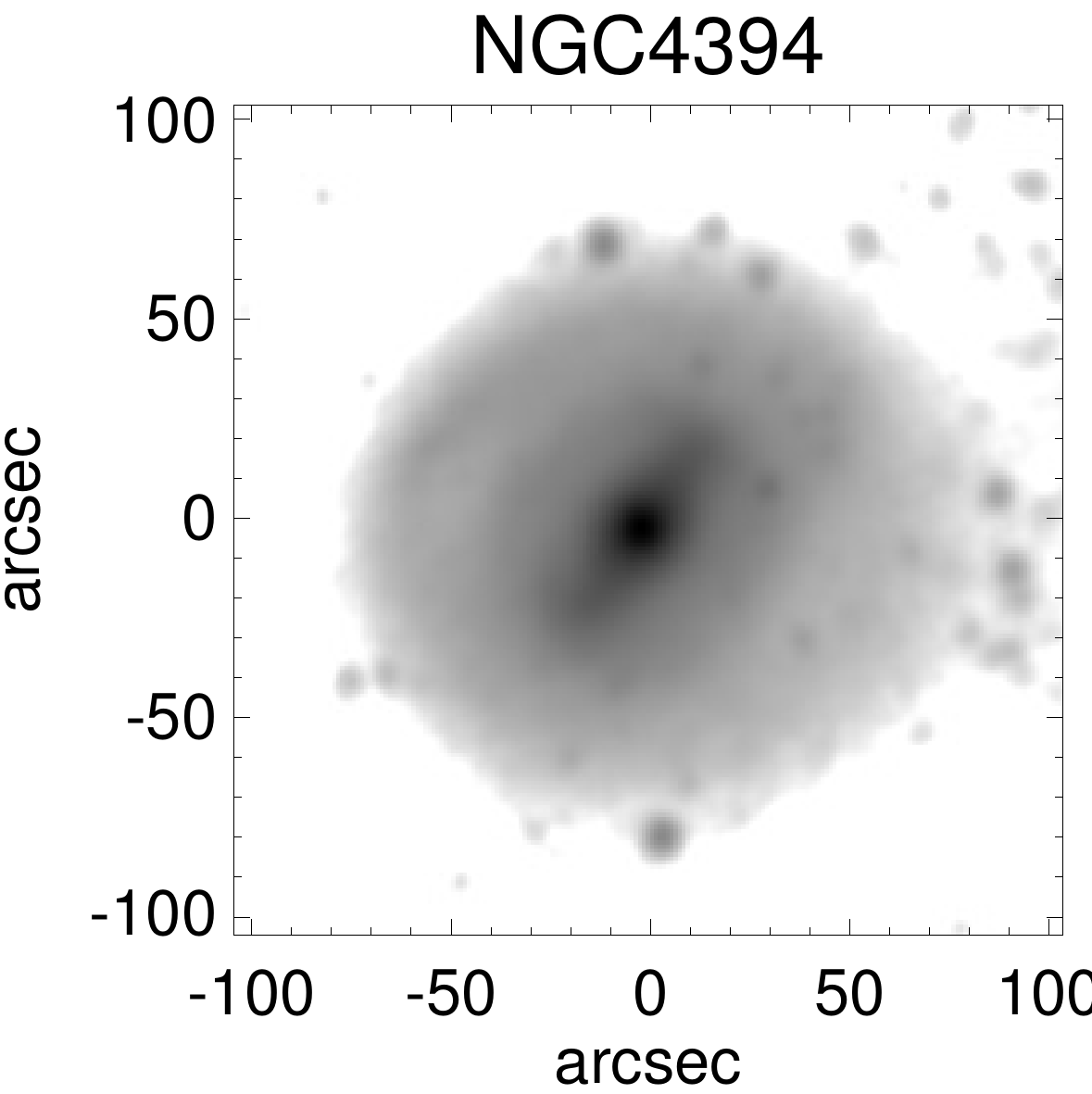}
\includegraphics[width=0.15\textwidth,trim={0 0 4cm 0},clip]{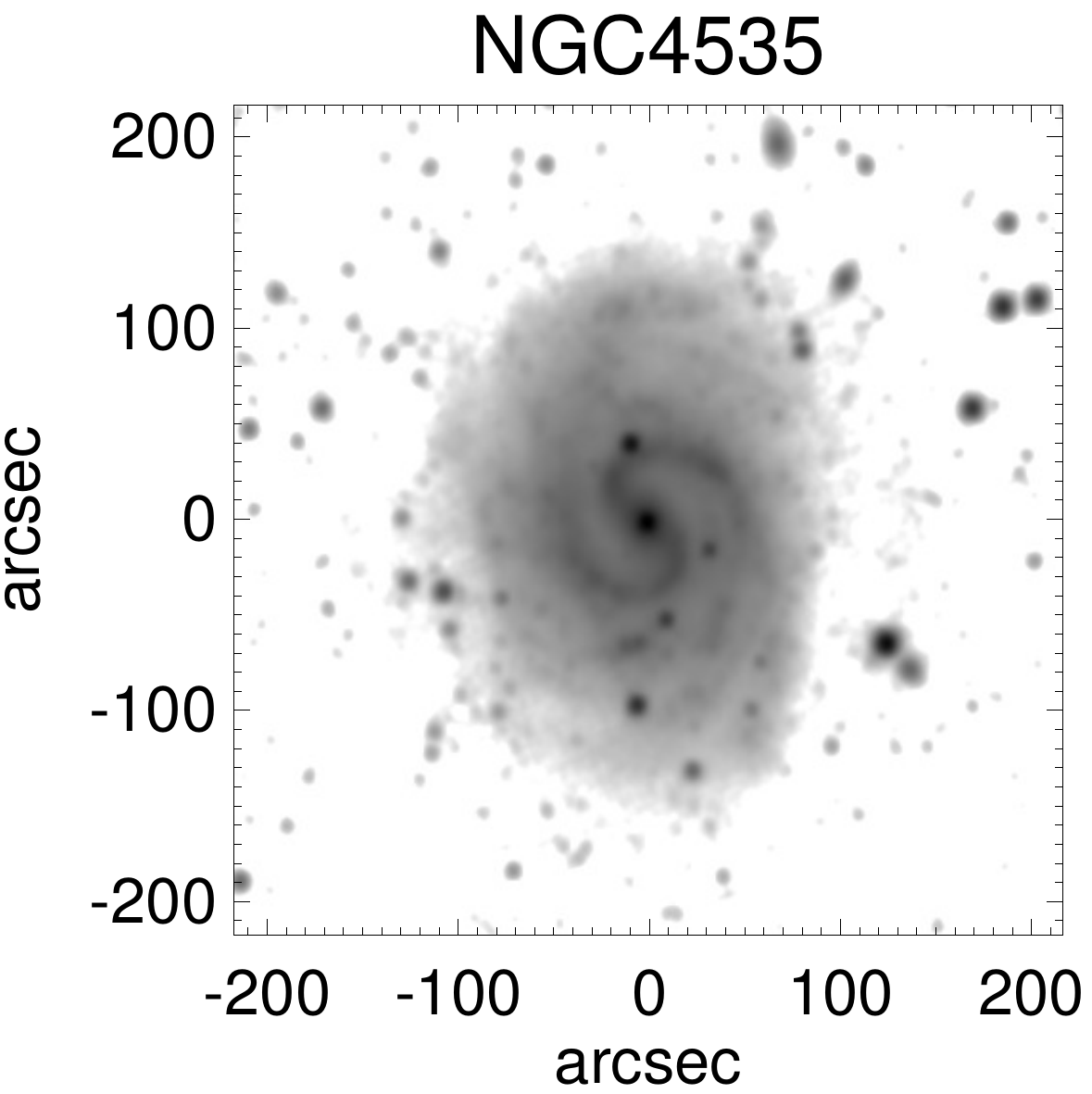}
\includegraphics[width=0.15\textwidth,trim={0 0 4cm 0},clip]{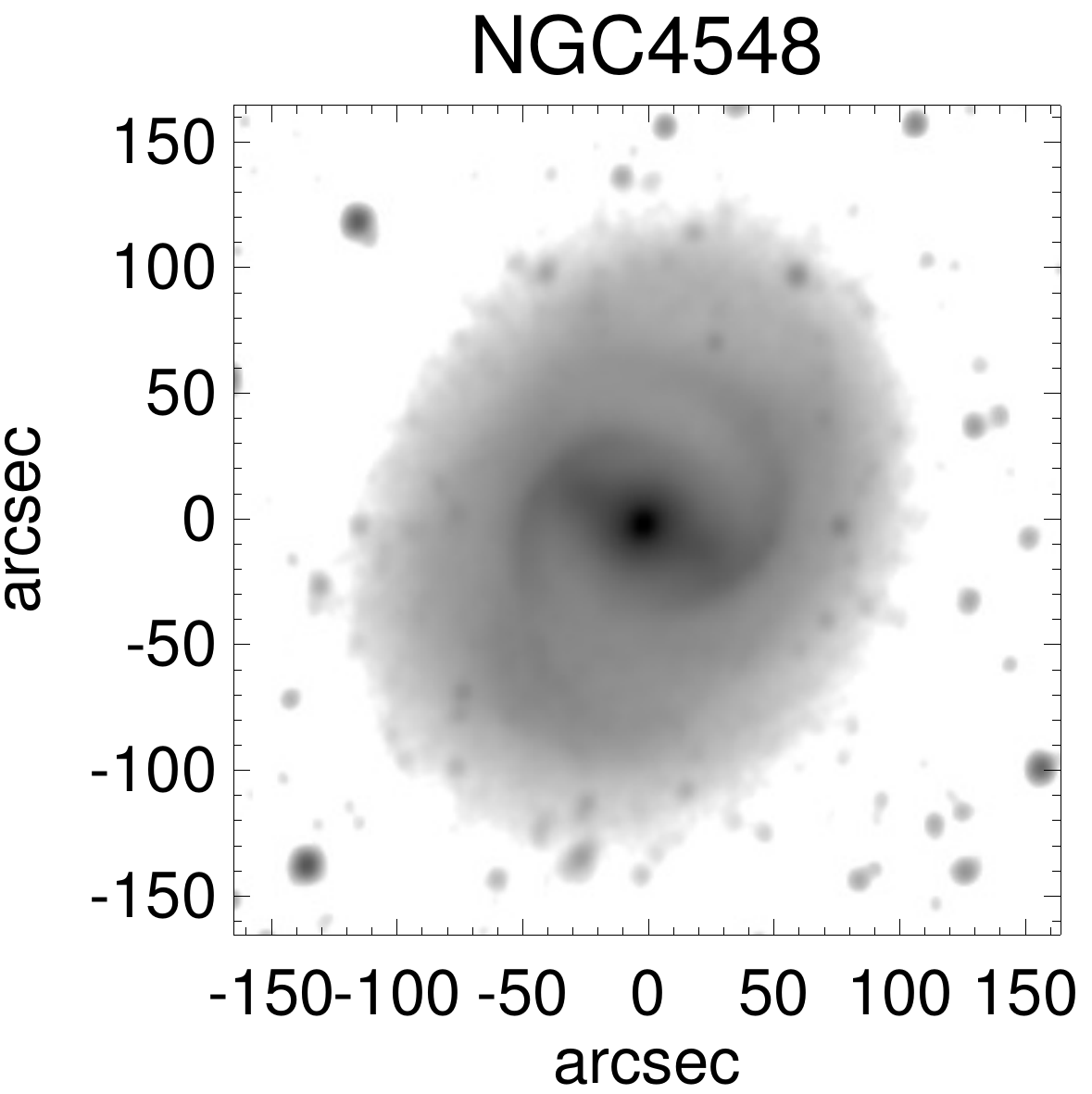}
\includegraphics[width=0.15\textwidth,trim={0 0 4cm 0},clip]{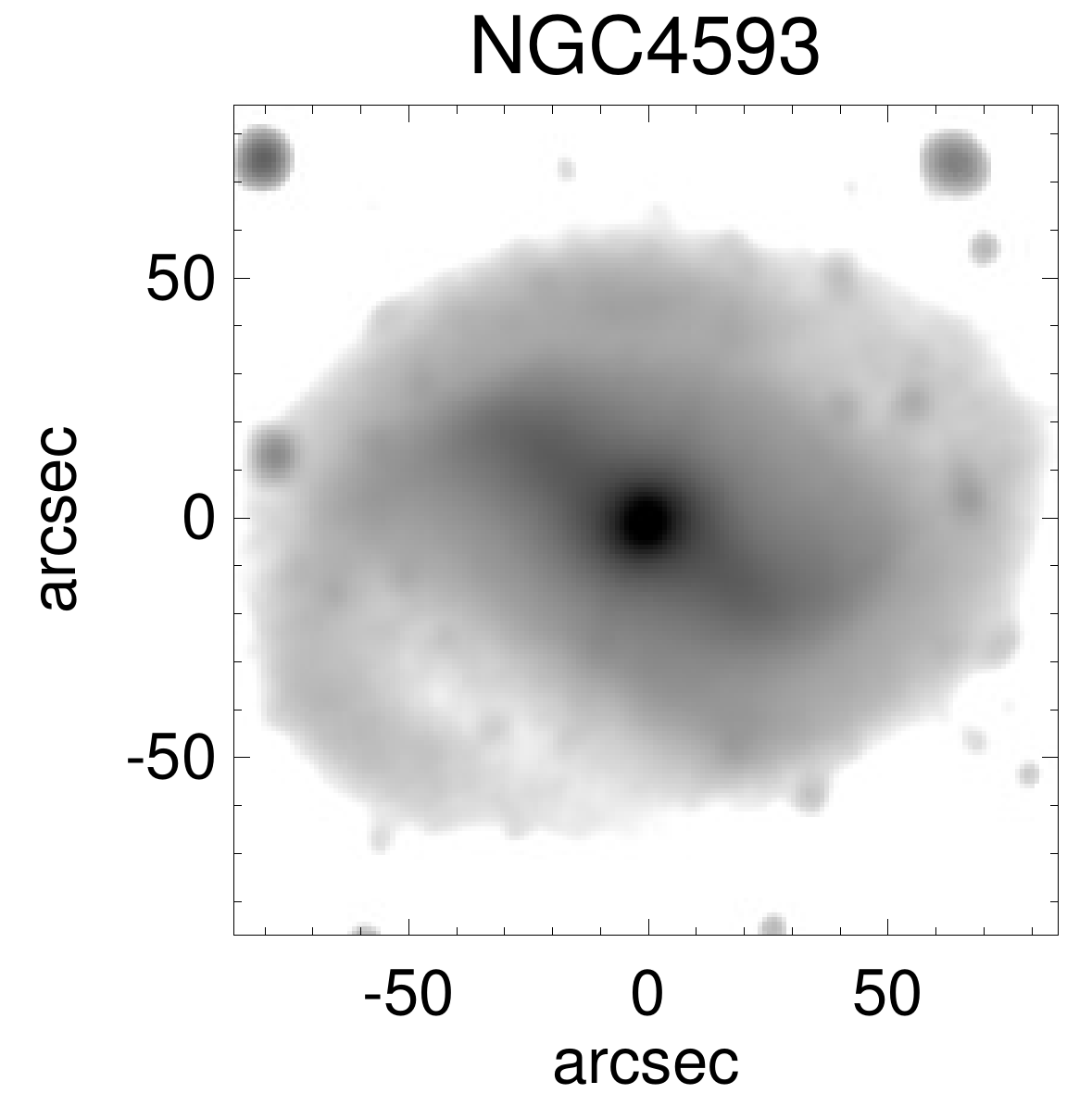}\\
\includegraphics[width=0.15\textwidth,trim={0 0 4cm 0},clip]{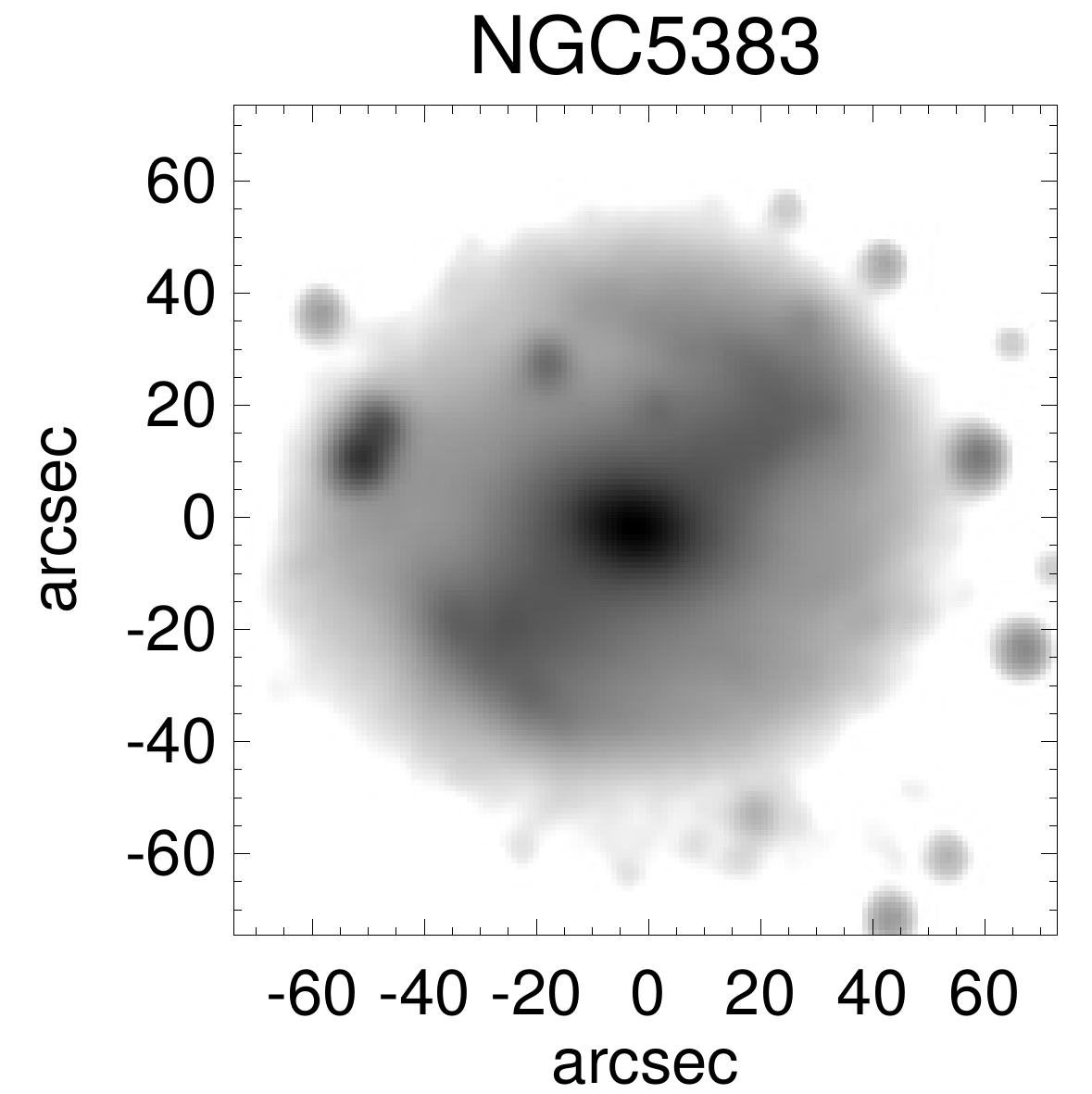}
\includegraphics[width=0.15\textwidth,trim={0 0 4cm 0},clip]{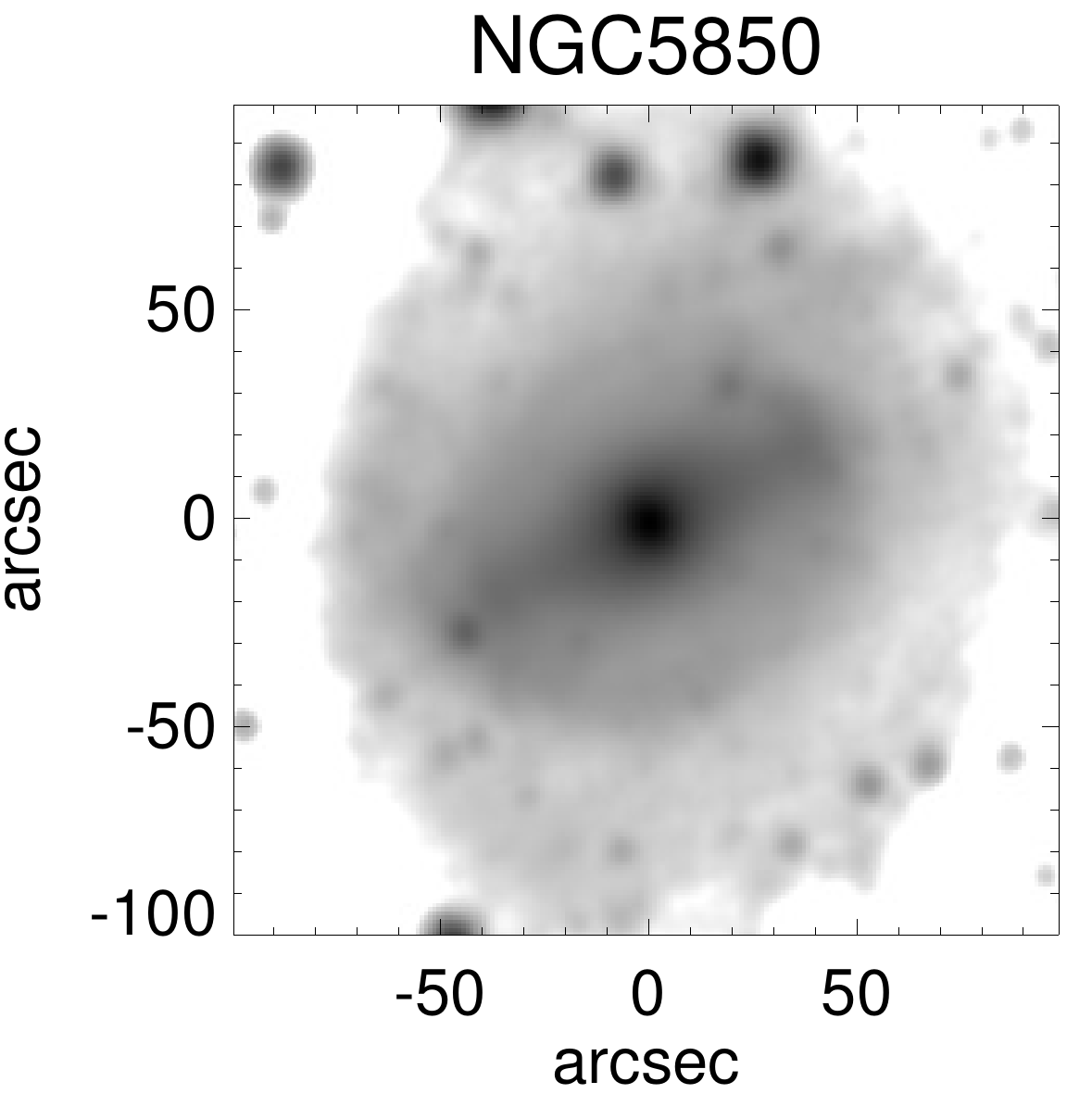}
\includegraphics[width=0.15\textwidth,trim={0 0 4cm 0},clip]{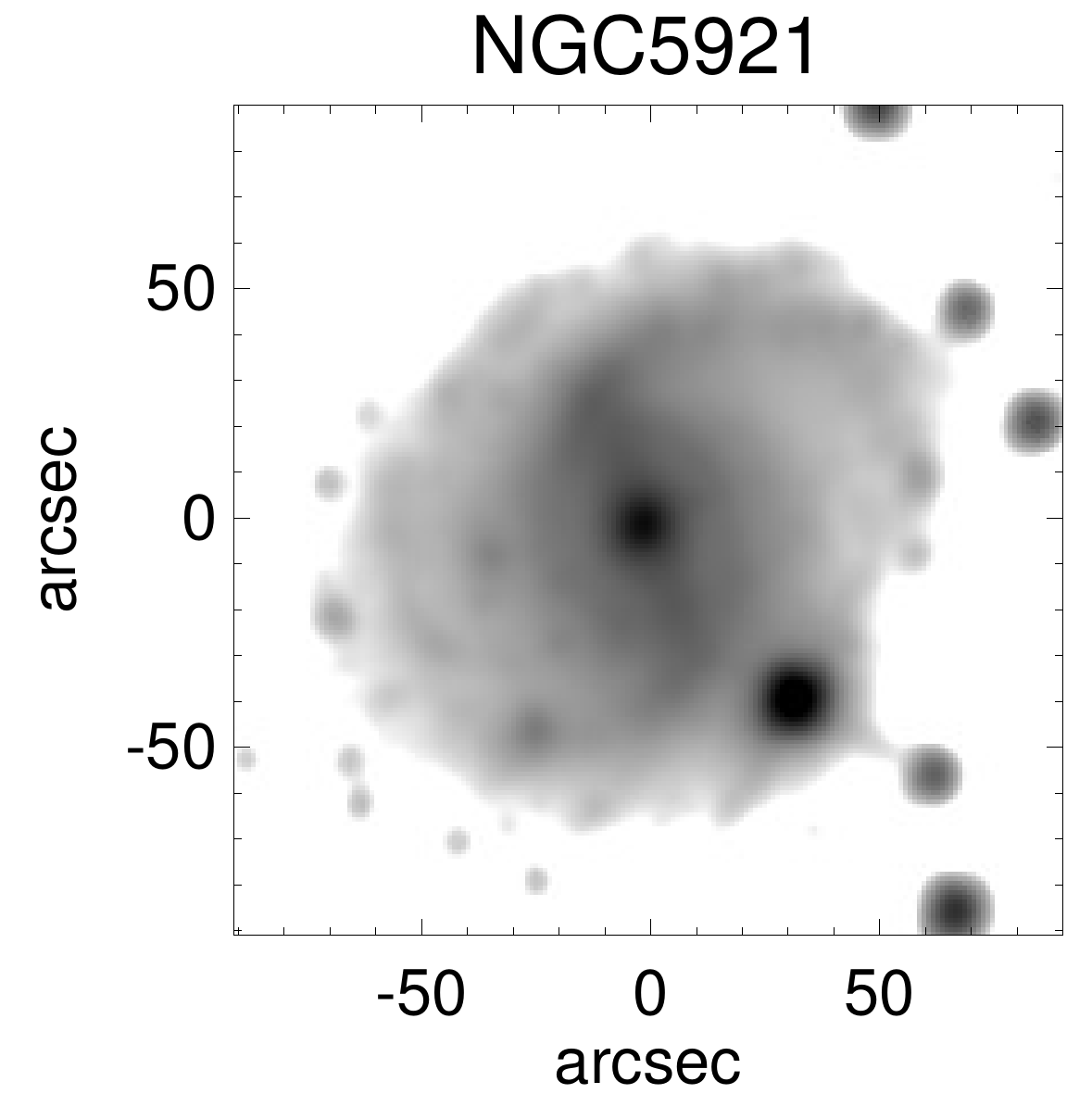}
\includegraphics[width=0.15\textwidth,trim={0 0 4cm 0},clip]{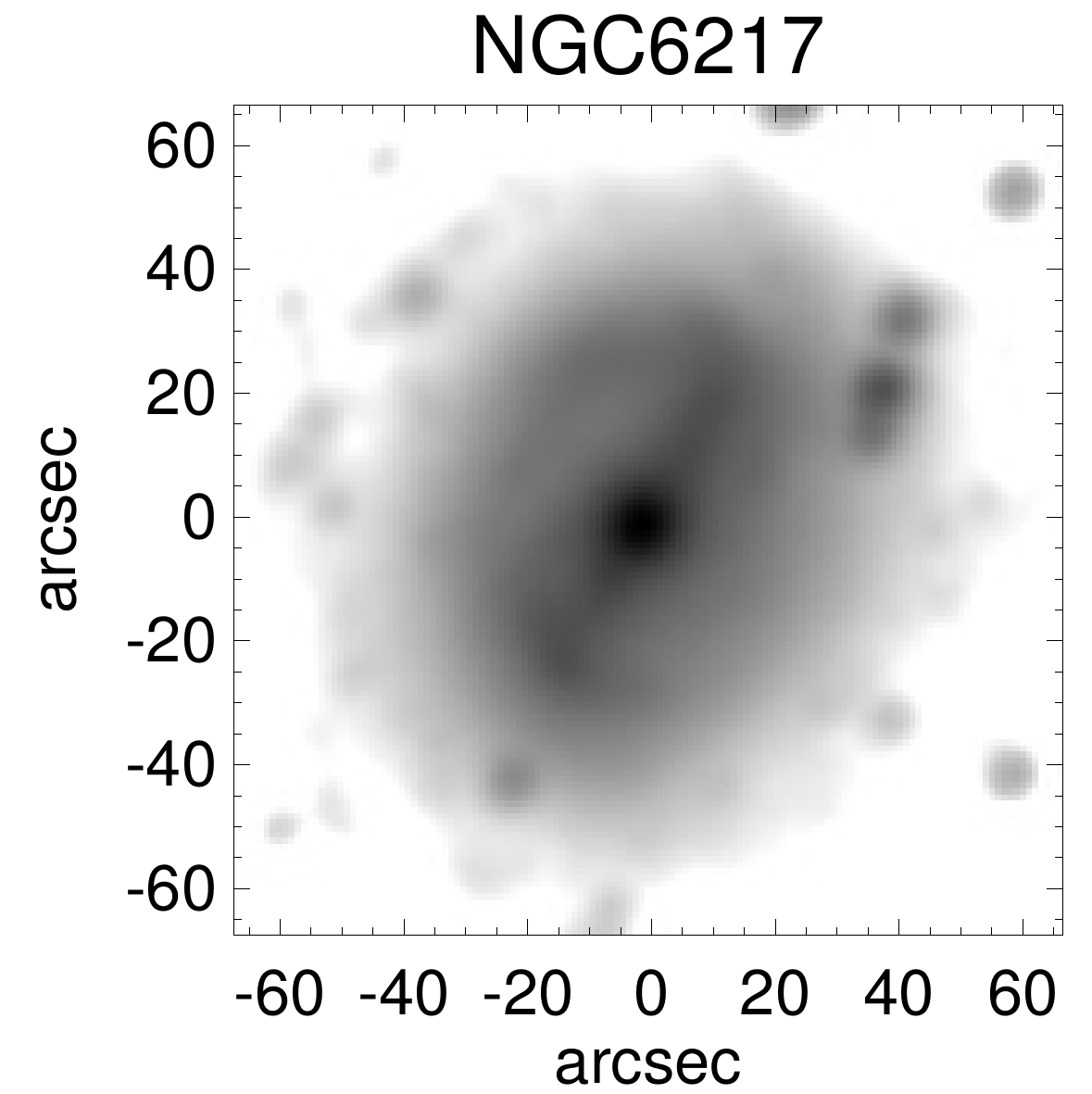}
\includegraphics[width=0.15\textwidth,trim={0 0 4cm 0},clip]{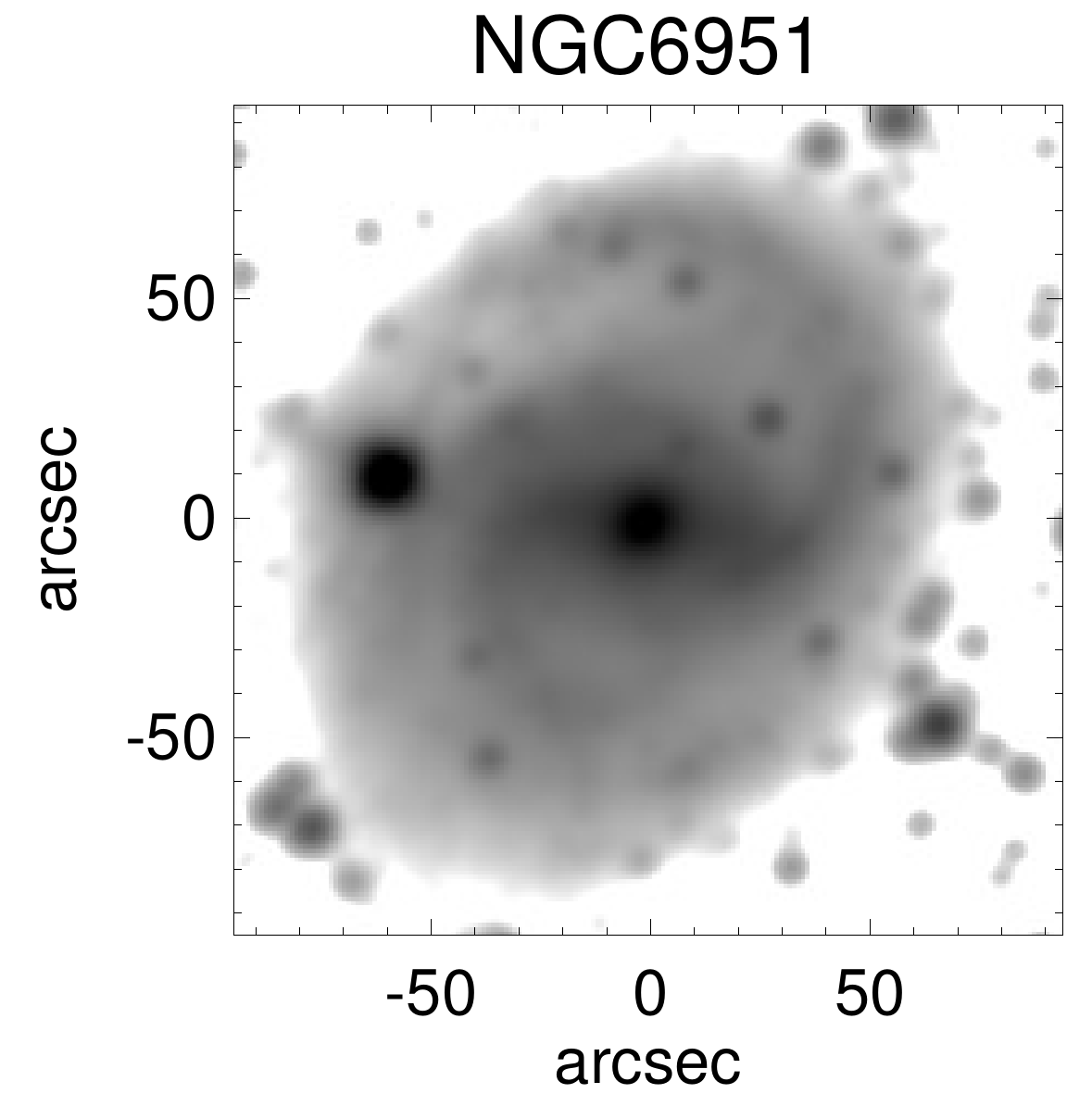}
\includegraphics[width=0.15\textwidth,trim={0 0 4cm 0},clip]{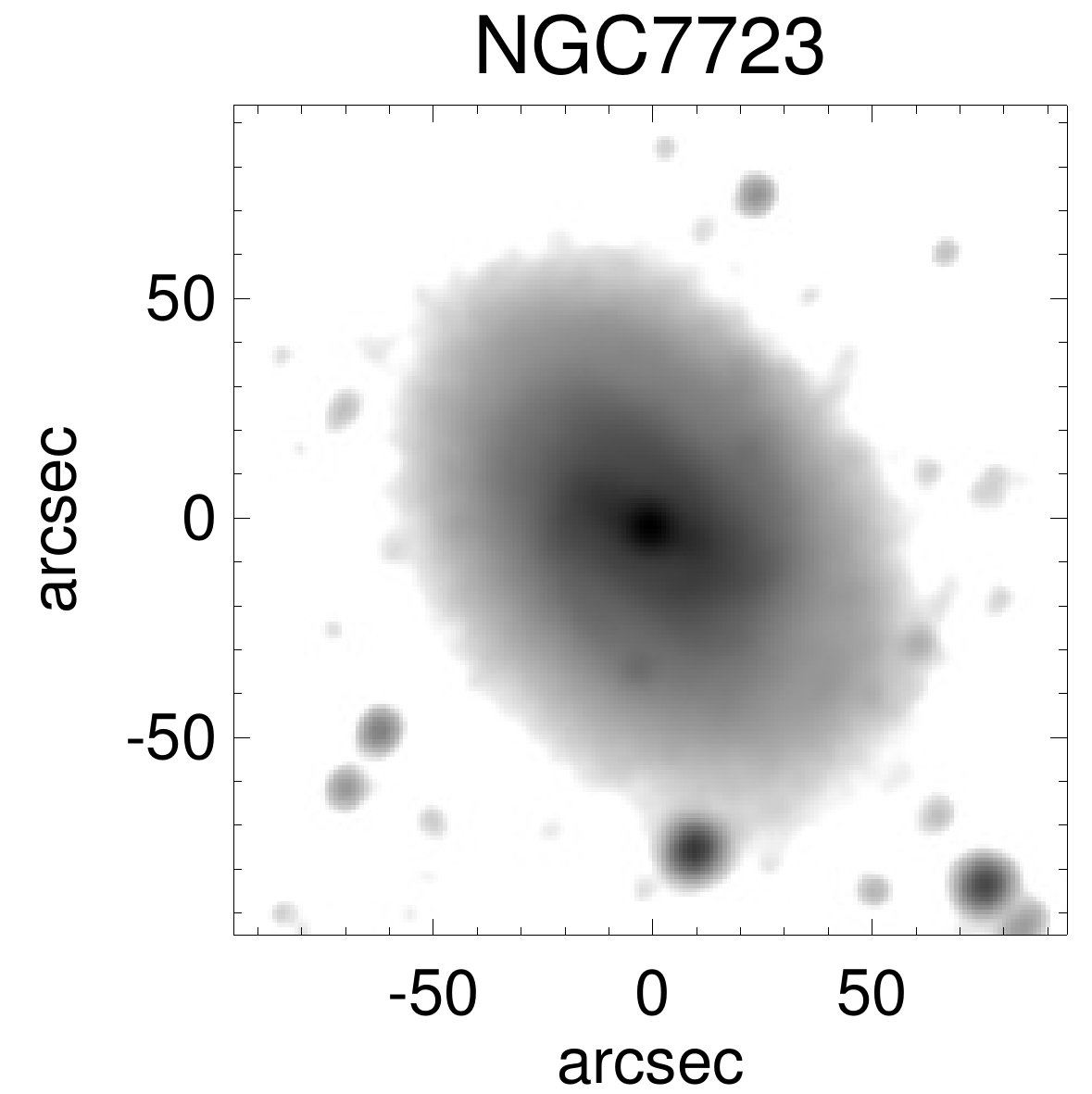}
\caption{
Sky-subtracted WISE 1 images of the 12 galaxies in our sample, 
in magnitude scale with surface brightness ($\mu_{3.4\mu \rm m}$, AB) in the range $[17-25] \, {\rm mag \, arcsec}^{-2}$.
}
\label{all_gals_fig}
\end{figure*}
%
%
\begin{figure*}
\centering
\includegraphics[width=0.49\textwidth]{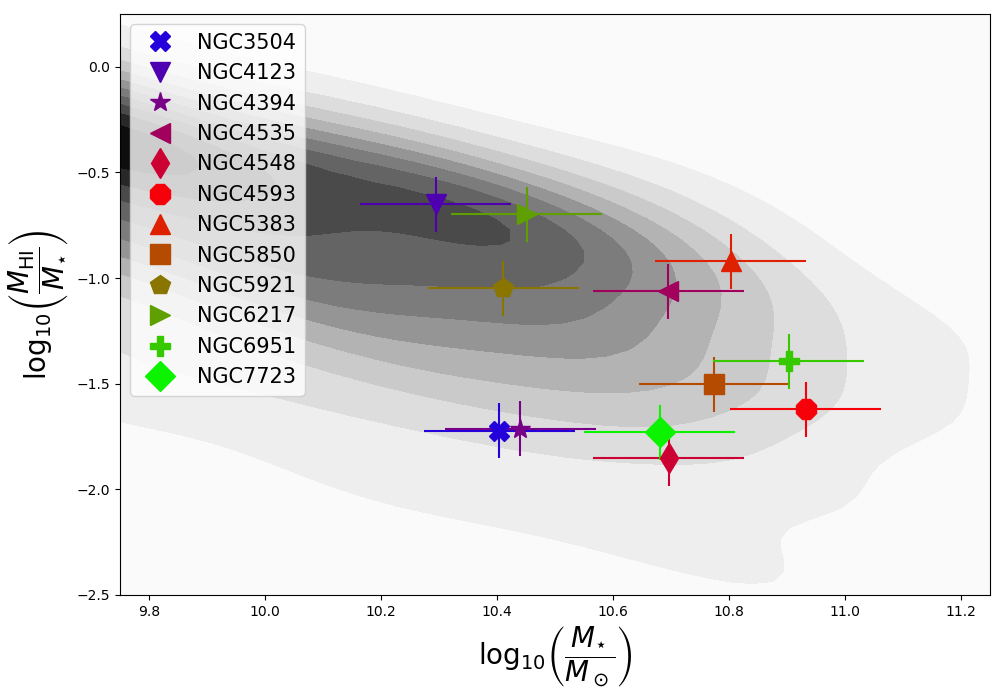}
\includegraphics[width=0.49\textwidth]{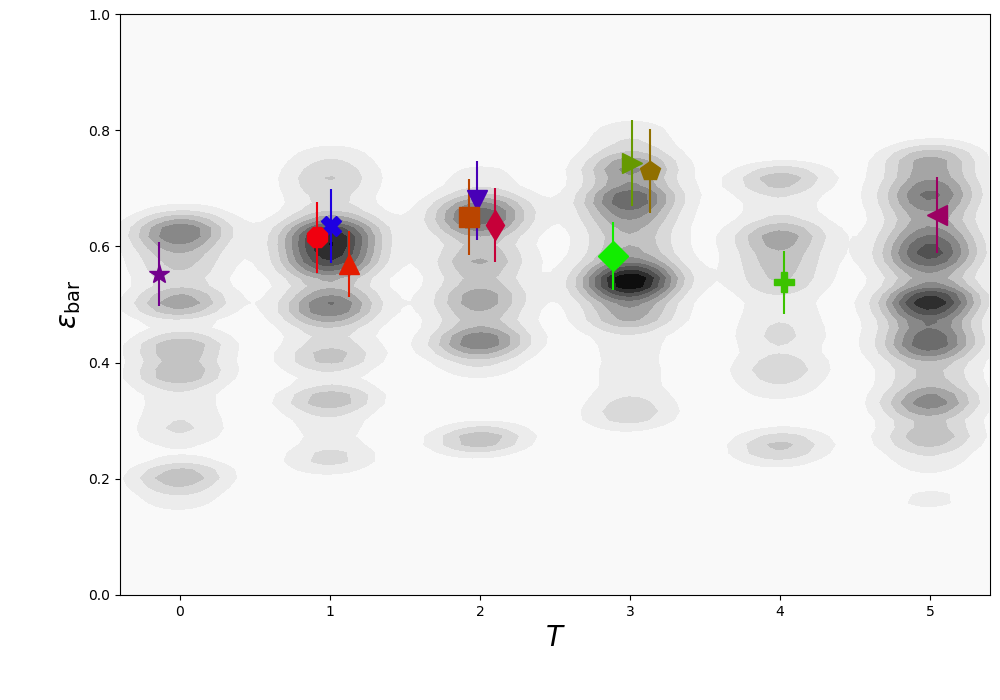}\\
\includegraphics[width=0.49\textwidth]{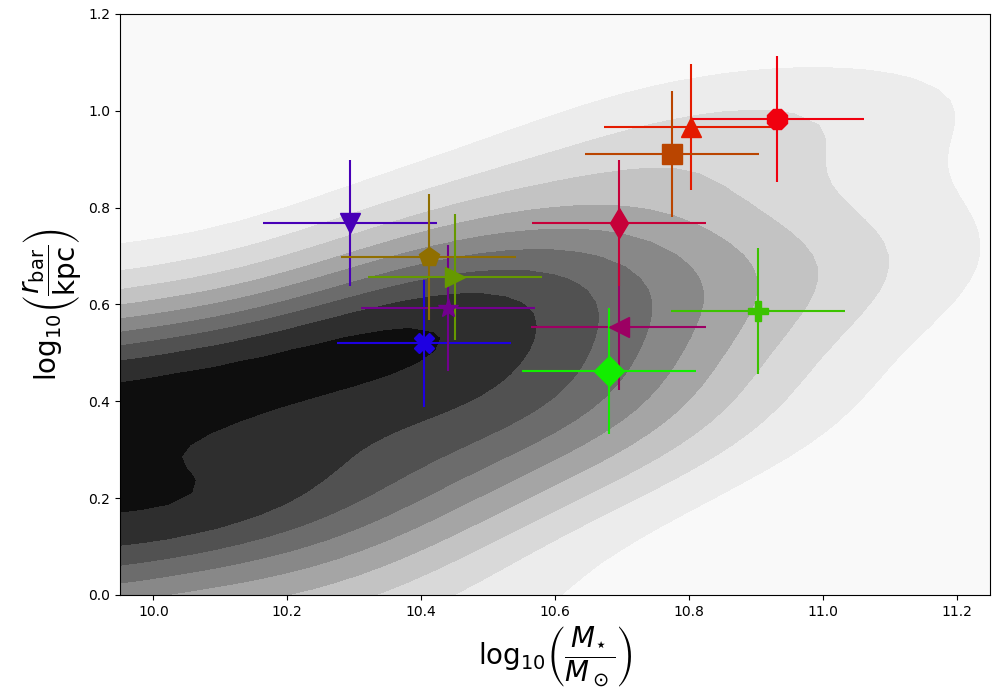}
\includegraphics[width=0.49\textwidth]{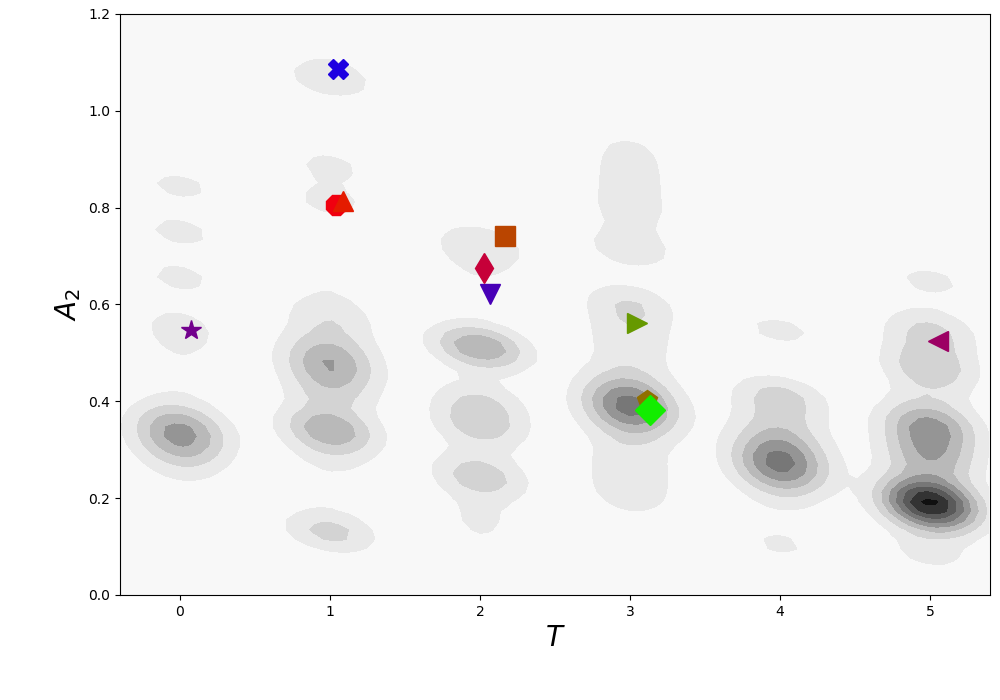}
\caption{
Properties of our sample of 12 barred galaxies (coloured symbols; see legend) and comparison with 508 barred galaxies in the 
S$^4$G survey (grey contours; disk inclinations lower than 50$^{\circ}$), 
as reported in \citet[][]{2016A&A...587A.160D} and \citet[][]{2019A&A...625A.146D}. 
\emph{Upper left panel:}
Gas fraction as a function of the total stellar mass. 
Error bars indicate the error on $M_{\ast}$ associated with 
a $30\%$ uncertainty on the mass-to-light ratio \citep[][]{2012AJ....143..139E}.
\emph{Lower left panel:} 
Bar length (in kiloparsecs) vs total stellar mass. 
Vertical error bars are associated with the distance uncertainty, which is typically $15 \%$ \citep[][]{2015ApJS..219....3M}. 
\emph{Upper right panel:} 
Deprojected bar ellipticity as a function of the revised Hubble stage. 
Error bars indicate the typical $10 \%$ uncertainty due to 2D deprojection effects for inclinations 
lower than 60$^{\circ}$ \citep[][]{2014ApJ...791...11Z}. 
\emph{Lower right panel:} 
Bar normalised $m=2$ Fourier amplitude (available for 11 galaxies) vs Hubble type. 
On the $x$-axes of the last two panels  small random offsets ($\lesssim 0.3$) were added to 
the $T$ values (integers) to avoid point overlapping.
}
\label{plot_pointing_sample}
\end{figure*}
%
%
\begin{figure}
\centering
\includegraphics[width=0.4\textwidth]{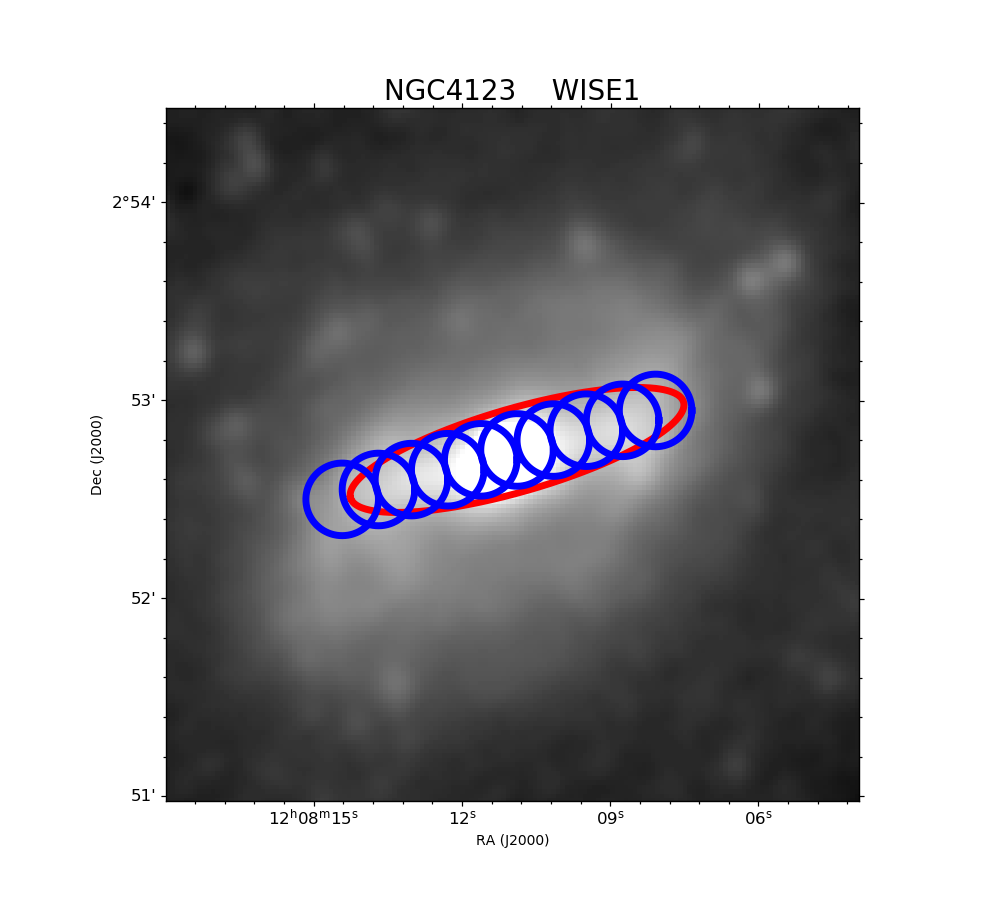}\\[-2ex]
\includegraphics[width=0.4\textwidth]{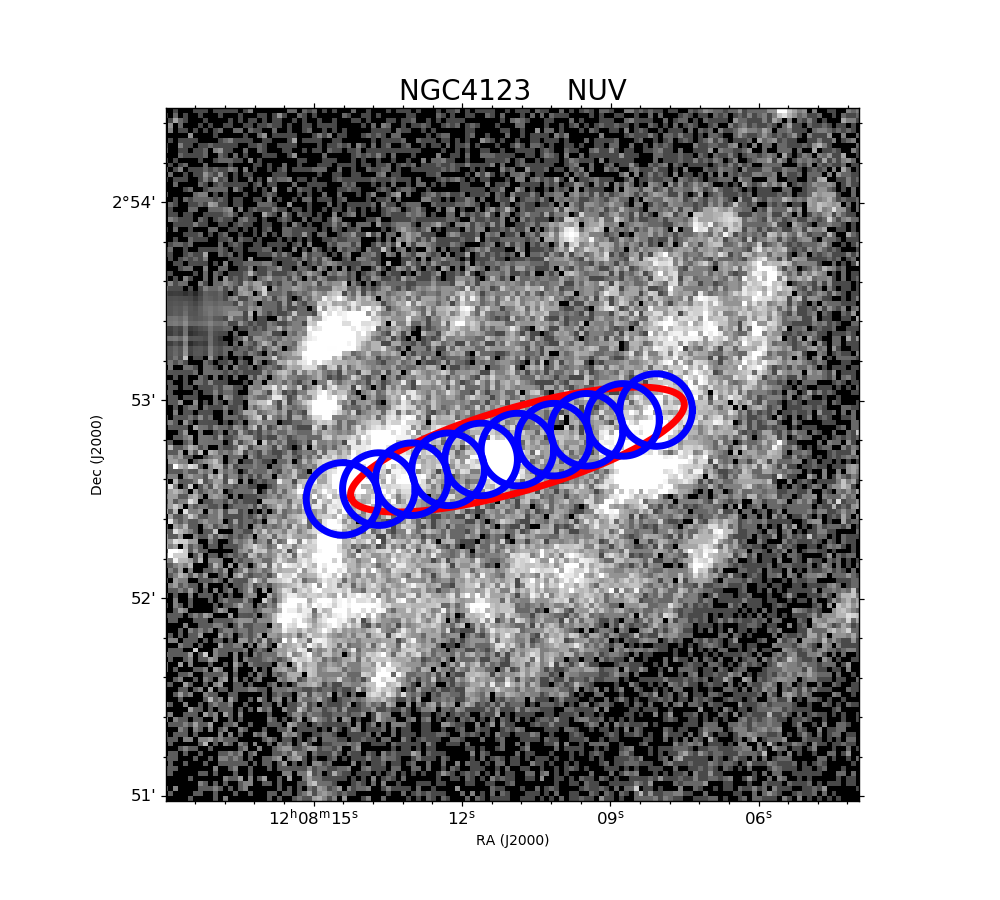}\\[-2ex]
\includegraphics[width=0.45\textwidth]{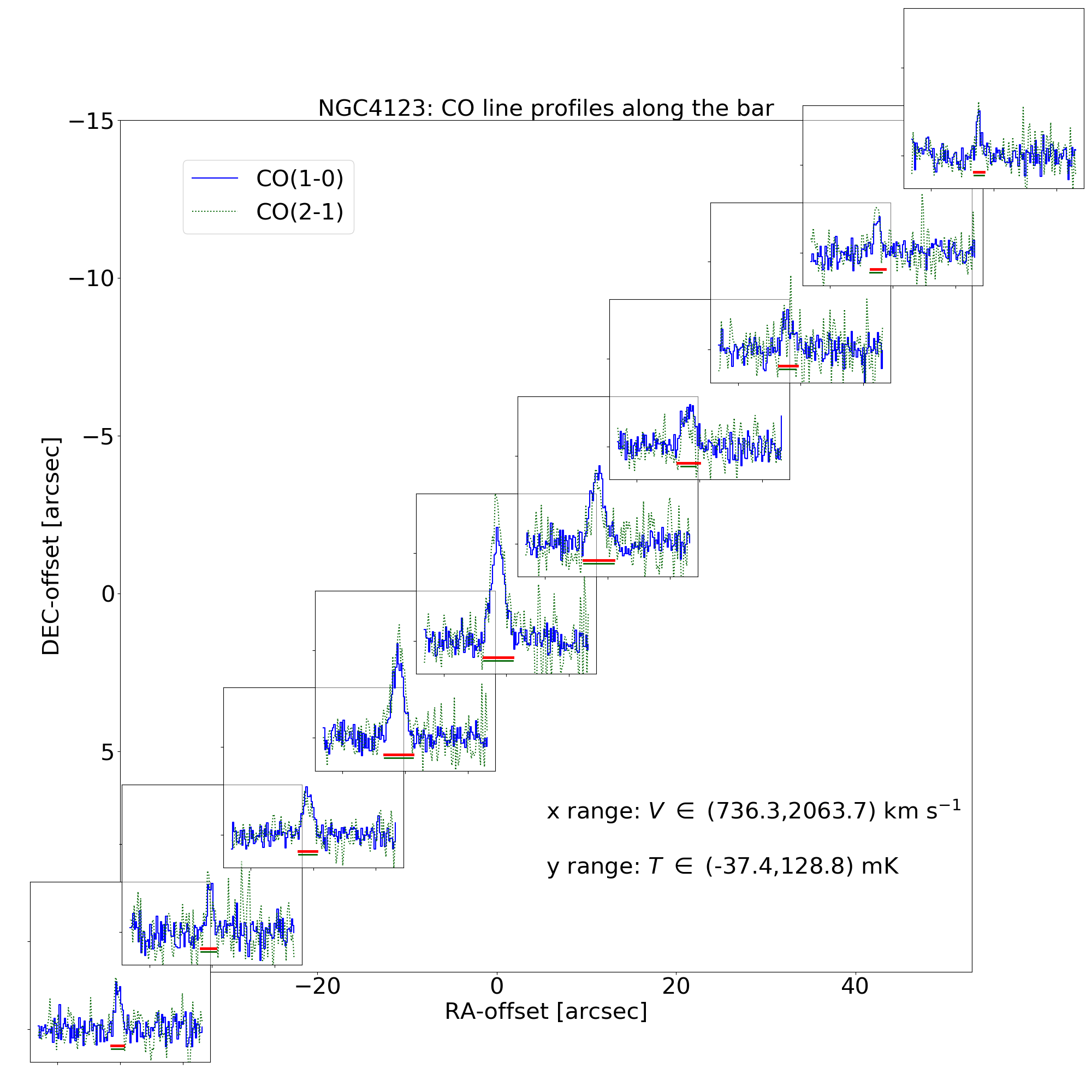}
\caption{
\emph{WISE} 1 (\emph{top panel}) and \emph{GALEX} far-UV (\emph{middle panel}) images of NGC~4123. 
The bar isophote, based on ellipse fitting of 3.6~$\mu$m images from \citet[][]{2015A&A...582A..86H} and \citet[][]{2016A&A...587A.160D}, 
is highlighted with a red ellipse. We show the circular beams where aperture photometry was performed (IRAM-30m 22$"$ pointings) in blue, 
spaced by 11$"$ and covering the whole bar. \emph{Bottom panel:} CO emission-line spectra along the bar of NGC~4123. 
The blue and green lines show the CO(1-0) and CO(2-1) spectra, respectively. 
The spectra are located on the plot in such a way that their centre corresponds to the offset of the pointing, relative to the central one. 
The red and green horizontal segments respectively show the zero-level line width of the CO(1-0) and CO(2-1) lines adopted for 
the determination of the velocity integrated intensities. 
The velocity (optical convention and in the Local Standard of Rest frame, in units of km s$^{-1}$) 
and intensity ranges of the spectra (in the main beam temperature scale, $T_{\rm mb}$, in units of mK) 
are also indicated in the lower right corner.}
\label{plot_pointing_example}
\end{figure}
%
%

This paper is organised as follows. In Sect.~\ref{sample_data} we present the sample and the  set of 
UV, optical, near- and mid-infrared, and millimetre data used. 
For the millimetre data we explain the observing strategy, data acquisition, and reduction of CO spectra; the spectra are shown in Appendix~\ref{app_30m_all}. 
In Sect.~\ref{MH2_CO} we describe the methodology used to infer molecular gas masses from CO(1-0) spectra (see Appendix~\ref{H2_tables}); 
$\Sigma_{\rm mol}$ profiles derived from CO(2-1) spectra are analysed in Appendix~\ref{CO_2_1_masses}. 
Section~\ref{ap_photometry} describes the use of aperture photometry to compute SFRs within the same regions covered by the IRAM-30m pointings 
(see also Appendix~\ref{SF_hybrid_non_hybrid}); the resulting values are tabulated in Appendix~\ref{Aperture_photometry_fluxes}. 
In Sect.~\ref{SFE_sect} we analyse the distribution of $\Sigma_{\rm SFR}$ and $\Sigma_{\rm mol}$ along the stellar bars. 
In Sect.~\ref{discussion} we discuss the physical properties that determine the presence or absence of star formation in galactic bars 
(see also Appendix~\ref{CO_2_1}). Finally, Sect.~\ref{summarysection} summarises the most important results of this work.
%
%
\section{Sample and data}\label{sample_data}
%
%
Our sample of 12 galaxies 
was selected from the extragalactic database HyperLeda
\footnote{\href{http://leda.univ-lyon1.fr}{http://leda.univ-lyon1.fr}}\citep[][]{2003A&A...412...45P} and met the following criteria:
%
%
\begin{itemize}
 \item they host prominent bars, according to the visual classification from ancillary optical images; 
 \item they probe both star-forming and quiescent bars, 
 based on the perusal of archival H$\alpha$ or mid-IR imaging; 
\item they have  morphological types spanning the range S0/a-Sc (i.e. Hubble types $T$ $\in [0,5])$;
 \item they are nearby, with recessional velocities  $<3000$ km s$^{-1}$; 
\item they have low disk inclinations ($i < 50^{\circ}$), enabling the study of surface densities (of SF and gas masses) 
 and minimising the effect of dust absorption in optical and near-IR wavelengths;
\item they have galactic declinations $\delta$ $>$ -15$^{\circ}$, avoiding high air masses during IRAM-30m observations;
\item they have $D_{25} > 2 \arcmin$, where $D_{25}$ is the length of the projected major axis 
of a galaxy at the isophotal level 25 mag arcsec$^{-2}$ in the $B$-band, ensuring good spatial resolution.
\end{itemize}
%
%

The galaxy sample is listed in Table~\ref{sample_12_props} together with some of the main galaxy properties. 
Their WISE 3.4 $\mu$m images are shown in Fig.~\ref{all_gals_fig}. 
We use the mean of the redshift-independent distances ($D$) available in the NASA/IPAC Extragalactic Database 
(NED)\footnote{\href{http://ned.ipac.caltech.edu}{http://ned.ipac.caltech.edu}}. 
Total stellar masses are taken from \citet[][]{2015ApJS..219....3M}, calculated from 3.6~$\mu$m imaging 
using the calibration of the mass-to-light ratio ($M/L$) by \citet[][]{2012AJ....143..139E}, 
which is also the $M/L$ applied in this work to estimate stellar densities within bars (Sect.~\ref{ap_photometry}), and 
from \citet[][]{2017MNRAS.466.1491H} in the case of NGC~6951 \citep[using the $K$ band and $M/L=0.6$;][]{2001ApJ...550..212B}.

According to the catalogue of quasars and active galactic nuclei by \citet[][]{2010A&A...518A..10V}, 
NGC~4593 is classified as Seyfert 1.0, NGC~4548 is a Seyfert 3.0 or hosts a 
low-ionisation nuclear emission-line region (LINER), and NGC~6217 and NGC~3504 harbour nuclear H{\sc\,ii} regions. 
NGC~6951 has been reported to host an active galactic nucleus (AGN) \citep[][]{1997ApJS..112..391H,1999ApJ...511..157K,2000A&A...353..893P}. 
The rest of the galaxies in our sample are not active.

\citet[][]{2020A&A...644A..38D} classified the distribution of massive SF in bars in the 
\emph{Spitzer} Survey of Stellar Structure in Galaxies \citep[S$^4$G;][]{2010PASP..122.1397S}, 
using continuum-subtracted H$\alpha$ and far-UV images, based on the detection of H{\sc\,ii} knots, clumps, or filaments along the major axis. 
For NGC~6951 (not in S$^4$G) we inspected the \emph{GALEX} far-UV image. 
One-half of our galaxies (NGC~3504, NGC~4535, NGC~4593, NGC~5383, NGC~5921, NGC~7723) 
harbour H{\sc\,ii} regions along the bar major axis, whereas in the other galaxies these regions are scarce. 
We note that in the case of NGC~6217 the distribution of SF is asymmetric and only presents a clear SF gap in the north-eastern part of the bar.

In Table~\ref{sample_12_props} we also show the morphological classification by \citet[][]{2015ApJS..217...32B} (S$^4$G) and/or NED (column 2). 
Eleven of the galaxies in our sample host inner features, the exception being NGC~4535. 
Of these, the majority have inner pseudo-rings (i.e. made of tightly wrapped spiral arms) and only NGC~5850 harbours a closed inner ring. 
This is relevant for our work since the presence of inner rings and the abundance of SF fuel along 
the bar can be closely connected \citep[][and references therein]{2019A&A...627A..26N,2020A&A...644A..38D}.

None of the 12 studied galaxies are currently merging. 
The degree of isolation of the S$^4$G galaxies in our sample was quantified by \citet[][]{2014MNRAS.441.1992L}; 
Col. 8 of Table~\ref{sample_12_props} lists the projected surface density of galaxies,
\begin{equation}
\Sigma_{3}^{A}={\rm log}_{10}\left(\dfrac{3}{\pi \, R_{3}^2}\right), 
\end{equation}
where $R_{3}$ is the projected distance to the third nearest neighbour galaxy, given in Mpc. 
Among the galaxies in our sample with the highest values of $\Sigma_{3}^{A}$, 
NGC~4548 and NGC~5850 are known to host signs of recent interactions with a nearby 
companion \citep[e.g.][]{1998AJ....115...80H,1999A&A...349..411V}, 
and NGC~4394 likely belongs to the M$\,$85 subgroup of the Virgo Cluster \citep[e.g.][]{2002A&A...384...24V}.

The selection of galaxies hosting high-amplitude bars biases our sample towards massive galaxies, 
spanning $10.2 < {\rm log_{10}} (M_{\star}/M_{\odot}) < 11$. 
This is shown in the upper left panel of Fig.~\ref{plot_pointing_sample}, 
where we display the gas fraction as a function of the total stellar mass. 
Total atomic gas (H{\sc\,i}) masses are estimated following \citet[][]{1988gera.book..522G} 
and \citet[][]{2018MNRAS.474.5372E}, 
\begin{equation}\label{gas}
M_{\rm HI}=2.356 \cdot 10^5 \cdot D^2 \cdot 10^{0.4 \cdot (17.4-m21c)},
\end{equation}
where $m21c$ is the corrected 21 cm line flux in magnitude from HyperLeda. 
For direct comparison with the physical properties of barred galaxies in the local Universe, in Fig.~\ref{plot_pointing_sample} 
we display in grey the same values for the S$^4$G sample \citep[][]{2020A&A...635A.197D} after imposing an inclination cut-off 
of 50$^{\circ}$, which shows good agreement. Our galaxies have H{\sc\,i} gas masses in the range $10^{8.7}-10^{9.9}M_{\odot}$, 
resulting in gas fractions ($M_{\rm HI}/M_{\star}$) spanning 1.5 dex.
%
%
\subsection{Bar structural parameters}
%
%
In order to characterise the basic properties of the bars in our sample, we use structural parameters derived from S$^4$G 3.6~$\mu$m images 
by \citet[][]{2015A&A...582A..86H} (bar position angles and sizes) and \citet[][]{2016A&A...587A.160D} (bar strengths). 
The disk orientation parameters are   from \citet[][]{2015ApJS..219....4S}. 
The bar parameters were deprojected to the disk plane following the same approach as in \citet[][]{2007MNRAS.381..943G}. 
Bar sizes and position angles were calculated from the maximum of the ellipticity profile, 
combined with the constancy of the position angle, at the bar region \citep[][]{1995A&AS..111..115W}. 

As for NGC~6951, which is not in the S$^4$G, we also ran ellipse fitting using the IRAF procedure \emph{ellipse} \citep[][]{1987MNRAS.226..747J} 
over the $i$-band image from \citet[][]{1999A&AS..134..333D}. The disk inclination and position angle are taken from  \citet[][]{2003AJ....126.1148B}: 
$i=39^{\circ}.4$ and PA$=143^{\circ}.1$ \citep[which is consistent with the optimal photometric estimation by][]{1993AJ....105.2090M}. 
The bar parameters are listed in Table~\ref{sample_12_props}.

The right panels of Fig.~\ref{plot_pointing_sample} show the distribution in the Hubble sequence of the 
intrinsic ellipticity of the bars ($\epsilon_{\rm bar}$) and their normalised $m=2$ Fourier amplitude ($A_{2}$). 
The grey points again correspond to the values in the S$^4$G ($i<50^{\circ}$) reported by \citet[][]{2016A&A...587A.160D} (see their Table 3). 
The lower left panel of Fig.~\ref{plot_pointing_sample} shows the bar radii 
in kiloparsecs as a function of the total stellar masses of the host galaxies. 
The bar parameters in our sample lie well within typical ranges for nearby disk galaxies of similar morphological types 
found in the literature \citep[see also][]{1995AJ....109.2428M,2005MNRAS.364..283E,2009A&A...495..491A,2012A&A...540A.103P}. 
Our sample of galaxies comprises fairly strong bars, as determined from either visual \citep[][]{2015ApJS..217...32B} or 
quantitative ($A_2$, $\epsilon_{\rm bar}$)  measurements.
%
%
\subsection{IRAM-30m CO spectra}\label{CO_SPECTRA}
%
%
We observed the CO(1-0) emission line (2.6 mm) along the bars of the 12 galaxies in our sample 
with the IRAM-30m radiotelescope at a central frequency of 115 GHz. 
This allowed us to obtain molecular gas mass surface density profiles (Sect.~\ref{MH2_CO}). 

Observations were carried out between June and November 2008 
using the dual polarisation receivers configuration AB with the 512 $\times$ 1 MHz filterbanks. 
We used a wobbler switching mode with a wobbler throw of 200 arcsec in azimuthal direction. 
The pointing was monitored on nearby quasars, Mars, or Jupiter every 60~-~90  minutes. 
The weather conditions were good during the observation period, 
with pointing accuracy better than 4 arcsec 
(data with poorer pointings were excluded). The average system temperature was $\sim$~300 K at 115 GHz on the $T_{\rm A}^*$ scale. 
For the data reduction only observations of good quality were chosen (suitable weather conditions and showing a flat baseline). 
The spectra were summed over the individual positions and the baseline was then subtracted.

We carried out several pointings (depending on the bar length, see Table \ref{sample_12_props}) 
separated by 11 arcsec and aligned with the bar major axis. 
The full width at half maximum (FWHM) is 21.5\arcsec\ at the frequency of our observations, resolving 1-4 kpc depending on the galaxy 
(see column 13 in Table \ref{sample_12_props}). 
The spatial distribution of the regions where the CO was measured 
is displayed in the upper panels of Fig.~\ref{plot_pointing_example} for NGC~4123 
(the same plots can be found in Appendix~\ref{app_30m_all} for all the galaxies in our sample), 
based on the visual measurement of the bar size and position angle. 
We show the \emph{WISE} 1 and far-UV images, with the CO pointings and the bar ellipse superimposed. 
The integration times per pointing ranged between 10 and 115 min and was divided into individual observing scans of 4-6 min duration.
The resulting CO(1-0) spectra are presented in the lower panel of Fig.~\ref{plot_pointing_example} and in Appendix~\ref{app_30m_all}. 
The $x$- and $y$-axes show the offset of the pointings (in RA and DEC) with respect to the centre of the galaxy.

The  CO spectra and intensities are expressed on the main beam temperature scale ($T_{\rm mb}$), which is computed from the forward and beam efficiency: 
$T_{\rm mb}$ = $T_{\rm A}^{*} \cdot F_{\rm eff}/B_{\rm eff}$ (at 115 GHz, $F_{\rm eff}$ = 0.95, and $B_{\rm eff}$ = 0.75 for the IRAM-30m). 
The velocity resolution varies between 10.4 and 10.5 km$\,$s$^{-1}$, 
which is adequate for our purposes because it is similar to the typical velocity dispersion of molecular clouds \citep[e.g.][]{1984ApJ...281..624S}. 
We simultaneously observed CO(2-1) (230 GHz); these data are suitable for studying the gas excitation (see Appendix~\ref{CO_2_1}) 
and molecular gas masses within a beam of size 0.9-2.0 kpc (${\rm FWHM}\approx 10.75$\arcsec), which is  two times better resolution than CO(1-0). 
The obtained CO(1-0) and CO(2-1) spectra are available online at the CDS associated with this publication.

In addition to the statistical error of the velocity-integrated line intensities, a typical flux calibration error  has to be taken into account. 
We adopt the errors determined in \citet[][]{2019A&A...627A.107L}, $15\%$ for CO(1-0) and $30\%$ for CO(2-1), derived 
from observations of several sources on different days. Despite the higher resolution achieved using CO(2-1),  here we opt to use CO(1-0) to 
estimate molecular gas masses, given the higher calibration error and poorer quality of the CO(2-1) spectra. 
Even so, in Appendix~\ref{CO_2_1_masses} we show that the results in this paper are consistently similar when 
CO(2-1)-based $\Sigma_{\rm mol}$ estimates are used, 
confirming that the resolution is not a big issue (on kiloparsec scales) in the statistical trends presented in this paper.

The current availability of homogeneous and state-of-the-art UV and IR datasets (Sects. \ref{UV_data} and \ref{IR_data}) 
makes the exploitation of these CO(1-0) data especially timely, as it allows   accurate measurements of 
star formation efficiencies (Sect.~\ref{SFE_sect}). 
While the use of interferometric data would maximise the spatial resolution (which is a caveat of this work), 
single-dish observations guarantee the recovery of all the CO flux along the bar major-axis.
%
%
\subsection{GALEX ultraviolet imaging}\label{UV_data}
%
%
We use the sky-subtracted and masked images from the \emph{GALEX/S$^4$G UV-IR catalogue} by \citet[][]{2018ApJS..234...18B}, 
that had been gathered from the \emph{GALEX} GR6/7 Data Release\footnote{
\href{http://galex.stsci.edu/GR6/}{http://galex.stsci.edu/GR6/}
} 
and reduced following \citet[][]{2007ApJS..173..185G}. 
Specifically, we use near-UV (NUV; $\lambda_{\rm eff}=2267\,\AA$) and far-UV (FUV; $\lambda_{\rm eff}=1516\,\AA$) photometry. 
Following \citet[][]{2007ApJS..173..185G}, we adopt a $14.8\%$ calibration error (0.15 mag) for both the FUV and NUV bands, 
but we note that \citet[][]{2007ApJS..173..682M} give smaller errors ($4.7\%$ and $2.8\%$ for FUV and NUV, respectively).

For NGC~6951 (not in S$^4$G) we obtained and processed the \emph{GALEX} NUV GR6/7 image ourselves; 
to our knowledge no usable FUV image exists for this galaxy. Emission at FUV wavelengths trace recent SF, 
of the order of several tens to $100$ Myr, while NUV traces $\lesssim$ 300 Myr populations \citep[][]{1998ARA&A..36..189K, 2014A&A...571A..72B}. 

Unless stated otherwise, we hereafter use NUV as the ultraviolet SF tracer for the sake of maximising the sample coverage. 
In Appendix~\ref{SF_hybrid_non_hybrid} we show that FUV and NUV yield similar results; this is also discussed in the following sections.
%
%
\subsection{WISE IR data}\label{IR_data}
%
%
Dust re-radiates the absorbed UV in mid-IR wavelengths. 
In order to trace the dust-enshrouded SF (Sect.~\ref{ap_photometry}) we use data from the Wide-field Infrared Survey Explorer 
\citep[][]{2010AJ....140.1868W} in the \emph{WISE} 3 ($\sim 12$~$\mu$m) and \emph{WISE} 4 ($\sim 22$~$\mu$m) passbands, 
that exist for all the spiral galaxies in our sample \citep[][]{2011ApJ...735..112J}. Stellar masses ($M_{\star}$) within the \text{CO} pointings are likewise 
computed using \emph{WISE} 1 ($\sim 3.4$~$\mu$m), while \emph{WISE} 2 ($\sim 4.6$~$\mu$m) and \emph{WISE} 3 are used to calculate IR colours 
and get a hint of non-stellar emission and AGN activity. Hereafter, we refer to the \emph{WISE} passbands with the letter `W'.

The \emph{WISE} images were downloaded from the NASA/IPAC Infrared Science archive\footnote{
\href{https://irsa.ipac.caltech.edu/applications/wise/}{https://irsa.ipac.caltech.edu/applications/wise/}
}. 
We subtracted the global background using the Python package \emph{photutils}. 
In this process, we masked emission that was above 2$\sigma$ for at least five connected pixels as belonging to the source. 
Both the mask making and the background subtraction use sigma-clipped statistics to calculate the median and mean values. 
After subtracting the global background, the \emph{WISE} maps were converted from digital numbers into janskys (Jy) using the photometric zero points given in 
Table 1 of the Explanatory Supplement to the \emph{WISE} All-Sky Data Release 
Products\footnote{
\href{https://wise2.ipac.caltech.edu/docs/release/allsky/expsup/sec2_3f.html}{https://wise2.ipac.caltech.edu/docs/release/allsky/expsup/sec2$\_$3f.html}
}. 

Following \citet[][see their Sect.~3.5]{2013AJ....145....6J}, we performed three corrections for extended source photometry. 
First, we applied an aperture correction that accounts for the point spread function (PSF) profile fitting used in the \emph{WISE} 
absolute photometric calibration. 
The corrections are 0.034, 0.041, 0.030, and 0.029 mag for the W1, W2, W3, and W4 bands, respectively, and the uncertainty is 1\%. 
The second correction is a colour correction that accounts for the spectral signature of the source convolved with the \emph{WISE} 
relative system response. The colour correction is given in 
\citet[][]{2010AJ....140.1868W} and in the \emph{WISE} Explanatory Supplement 21, Section IV.4.h. For bands W1, W2, and W4 
this corrections is always very small, $\lesssim$ 1\%, and we neglect it, but for W3 it can be as large as $\sim 10\%$. 
The third correction, related to a discrepancy in the calibration between the \emph{WISE} photometric standard blue stars and red galaxies, 
only applies to the W4 images and accounts for a factor of 0.92 \citep[][]{2011ApJ...735..112J} 
for non-bulge dominated objects (W2-W3 > 1.3 mag, which is the case for our objects). 
The calibration accuracy of the \emph{WISE} maps is 2.4\%, 2.8\%, 4.5\%, and 5.7\% for the W1, W2, W3, and W4 images, 
respectively \citep[][]{2011ApJ...735..112J}.
%
%
\section{Integrated CO spectra and molecular gas mass}\label{MH2_CO}
%
%
\begin{figure}
\centering
\includegraphics[width=0.47\textwidth]{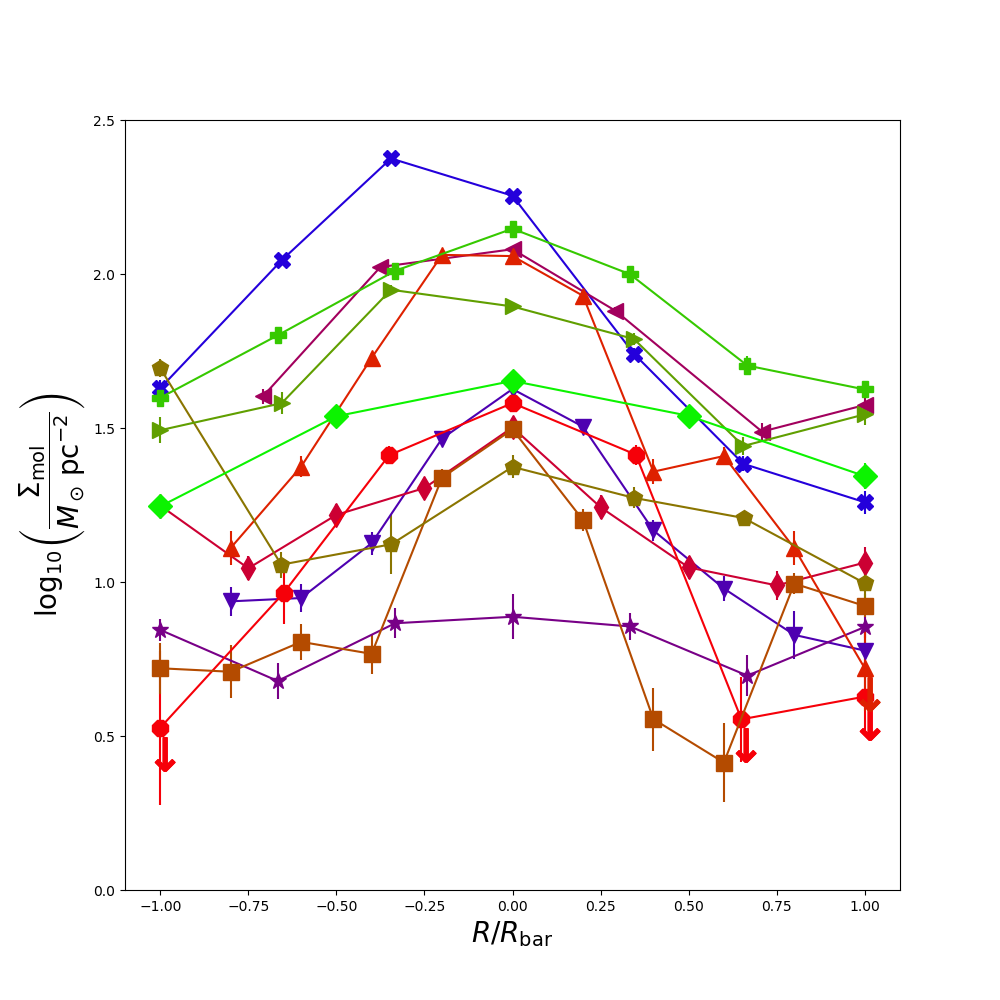}\\[-3ex]
\includegraphics[width=0.47\textwidth]{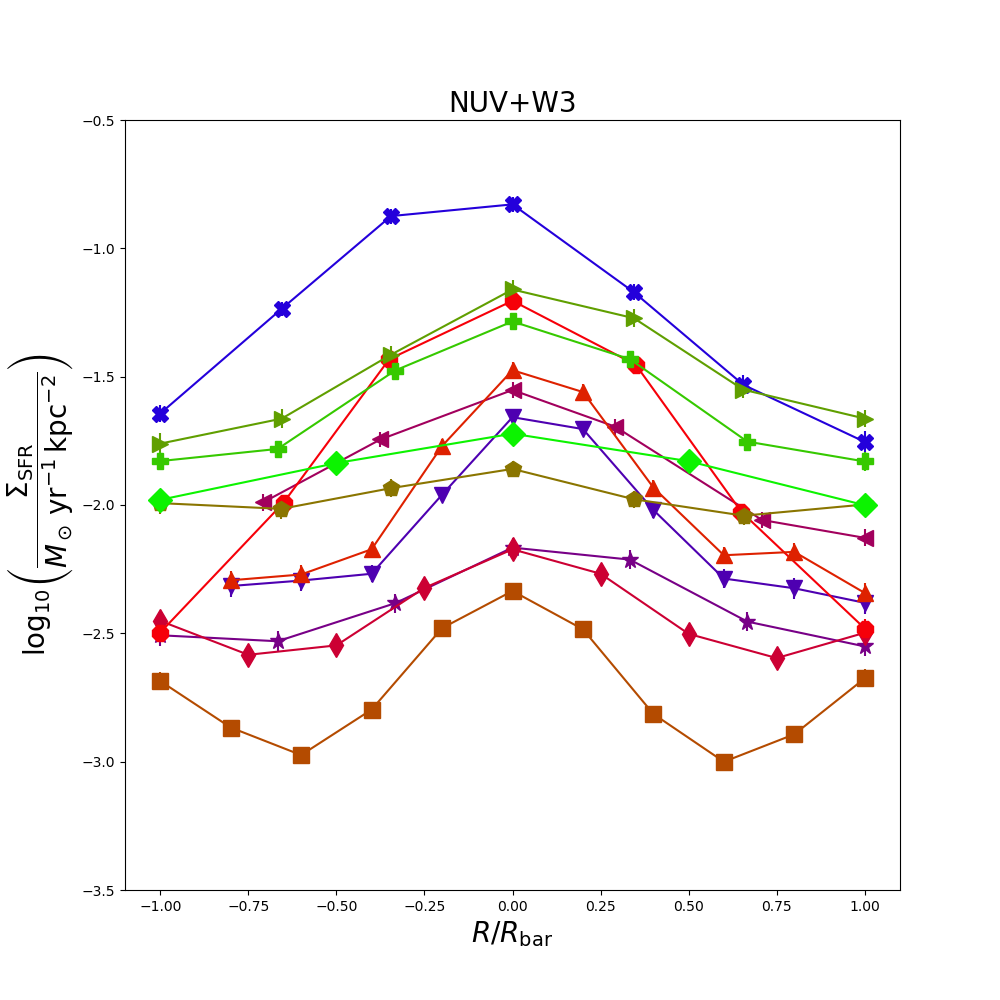}\\[-3ex]
 \includegraphics[width=0.47\textwidth]{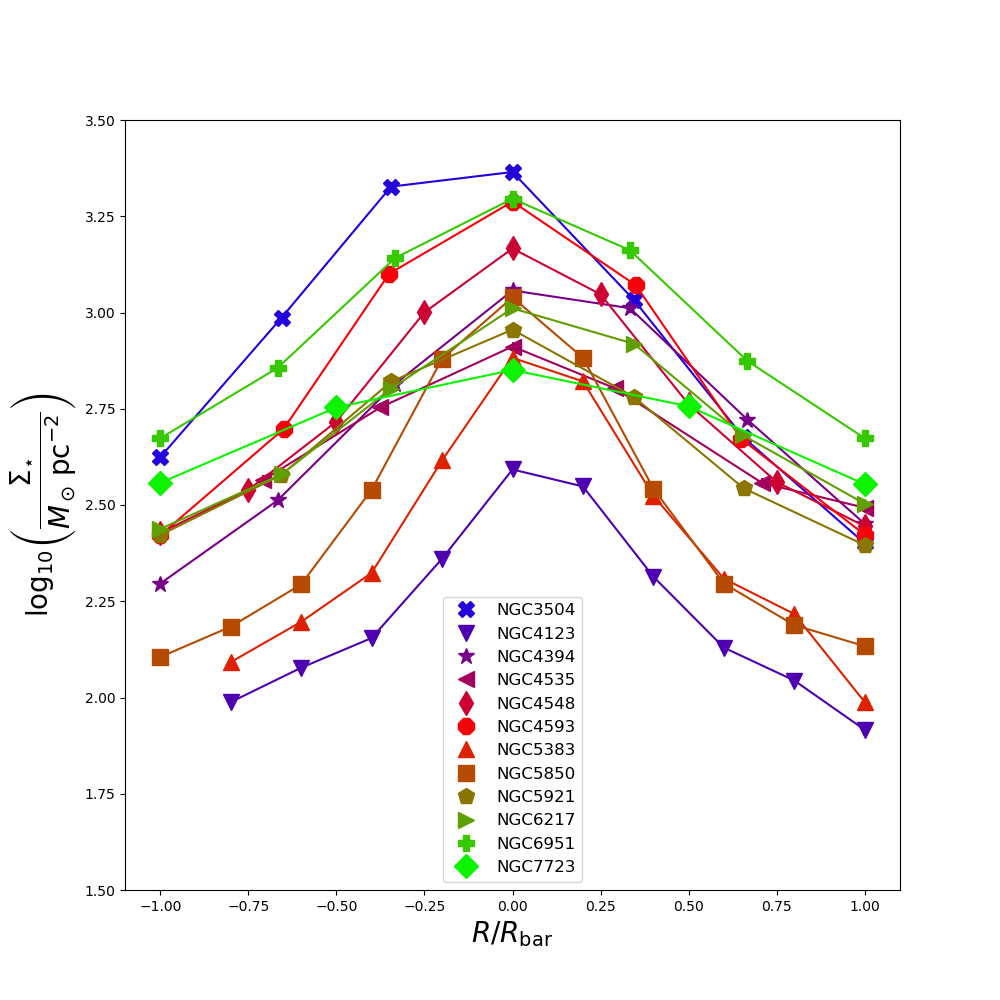}
\caption{
Profiles of molecular gas mass (\emph{top}), star formation rate (\emph{middle}), 
and stellar surface densities (\emph{bottom}) along the bar major axis, 
two-folded so that negative and positive numbers indicate the west and east sides of the bar, respectively. 
In the upper panel the downward-pointing arrows correspond to the pointings with no CO(1-0) detection (Eq. \ref{eq_upper_limit}). 
}
\label{PLOTS_ALLTOGETHER_A}
\end{figure}
%
%
From the CO(1-0) spectra measured along the bars (Sect.~\ref{CO_SPECTRA}), we obtain the velocity-integrated intensities (in K km s$^{-1}$)
\begin{equation}
I_{\rm CO(1-0)} = \int T dV
\end{equation}
by summing the spectra over the visually determined zero-level velocity width, 
using the Python \emph{SciPy} package for the Simpson's rule 
(i.e. approximating each part of the line with parabolas to perform the numerical integration). 
We estimate the uncertainty as
\begin{equation}\label{error_Ico}
{\rm Error} (I_{\rm CO(1-0)}) = \sigma \sqrt{{\Delta V}{\delta V}},
\end{equation}
where $\delta V$ is the channel width (10.45 km s$^{-1}$), $\Delta V$ is the total line width, 
and $\sigma$
is the root mean square (rms) noise of the spectrum, calculated outside the line emission. 
We checked that increasing $\Delta V$ by $25\%$ and $50\%$ results in a median difference 
in $I_{\rm CO(1-0)}$ of $3.2\%$ and $4.2\%$, respectively, 
which is lower than the median error ($\approx 8\%$) on $I_{\rm CO(1-0)}$ (Eq.~\ref{error_Ico}). 
We also confirmed that the velocity-integrated intensities determined within the line width agree 
well with those obtained from Gaussian fits, the difference being $5.5\%$  \citep[consistent with][]{2016ApJ...827..106A}. 
We hence conclude that the measurement procedure is not a source of major uncertainty in the $I_{\rm CO(1-0)}$ estimates.

In the event of no detection (which is only the case for four spectra of the  outermost pointings of 
NGC~4593 and NGC~5383) we calculated the upper limit as
\begin{equation}\label{eq_upper_limit}
I_{\rm CO(1-0)}=3 \sigma  \sqrt{\Delta V \delta V}.
\end{equation}
For CO(1-0) we adopted a value of $\Delta V = 200$ km s$^{-1}$, which is the median line width of CO(1-0) in this work. 
These pointings are shown as arrows in the plots. 
For CO(2-1) we followed the same steps as in the case of CO(1-0). We 
adopted $\Delta V =  \Delta V$ (CO(1-0)) if CO(1-0) was detected, and $\Delta V = 200$ km s$^{-1}$ otherwise.

We calculate the molecular gas mass ($M_{\rm mol}$) from the CO(1-0) 
luminosity ($L^\prime_{\rm CO}$) following \citet[][]{1997ApJ...478..144S},
\begin{equation}
L^\prime_{\rm CO} [{\rm K \, km\, s^{-1} pc^{-2}}]= 3.25 \times 10^7\, S_{\rm CO,tot} \nu_{\rm rest}^{-2} D^{2} (1+z)^{-1},
\label{eq:lco}
\end{equation}
where $S_{\rm CO, tot}$ is the CO line flux (in Jy$\,$km$\,$s$^{-1}$), 
$D$ is the distance in Mpc, $z$ the redshift, 
and $\nu_{\rm rest}$ is the rest frequency of the line in GHz, and then
\begin{equation}\label{eq_mh2}
M_{\rm mol} [M_\odot]= 1.36 \, \alpha_{\rm CO} \, L^\prime_{\rm CO}.
\end{equation}
We adopt the Galactic value 
$\alpha_{\rm CO}= \alpha_{\rm CO, gal} = {3.2}\,M_\odot$\,(K\,km\,s$^{-1}$\,pc$^{-2}$)$^{-1}$ 
\citep[][]{2013ARA&A..51..207B}. 
This conversion factor corresponds to 
$X=N_{\rm H_{2}}/I_{\rm CO} = 2 \cdot 10^{20} \, \rm cm^{-2} \, {\rm (K \, km \, s}^{-1})^{-1}$ \citep[][]{1986ApJ...309..326D,2011ApJ...730L..13B}. 
The factor $1.36$ includes the mass of helium and heavy metals \citep[as in e.g.][]{2011ApJ...730L..13B,2013ARA&A..51..207B}. 
Molecular gas surface density profiles ($\Sigma_{\rm mol}$, in units of M$_\odot$ pc$^{-2}$) are obtained as
\begin{equation}\label{eq_sigma_h2}
\Sigma_{\rm mol}= 1.36 \, \alpha_{\rm CO} \, I_{\rm CO} \, {\rm cos}(i),
\end{equation}
where $i$ corresponds to the disk inclination. 
Inclination corrections (see also Eqs. \ref{Eq_NUV_WISE3_paper_2} and \ref{eq_sigma_mass}) are performed following 
\citet[][]{2008AJ....136.2782L} and \citet[][]{2019ApJS..244...24L}, among others. 
We do not take into account the 3D geometry of the bar or bulge; 
this parametrisation is beyond the scope of this paper.

The $\Sigma_{\rm mol}$ profiles along the bar major axes are shown in the upper panel of Fig.~\ref{PLOTS_ALLTOGETHER_A}. 
The molecular gas is typically concentrated in the centre of galaxies, 
and in some cases we detect peaks at one or both edges of the bars (e.g. NGC~4394, NGC~5850, NGC~4548, NGC~5921). 
In a few cases (e.g. NGC~3504 or NGC~5921) the profiles are rather asymmetric, 
while in NGC~7723 it is shallower than in the rest of the sample. 
In Appendix~\ref{H2_tables} we list, for each of the pointings, 
the corresponding J2000 equatorial coordinates, velocity-integrated intensities, molecular gas masses, and surface densities. 

We report $\Sigma_{\rm mol}$ values that are consistent with those in the literature. 
\citet[][]{2008AJ....136.2846B} find average values of $\Sigma_{\rm mol}$ for 
disks of local non-barred spiral galaxies of $\sim$10-15 $M_{\odot}$ pc$^{-2}$. 
At the bar region, $\Sigma_{\rm mol}$ is typically in the range $\approx$ 20-90 $M_{\odot}$ pc$^{-2}$ 
\citep[][]{1998A&A...337..671R,1999ApJ...526...97R,2007PASJ...59..117K,2016PASJ...68...89M,2018PASJ...70...37M,2019PASJ...71S..13Y,2020MNRAS.495.3840M}, 
under the assumption of a common $\alpha_{\rm CO}=3.2 M_{\odot}$ (K\,km\,s$^{-1}$\,pc$^{-2}$)$^{-1}$, as used here. 
Our use of a bigger sample reveals a somewhat larger range: $\Sigma_{\rm mol}$ varies by $\approx 1.5-2$ orders of 
magnitude between the probed galaxies. 
All apertures along the bar of NGC~4394 have $\Sigma_{\rm mol}<10\, M_\odot$ pc$^{-2}$. 
The central peaks of NGC~4535, NGC~5383, and NGC~6951 are larger than 50 $M_\odot$ pc$^{-2}$ (see Fig.~\ref{PLOTS_ALLTOGETHER_A});  in the case of NGC~3504  $\Sigma_{\rm mol}$ goes beyond 100 $M_\odot$ pc$^{-2}$ and is rather asymmetric, which  is likely explained by the efficiency of these strong bars in sweeping the disk gas and enhancing its central concentration 
\citep[e.g.][and references therein]{1993RPPh...56..173S}.
%
%
\section{Star formation rates and stellar masses}\label{ap_photometry}
%
%
\begin{figure}
\centering
\includegraphics[width=0.49\textwidth]{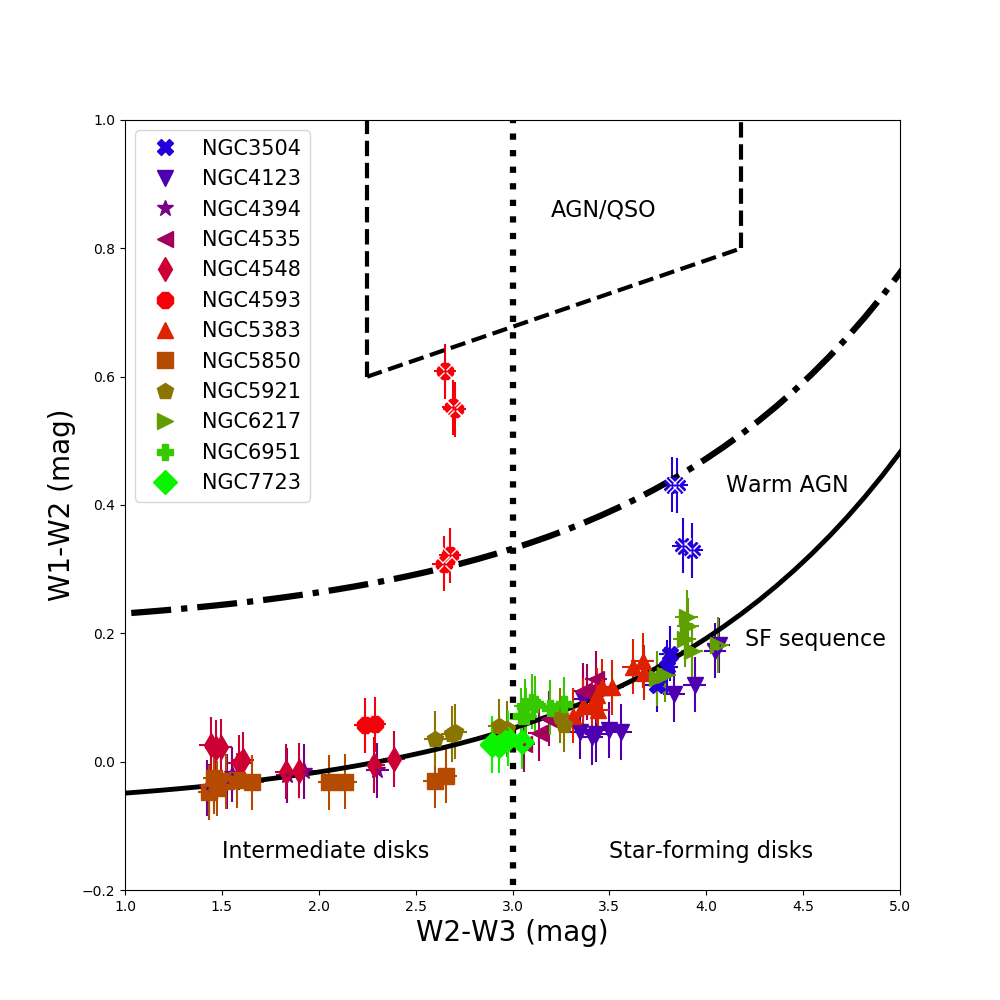}
\caption{
\emph{WISE} colour-colour magnitude plot within each IRAM pointing. 
A solid black line indicates the fit of the star formation sequence 
from \citet[][see their Fig.~10 and Eq. 1]{2019ApJS..245...25J}, 
while the vertical dotted line  separates the  intermediate and star-forming disks. 
The  polygon at the top of the figure, outlined with a dashed line, 
maps the region of \emph{WISE}-obscured  AGNs, QSOs,  LINERs, and 
ULIRGs. The dash-dotted line indicates the region where 
low-power Seyferts and LINERs reside \citep[see also][]{2011ApJ...735..112J,2020ApJ...903...91Y}, 
which is where the central pointings of NGC~4593 and NGC~3504 (x symbols) lie.
}
\label{plot_AGN}
\end{figure}
%
%
\begin{figure*}
\centering
\includegraphics[width=0.49\textwidth]{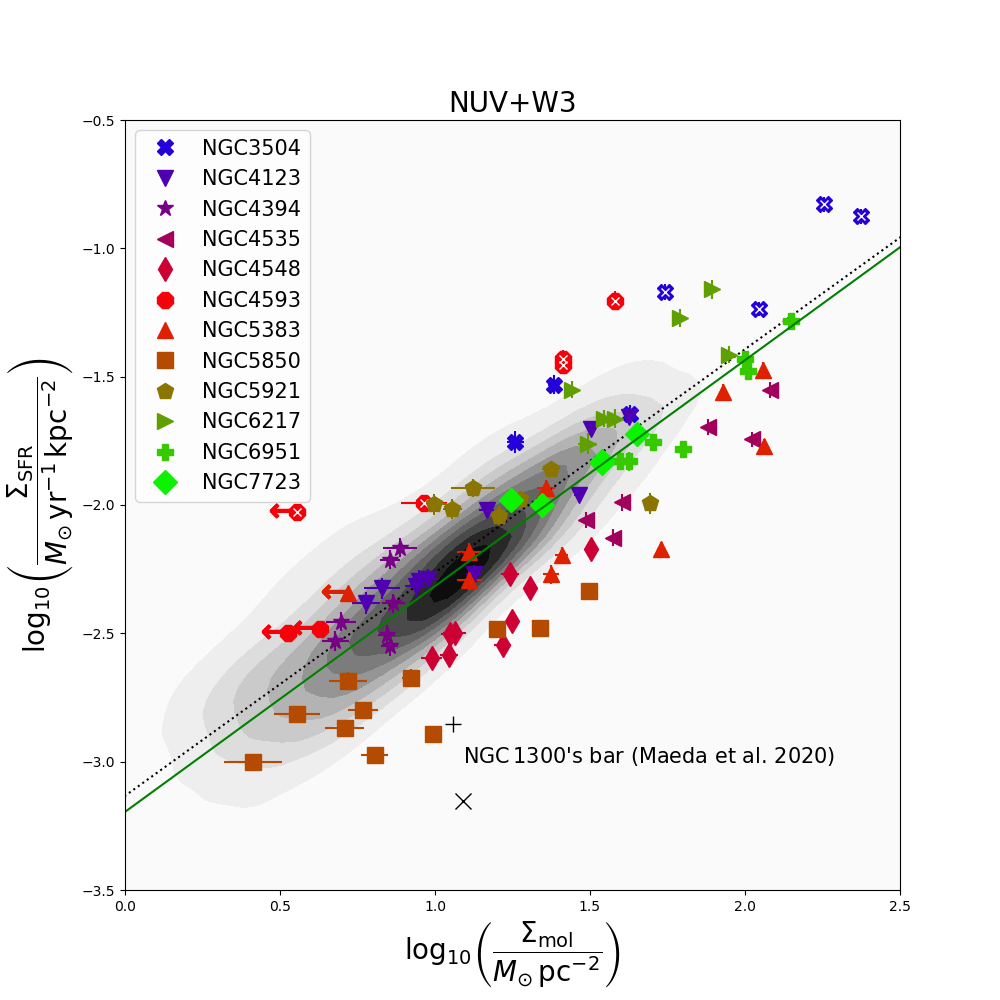}
\includegraphics[width=0.49\textwidth]{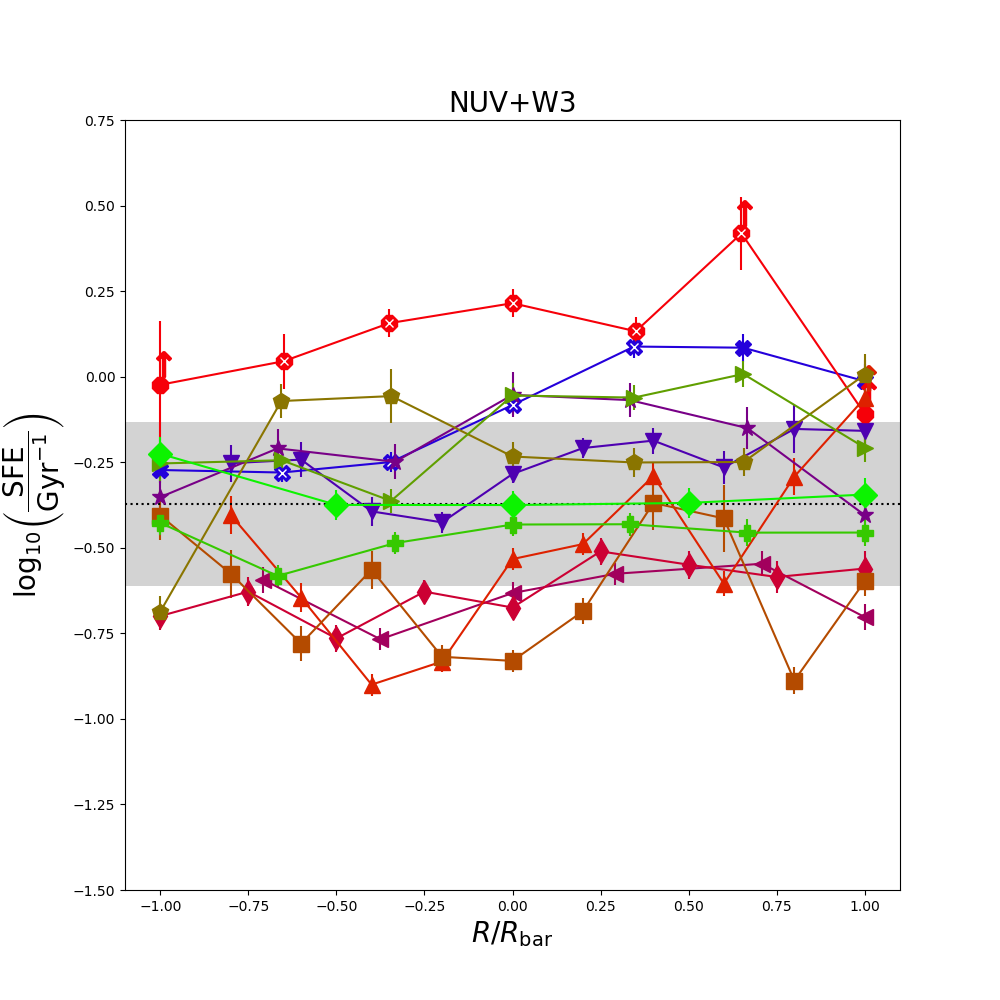}
\includegraphics[width=0.49\textwidth]{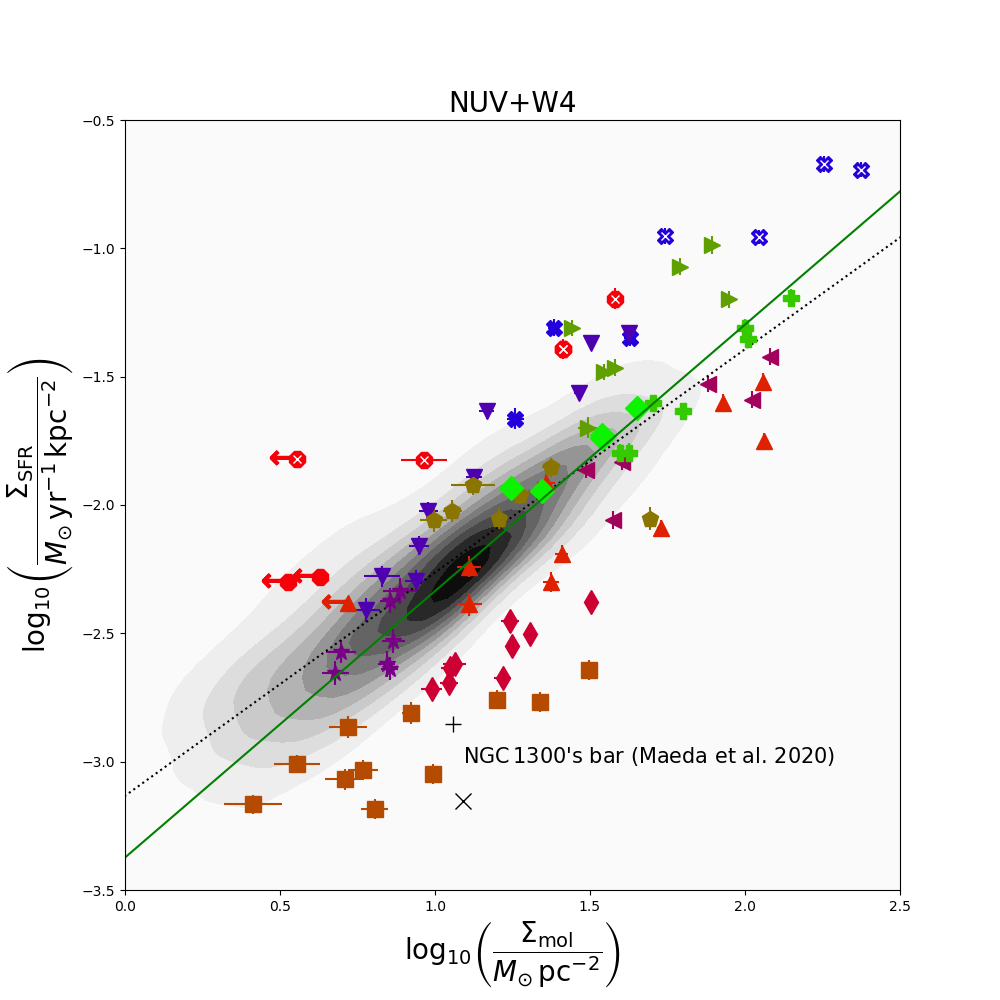}
\includegraphics[width=0.49\textwidth]{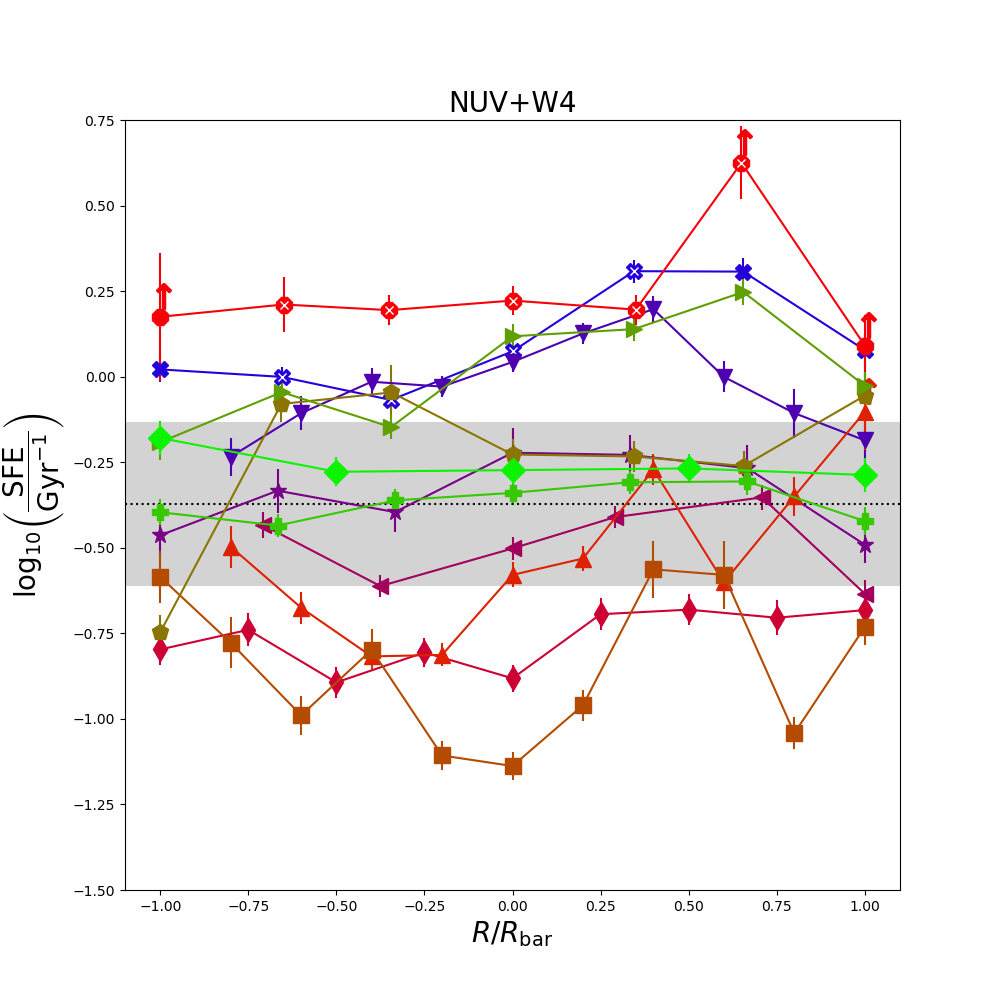}
\caption{
Kennicutt-Schmidt law and star formation efficiency within bars.
\emph{Left panels:} 
Star formation rate surface density traced from NUV combined with \emph{WISE} 3 (\emph{upper panels}) and 
\emph{WISE} 4 (\emph{lower panels}) vs molecular gas surface density,  traced from the CO(1-0) emission. 
The colour palette and symbols indicate different galaxies (see legend). 
The grey contours show the measurements from \citet[][]{2011ApJ...730L..13B}, obtained from FUV and 22~$\mu$m images, 
whose linear fit is also shown with a black dotted line. 
The white x symbols indicate those points with high W2-W1 (Fig \ref{plot_AGN}) that may correspond to less reliable SFR estimates due to AGNs. 
The leftward arrows correspond to the pointings with no CO(1-0) detection (Eq. \ref{eq_upper_limit}). 
The green line corresponds to the linear fit to the cloud of points. 
The grey cross and x symbol indicate the $\Sigma_{\rm SFR}$ and $\Sigma_{\rm mol}$ values at the bar region of NGC~1300 
reported by \citet[][]{2020MNRAS.495.3840M} from the combination of FUV and W4 (`Bar-A' and `Bar-B' in their Table 3). 
\emph{Right panels:} Profiles along the bar major axis of star formation efficiency. 
The grey rectangle traces the median SFE (black dotted line) plus or minus 1$\sigma$ from \citet[][]{2011ApJ...730L..13B}.
}
\label{PLOTS_ALLTOGETHER_B}
\end{figure*}
%
%
In order to calculate the SFR and $M_{\star}$ within each IRAM pointing, 
we perform aperture photometry over the \emph{GALEX} FUV and NUV 
and \emph{WISE}~1-2-3-4 images with the aid of Python's \emph{photutils} and \emph{astropy} packages. 
We proceed in the following way. We multiply the corresponding background-subtracted \emph{GALEX}and \emph{WISE} images, $I(x,y)$, with the IRAM beam 
pattern (approximated as a normalised Gaussian beam) placed at the position where the CO beam 
was pointed during the IRAM-30m CO(1-0) observations ($x_0,y_0)$:
\begin{equation}
I_{\rm beam} (x,y) = I(x,y) \cdot \exp{\left(-\frac{(x-x_0)^2+(y-y_0)^2}{2\sigma^2}\right)}.
\label{eq:multipy-beam}
\end{equation}

The Gaussian standard deviation ($\sigma$) is related to the FWHM 
as $\sigma$ = FWHM/$(2\sqrt{2\ln(2)})$ =  FWHM/2.35. 
From the resulting maps, $I_{\rm beam}(x,y)$, we then measure the total fluxes (in Jy). 
The resulting values are listed in Appendix~\ref{Aperture_photometry_fluxes}, together with the errors, 
taken as the quadratical sum of the calibration error (which is dominant) and the photometric error. 
We then compute the luminosities $L_{\nu}$ (in erg\,s$^{-1}$\,Hz$^{-1})$ 
and $L = \nu L_{\nu}$ (in erg\,s$^{-1}$).

We compute dust attenuation-corrected SFRs combining \emph{GALEX} NUV 
emission-line and \emph{WISE} infrared continuum measurements at 12~$\mu$m (W3) and 22~$\mu$m (W4). 
We use the linear `hybrid' SFR calibrations from the \emph{GALEX}-SDSS-\emph{WISE} Legacy Catalog \citep[][]{2016ApJS..227....2S,2018ApJ...859...11S}, 
as reported by \citet[][see their Table.~7]{2019ApJS..244...24L}:
\begin{equation}\label{Eq_NUV_WISE3_paper}
{\rm SFR}_{\rm NUV,W3}[M_{\odot} \, {\rm yr}^{-1}] = \frac{L_{NUV}{\rm [erg\,s^{-1}]}}{10^{43.24}}+\frac{L_{W3}{\rm [erg\,s^{-1}]}}{10^{42.86}},
\end{equation}
\begin{equation}\label{Eq_NUV_WISE4}
{\rm SFR}_{\rm NUV,W4}[M_{\odot}\, {yr^{-1}}] = \frac{L_{NUV}{\rm [erg\,s^{-1}]}}{10^{43.24}}+\frac{L_{W4}{\rm [erg\,s^{-1}]}}{10^{42.79}}.
\end{equation}
W3 and W4 are robust mid-IR SF tracers used to trace dust-enshrouded SF \citep[e.g.][]{2017ApJ...850...68C,2018ApJ...865..154H}. 
The W3 images are an order of magnitude more sensitive to SF than the W4 \citep[e.g.][]{2019ApJS..244...24L}. 
Nevertheless, \citet[][]{2019ApJS..244...24L}  argue that W3 suffers from larger systematic uncertainties and report variations in 
the W3-to-W4 ratio due to PAHs \citep[][]{2005ApJ...628L..29E}. The centre of the W3 band is indeed close to the 11.3 $\mu$m PAH emission, 
but is still dominated by continuum because of its relatively large bandwidth \citep[see discussion in][]{2017ApJ...850...68C}. 
PAH abundances are high in regions of active SF and are likely produced in molecular clouds \citep[][]{2010ApJ...715..701S}; 
PAHs can be destroyed near stars and/or AGNs where the radiation field is very intense.
 
In order to make sure that our analysis is not biased by the choice of the SF tracer, 
we tested different SF recipes (see Appendix~\ref{SF_hybrid_non_hybrid}), 
including estimates from NUV, FUV, and W3-4, and different hybrid combinations. 
We checked and confirmed that the trends presented in this paper are qualitatively the same when FUV is used instead of NUV, 
but we found that the SFEs and the scatter of the KS law are somewhat sensitive to wavelength (see discussion in Sect.~\ref{SFE_sect}). 
Interestingly enough, the bulk of the SFRs in our galaxies are traced by \emph{WISE} (Fig.~\ref{COMPARISON_TRACERS_GALS_ALL_2}). 
Therefore, whether we use NUV or FUV as a SF tracer does not make a big difference in the results presented in the next sections 
as long as they are combined with W3 or W4 (see Fig.~\ref{COMPARISON_TRACERS_GALS_ALL}). 
Unless stated otherwise, we estimate SFRs from NUV and W3 combined.

The star formation rate surface density ($\Sigma_{\rm SFR}$), corrected for inclination (i), in each aperture is calculated as
\begin{equation}\label{Eq_NUV_WISE3_paper_2}
\Sigma_{\rm SFR}[M_{\odot} \, {\rm yr}^{-1} \, {\rm pc}^{-2}]=\frac{{\rm SFR} \cdot {\rm cos}(i)}{1.13 \cdot \rm (FWHM_{pc})^2},
\end{equation}
where the denominator corresponds to the 2D integral of a Gaussian beam, 
which we take as the FWHM (21.5\arcsec) of the IRAM beam (in units of pc) at the distance of the object, FWHM$_{\rm pc}$.
This expression, together with Eq.~\ref{eq:multipy-beam}, means that we calculate the weighted mean value of the 
SFR with the IRAM beam, where the weighting function is the IRAM beam shape.

We calculate the stellar mass from the W1 fluxes within each IRAM pointing ($I_{3.4 \mu \rm m, beam}$) adopting a constant mass-to-light 
ratio $\Upsilon_{\star}^{3.6}=0.53$ \citep[in solar units at $3.6$ $\mu$m,][]{2012AJ....143..139E} 
and neglecting the small difference between 3.6~$\mu$m (the \emph{Spitzer}-IRAC 1 central wavelength) 
and 3.4~$\mu$m (the W1 central wavelength), as done in  \citet[][]{2019ApJS..244...24L}, among others. 
With this we derive the inclination-corrected stellar mass surface density as
\begin{equation}\label{eq_sigma_mass}
\begin{split}
\Sigma_{\star} [M_{\odot} \, {\rm pc}^{-2}] & = \frac{L_{\rm 3.4, \odot}  \cdot \Upsilon_{\star}^{3.6} \cdot {\rm cos}(i)}{1.13\cdot {\rm FWHM}^2} \\ 
&= 1.88 \cdot 10^5 \cdot I_{\rm 3.4 \mu{\rm m}, beam}[{\rm Jy}] \cdot \Upsilon_{\star}^{3.6} \cdot {\rm cos}(i),
\end{split}
\end{equation} 
%
%
\input{table_SFE_bars.txt}
%
%
where $L_{\rm 3.4, \odot}$ is the luminosity $\nu L_\nu$ at 3.4~$\mu$m in units of solar 
luminosity at 3.6~$\mu$m \citep[$1.4 \times 10^{32}$ erg s$^{-1}$,][]{2014MNRAS.445..881C}. 
W1-based stellar masses are likewise utilised to normalise SFRs and obtain specific star formation rates (sSFRs) across bars. 
We opted not to estimate $M_{\star}$ from \emph{WISE} 1 and 2 colours \citep[][]{2016ApJ...821..113Z}; 
the resulting $\Sigma_{\star}$ profiles were not reliable (e.g. minima of $\Sigma_{\star}$ in the centre of 
NGC~4593), quite possibly because of non-stellar contaminants such as hot dust and PAHs \citep[see e.g.][]{2011MNRAS.417..812Z,2012ApJ...744...17M}.

In order to assess whether AGNs might have a contribution to the computed \emph{WISE} fluxes, 
we construct \emph{WISE} colour-colour magnitude diagrams (W1-W2 versus W3-W2) for all galaxies and within 
each aperture (Fig.~\ref{plot_AGN}) 
and include the classifications of \citet[][]{2017ApJ...836..182J}, presented in \citet[][see their Fig.~10]{2019ApJS..245...25J} 
into star-forming galaxies and AGNs or quasars (QSOs). 
The majority of the sampled regions in our bars belong to the SF sequence;   roughly half belong to the `intermediate galaxy' class 
\citep[W2-W3<3, mostly quiescent bars as reported by][]{2020A&A...644A..38D} 
and the other half to active star-forming disks (W2-W3>3). 
There are, however, a few points that lie above this SF sequence, corresponding to the central 
regions of NGC~4593 and NGC~3504. NGC~4593 is classified a Seyfert 1 by 
\citet[][]{2010A&A...518A..10V} which can explain the elevated W1-W2 values that bring 
it close to the AGN--QSO region. The inferred SFRs and stellar masses within those 
apertures are thus less reliable. \citet[][]{2010A&A...518A..10V} report nuclear H{\sc\,ii} regions in NGC~3504, 
which show high W2-W3 and W1-W2 values. Interestingly, the \emph{WISE}  colours of NGC~4548, which is classified as a LINER 
\citep{2010A&A...518A..10V}, are not affected by this nuclear activity and are entirely compatible with the SF sequence. 
This is also the case of the active galaxy NGC~6951 \citep[][]{1997ApJS..112..391H,2000A&A...353..893P}. 
We note that in some cases the identification of AGN footprints may be hindered by the large size of the  beams used. 
In Fig.~\ref{plot_AGN} and the following figures we use x symbols to indicate pointings whose elevated W2-W1 colours are suggestive of the presence of AGNs.

The profiles along bars of star formation rate surface density 
and stellar surface density are shown in the central and lower panels of Fig.~\ref{PLOTS_ALLTOGETHER_A}, respectively. 
Their shapes are quite similar to those of molecular gas surface density, shown in the upper panel. 
All $\Sigma_{\rm SFR}$ profiles are centrally peaked; some of these peaks are 
associated with the presence of nuclear rings (e.g. NGC~5383 and NGC~5850), 
including the case of NGC~6951 \citep[e.g.][]{1999ApJ...511..157K}. 
A few galaxies (e.g. NGC$\,4548$ and NGC$\,5850) $ show humps at the bar ends. 
In general, three galaxies in our sample (NGC~4394, NGC~4548, and NGC~5850) show low levels of SF along the bar major-axis. 
The visual inspection of the FUV images of NGC~4123, NGC~6951, and NGC~6217 (only in the north-east) 
also reveal SF gaps along the bar \citep[see e.g.][]{2020A&A...644A..38D}; 
these gaps are typically smaller than the IRAM-30m pointings and cannot be resolved. 
The $\Sigma_{\rm SFR}$ values directly obtained from FUV in this work (Appendix~\ref{SF_hybrid_non_hybrid}) 
for S0/a-Sc galaxies are consistent with the average profiles along bars presented by \citet[][]{2020A&A...644A..38D} for $T<5$. 
They report a standard deviation of $1-1.5$ mag of the FUV emission in the inner parts of galaxies with log$_{10}(M_{\star}/M_{\odot})=10-11$.

None of the $\Sigma_{\star}$ profiles show enhancements at the bar ends. 
This is not surprising as these galaxies do not have ansae structures \citep[stellar blobs at the end of the bars;][]{1965AJ.....70..501D} 
even though most ansae are detected in early-type spiral galaxies, 
like some of the ones probed in this work \citep[e.g.][]{2007AJ....134.1863M,2007MNRAS.381..401L}. 
Our galaxies are characterised by large central mass concentrations. 
At least four galaxies (NGC~4394, NGC~4548, NGC~4593, and NGC~5850) host 
barlenses \citep[which are lens-like stellar structures embedded in bars postulated to be the face-on counterparts 
of boxy or peanut bulges; e.g.][]{2014MNRAS.444L..80L,2015MNRAS.454.3843A}, 
nuclear rings and bars (NGC~5850), or nuclear lenses (NGC~3504 and NGC~4394), according to \citet[][]{2015ApJS..217...32B} 
and/or the NED, as shown in Table \ref{sample_12_props}.

The resulting values of $\Sigma_{\rm SFR}$ and $\Sigma_{\rm \ast}$ are tabulated in Appendix~\ref{Aperture_photometry_fluxes}. 
The relationship between the values of SFRs and $M_{\star}$, presented here, and 
molecular gas masses is analysed in the next section.
%
%
\section{Star formation efficiency along bars}\label{SFE_sect}
%
%
\begin{figure*}
\centering
\includegraphics[width=0.49\textwidth]{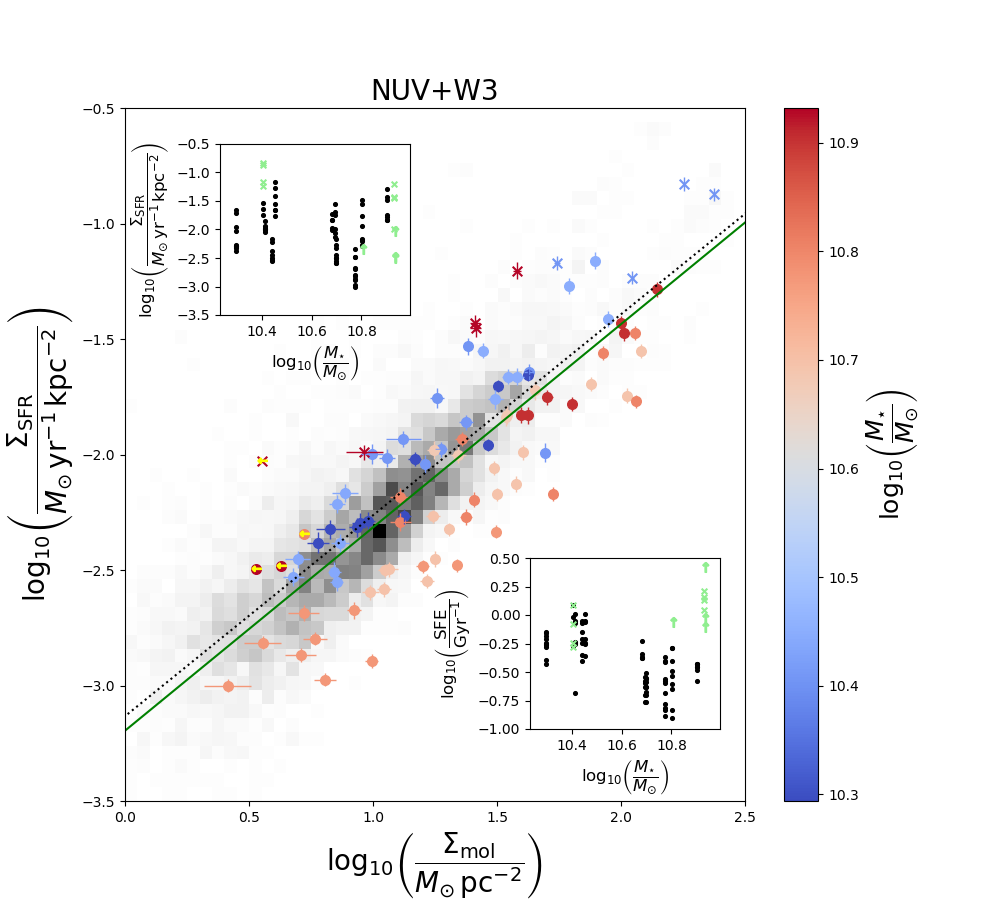}
\includegraphics[width=0.49\textwidth]{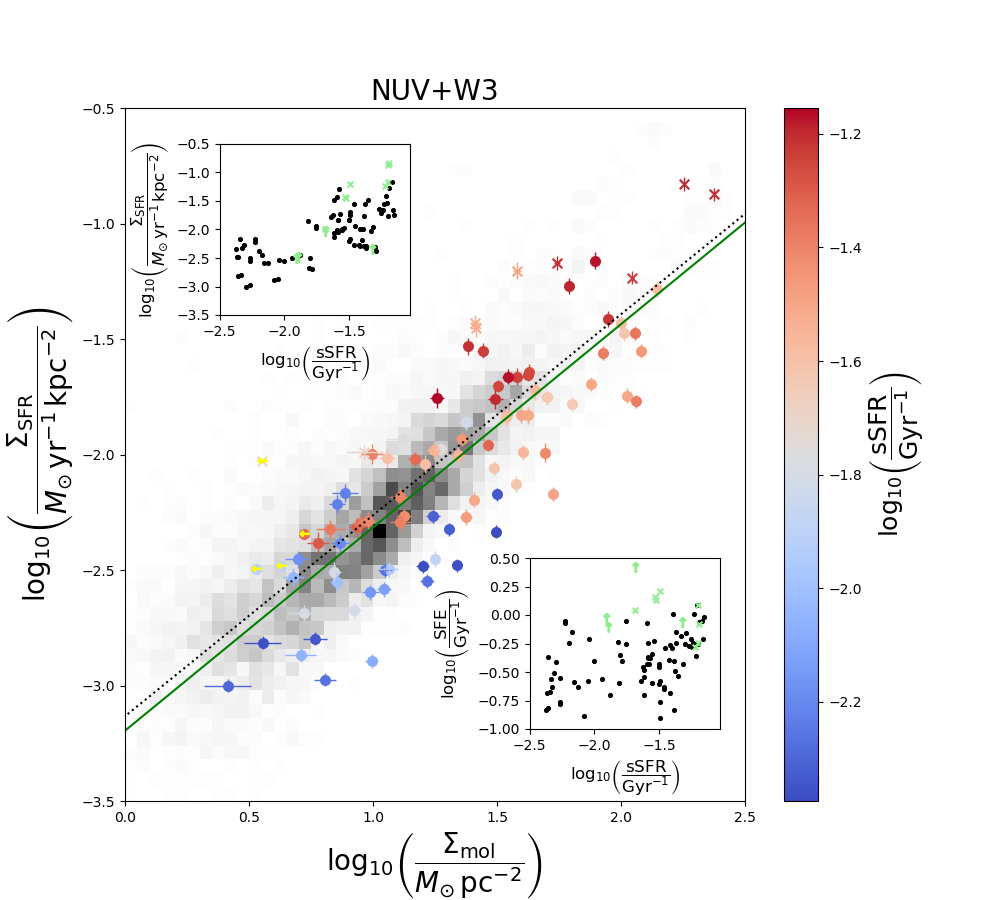}
\includegraphics[width=0.49\textwidth]{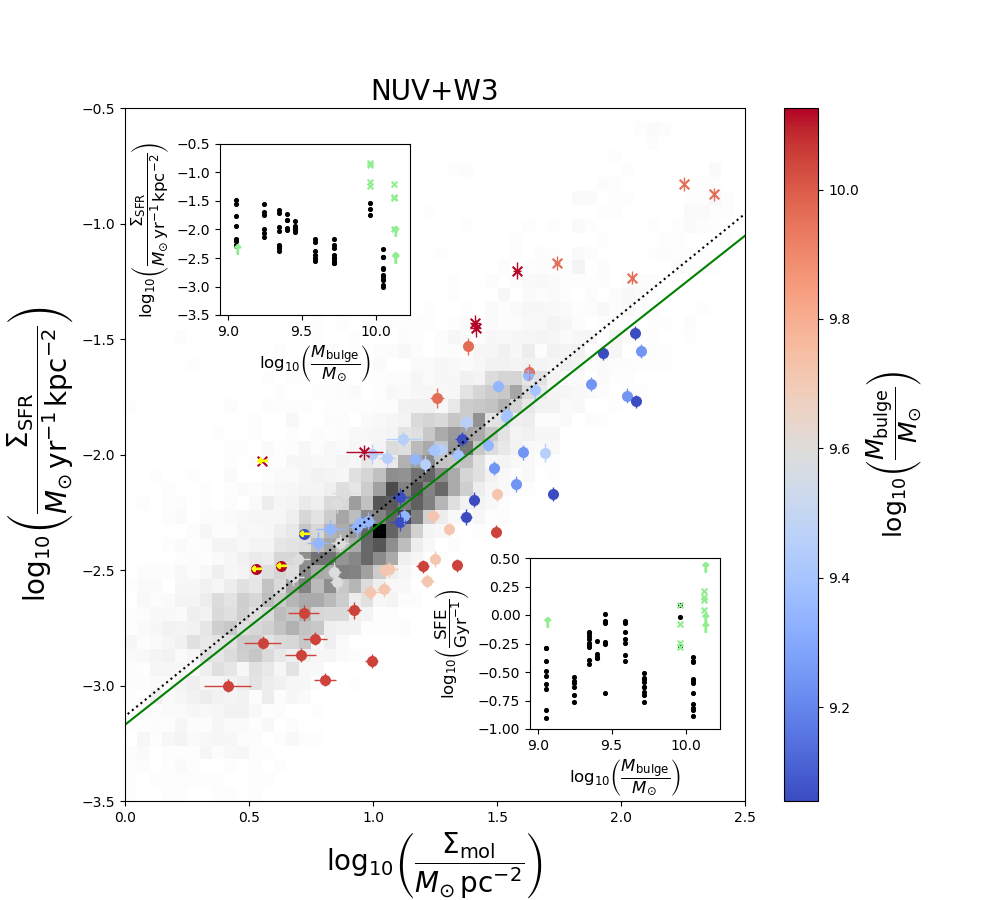}
\includegraphics[width=0.49\textwidth]{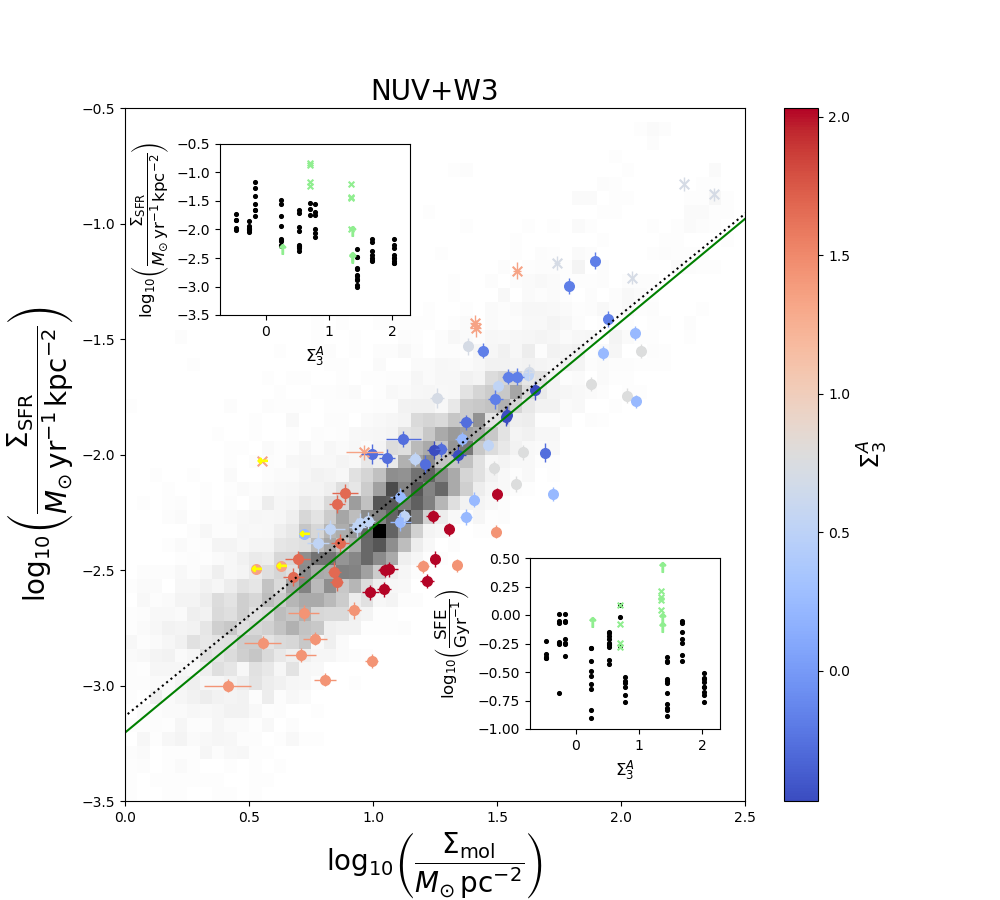}
\caption{
Same as  left panel of Fig.~\ref{PLOTS_ALLTOGETHER_B}, 
but the  colour-coding (see colour bars on the right)  indicates the total stellar mass of the host galaxy (\emph{top left}), 
the specific star formation of each aperture along the bars (\emph{top right}), the bulge stellar mass (\emph{bottom left}), 
and the projected surface density to the third-nearest neighbour galaxy (\emph{bottom right}). 
The background black and grey density plots trace the measurements from \citet[][]{2011ApJ...730L..13B}. 
The x symbols indicate pointings whose elevated W2-W1 colours suggest the presence of AGNs. 
The insets  show $\Sigma_{\rm SFR}$ (upper left corners) and SFE (lower right corners) 
vs the parameters used for the colour-coding of the main plot; 
black solid points correspond to the pointings with CO(1-0) detection whose WISE colours lie in the SF sequence, 
while non-detections and AGN candidates are respectively highlighted with green x symbols and arrows (i.e. less reliable SFEs and SFRs).
}
\label{BAR_VS_GLOBAL_2}
\end{figure*}
%
%
\begin{figure*}
\centering
\includegraphics[width=0.49\textwidth]{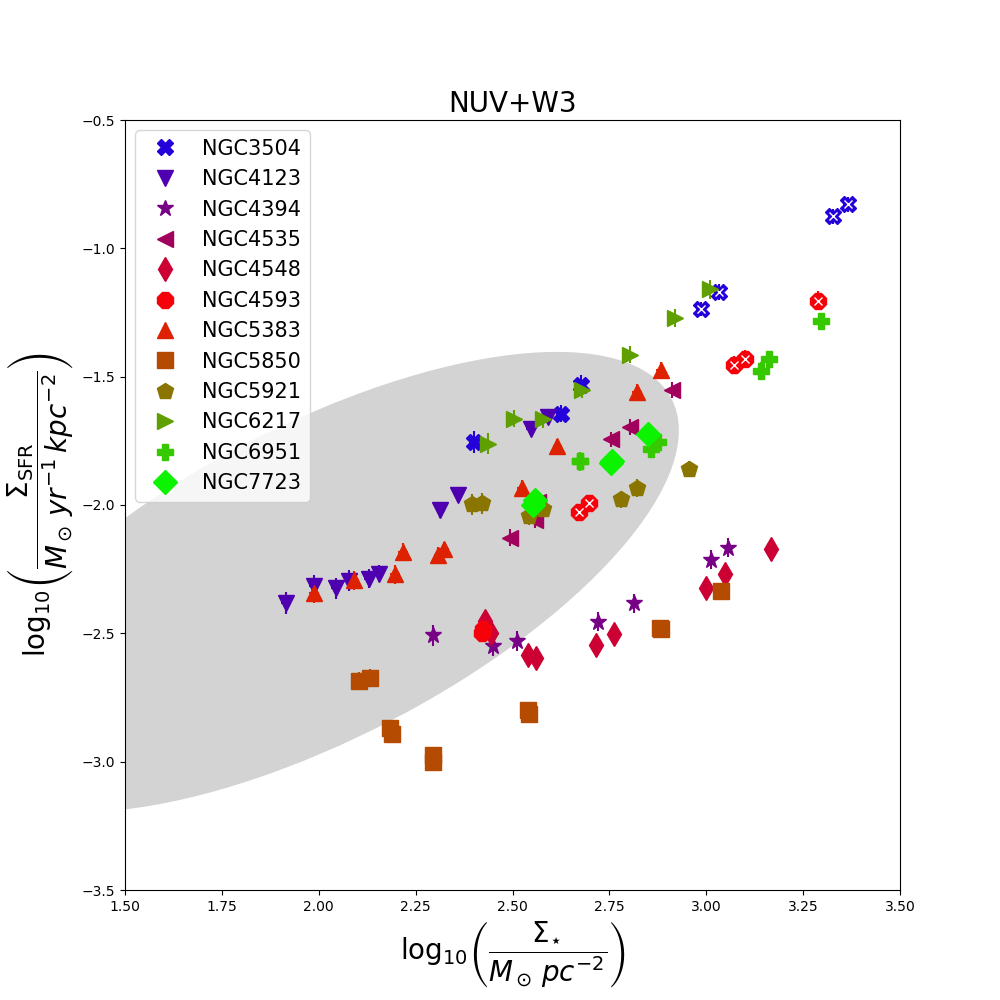}
\includegraphics[width=0.49\textwidth]{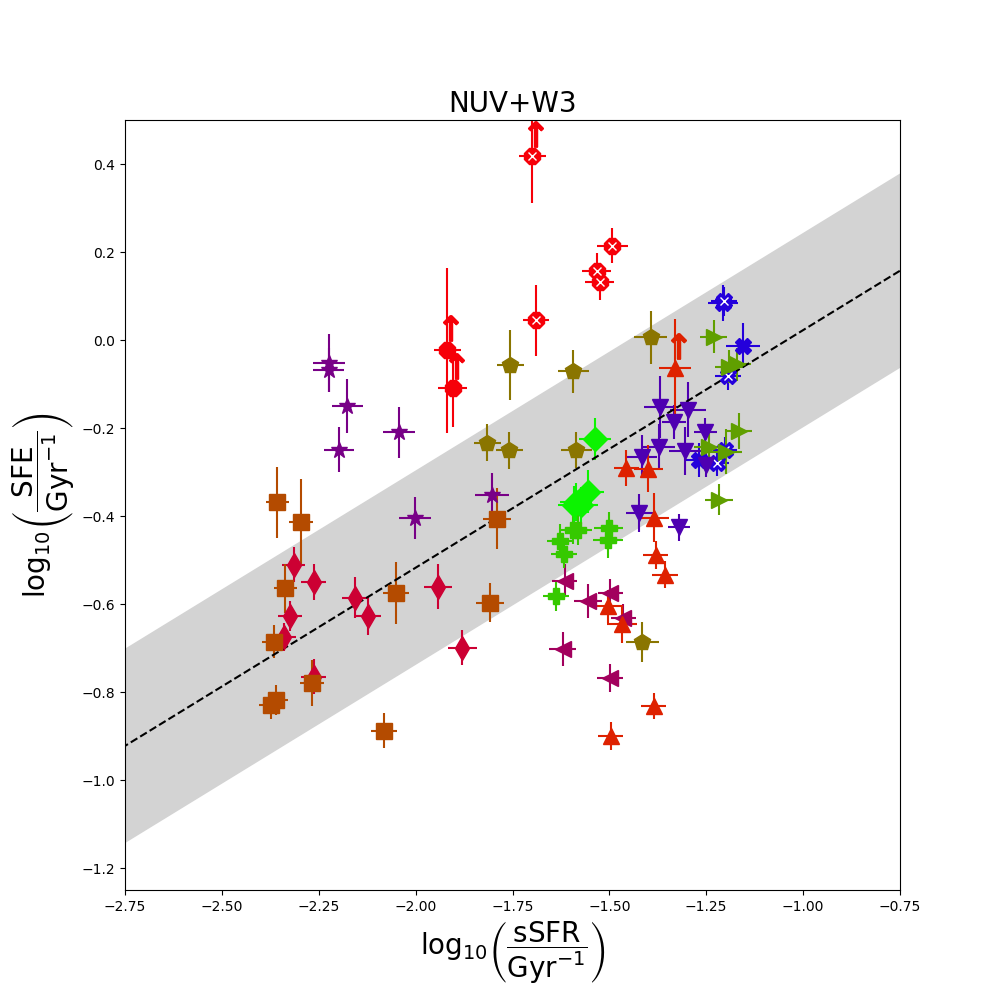}
\caption{
Scaling relations involving surface densities of stellar mass, molecular gas mass, and star formation rate.
\emph{Left:}
Star-forming main sequence measured within the CO pointings covering the bars, 
for each galaxy in our sample (see legend). Star formation rates are calculated 
from the combination of NUV and \emph{WISE} 3 images (Eq. \ref{Eq_NUV_WISE3_paper}), 
while stellar masses are estimated from \emph{WISE} 1 images (Eq. \ref{eq_sigma_mass}). 
The grey ellipse corresponds to the spatially resolved main sequence reported by \citet[][]{2018MNRAS.474.2039E}, 
traced from their Fig.~4 \citep[see][]{2019A&A...623A.154R}, based on Mapping Nearby Galaxies at APO \citep[MaNGA;][]{2015ApJ...798....7B}.
\emph{Right:} Star formation efficiency vs  specific star formation. The grey area corresponds to the best fit (black dotted line) 
reported by \citet[][COLD GASS, see their Table 1; we only use their secure detections]{2011MNRAS.415...61S} plus or minus 1$\sigma$. 
For the sake of consistency, we add the mass of helium and heavy metals ($\times 1.36$) to the molecular gas masses 
used to calculate SFEs in COLD GASS. The colours and symbols are the same as in Fig.~\ref{PLOTS_ALLTOGETHER_B}.
}
\label{Fig_MS}
\end{figure*}
%
%
Here we study the relationship between $\Sigma_{\rm SFR}$ and $\Sigma_{\rm mol}$ in bars, 
so that we can determine whether the SF activity is mainly controlled by the H$_{2}$ mass 
(i.e. GMCs and diffuse molecular gas) or by distinct physical conditions in bars. 
We quantify the link between star formation and molecular gas by measuring the star formation efficiency SFE = $\Sigma_{\rm SFR} / \Sigma_{\rm mol}$. 
SFE is the inverse of the gas depletion time ($\tau_{\rm dep}$), which  is  the time needed to use up the existing gas reservoirs at a given SFR. 

The SFEs are calculated within each of the pointings sampling the bars in our sample. 
$\Sigma_{\rm SFR}$ and $\Sigma_{\rm mol}$ are correlated even within bars (see Fig.~\ref{PLOTS_ALLTOGETHER_B}). 
This means that, on average, the presence of ionised regions in strong bars is to a large extent controlled by the availability of H$_{2}$ gas, 
even though the timescales for SF might well be larger than those of the inflow of GMCs along the bars 
\citep[see e.g. the models by][]{2006A&A...454..165P}. The local KS law depends on the spatial resolution of the data; for example, 
a higher slope of the scaling relation has been found for a coarser spatial resolution \citep[e.g.][]{2010A&A...510A..64V}. 
Therefore it is instructive to compare these results to those obtained from CO(2-1) which has a two times higher angular resolution. 
In Appendix~\ref{CO_2_1_masses} we show that the reported $\Sigma_{\rm SFR}$-$\Sigma_{\rm mol}$ 
relation is similar when CO(2-1) spectra (beam size of $\sim 1-2$ kpc) are used to 
compute $\Sigma_{\rm mol}$ (see Fig.~\ref{CO_2_1_KS}) instead of CO(1-0). 
On the other hand, our pointings are sufficiently large to probe physical areas ($\gtrsim 500$ pc) where 
the molecular gas mass and the star formation rates are tightly correlated, as increasing resolution can 
eventually wash out the scaling reported on large scales and the KS law breaks down \citep[][]{2010ApJ...722.1699S,2014MNRAS.439.3239K} 
\citep[for a similar discussion based on the analysis of the spatially resolved star-forming main sequence, see][and references therein]{2018ApJ...865..154H}.

The SFEs vary between different galaxies by up to an order of magnitude (Fig.~\ref{PLOTS_ALLTOGETHER_B}), 
ranging for $0.1\lessapprox$ SFE/Gyr$^{-1}$ $\lessapprox 1.8$ 
when $\Sigma_{\rm SFR}$ is estimated from the combination NUV and W3.  
For a given galaxy there is no systematic trend in SFE along the bar major axis 
(right panel of Fig.~\ref{PLOTS_ALLTOGETHER_B}).
In Table.~\ref{table_SFE_bars} we show the mean and dispersion of SFE as a function of $R/R_{\rm bar}$ 
for the different hybrid SF recipes used in this work (Sect.~\ref{ap_photometry}). 
The mean SFE in the central region ($|R/R_{\rm bar}|$<1/4) and bar-ends ($|R/R_{\rm bar}|>3/4$) is $\sim 0.4$ Gyr$^{-1}$, 
which is roughly 0.1 dex larger than the average in the mid-part of the bar ($1/4<|R/R_{\rm bar}|<3/4$). 
However, this difference is smaller than the dispersion of SFE across bars ($\sim 0.27$ dex) and SFE gradients are 
not obvious in individual galaxies; we thus conclude that SFEs in the bar are not remarkably different than those in the central regions. 
While the SFE depends on the adopted SF tracer and recipe to some extent 
(depending on the choice the mean integrated SFE within bars varies by a maximum of 0.08 dex; Table~\ref{table_SFE_bars}), 
the discussed radial trends versus radius are qualitatively the same.

We note that a smaller scatter in the $\Sigma_{\rm SFR}-\Sigma_{\rm mol}$ relation is obtained when FUV or NUV 
are combined with W3, instead of W4, to account for the dust (Fig.~\ref{PLOTS_ALLTOGETHER_B} and 
Appendix~\ref{SF_hybrid_non_hybrid}). This is in agreement with \citet[][]{2017ApJ...850...68C}, 
who analysed the relation between the luminosity at W3 and W4 passbands and the SF estimated from the total infrared luminosity: 
they conclude that the W3 SFR relation has a 1$\sigma$ scatter of 0.15, while the W4 SFR relation shows more scatter (0.18 dex). 
In this work we obtain a 1$\sigma$ scatter of 0.27 dex in the KS law when using SFR$_{NUV,W3}$; 
the scatter increases to 0.37 dex for SFR$_{NUV,W4}$. 
In addition, the 24 $\mu$m diffuse emission within the IRAM-30m pointing may not be related to SFR, but rather 
to the circumstellar dusty envelopes of AGB stars \citep[e.g.][]{2009A&A...493..453V}; this can be another physical reason 
for the above-mentioned scatter when using W4.

On average, our SFE within bars are consistent with median values SFEs obtained in the literature for spiral galaxies. 
In particular, we compare our measurements with those reported by \citet[][]{2011ApJ...730L..13B} (median SFE 0.43 Gyr$^{-1}$, $\sigma=0.24$ dex, 
under the assumption of the same $\alpha_{\rm CO}$ applied in this work), shown in black and grey in the left panels 
of Fig.~\ref{PLOTS_ALLTOGETHER_B} (see also Figs.~\ref{COMPARISON_TRACERS_GALS_ALL_2}~and~\ref{COMPARISON_TRACERS_GALS_ALL}). 
There is good agreement between the linear fits of the KS law in this work and that of 
\citet[][]{2011ApJ...730L..13B} for SFR$_{NUV, W3}$, 
shown with solid green and black dotted lines in Fig.~\ref{PLOTS_ALLTOGETHER_B}, respectively. 
We again report differences between W3 and W4: the slope of the KS law is $\sim 15 \%$ larger for 
SFR$_{NUV, W4}$ than for SFR$_{NUV, W3}$ (see Table~\ref{table_slopes_scatter} in Appendix~\ref{SF_hybrid_non_hybrid}). 

The SFEs are almost a factor of 2 lower than the average in a few of our galaxies, 
such as NGC~5850 (<SFE>$=0.25 \pm 0.03$ Gyr$^{-1}$) and NGC~4548 (<SFE>$=0.24 \pm 0.01$ Gyr$^{-1}$); 
the resulting SFEs are even lower (by a factor of $\sim 1.5$) when using SFR$_{NUV, W4}$. 
This is also the case of NGC~4535 (<SFE>$=0.23 \pm 0.02$ Gyr$^{-1}$), which hosts H{\sc\,ii} regions in the mid-parts of the bars. 
NGC~3504 and NGC~4593 are the galaxies in the sample with the highest SFE values (even larger than 1 Gyr$^{-1}$); 
we note, however,  that the measurement in the central pointings may be affected to some extent by AGN contamination 
(x symbols in Figs. \ref{PLOTS_ALLTOGETHER_B} and \ref{BAR_VS_GLOBAL_2}), as discussed in Sect.~\ref{ap_photometry}. 

It is not trivial to explain the differences in SFE between different galaxies as they are all grand design hosting high-amplitude bars. 
Interestingly enough, we find a segregation as a function of the total stellar mass of the host galaxy (Fig.~\ref{BAR_VS_GLOBAL_2}): 
the massive galaxies in our sample ($10^{10.7} \lesssim M_{\star}/M_{\odot} \lesssim 10^{11}$) 
have lower SFEs than their fainter counterparts ($10^{10.2} \lesssim M_{\star}/M_{\odot} \lesssim 10^{7}$). 
In Fig.~\ref{BAR_VS_GLOBAL_2} we also show that the bars with the lowest $\Sigma_{\rm SFR}$ and sSFR 
values  cohabit with massive bulges, according to 2D  photometric decompositions by \citet[][]{2015ApJS..219....4S}, and higher $\Sigma_{3}^{A}$ 
(Table~\ref{sample_12_props}, Sect.~\ref{sample_data}). 
Naturally, these statistical trends may imply that environmental and internal effects control the global gas availability and 
KS law, in the bars and elsewhere in the disk. 
These trends are also clear when we use FUV or \emph{WISE} 4 for the SFR estimates (Fig.~\ref{BAR_VS_GLOBAL_2_W4}). 
This is further discussed in Sect.~\ref{SFE_bars_discussion}.

One possible source of uncertainty in our analysis of SFEs in bars is the assumption of a constant CO-to-H$_2$ conversion factor, 
using the same value as found in our Galaxy \citep[][]{2013ARA&A..51..207B}. It is well accepted that $\alpha_{\rm CO}$ 
can vary in different types of galaxies, 
and within galaxies \citep[e.g.][]{2013ApJ...777....5S,2013ApJ...764..117B}, 
mainly due to its dependence on metallicity and star formation activity \citep[see][and references therein] {2013ARA&A..51..207B}. 
In particular, $\alpha_{\rm CO}$ can be considerably higher for low-metallicity galaxies 
\citep[below 12 + log(O/H) $\sim$  8.4, e.g.] []{2011ApJ...737...12L,2013ARA&A..51..207B, 2015A&A...583A.114H} 
and lower by a factor of $3-10$ in extreme starbursts, as in ultraluminous infrared galaxies (ULIRGs) \citep[][]{1998ApJ...507..615D,2003ApJ...582...37D}. 
Our sample does not contain any low-metallicity galaxies nor  does it contain extreme starburst galaxies, as we can see from Fig.~\ref{Fig_MS} (left panel) showing that all our pointings 
follow the spatially resolved star-forming main sequence \citep[][see their Fig.~4]{2018MNRAS.474.2039E} 
or lie below it \citep[interestingly in the case of NGC~4394, NGC~4548, and NGC~5850, which lack H{\sc\,ii} 
knots and clumps in the mid-part of the bar;][]{2020A&A...644A..38D}. 
This means that we do not expect large variations in $\alpha_{\rm CO}$ 
due to a low metallicity or a high SFR within our sample.

A lower CO-to-H$_2$ conversion factor in the bar region than in the rest of the disk has been suggested in previous observational work 
\citep[e.g. the case of NGC~3627, studied by][]{2015PASJ...67....2M}, but remains a matter of debate \citep[see also][]{2011MNRAS.411.1409W}. 
\citet[][]{2012PASJ...64...51S} argues that SFEs in the bar of Maffei 2 can be underestimated by a factor of 0.5-0.8 by 
assuming a constant $\alpha_{\rm CO}$, which is linked to the presence of diffuse, non-bound, and non-optically thick molecular gas 
\citep[see also the analysis of NGC~1300 by][]{2020MNRAS.495.3840M}. 
We cannot entirely exclude this possibility based on our data, and we thus acknowledge that part of the dispersion of SFE along bars might 
be a consequence of the $\alpha_{\rm CO}$ uncertainty. 
However, even if $\alpha_{\rm CO}$ were lower, we can still conclude that SFE is not systematically low along the bars probed in this work. 
We also note that the central pointings of AGN hosts can have more uncertain $\alpha_{\rm CO}$ \citep[][]{1999ApJ...516..114P,2013ARA&A..51..207B}. 
Improving the $\alpha_{\rm CO}$ calibration is beyond the scope of this paper and is not within reach with the data presented here. 

Finally, in order to illustrate the consistency of our results with COLD GASS \citep[][]{2011MNRAS.415...61S}, 
in the right panel of Fig.~\ref{Fig_MS} we show the star formation efficiency versus specific star formation rate within bars. 
The grey area corresponds to the best fit ($\pm 1\,\sigma$) reported in COLD GASS based on 
global integrated measurements of the CO(1-0) line from IRAM-30m for 222 galaxies (with and without bars); 
the authors show the consistency in their reported H$_2$ depletion times with 
integrated and resolved measurements in HERACLES \citep[][]{2009AJ....137.4670L}.
Altogether, the slope of the relationship in this work coincides with the best fit reported by \citet[][]{2011MNRAS.415...61S}.
%
%
\subsection{Link between gravitational torques, shear, and star formation}\label{SF_SHEAR}
%
%
The shear ($\Gamma$) in a galaxy is expected to be correlated with the star formation rate \citep[][]{2005MNRAS.361L..20S}. The value of  
$\Gamma$ can be estimated from the slope of the rotation curves ($V$), 
\begin{equation}
\Gamma=-d {\rm ln} \Omega/d {\rm ln} r,
\end{equation} 
where $\Omega(r)=V/r$ is the angular velocity \citep[e.g.][]{2018MNRAS.477.1451F} 
at a given radius $r$, and $\Gamma=1$ in the flat regime. However,   
$\Gamma$ cannot be directly estimated for the galaxies in our sample as we lack ancillary high-resolution rotation curves or 2D 
integral field unit (IFU) or Fabry-P\'erot (FP) dynamics. However we have at our disposal the stellar component (bulge+disk) of the circular velocity, 
from which we here calculate the shear ($\Gamma_{\star}$). 
Specifically, \citet[][]{2016A&A...587A.160D} estimated the radial force field ($F_{\rm R}$) 
by applying the NIR-QB code \citep[][]{1999AJ....117..792S,2002MNRAS.337.1118L} to 3.6~$\mu$m S$^4$G images, 
and calculated the stellar contribution to the circular velocity as 
\begin{equation}
V_{\rm disk+bulge}(r)=\sqrt{\Upsilon_{3.6 \rm \mu m}\left<F_{\rm R}(r)\right> r}, 
\end{equation}
where $r$ is the galactocentric radius, $F_{\rm R}$ corresponds to the radial force obtained for $M/L$=1, 
and $\Upsilon_{3.6 \rm \mu m}=0.53$ is the mass-to-light ratio at 3.6~$\mu$m obtained by \citet[][]{2012AJ....143..139E}, 
which is assumed to be constant throughout the disk. Here we evaluate $\Gamma_{\star}$ in the same radial ranges where the IRAM-30m dish was pointed. 
The value of  $\Gamma_{\star}$ is a lower bound of $\Gamma$ as it does not include the contribution of dark matter and gas components to the potential well, 
nor the contribution of non-circular motions, and thus the conclusions in this section on the $\Gamma$-SFR connection are not definitive.

\begin{figure}
\centering
\includegraphics[width=0.49\textwidth]{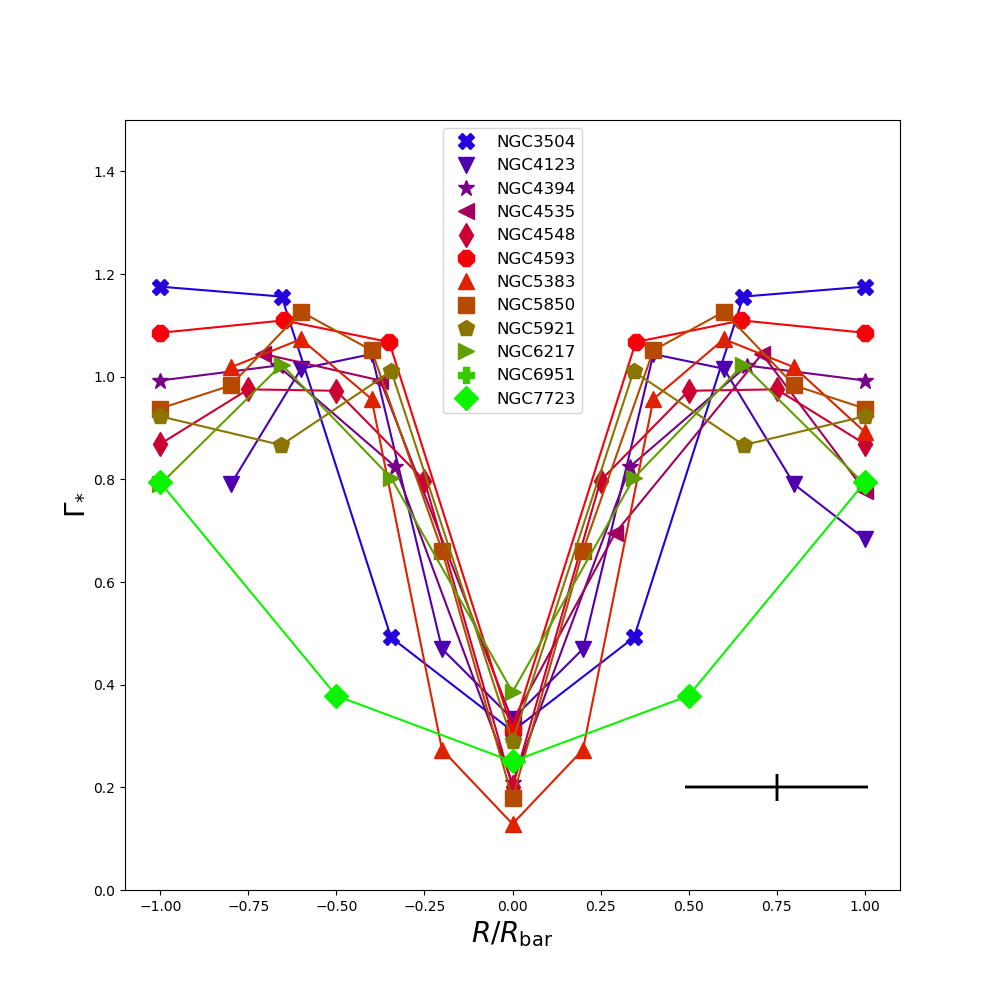}\\
\includegraphics[width=0.49\textwidth]{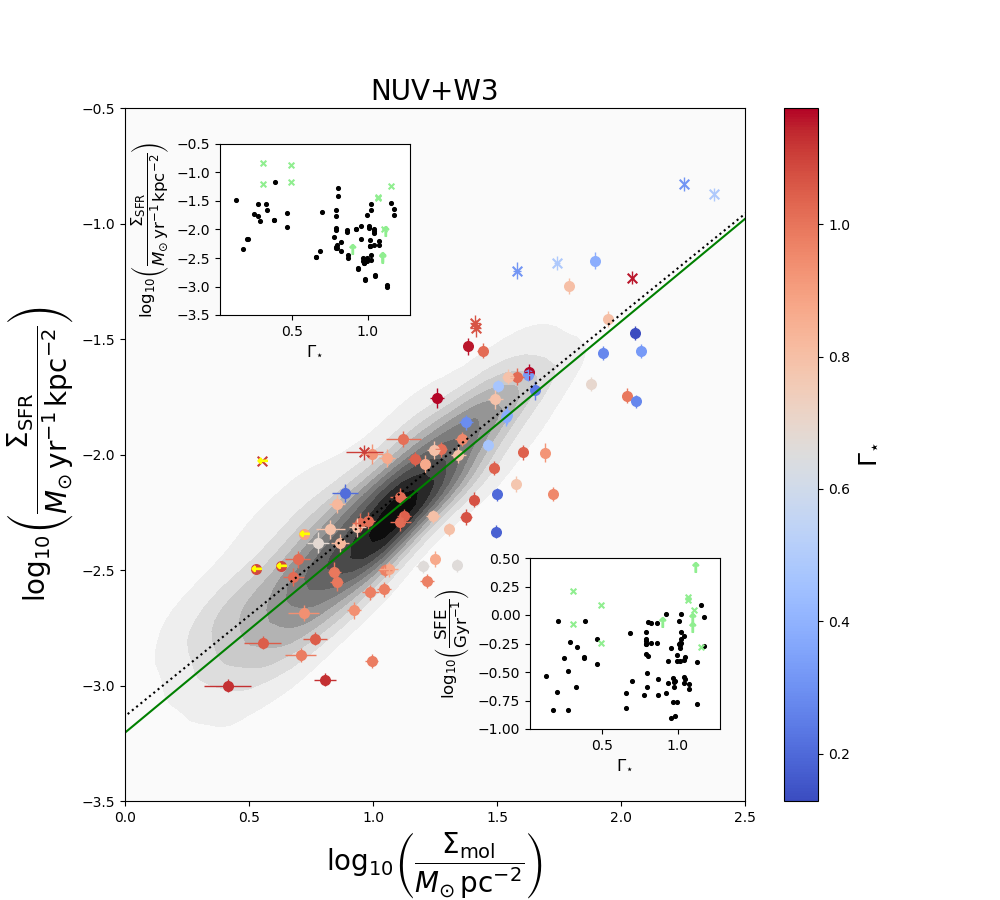}\\
\caption{
Shear and its connection to star formation.
\emph{Upper panel:} 
Shear evaluated from the stellar component of the circular velocity (see text) within each IRAM-30m aperture 
as a function of the radius along the bar major axis. 
The black cross indicates the mean bar range covered by the IRAM-30m pointings ($x$-axis, i.e. FWHM/$R_{\rm bar}$) 
and error on $\Gamma_{\star}$. 
The latter corresponds to a $13 \%$ typical uncertainty on the slope of the disk+bulge rotation curve \citep[][]{2016A&A...587A.160D}, 
which is associated with the uncertainty on the disk thickness determination.
\emph{Lower panel:} 
As in Fig.~\ref{BAR_VS_GLOBAL_2}, but colour-coded by the stellar contribution to the shear. 
The grey contours show the measurements from \citet[][]{2011ApJ...730L..13B}.
}
\label{PLOTS_ALLTOGETHER_GRAV_POT}
\end{figure}

As expected, the shear rate parameter is lowest in the central regions of the galaxies 
(Fig.~\ref{PLOTS_ALLTOGETHER_GRAV_POT}, upper panel) where the inner slope of the rotation curve is highest. 
We find that the SFR surface density is highest in the pointings with lowest shear rates 
(blue points in the lower panel of Fig.~\ref{PLOTS_ALLTOGETHER_GRAV_POT}). 
The implications of the connection between $\Gamma$ and SFR are addressed in Sect.~\ref{SFE_bars_discussion}. 
Having characterised the SFEs, SFRs, and molecular gas masses within bars, we next discuss the results in light of galaxy evolution.
%
%
%
%
\section{Discussion: Physical mechanisms that drive star formation in galactic bars}\label{discussion}
%
%
Bars in galaxies have been studied intensively during the last decades \citep[e.g.][and references therein]{2013seg..book..305A,2013seg..book....1K,2013seg..book..155B}, 
mainly with the motivation to investigate their role in the transport of material towards 
the central regions of galaxies \citep[e.g.][]{2004ARA&A..42..603K,2005ApJ...626..159B,2009A&A...495..775P,2015MNRAS.451..936S,2021MNRAS.500.2380Z}. 
The fraction of dense gas measured by  HCN emission, for example, can be enhanced in bars \citep[such as in NGC~2903;][]{2008A&A...491..703L}. 
In some cases the funnelled gas is known to enable  episodes of SF and nuclear activity 
\citep[e.g.][]{2000ApJ...529...93K,2011MNRAS.416.2182E,2012ApJS..198....4O,2013ApJ...776...50C,2015MNRAS.446.2468E,2015A&A...584A..88F,2016A&A...595A..63V,2017ApJ...848...87C,2020MNRAS.499.1406L} 
and can increase the central stellar mass \citep[e.g.][]{2016A&A...596A..84D}.

The relation between H$_2$ and SF within bars has so far received much less attention, and is not well understood yet. 
This is of particular interest given the extreme physical conditions of bars as places with strong shear, shocks, and non-circular 
velocities \citep[e.g.][]{1984MNRAS.207....9P,2004A&A...413...73Z} or significant magnetic field strengths \citep[e.g.][]{2002A&A...391...83B}. 
These make them ideal testing rooms for exploring the parameters that trigger or inhibit SF 
since there are no obvious differences between the H{\sc\,ii} regions in bars and those in the outer disks \citep[e.g.][]{1999A&A...346..769M}. 
We   traced the molecular gas (Sect.~\ref{MH2_CO}) and SF (Sect.~\ref{ap_photometry}) along the bars of 12 spiral galaxies (Sect.~\ref{sample_data}). 
These galaxies host bars with different degrees of SF, with no prior selection on gas content. 
We took a different approach in comparison to previous studies that either observed individual objects or a small set of galaxies 
\citep[see e.g.][]{2016PASJ...68...89M,2019PASJ...71S..13Y}, 
often reporting a low SFE in bars \citep[see][and references therein]{2020MNRAS.495.3840M}. 
Instead, we used a fairly large sample of 12 galaxies and show that, on average, the SFE is not remarkably diminished in bars 
relative to the typical values found in spiral galaxies (for further details, see Sect.~\ref{SFE_bars_discussion}).
%
%
\subsection{Spatial distribution of SF in bars}\label{SF_distribution}
%
%
Different spatial distributions for the recent SF in bars, as traced by the H$\alpha$ emission, have been reported 
\citep[e.g.][]{2002AJ....124.2581S,2008A&A...485....5Z,2019A&A...627A..26N,2020MNRAS.495.4158F,2020A&A...644A..38D}. 
When any H$\alpha$ is detected in the bar it seems to come  (1) from H{\sc\,ii} regions distributed along the bar, 
(2) from the nuclear or circumnuclear region with little or no emission from the bar, or (3) from the bar 
and the nuclear region (i.e. an intermediate case between 1 and 2) \citep[][]{1997A&A...326..449M,2007A&A...474...43V}. 
All of the strongly barred galaxies studied in this work have prominent circumnuclear SF relative to the underlying disk, 
and show different degrees of SF at the mid-part of the bar 
(Sect.~\ref{ap_photometry}). 

The different H$\alpha$ bar morphologies are interpreted as stages of an evolutionary sequence of the bar 
by \citet[][]{2007A&A...474...43V} based on the AMIGA project \citep[][]{2005A&A...436..443V} \citep[see also earlier work by][]{1997A&A...326..449M}. 
The sequence starts with SF distributed along the bar. 
The gas is then progressively depopulated from the bar through gas inflow towards the centre of the 
galaxy until H$\alpha$ emission is only seen in the nuclear or circumnuclear region. 
The later stage would occupy most of the bar life times, given the observed frequency. 
In this scenario the absence of SF in many barred galaxies is explained by the lack of gas. 
Furthermore, \citet[][]{2020A&A...644A..38D} show that strongly 
barred early-type spiral galaxies are characterised by a $\sim$0.5 mag  brighter central UV emission (i.e. $\gtrsim 50 \%$ larger $\Sigma_{\rm SFR}$), 
compared to their weakly barred counterparts. 
This scenario can indeed be explained by the efficiency of a strong bar potential at inducing central gas concentration and starbursts. 
Many of the barred galaxies in our sample have central $\Sigma_{\rm SFR}$ values that are substantially larger than 
those in \citet[][]{2011ApJ...730L..13B} (Fig.~\ref{PLOTS_ALLTOGETHER_B}) or \citet[][based on MaNGA DR13 datacubes]{2018MNRAS.474.2039E} 
(Fig.~\ref{Fig_MS}), and those correspond to places with high specific SFRs (Fig.~\ref{BAR_VS_GLOBAL_2}). 

In addition, \citet[][]{2020A&A...644A..38D} showed that, on average, inner-ringed galaxies are 
characterised by a UV and H$\alpha$ deficit in the central parts \citep[see also][and references therein]{2019A&A...627A..26N} 
in both barred and non-barred systems. 
This picture can be partially explained by gas being trapped at 
the 1/4 ultraharmonic resonance \citep[][]{1984MNRAS.209...93S,1996FCPh...17...95B}, where inner rings tend to form 
\citep[e.g.][]{1993RPPh...56..173S,2000A&A...362..465R,2019A&A...625A.146D}, 
slowing down its migration to the nuclear regions \citep[e.g.][]{2019A&A...627A..26N}. 
In this work the almost ubiquitous presence of inner pseudo-rings in our sample 
($> 80 \%$, see Sect.~\ref{sample_data} and Table \ref{sample_12_props}) 
does not systematically imply a lack of molecular gas and SF along the bars. 
This suggests that either gas is not irreversibly caught at the 1/4 ultraharmonic resonance 
or that  only closed, well-defined rings have a strong effect in trapping the gas particles, unlike pseudo-rings. 
The only galaxy in our sample with a closed ring, 
NGC~5850, is indeed characterised by a low-amplitude centrally peaked $\Sigma_{\rm mol}$ profile that 
drops along the bar showing very little SF (Fig.~\ref{PLOTS_ALLTOGETHER_A}), and gets enhanced again at one of the bar ends, at the ring radius.

Distinct distributions of massive SF within bars in galaxies of different morphological types 
have been reported by \citet[][]{2020A&A...644A..38D} using both 
stacking techniques (NUV and FUV) and visual classifications (in both H$\alpha$ and FUV). 
They also find differences in the statistical distributions of star-forming 
and quiescent bars as a function of physical properties, such as cold gas fraction and tangential-to-radial force ratios; 
for a segregation in the SFR-$M_{\star}$ plane in the Mapping Nearby Galaxies at 
APO \citep[MaNGA;][]{2015ApJ...798....7B}, see \citet[][]{2020MNRAS.495.4158F}. 
S0 galaxies tend to host SF exclusively in the circumnuclear regions, but are not studied in this work. 
Star-forming bars are most common among late-type galaxies ($T > 5$); \citet[][]{2020A&A...644A..38D} 
argue that this is a consequence of low shear given their lower central masses, 
which in turn may favour the gravitational collapse of GMCs and the formation of stars \citep[e.g.][]{2005MNRAS.361L..20S} 
(see next Sect.~\ref{SFE_bars_discussion}).

Star formation at the bar ends, but scant along the bar, is typical for early- and intermediate-type spirals \citep[][]{2020A&A...644A..38D}. 
This most likely results from the interplay of gas flow, shocks, and enhanced shear in centrally concentrated galaxies with large bar amplitudes. 
Even so, the authors identify quite a few galaxies ($\sim 1/3$) with $0 \le T \le 5$ and H{\sc\,ii} regions along bars, 
and a key parameter missing from their analysis is the fuel for SF. 
In the present paper, that purposely targets star-forming as well as quiescent bars hosted by S0/a-Sc galaxies, 
we show that the degree of SF in bars correlates with the mass surface density of molecular gas, 
in agreement with the KS law reported for disks as a whole. 
We detect H$_2$ along the bars of all probed galaxies, regardless of their $T$-type, 
very clearly in the circumnuclear regions (Fig.~\ref{PLOTS_ALLTOGETHER_A}) and at the bar ends in some objects. The value of 
$\Sigma_{\rm mol}$ varies by almost 2 orders of magnitude between the probed galaxies.

Bar stacks by \citet[][]{2020A&A...644A..38D} show that the average FUV emission in the mid- and outer part of strong bars hosted by more than 
$100$ S0/a-Sc galaxies is $\sim 26.3$ mag arcsec$^{-2}$, which corresponds to 
$\Sigma_{\rm SFR} \approx 10^{-3.2} \, M_{\odot} \, {\rm yr}^{-1} \, {\rm kpc}^{-2}$ (see their Fig.~3). 
This value is consistent with the mean $\Sigma_{\rm SFR}$ of the 12 galaxies probed in this work, as shown in 
the upper panel of Fig.~\ref{COMPARISON_TRACERS_GALS_ALL_2}; in the case of NGC~4548 and NGC~5850 
$\Sigma_{\rm SFR}$, as traced from FUV, is 0.5-1 orders of magnitude lower. 
The quiescent bars in our sample (NGC~4394, NGC~4548, and NGC~5850), all having 
$\Sigma_{\rm SFR}\lesssim 10^{-2.25} \, M_{\odot}$ yr$^{-1}$ kpc$^{-2}$ when NUV is combined with W4 or W3, 
show low $\Sigma_{\rm mol}$. Interestingly, these three bars clearly cluster below the main sequence (Fig.~\ref{Fig_MS}) and 
are also characterised by low specific SFRs (see lower panel of Fig.~\ref{BAR_VS_GLOBAL_2}). Their atomic gas content is likely to be low as well 
\citep[e.g. this is clear from the inspection of the VIVA H{\sc\,i} map for NGC$\,4548$ in][]{2009AJ....138.1741C}, 
but the analysis of H{\sc\,i} is beyond the scope of this paper. Differences in SFEs between galaxies are discussed next.
%
%
\subsection{The SFE is not always diminished in bars}\label{SFE_bars_discussion}
%
%
In general, the SFEs in the bars of our galaxy sample (Fig.~\ref{PLOTS_ALLTOGETHER_B}) are 
similar to those reported by \citet[][]{2011ApJ...730L..13B} (0.43 Gyr$^{-1}$) 
from the analysis of 30 nearby disks at a resolution of 1 kpc \citep[see also][]{2008AJ....136.2846B}. 
Of the three quiescent bars that we resolve in this work, with 
$\Sigma_{\rm SFR}\lessapprox 10^{-2.25} \, M_{\odot} \, {\rm yr}^{-1} \, {\rm kpc}^{-2}$, 
NGC~4548 and NGC~5850 have a factor of $\sim$2 lower SFEs, 
while NGC~4394 (<SFE>$=0.64 \pm 0.07$ Gyr$^{-1}$) does not. 
On the other hand, NGC~4535, which hosts SF along the bar, has a fairly low SFE (<SFE>$=0.23 \pm 0.02$ Gyr$^{-1}$). 

Combining data from the Nobeyama 45m and ALMA 12m (with no sensitivity to diffuse gas), 
\citet[][]{2020MNRAS.495.3840M} study the star formation activity in NGC~1300 and report a lower SFE at the bar region than in 
the bar-end or spiral arms (see their Fig.~A.2). They find values of SFE at the bar region that are  
much lower than those in \citet[][]{2011ApJ...730L..13B}. 
They argue that molecular gas exists in strong bars with no clear H{\sc\,ii} regions \citep[][]{2018PASJ...70...37M}. 
Likewise, based on single-dish CO observations of NGC~2903 and NGC~4303, 
\citet[][]{2016PASJ...68...89M} and \citet[][]{2019PASJ...71S..13Y} conclude that the 
H$_{2}$ gas density in the bar regions is lower than that in the arms. 
Here, SFEs in the central regions are not systematically different than those in outer parts of the bar and at the beginning of the arms or rings.
\citet[][]{2020MNRAS.495.3840M} argue that the presence of a large amount of diffuse molecular gas across bars makes the SFE low in appearance. 
However, they conclude that the bar SFE remains low even when diffuse gas is excluded, and thus other mechanisms might explain this trend. 
In addition, \citet[][]{2020MNRAS.495.3840M} find a tight correlation between the SFE and $R_{21}=I_{\rm CO(2-1)}/I_{\rm CO(1-0)}$. 
$R_{21}$ is a proxy of the gas excitation \citep[e.g.][]{2012ApJ...761...41K}, 
which \citet[][]{2020MNRAS.495.3840M} report to be low in the bar region of NGC~1300. 
We do not confirm these trends in the bars in our sample (Fig.~\ref{plots_R21}; see Appendix~\ref{CO_2_1} for further discussion).

An explanation for the presence or absence of SF are the local conditions of the gas in bars. 
Cloud-cloud collisions, inner-cloud turbulence, or the bar tidal field, 
can tear apart GMCs \citep[e.g.][]{2000A&A...363...93H,2019PASJ...71S..13Y} 
and enhance the fraction of diffuse gas. In addition, as discussed by \citet[][]{1997AJ....114..965R} and \citet[][]{2002AJ....124.2581S}, 
GMCs can be disrupted, or their formation inhibited, by diverging streamlines on the trailing side of the bar dust lanes. 
\citet[][]{2000ApJ...532..221S}  argue, however,  that this effect can be neutralised by self-gravity. 
Recent models by \citet[][]{2018A&A...609A..60K} show that SF is less efficient in the turbulent interstellar medium, 
and point to the effect of bars quenching SF without needing to deplete the gas. 
One possible explanation for the low SFE in their galaxy (or in galaxies like NGC~4548 and NGC~5850 here) is indeed that GMCs 
in bars are gravitationally unbound \citep[e.g.][]{2013ApJ...779...45M,2013MNRAS.429.2175N}, 
but we argue that this is not the general picture for all strong bars. 

Evidence exists for the inhibition of SF based on CO observations \citep[e.g.][]{1998A&A...337..671R}. 
Fluid dynamic simulations of bars \citep[e.g.][]{1992MNRAS.259..345A,2000ASPC..221..243A} predict that the highest gas density 
loci are also the loci of strong shocks and high shear within strong bars. 
Observations of H$\alpha$ velocity gradients suggest that shear makes SF drop, whereas shocks enhance it in general \citep[][]{2004A&A...413...73Z}. 
In Sect.~\ref{SF_SHEAR} we showed that the pointings having the highest $\Sigma_{\rm SFR}$ are indeed those with  
the lowest shear rate parameters, calculated from the stellar contribution to the circular velocity \citep[from ][]{2016A&A...587A.160D}. 
In this paper we use a lower limit of the true shear since we neglect the contribution of dark matter, gas, and non-circular motions.

The occurrence of SF at the bar ends 
is witnessed in simulations as well \citep[e.g.][]{2015MNRAS.454.3299R}. It may result from a combination of kiloparsec-scale dynamics 
(gas flows) and parsec-scale turbulence and clouds collisions, under a low shear, as discussed in \citet[][]{2020MNRAS.495.4158F}. 
The same physics explain why H{\sc\,ii} regions are preferentially located on the leading side of the bars 
\citep[e.g.][]{1992MNRAS.259..345A,2000A&A...354..823R,2002AJ....124.2581S,2010A&A...521A...8P,2019A&A...627A..26N,2020A&A...644A..38D,2020MNRAS.495.4158F}: 
it is likely a consequence of shear and turbulence forces inhibiting SF everywhere but 
on the bar leading side \citep[][]{2015MNRAS.446.2468E,2015MNRAS.454.3299R,2020MNRAS.495.4158F} and at the bar ends. 

A possible connection between bar strength and the occurrence of SF has been discussed 
in the past \citep[e.g.][]{2002ApJ...570L..55J}, given the expectation that weaker bars have weaker shocks and shear \citep[][]{1992MNRAS.259..345A}. 
However, to our knowledge this is   not supported by any recent observational work with large unbiased samples. 
In some galaxies H{\sc\,ii} regions are indeed found in bars where strong shocks and shear are also detected 
\citep[e.g.][]{1997A&A...326..449M,2002AJ....124.2581S,2004A&A...413...73Z,2008A&A...485....5Z}. 
We searched for the existence of correlations of the SFE with various structural parameters, 
such as bar-to-total and bulge-to-total flux ratios \citep[from][]{2014ApJ...782...64K,2015ApJS..219....4S}, 
bar strength \citep[e.g. normalised $m=2$ Fourier amplitude or ellipticity, from][]{2016A&A...587A.160D}, 
or even tangential-to-radial forces evaluated locally \citep[e.g.][]{2019A&A...631A..94D} 
within the area covered by the IRAM-30m pointings; we found no evidence of any correlations (plots  not shown here).
This implies that either  the range of bar parameters covered is too narrow (our sample only contains strong bars, Fig.~\ref{plot_pointing_sample}) 
or that other, more fundamental, global dynamical and/or neutral gas properties can explain the differences in SFE. 

We find an interesting segregation as a function of the total stellar mass of the host galaxy (Fig.~\ref{BAR_VS_GLOBAL_2}): 
galaxies with highest $M_{\star}$ tend to host bars with low SFE. 
Even though we have used the largest sample to date for the stated scientific goals, 
the robustness of the analysis is still limited by sample size. 
Thus, we encourage this result to be verified elsewhere with larger surveys, 
such as EDGE-CALIFA \citep[][]{2017ApJ...846..159B,2021MNRAS.503.1615S} 
or Physics at High Angular Resolution in Nearby GalaxieS 
\citep[PHANGS; e.g.][]{2021MNRAS.502.1218R,2021MNRAS.501.3621A,2021arXiv210102855L}. 
The analysis of SF scaling relations at $\sim$100 pc from PHANGS was released by \citet[][]{2021arXiv210409536P} 
during the review process of this paper and will be analysed elsewhere.

\citet[][]{2010A&A...510A..64V} conclude that for M$\,$33 the stellar disk, which is gravitationally 
dominant with respect to the gaseous disk, plays a major role in driving the SFR, and the usual phase shift between gas and 
stellar components may also play a role to explain the aforementioned tendency. 
Whether the segregation of the KS law as a function of $M_{\star}$ hints at the SF suppression by 
large-amplitude bars \citep[typically hosted by massive galaxies;][]{2016A&A...596A..84D} or points to more fundamental dynamical differences 
remains to be confirmed. We note that \citet[][see their Sect.~4.1.2]{2013AJ....146...19L} already discussed a weak correlation relating 
gas depletion times to galaxy masses and average surface densities 
\citep[see their Fig.~7; see also][]{2011MNRAS.415...61S,2012ApJ...758...73S}.

To sum up, H$\alpha$ and far- and near-UV maps of disk galaxies show a massive SF along some bars, 
whereas there is a lack of SF in others. Among early- and intermediate-type galaxies, 
$\sim 2/3$ lack H{\sc\,ii} regions along the bar, but $\sim 1/3$ do not \citep[][]{2020A&A...644A..38D}. 
We argue, for S0/a-Sc galaxies, that the degree of SF in bars is not primarily controlled by the SFE. 
Similarly, \citet[][]{2020A&A...644A..79G} recently presented a multi-$\lambda$ analysis of bars in 
NGC~3351, NGC~4579, and NGC~4725, and concludes that bars that are devoid of SF (as traced from UV) 
are also devoid of molecular and neutral hydrogen. 
Our observations are not easy to reconcile with the theoretical expectation of a lower SFE 
in the central parts of gas-rich barred galaxies,  compared to their unbarred counterparts, 
or a flattening of the slope of the KS relation in the bar region \citep[][]{2018A&A...609A..60K}; 
we note,  however, that our sample is not limited to gas-rich galaxies (Sect.~\ref{sample_data}).
We conclude that gas depletion times are not always high along strong bars, either in the central, mid-, or end parts. 
Strong bars with lowest SFEs are typically hosted by massive galaxies ($M_{\star} \gtrsim 10^{10.7} M_{\odot}$). 
%
%
\subsection{Environmental versus internal quenching}
%
%
We have shown that NGC~4394, NGC~4548, and NGC~5850 host bars with remarkably low values of $\Sigma_{\rm SFR}$ and sSFR. 
What physical mechanisms, other than those discussed so far, could have caused the quenching of SF in their bars?

Given their high $\Sigma_{3}^{A}$ values (lower right panel of Fig.~\ref{BAR_VS_GLOBAL_2}), 
environmental effects are likely responsible for the low level of star formation in these systems. 
NGC~5850 is perturbed, as shown by \citet[][]{1998AJ....115...80H}, due to the encounter with the 
nearby massive elliptical NGC~5846. Using interferometric IRAM observations, \citet[][]{2000ASPC..197...61L} 
report CO emission from the nuclear ring, whose distribution is asymmetric due to the interaction. 
Likewise, NGC~4394 and NGC~4548 belong to the Virgo cluster. 
NGC~4548 shows signatures of a past interaction, such as an inner polar ring \citep[][]{2002AstL...28..207S}, 
and is perturbed and warped \citep[][]{1999A&A...349..411V}.

On the other hand, other internal mechanisms, besides those discussed in Sect.~\ref{SFE_bars_discussion}, 
may account for the low SFE in NGC~4548 or NGC~5850. 
According to numerical models by \citet[][]{2009ApJ...707..250M}, given that SF occurs in gravitationally unstable gas disks, 
massive bulges can morphologically quench SF in galaxies. 
The stellar mass of the bulges hosted by NGC~4548 and NGC~5850 are $10^{9.7}\,M_{\odot}$ and $10^{10.1}\,M_{\odot}$, respectively. 
This means that they are indeed two of the galaxies with the largest bulges in our sample 
(see the lower left panel of Figs. \ref{BAR_VS_GLOBAL_2} and \ref{BAR_VS_GLOBAL_2_W4}); 
however, we have checked that no correlation exists between the residuals of the KS law 
and bulge-to-total flux ratios. We conclude that both morphological and environmental quenching are likely 
responsible for the low $\Sigma_{\rm SFR}$ and SFE in NGC~4548 and NGC~5850.
%
%
\section{Summary and conclusions}\label{summarysection}
%
%
The principal aim of this work was to study the processes that enhance or inhibit star formation within strong stellar bars of spiral galaxies. 
Bars are excellent laboratories for conducting exploratory research on SF as they are characterised by extreme physical conditions, 
such as strong shear, shocks, non-circular motions, and magnetic fields. 
We  selected a sample of 12 nearby, practically face-on, strongly barred galaxies with total stellar masses 
${\rm log}_{10}(M_{\star}/M_{\odot})\in [10.2,11]$, Hubble types $T\in [0,5]$, and different degrees of star formation in bars. 

In order to trace the distribution of molecular gas within bars we obtained 
CO(1-0) spectra (21.5$\arcsec$ FWHM) and CO(2-1) spectra (10.75$\arcsec$ FWHM) with the IRAM-30m radiotelescope, 
performing several pointings along their major axes. At the distance to our galaxies ($\lesssim 40$ Mpc), 
the resolution elements associated with the IRAM pointings were in the range $1.7-3.9$ kpc for CO(1-0) and $0.85-1.95$ kpc for CO(2-1). 
The CO spectra are made available in electronic form at the CDS linked to this publication. CO emission was detected in the full set of galaxies. 
We computed the molecular gas surface density ($\Sigma_{\rm mol}$) from the velocity-integrated line intensity of the CO(1-0) spectra. 
A remarkable central enhancement of $\Sigma_{\rm mol}$ was observed in the whole sample, 
while four galaxies show secondary peaks at the bar ends. 

To trace the star formation (SF), we used \emph{GALEX} near- and far-UV and \emph{WISE} 12 and 22~$\mu$m photometry. 
We performed aperture photometry centred on the CO pointings and computed the SFRs. 
We find similar distributions of the surface density of SF rate ($\Sigma_{\rm SFR}$) in bars; 
the 12 galaxies host central starbursts, which is not surprising as strong bar torques are known to actively funnel 
the gas towards the central regions \citep[e.g.][]{1993RPPh...56..173S}. 
Three quiescent bars in our sample (NGC~4394, NGC~4548, and NGC~5850) 
have $\Sigma_{\rm SFR}\lesssim 10^{-2.25} \, M_{\odot}$ yr$^{-1}$ kpc$^{-2}$ 
as quantified from the combination of NUV and WISE 3 and WISE 4 emission along the bar major axis; 
interestingly, these are the galaxies in the densest environments (highest surface density to the third-nearest neighbour galaxy).

$\Sigma_{\rm SFR}$ and $\Sigma_{\rm mol}$ are tightly correlated within bars. 
This means that  the SF activity in bars of strongly barred galaxies is primarily controlled by the content of molecular gas. 
We   calculated the star formation efficiencies (SFE = $\Sigma_{\rm SFR} / \Sigma_{\rm mol}$). The
SFEs within bars vary between different galaxies by up to an order of magnitude, 
spanning $0.1\lessapprox$ SFE/Gyr$^{-1}$ $\lessapprox 1.8$ when measured from NUV and WISE 3 combined, 
but are roughly constant along the bar major axes. 

The derived SFRs and SFEs follow the typical values found in disks of spirals galaxies \citep[e.g.][]{2011ApJ...730L..13B}. 
We do not find remarkable differences in the KS law in 
the central regions of barred and non-barred galaxies, as predicted from numerical models by \citet[][]{2018A&A...609A..60K} 
for gas-rich galaxies. 
However, in some galaxies the SFE within bars can be a factor of $\sim 2$ lower than the average in spirals. 
Interestingly, the two with lowest SFE and $\Sigma_{\rm SFR}$ 
(NGC~4548 and NGC~5850) are also those hosting massive stellar bulges, while having signatures of recent interactions; 
both environmental and internal quenching are plausible explanations for their low level of star formation.

In general, whether or not a strong bar hosts star-forming regions is not only determined by the degree of SFE, as has been previously claimed, 
and thus other physical factors must also come into play. 
Interestingly, the higher the total stellar mass of the galaxies, the lower the SFE within their bars. 
The latter needs to be verified with larger samples that include faint systems as well as weak bars 
as it might link the SF suppression to large-amplitude bars \citep[typically hosted by massive galaxies, e.g.][]{2016A&A...596A..84D}.

The novelty of this work is that it presents a blind study of the molecular gas in bars for galaxies 
with different star formation distributions and properties, with the aid of multi-$\lambda$ data. 
In most previous works the emphasis was placed on  objects for which molecular gas was present, 
or objects harbouring a lot of star formation in their bars. We have shown that the star formation efficiency is not uniformly 
inhibited in strong bars using a fairly large sample of 12 galaxies.
%
%
\begin{acknowledgements}
%
%
%
We thank the anonymous referee for a constructive and detailed report. 
This project has received funding from the European Union's Horizon 2020 research and innovation programme 
under the Marie Sk$\l$odowska-Curie grant agreement No 893673. 
We acknowledge financial support from the European Union's Horizon 2020 research and innovation programme under 
Marie Sk$\l$odowska-Curie grant agreement No 721463 to the SUNDIAL ITN network, 
from the State Research Agency (AEI-MCINN) of the Spanish Ministry of Science and Innovation 
under the grant "The structure and evolution of galaxies and their central regions" 
with reference PID2019-105602GB-I00/10.13039/501100011033, and from IAC project P/300724, 
financed by the Ministry of Science and Innovation, through the State Budget and by the 
Canary Islands Department of Economy, Knowledge and Employment, through the Regional Budget of the Autonomous Community. 
SDG acknowledges support from the Spanish Public Employment Service (SEPE). 
Furthermore, we acknowledge support by the research project  AYA2017-84897-P from the Spanish Ministerio 
de Econom\'ia y Competitividad, from the European Regional Development Funds (FEDER) and the Junta de Andaluc\'ia (Spain) grants FQM108. 
DE acknowledges support from a Beatriz Galindo senior fellowship (BG20/00224) from the Ministry of Science and Innovation. 
LVM acknowledges financial support from the grants AYA2015-65973-C3-1-R and RTI2018-096228- B-C31 (MINECO/FEDER, UE), 
as well as from the State Agency for Research of the Spanish MCIU through the Center of Excellence Severo Ochoa award to the 
Instituto de Astrof\'isica de Andaluc\'ia (SEV-2017-0709). This research makes use of python (\href{http://www.python.org}{http://www.python.org}), 
Matplotlib \citep[][]{Hunter2007}, and Astropy \citep[][]{2013A&A...558A..33A,2018AJ....156..123A}. 
We acknowledge the usage of the HyperLeda database (\href{http://leda.univ-lyon1.fr}{http://leda.univ-lyon1.fr}). 
We thank Alexandre Bouquin for providing us with the \emph{GALEX} FUV and NUV images used in this work. 
We thank St\'ephane Courteau, Estrella Florido, Ra\'ul Infante-Sainz, Tom Jarrett, Johan H. Knapen, Heikki Salo, 
and Miguel Querejeta for useful discussions. We thank S\'ebastien Comer\'on and Facundo D. Moyano for valuable comments on the manuscript.
%
%
{\it Facilities}: GALEX, WISE, \emph{Spitzer} (IRAC).
%
%
\end{acknowledgements}
%
%
\bibliographystyle{aa}
\bibliography{bibliography}
\clearpage
%
%
\begin{appendix}
%
%
%
%
\section{IRAM-30m pointings and CO spectra}\label{app_30m_all}
\begin{figure}
\centering
\includegraphics[width=0.45\textwidth]{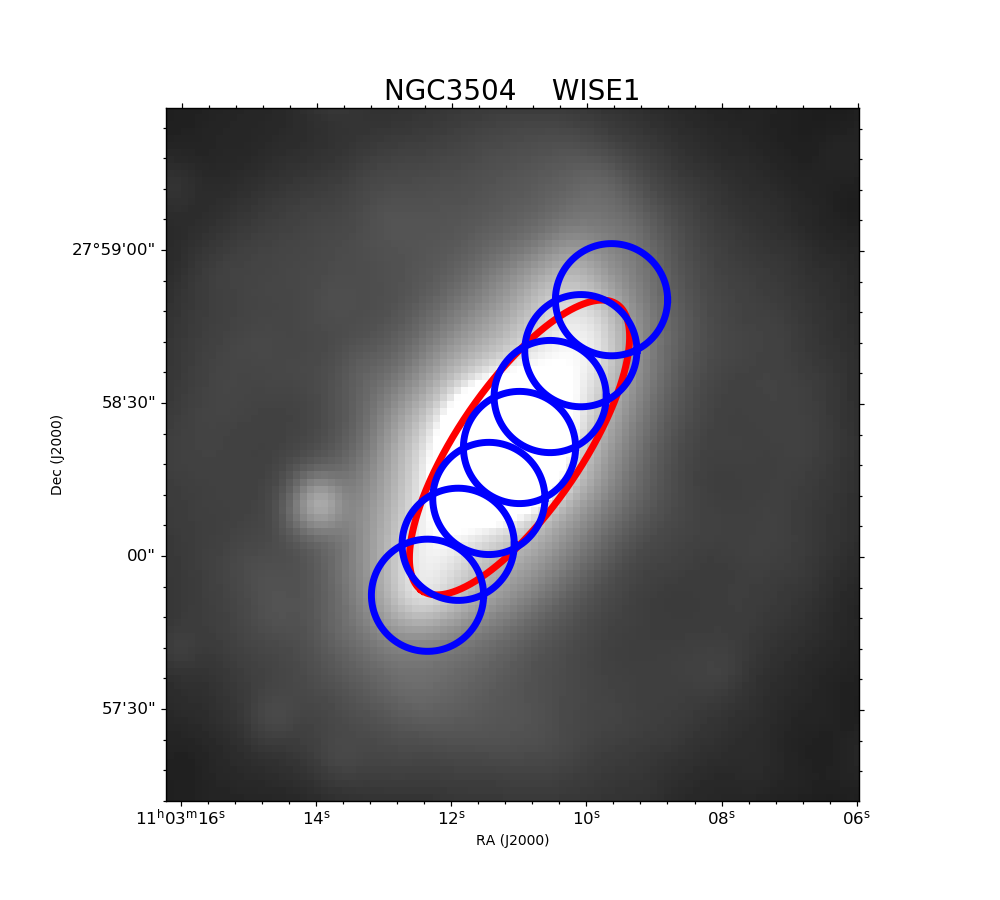}\\[-3ex]
\includegraphics[width=0.45\textwidth]{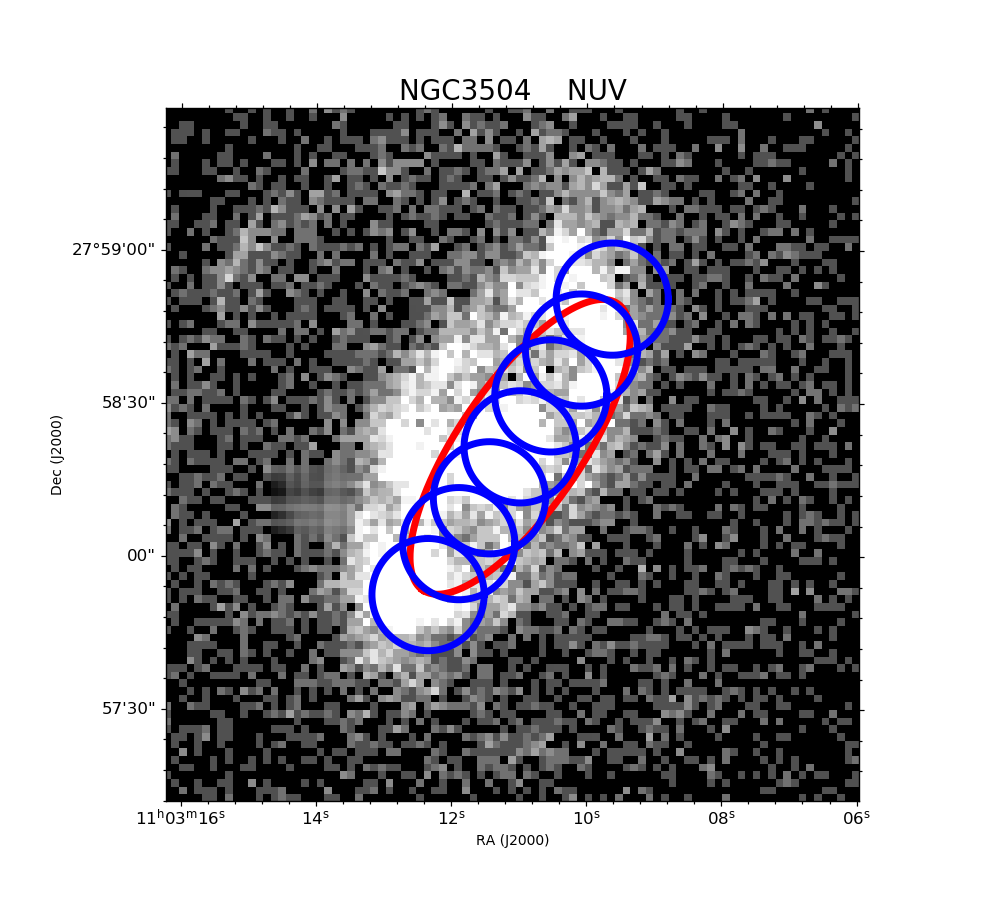}\\[-3ex]
\includegraphics[width=0.49\textwidth]{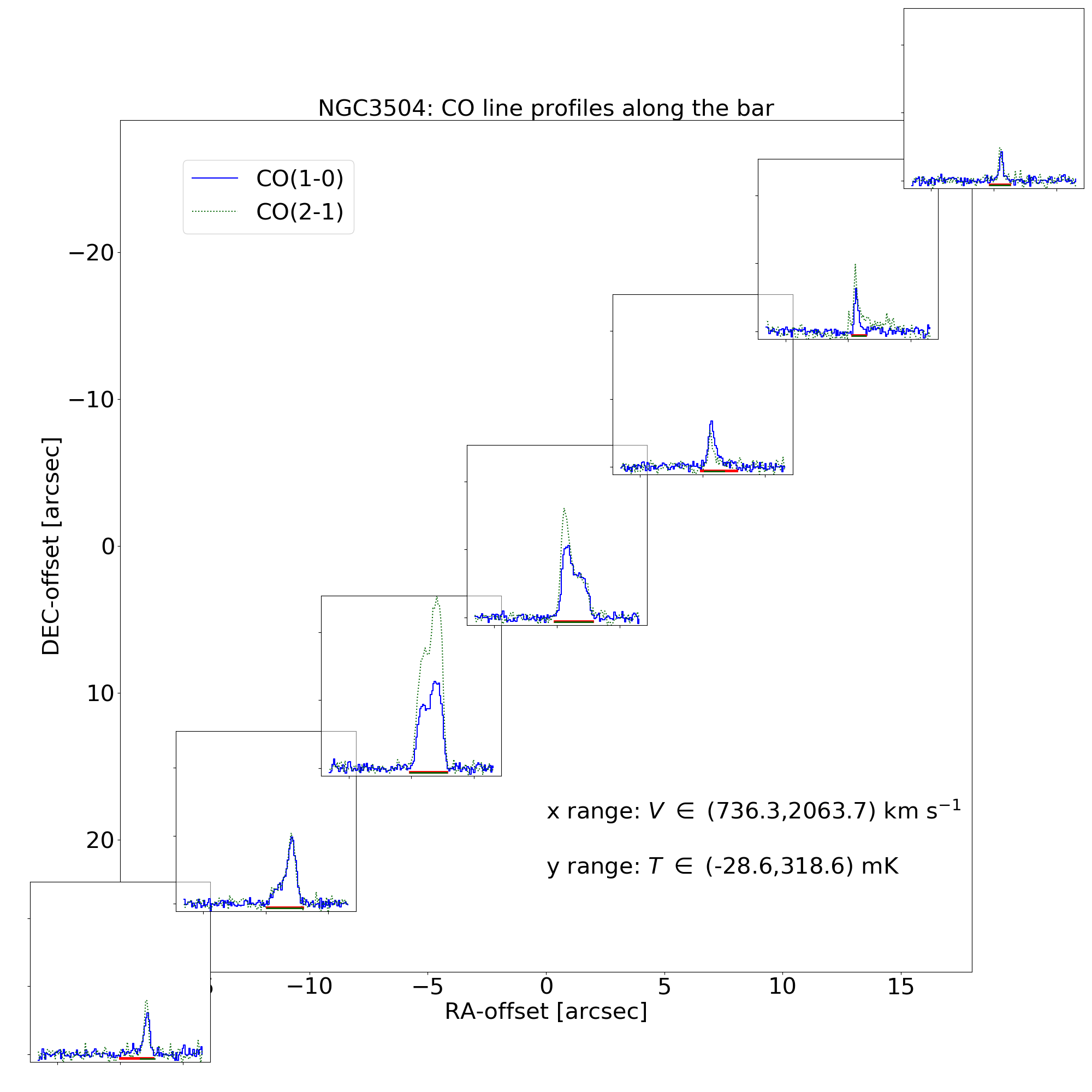}
\caption{
As in Fig.~\ref{plot_pointing_example}, but for NGC~3504.
}
\label{plots_spectra_app_0}
\end{figure}

\begin{figure}
\centering
\includegraphics[width=0.45\textwidth]{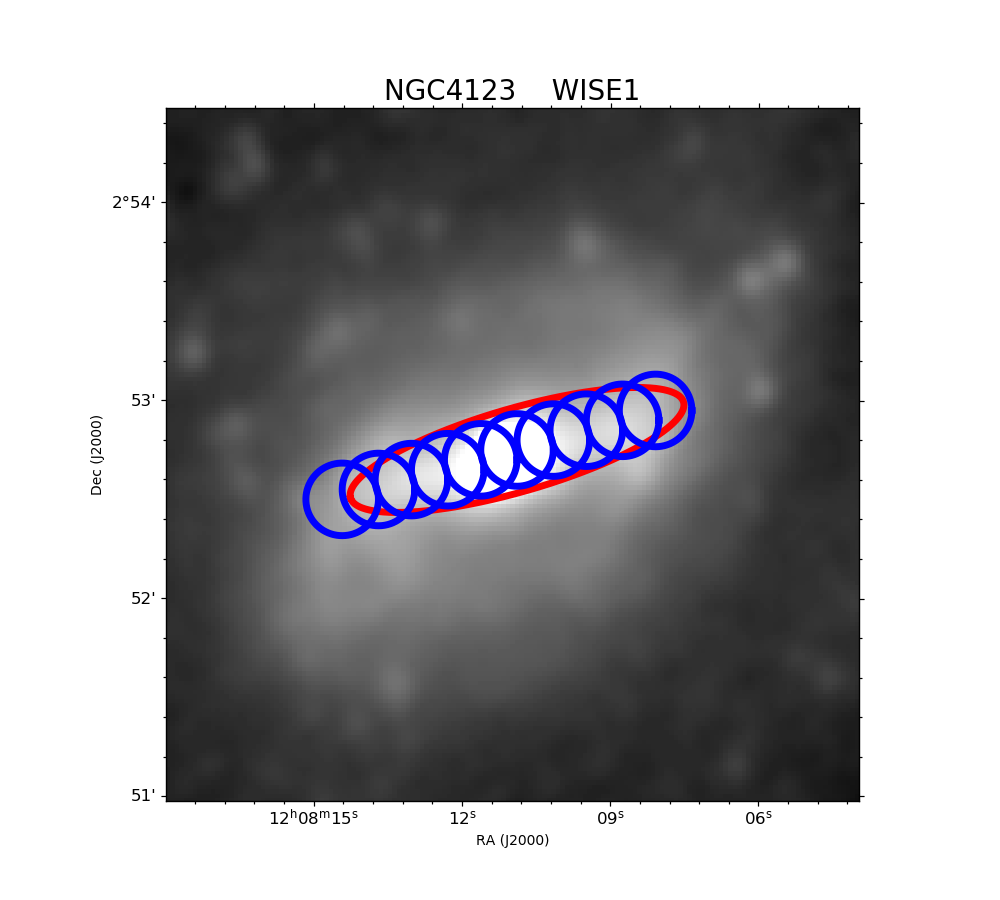}\\[-3ex]
\includegraphics[width=0.45\textwidth]{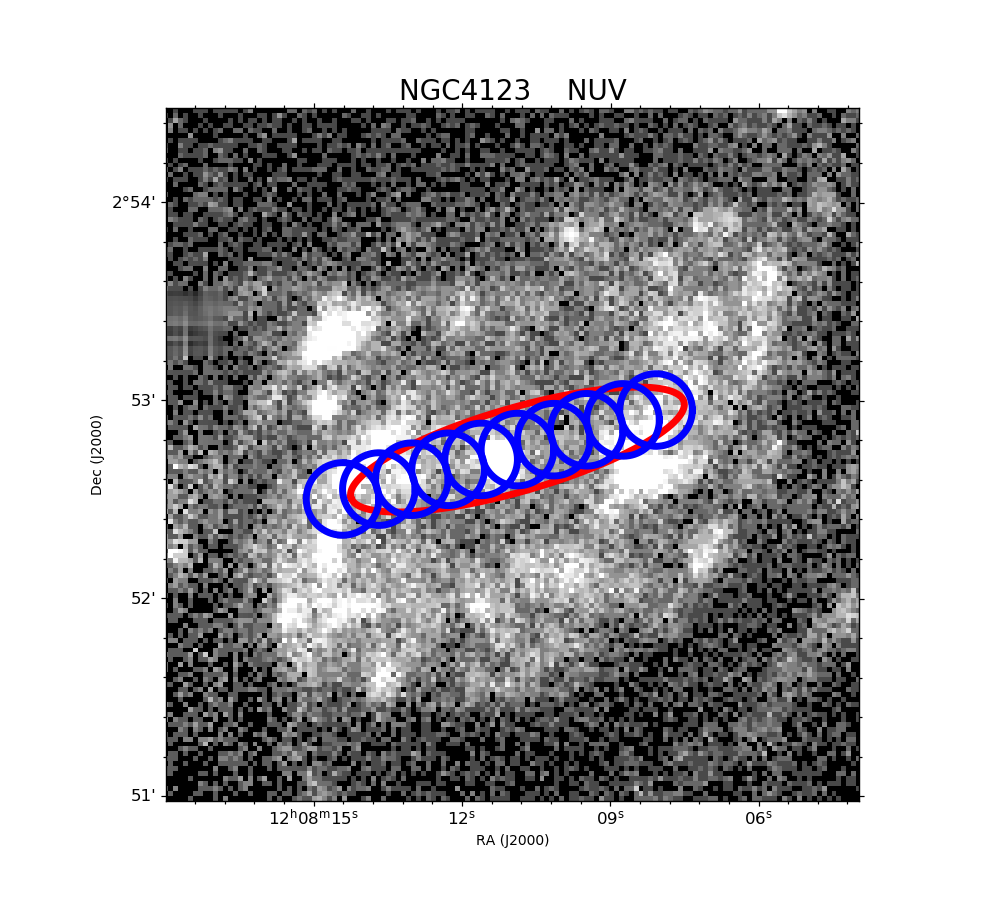}\\[-3ex]
\includegraphics[width=0.49\textwidth]{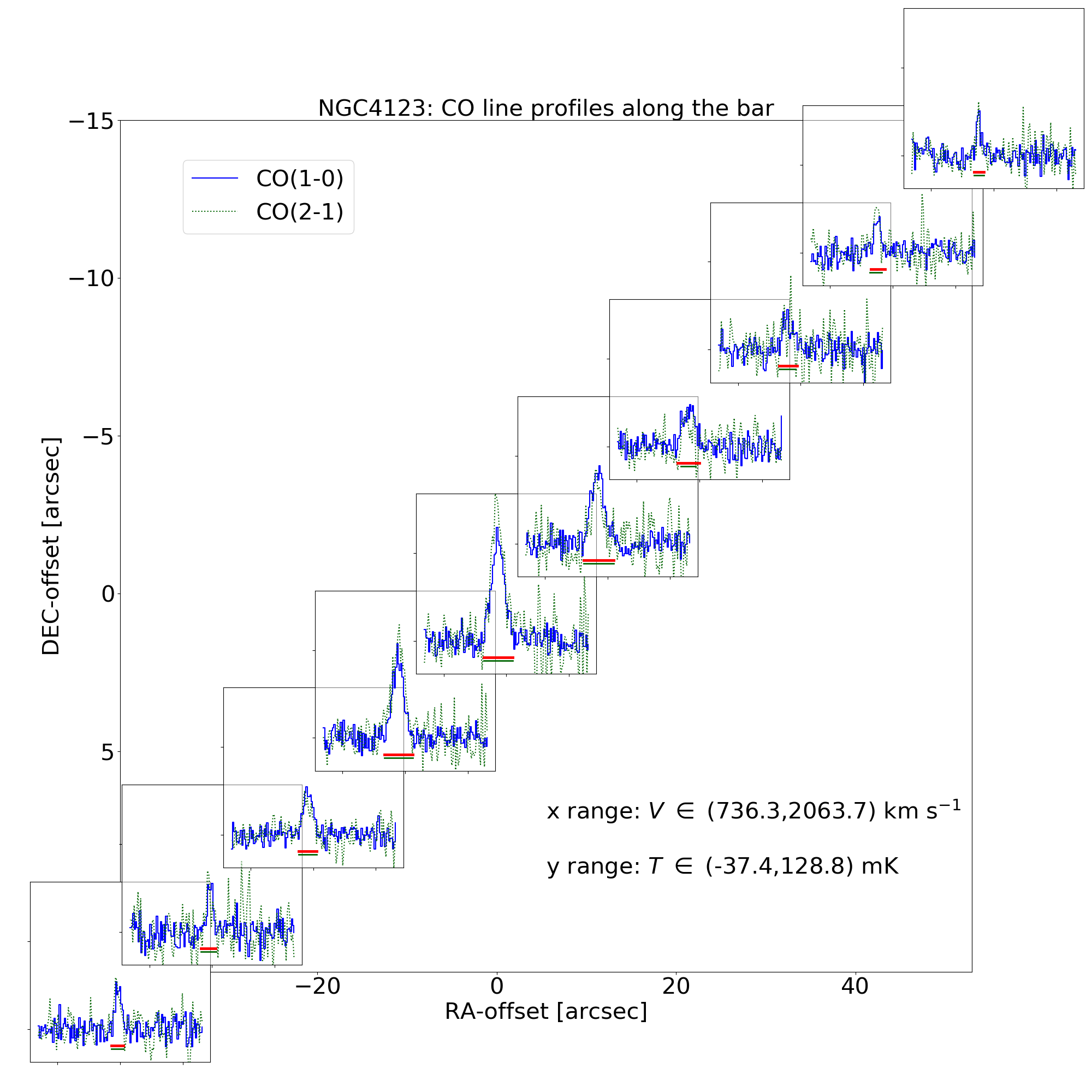}
\caption{
As in Fig.~\ref{plot_pointing_example}, but for NGC~4123.
}
\label{plots_spectra_app_1}
\end{figure}

\begin{figure}
\centering
\includegraphics[width=0.45\textwidth]{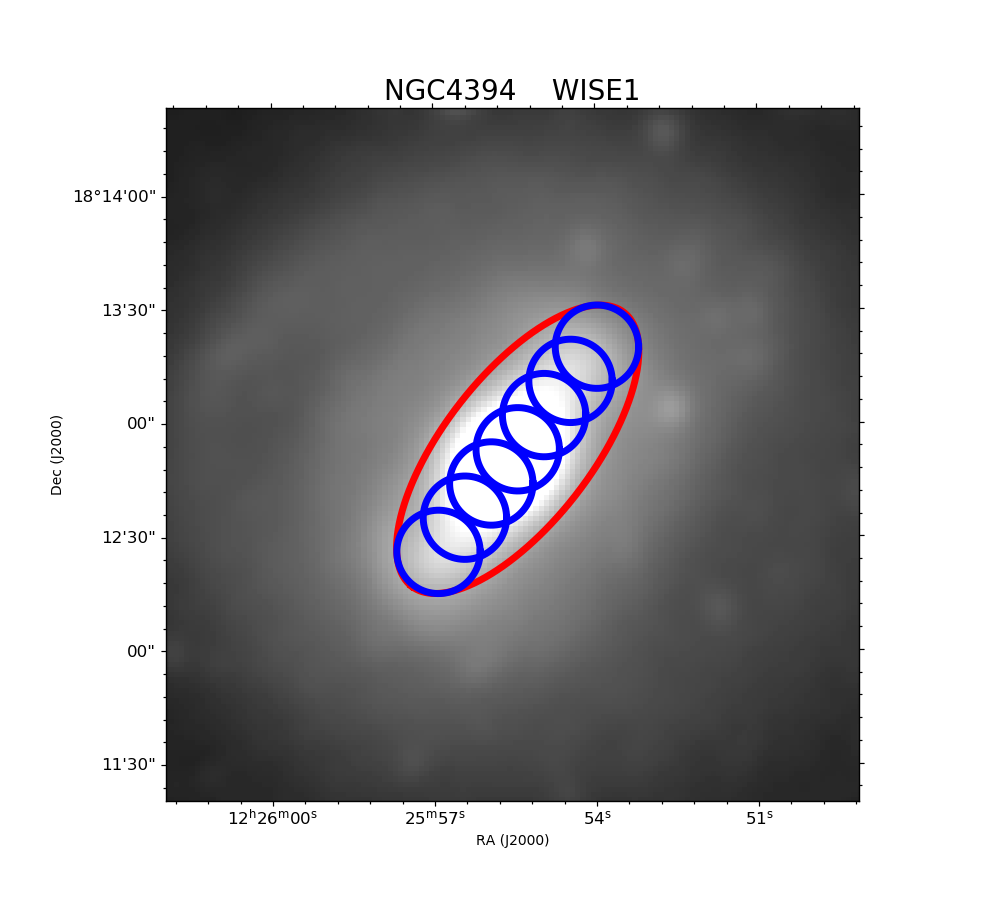}\\[-3ex]
\includegraphics[width=0.45\textwidth]{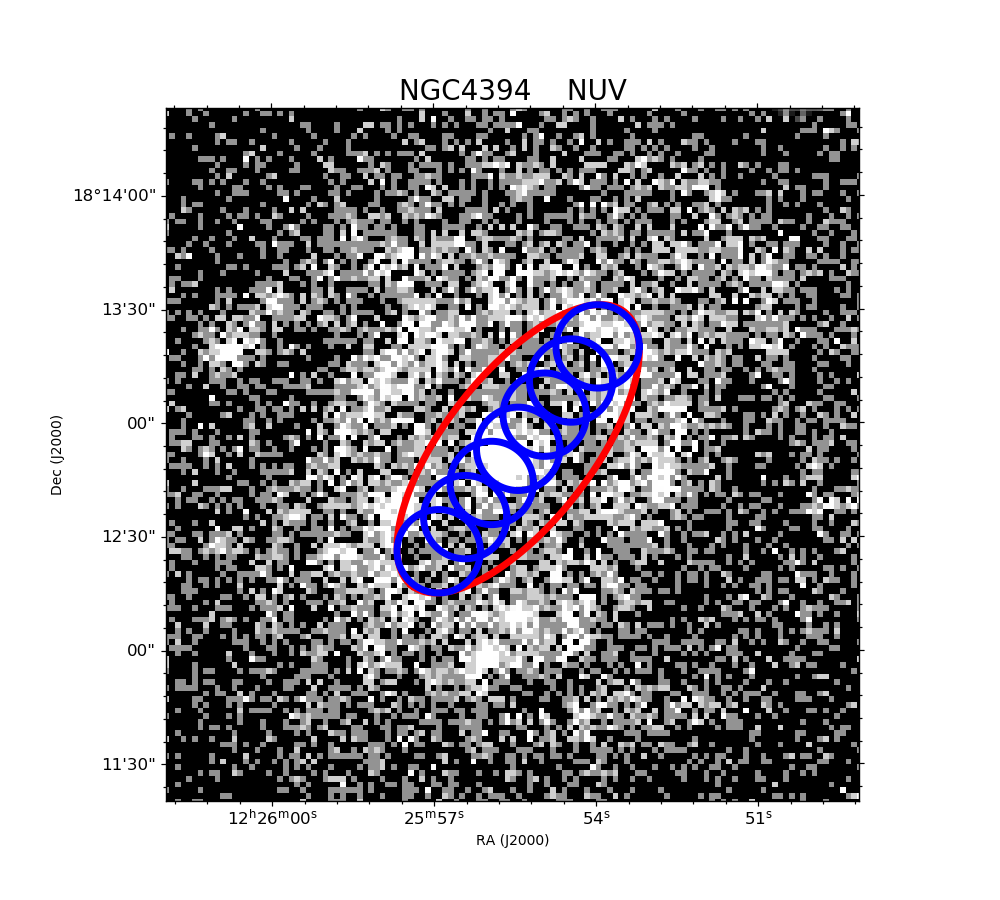}\\[-3ex]
\includegraphics[width=0.49\textwidth]{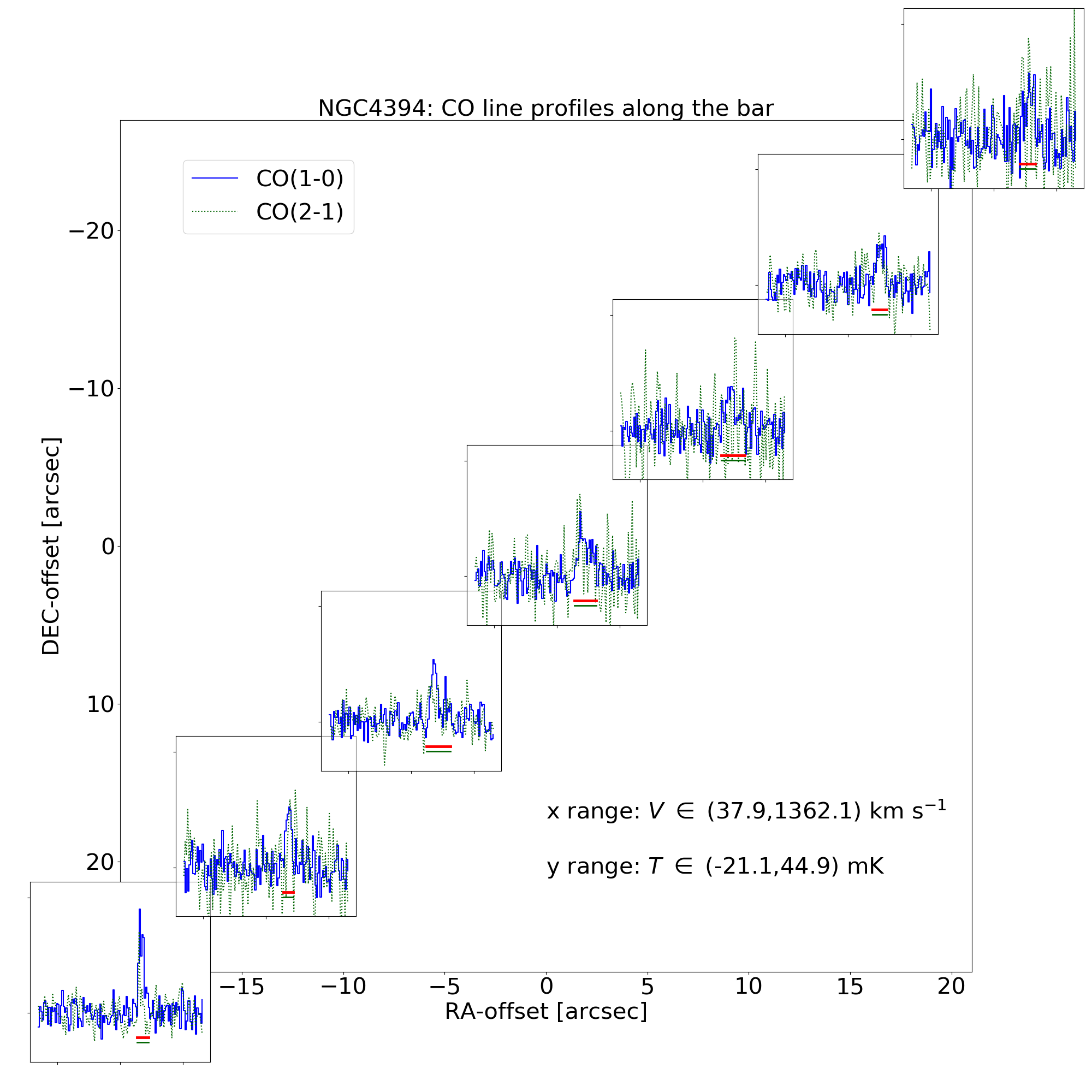}
\caption{
As in Fig.~\ref{plot_pointing_example}, but for NGC~4394.
}
\label{plots_spectra_app_3}
\end{figure}

\begin{figure}
\centering
\includegraphics[width=0.45\textwidth]{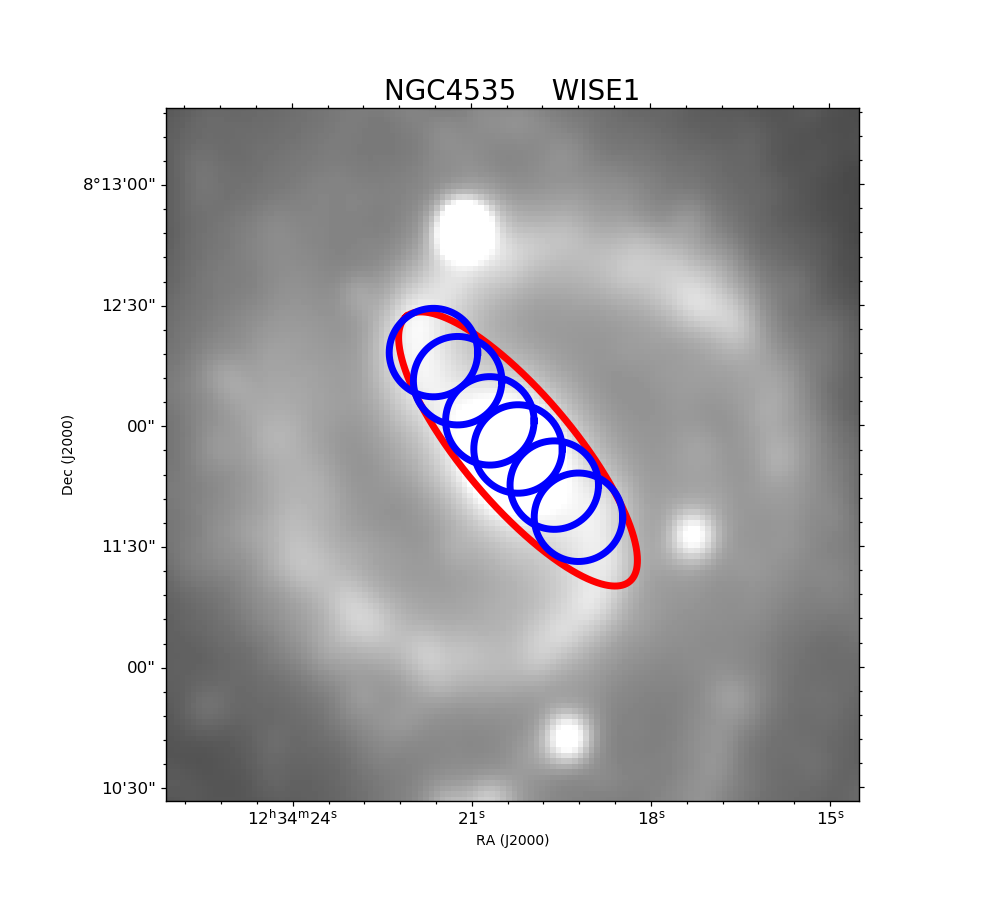}\\[-3ex]
\includegraphics[width=0.45\textwidth]{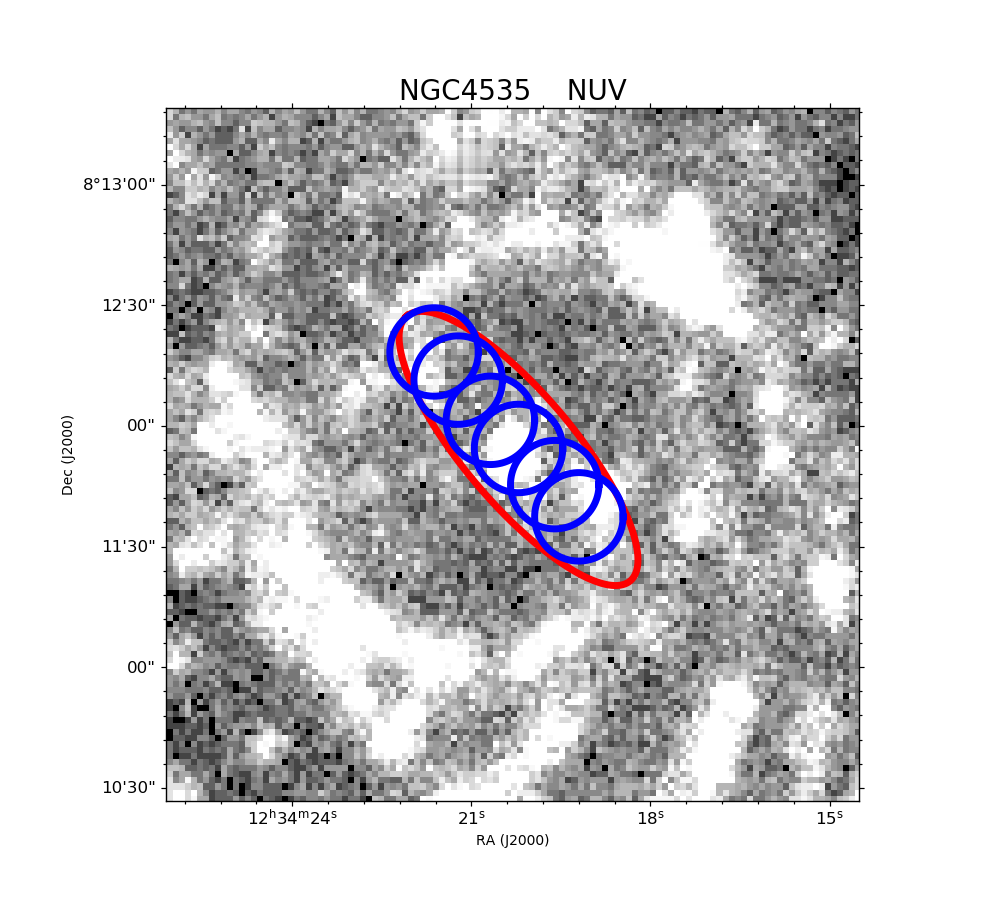}\\[-3ex]
\includegraphics[width=0.49\textwidth]{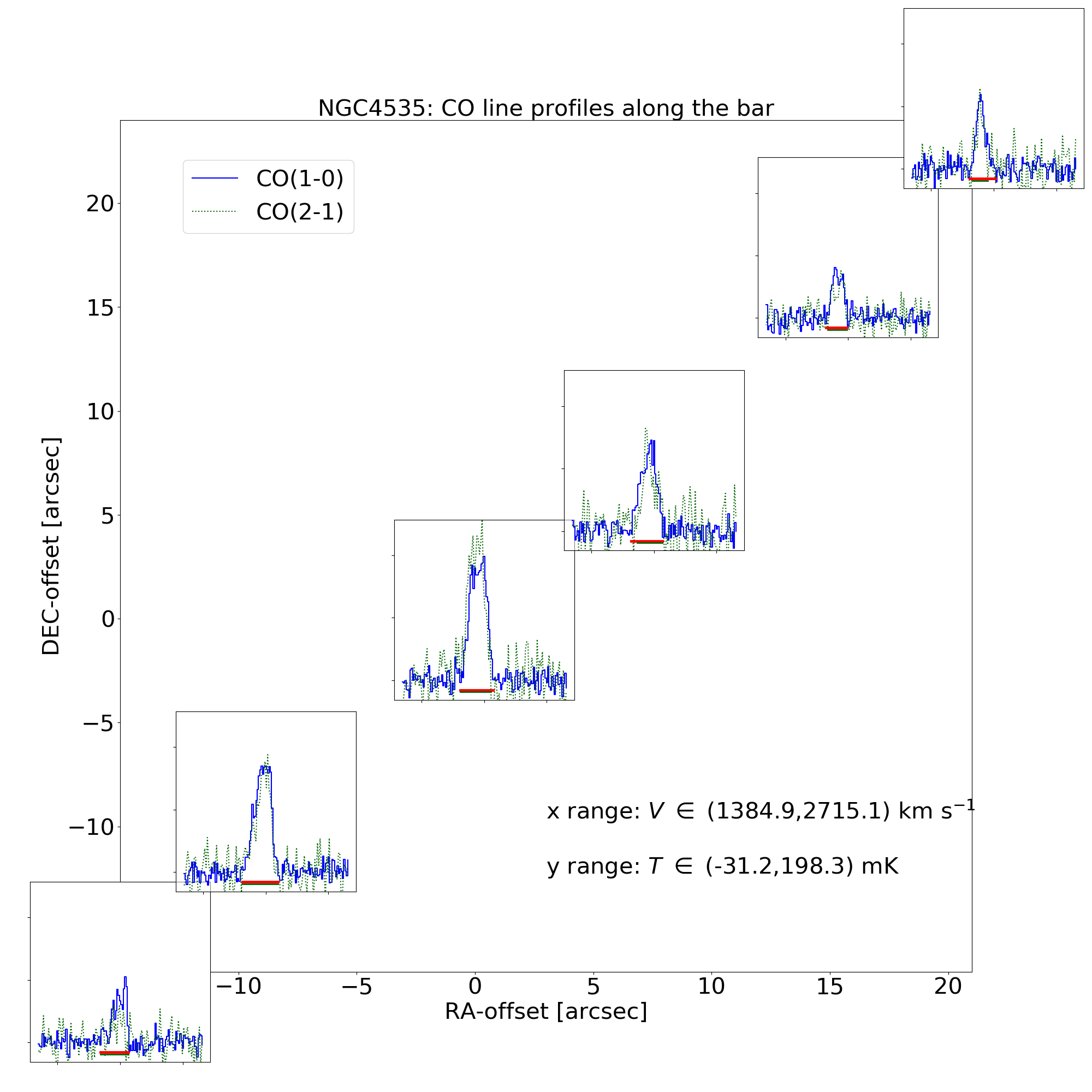}
\caption{
As in Fig.~\ref{plot_pointing_example}, but for NGC~4535.
}
\label{plots_spectra_app_4}
\end{figure}

\begin{figure}
\centering
\includegraphics[width=0.45\textwidth]{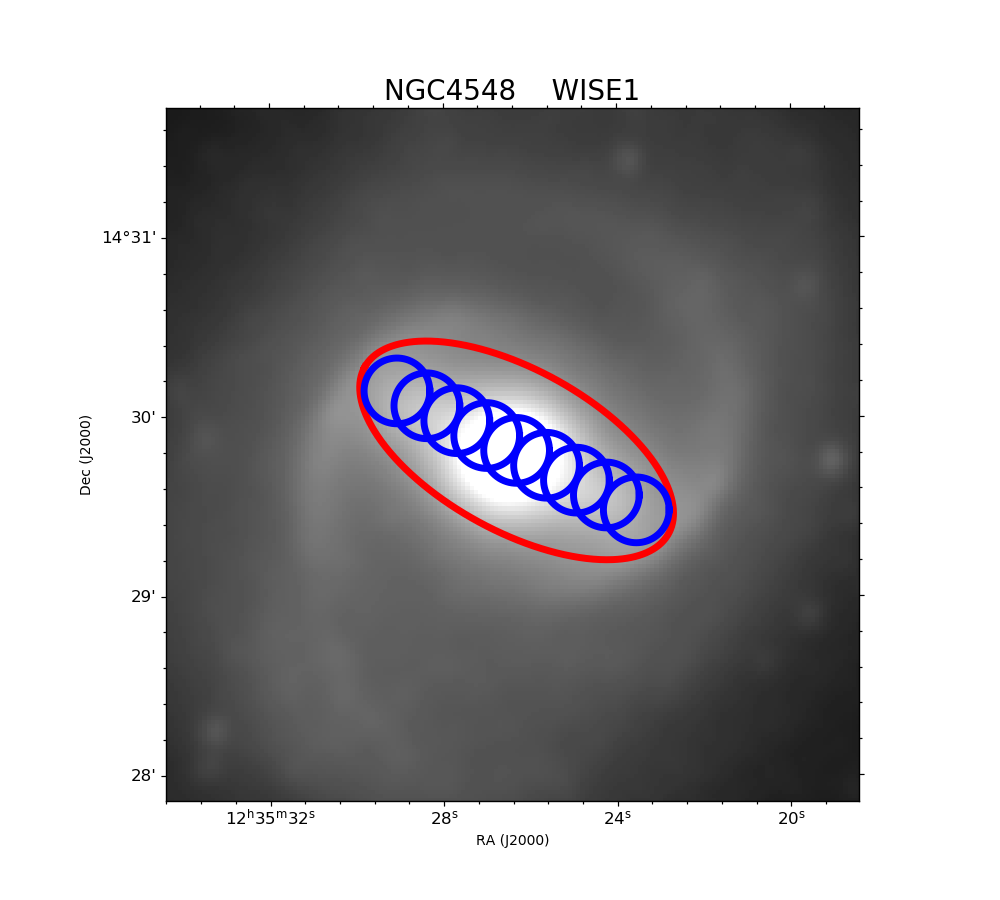}\\[-3ex]
\includegraphics[width=0.45\textwidth]{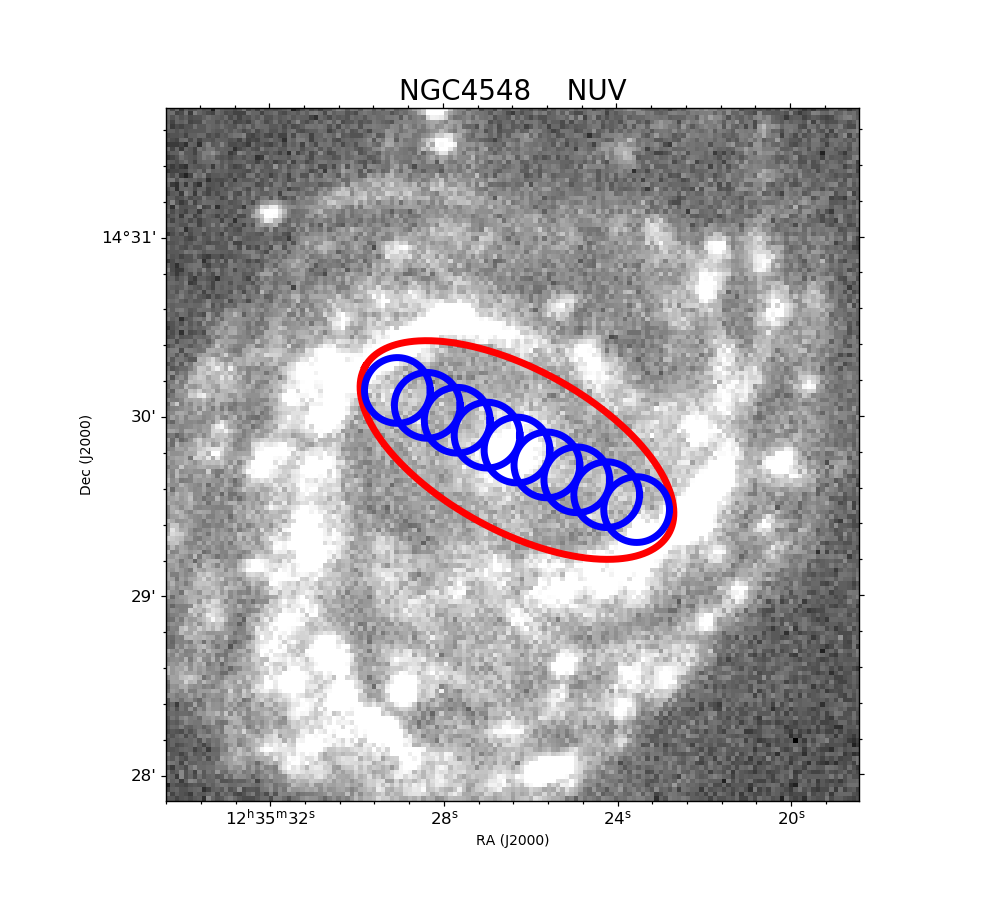}\\[-3ex]
\includegraphics[width=0.49\textwidth]{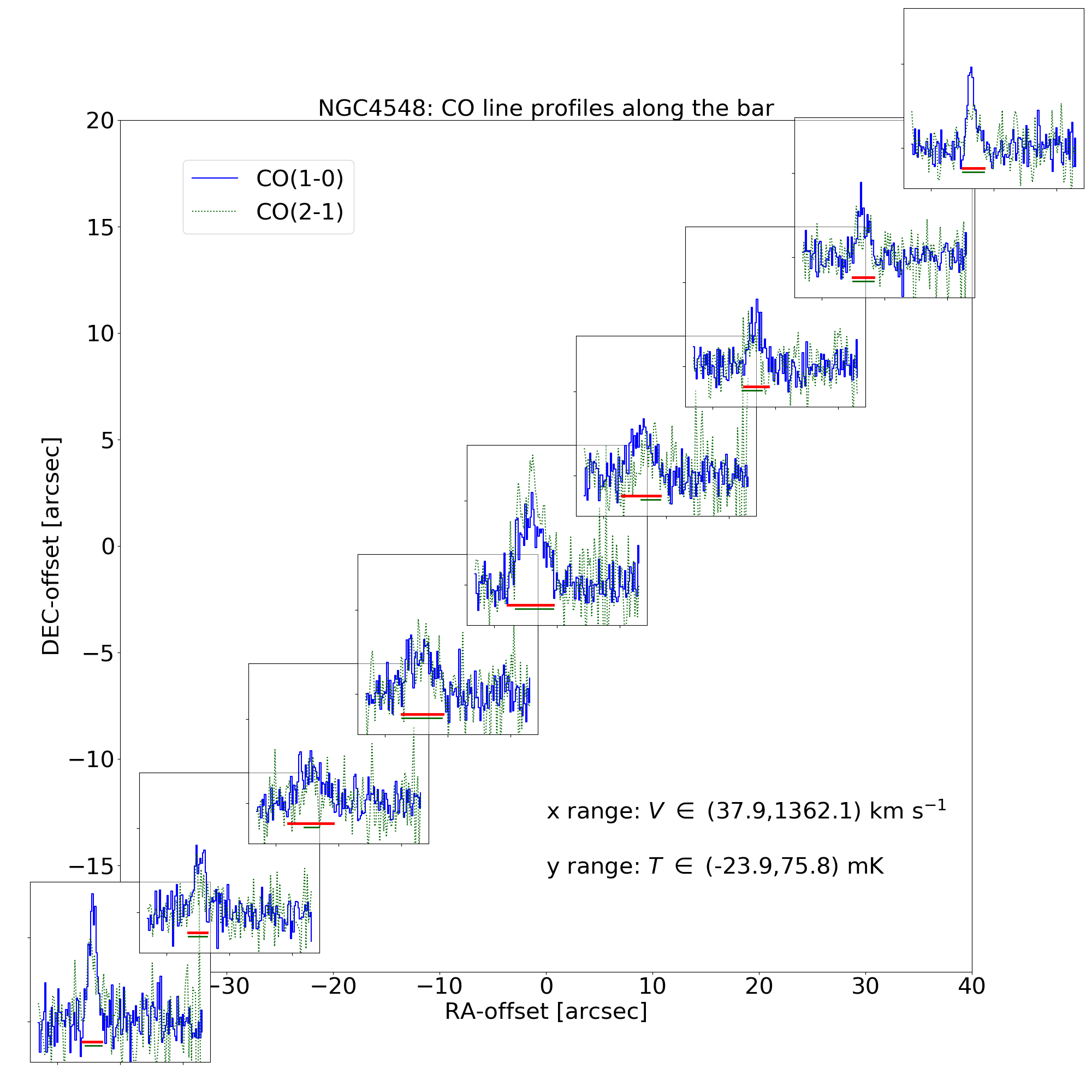}
\caption{
As in Fig.~\ref{plot_pointing_example}, but for NGC~4548.
}
\label{plots_spectra_app_5}
\end{figure}

\begin{figure}
\centering
\includegraphics[width=0.45\textwidth]{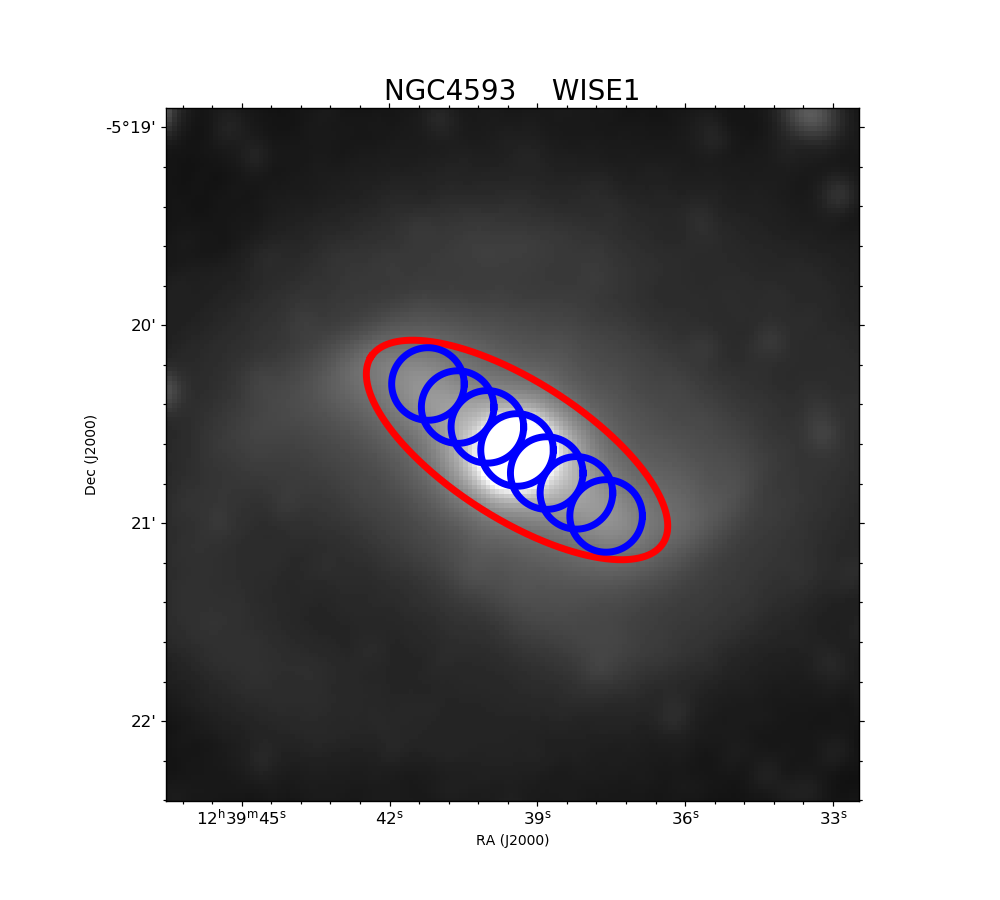}\\[-3ex]
\includegraphics[width=0.45\textwidth]{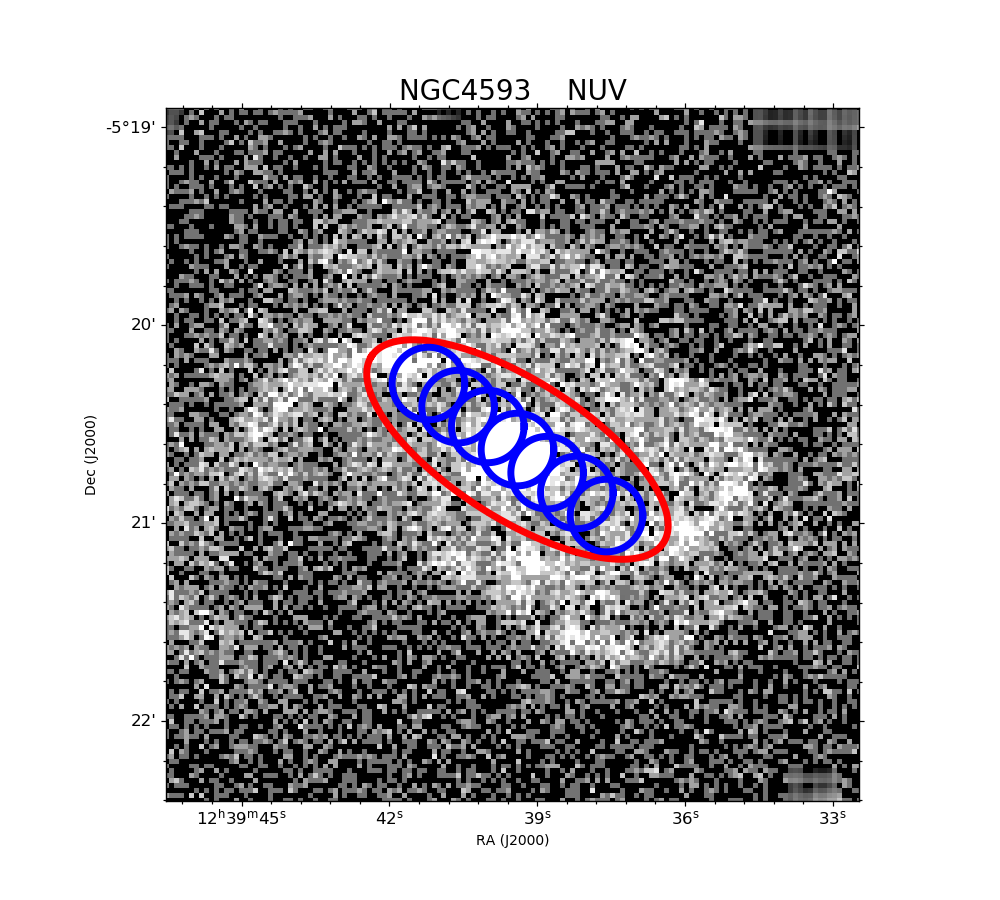}\\[-3ex]
\includegraphics[width=0.49\textwidth]{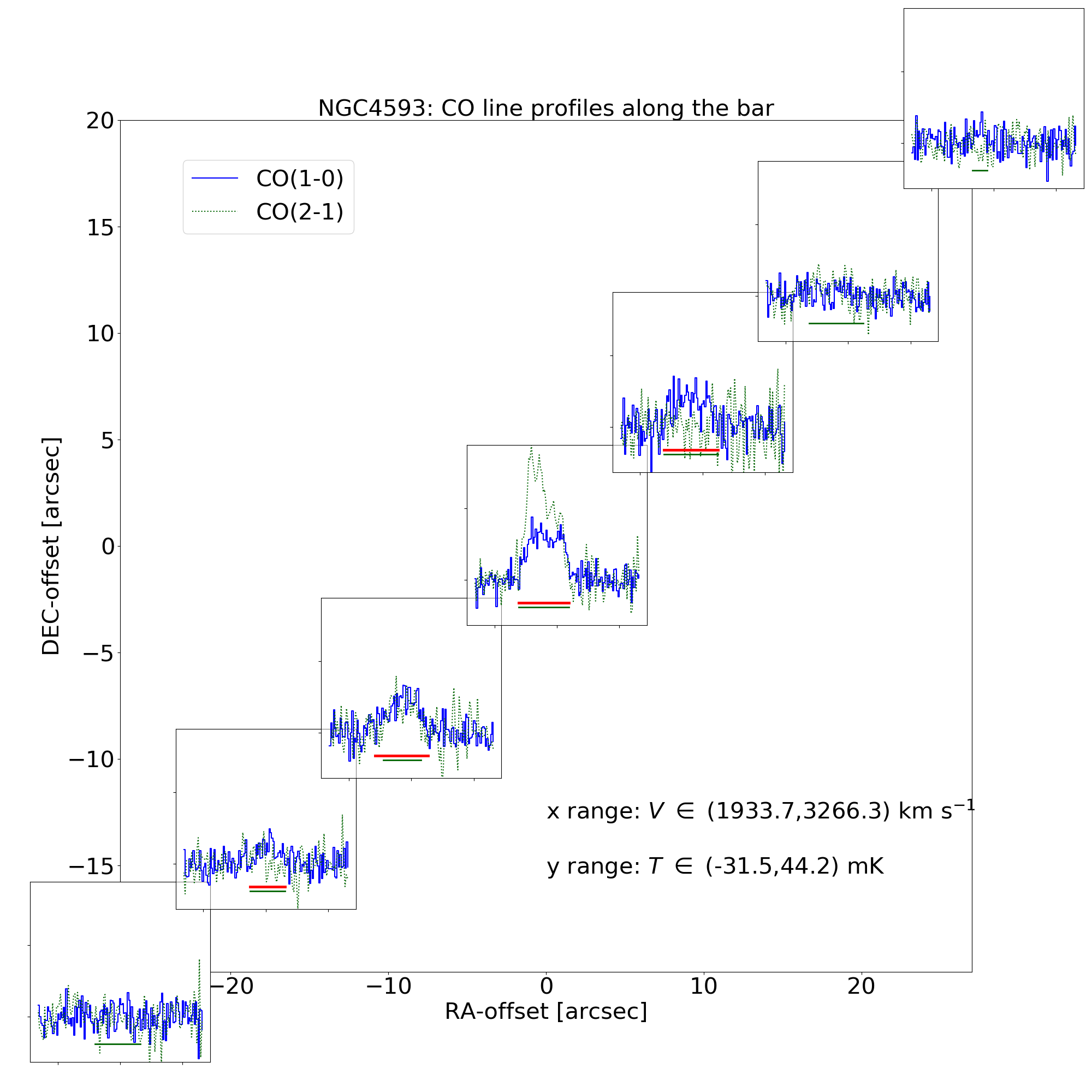}
\caption{
As in Fig.~\ref{plot_pointing_example}, but for NGC~4593.
}
\label{plots_spectra_app_6}
\end{figure}

\begin{figure}
\centering
\includegraphics[width=0.45\textwidth]{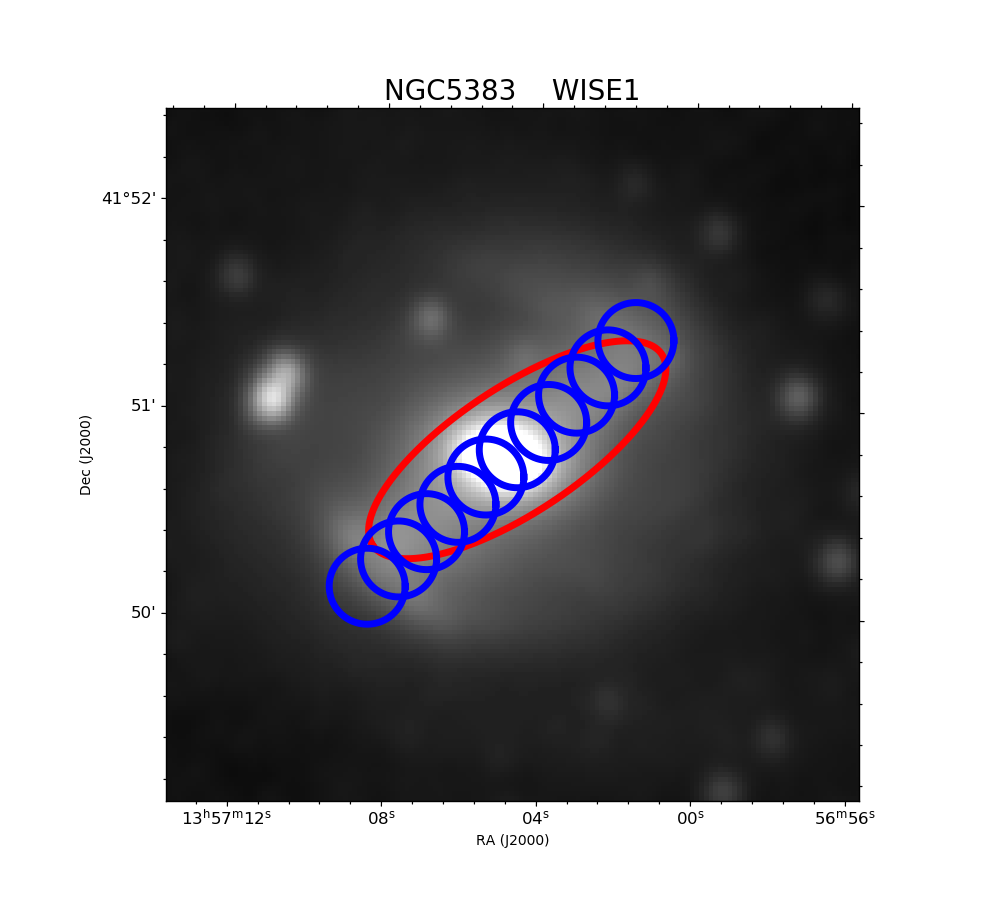}\\[-3ex]
\includegraphics[width=0.45\textwidth]{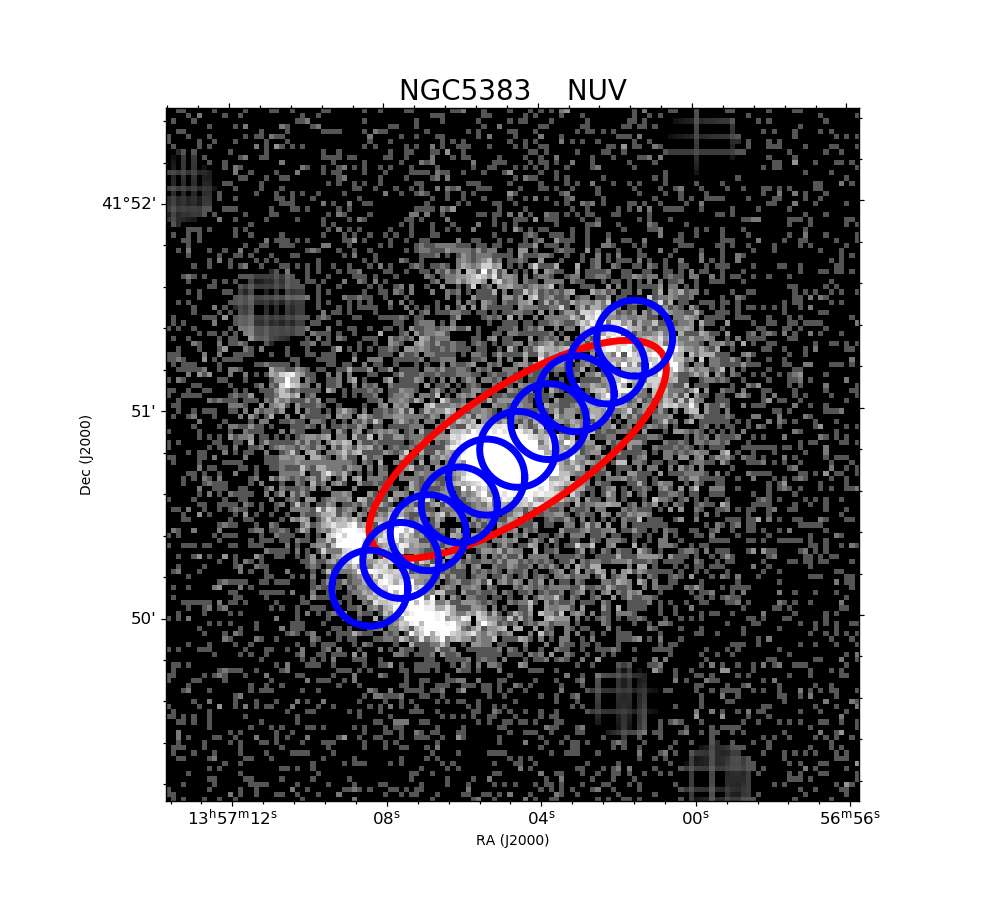}\\[-3ex]
\includegraphics[width=0.49\textwidth]{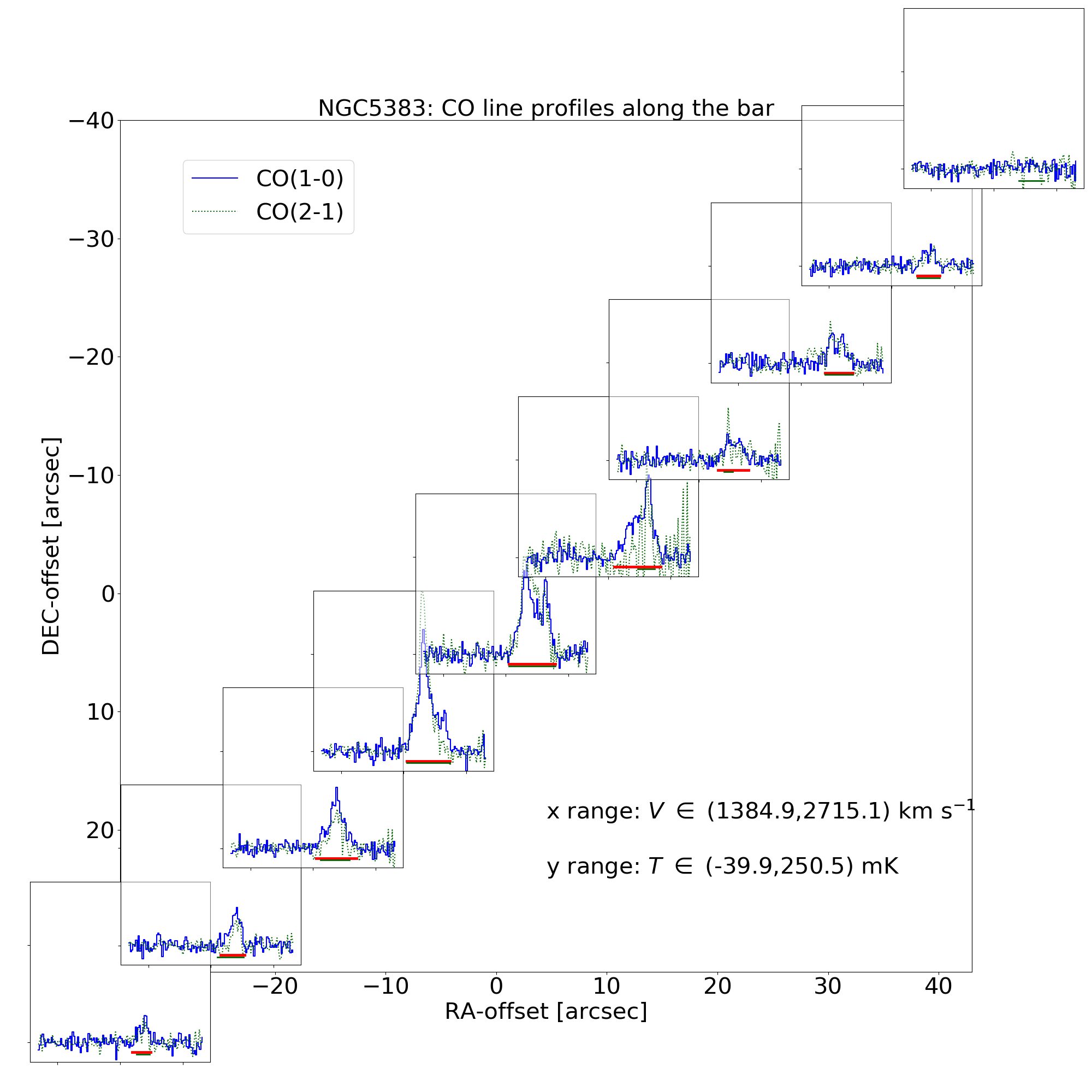}
\caption{
As in Fig.~\ref{plot_pointing_example}, but for NGC~5383.
}
\label{plots_spectra_app_7}
\end{figure}

\begin{figure}
\centering
\includegraphics[width=0.45\textwidth]{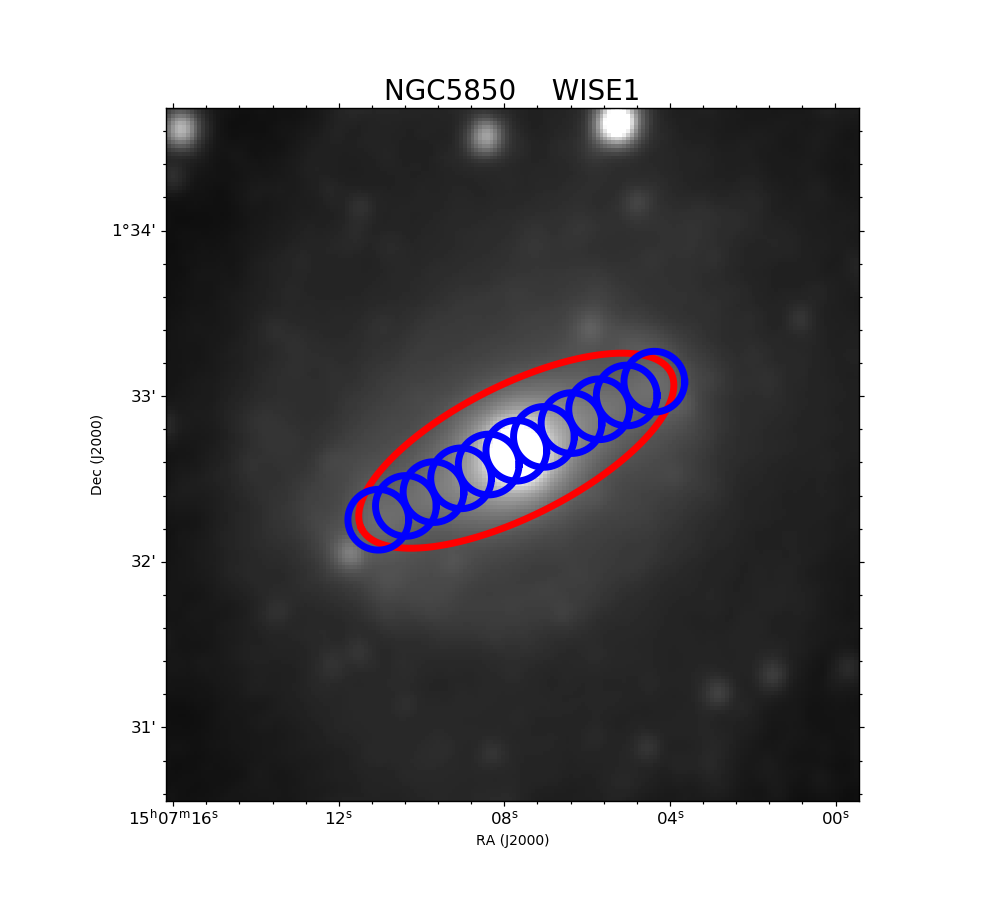}\\[-3ex]
\includegraphics[width=0.45\textwidth]{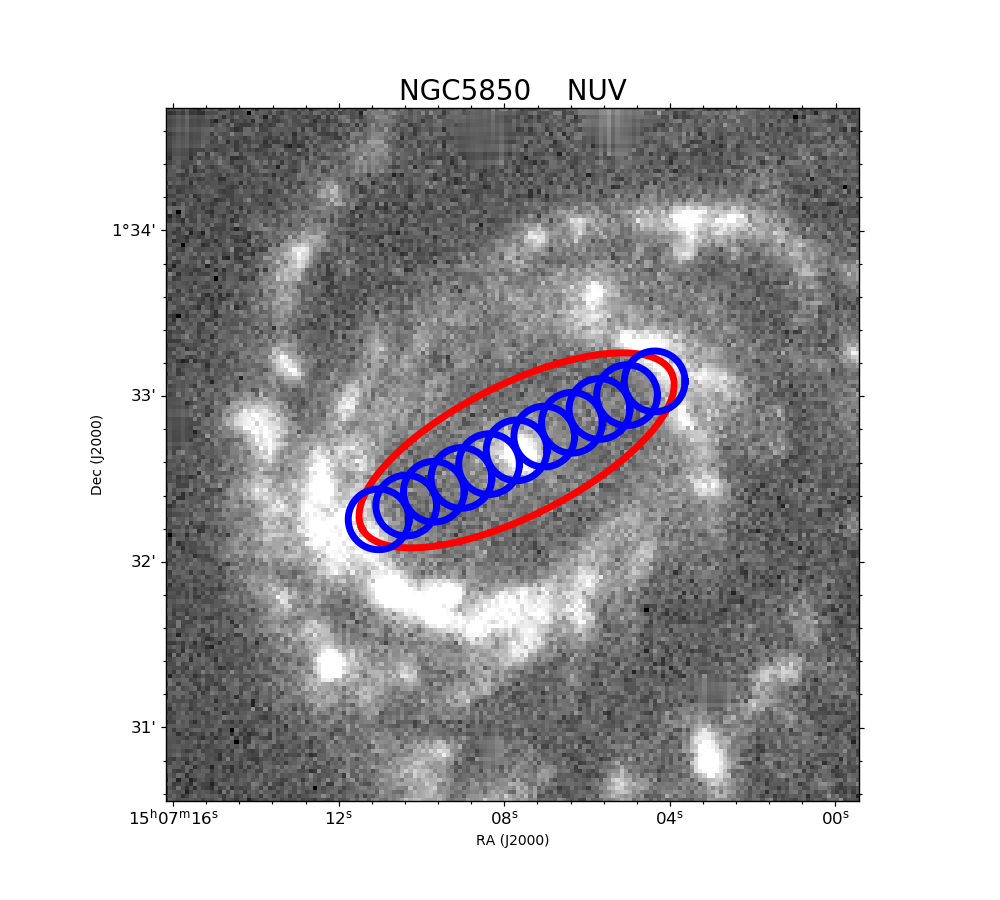}\\[-3ex]
\includegraphics[width=0.49\textwidth]{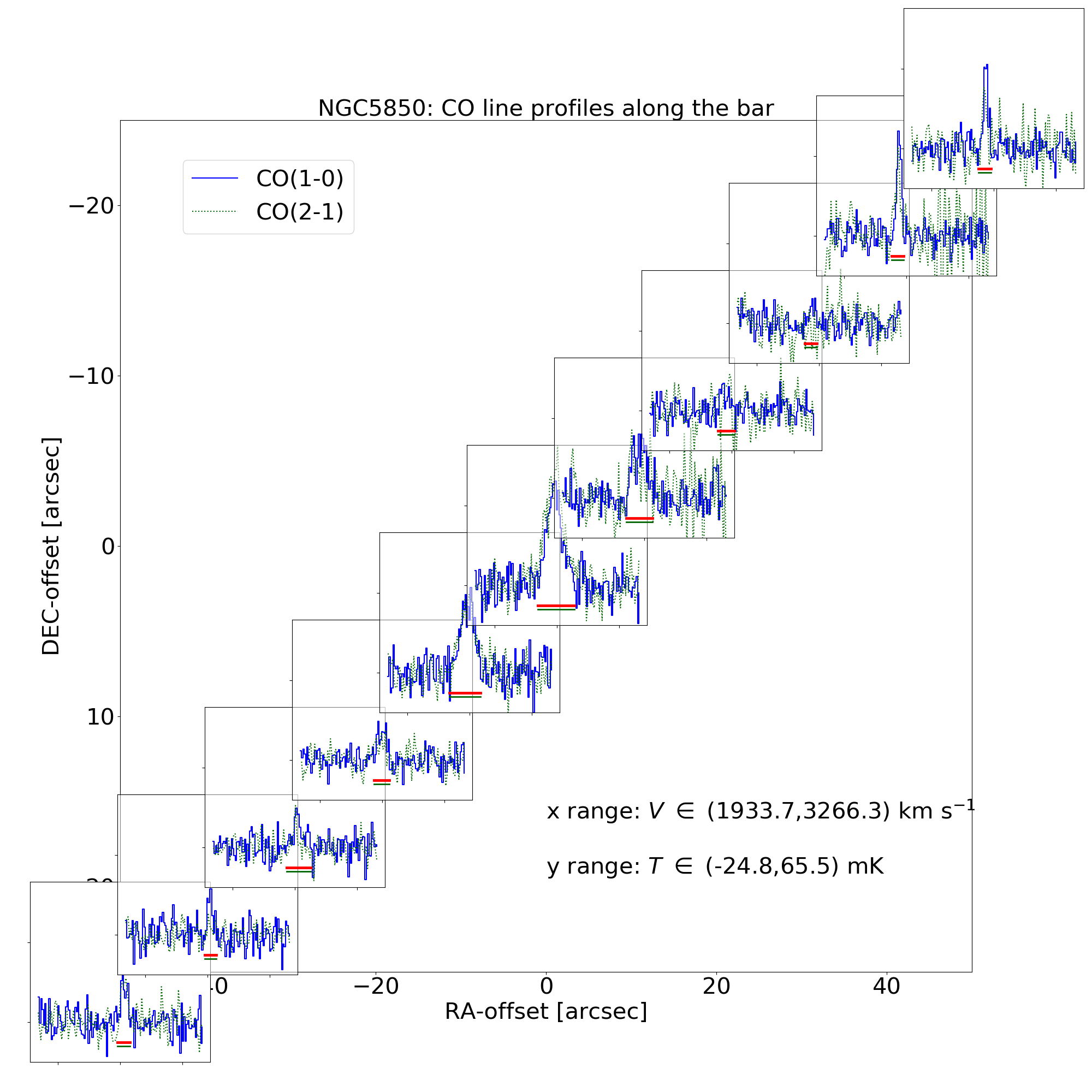}
\caption{
As in Fig.~\ref{plot_pointing_example}, but for NGC~5850.
}
\label{plots_spectra_app_8}
\end{figure}

\begin{figure}
\centering
\includegraphics[width=0.45\textwidth]{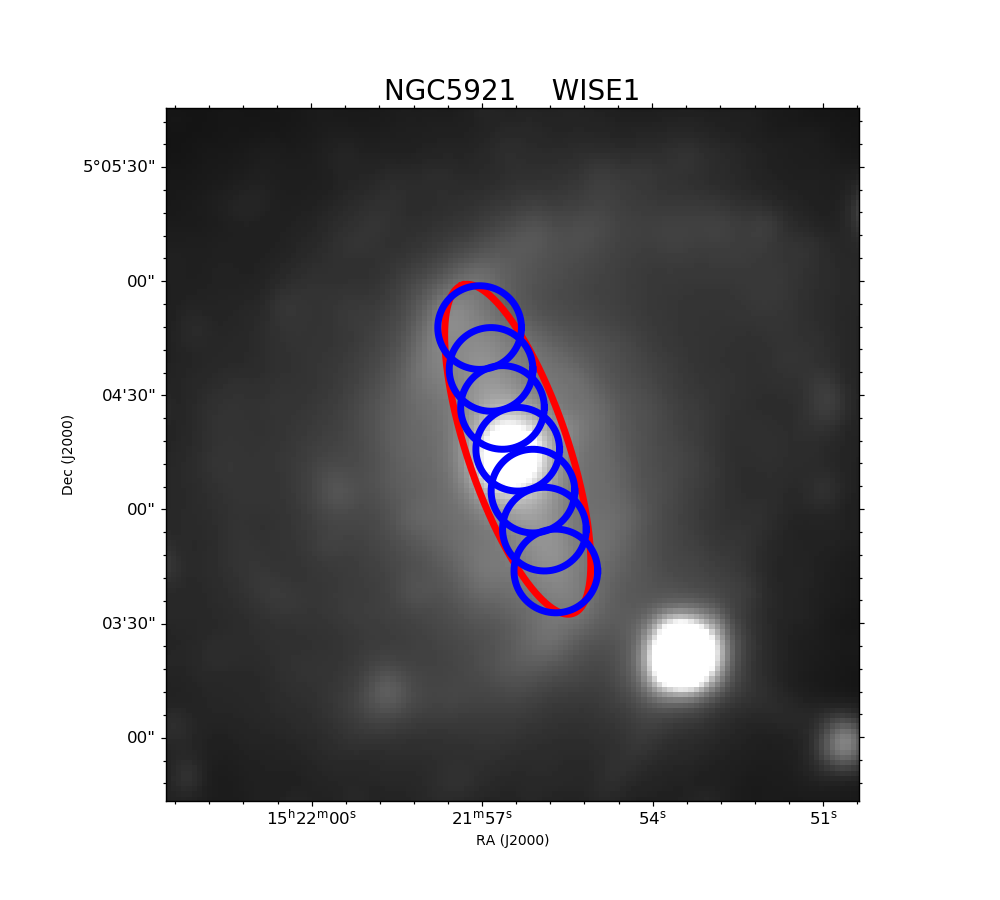}\\[-3ex]
\includegraphics[width=0.45\textwidth]{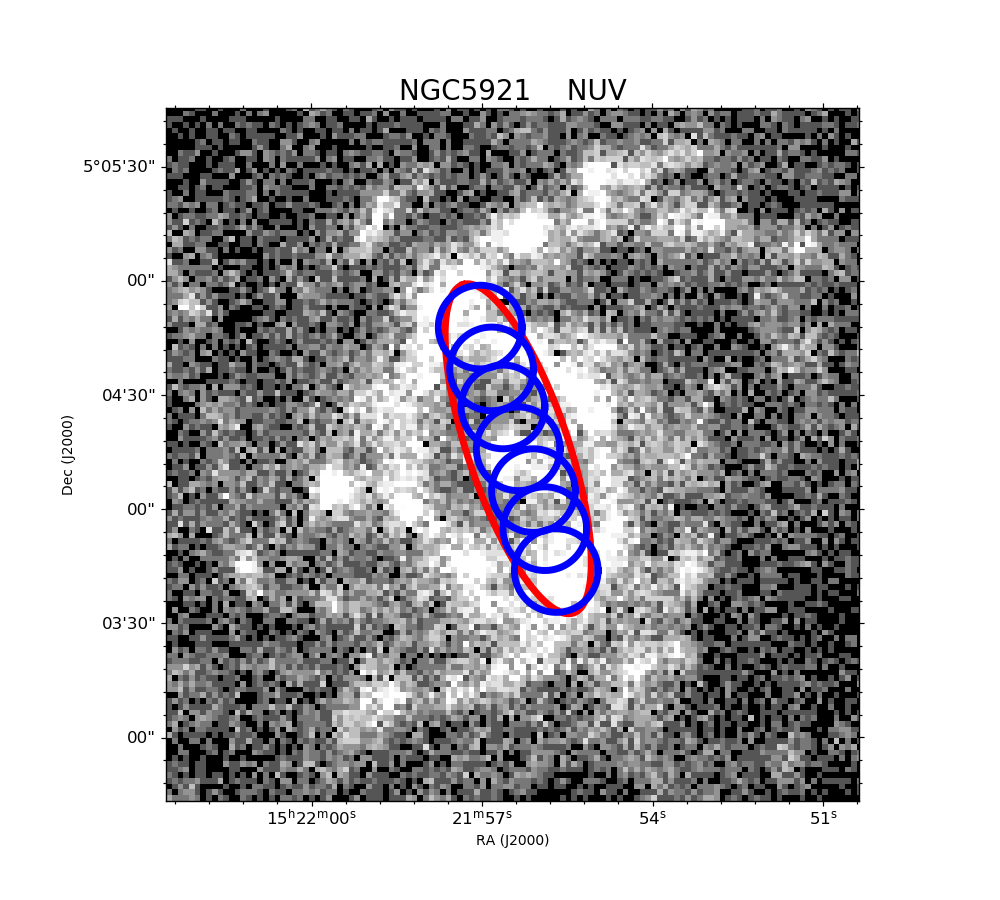}\\[-3ex]
\includegraphics[width=0.49\textwidth]{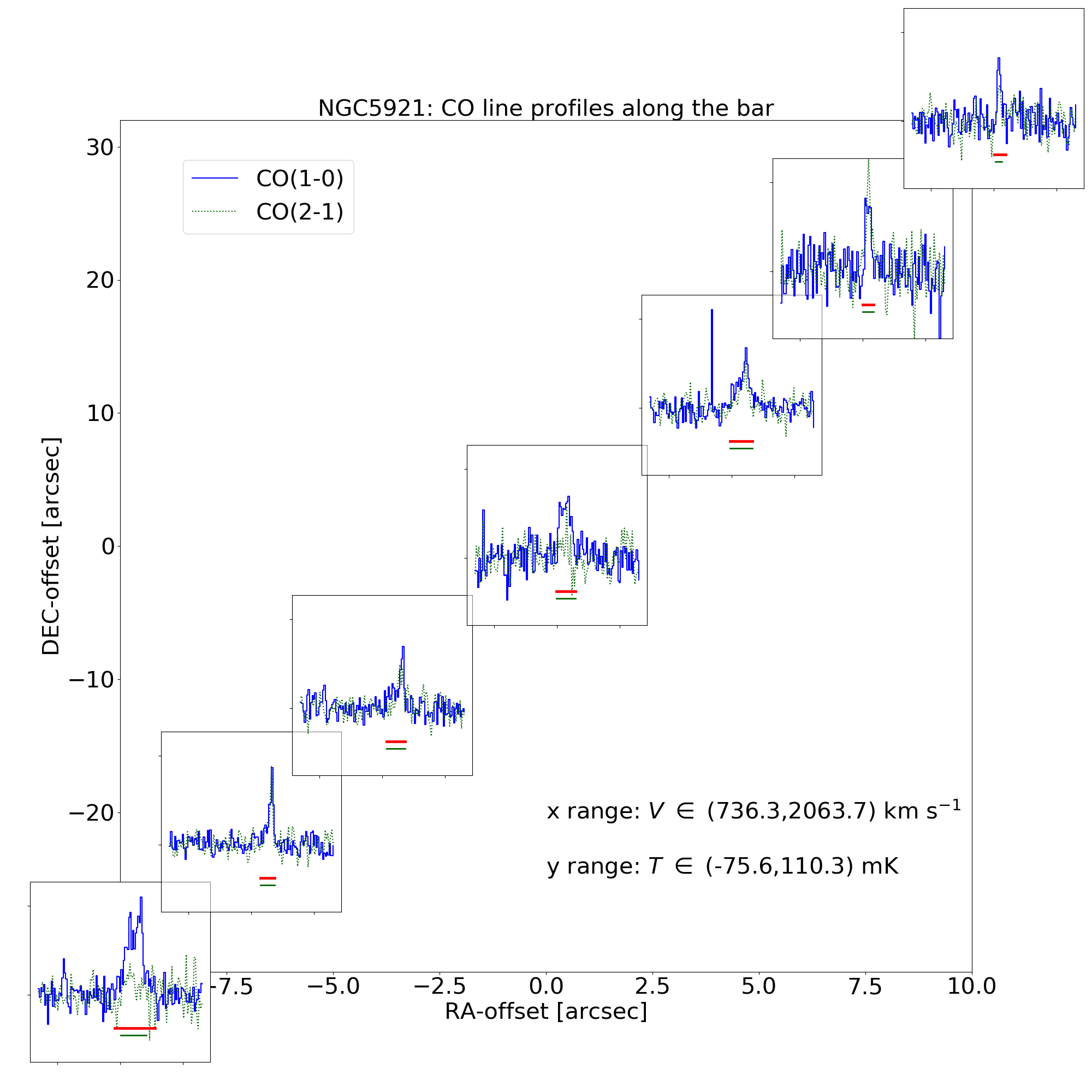}
\caption{
As in Fig.~\ref{plot_pointing_example}, but for NGC~5921.
}
\label{plots_spectra_app_9}
\end{figure}

\begin{figure}
\centering
\includegraphics[width=0.45\textwidth]{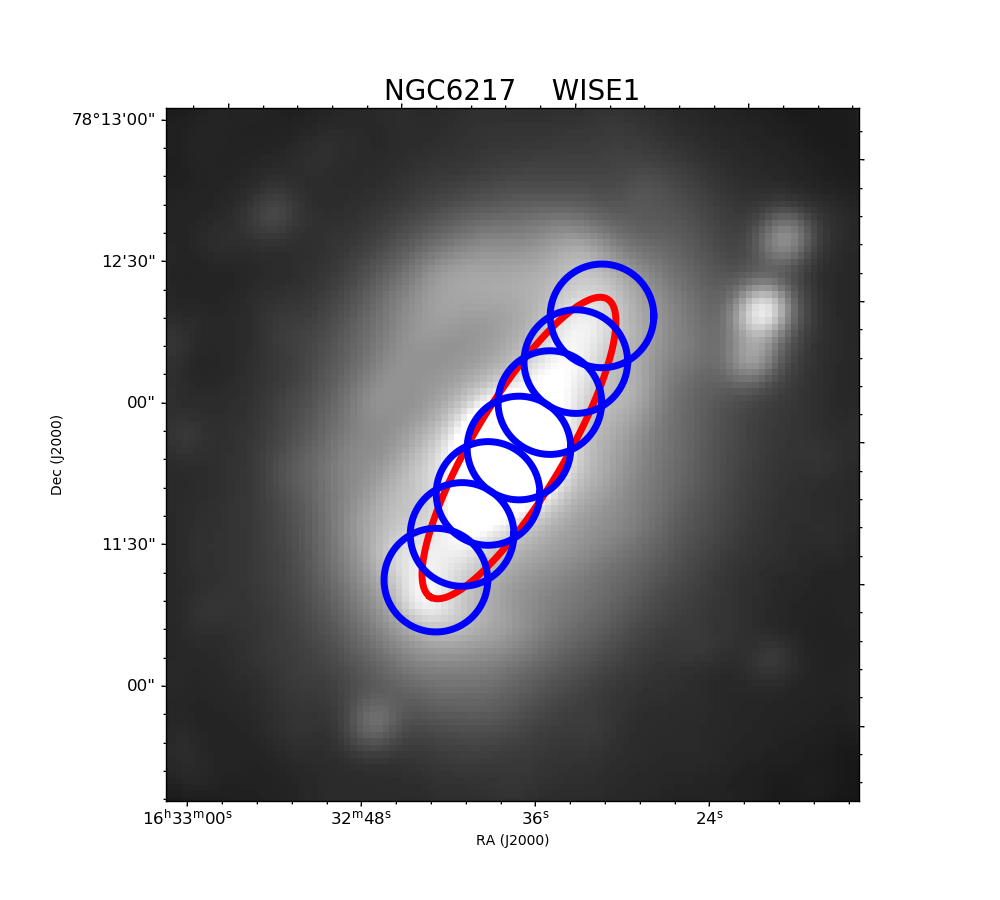}\\[-3ex]
\includegraphics[width=0.45\textwidth]{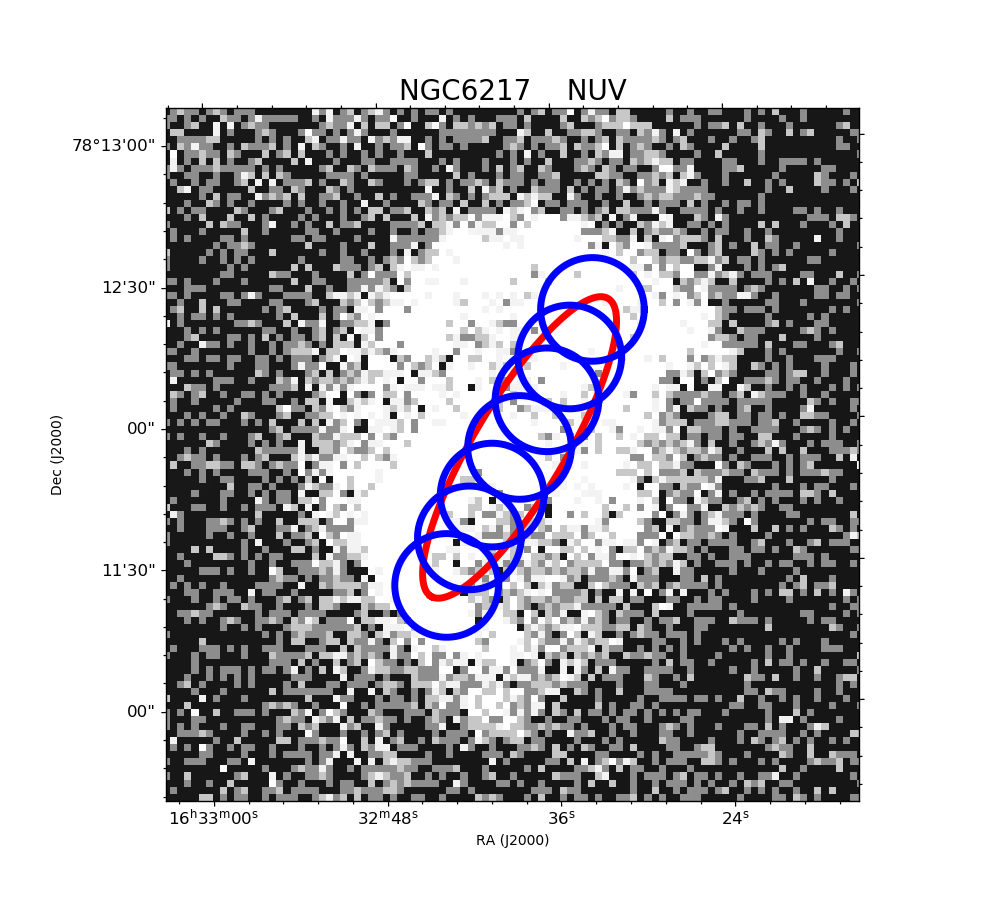}\\[-3ex]
\includegraphics[width=0.49\textwidth]{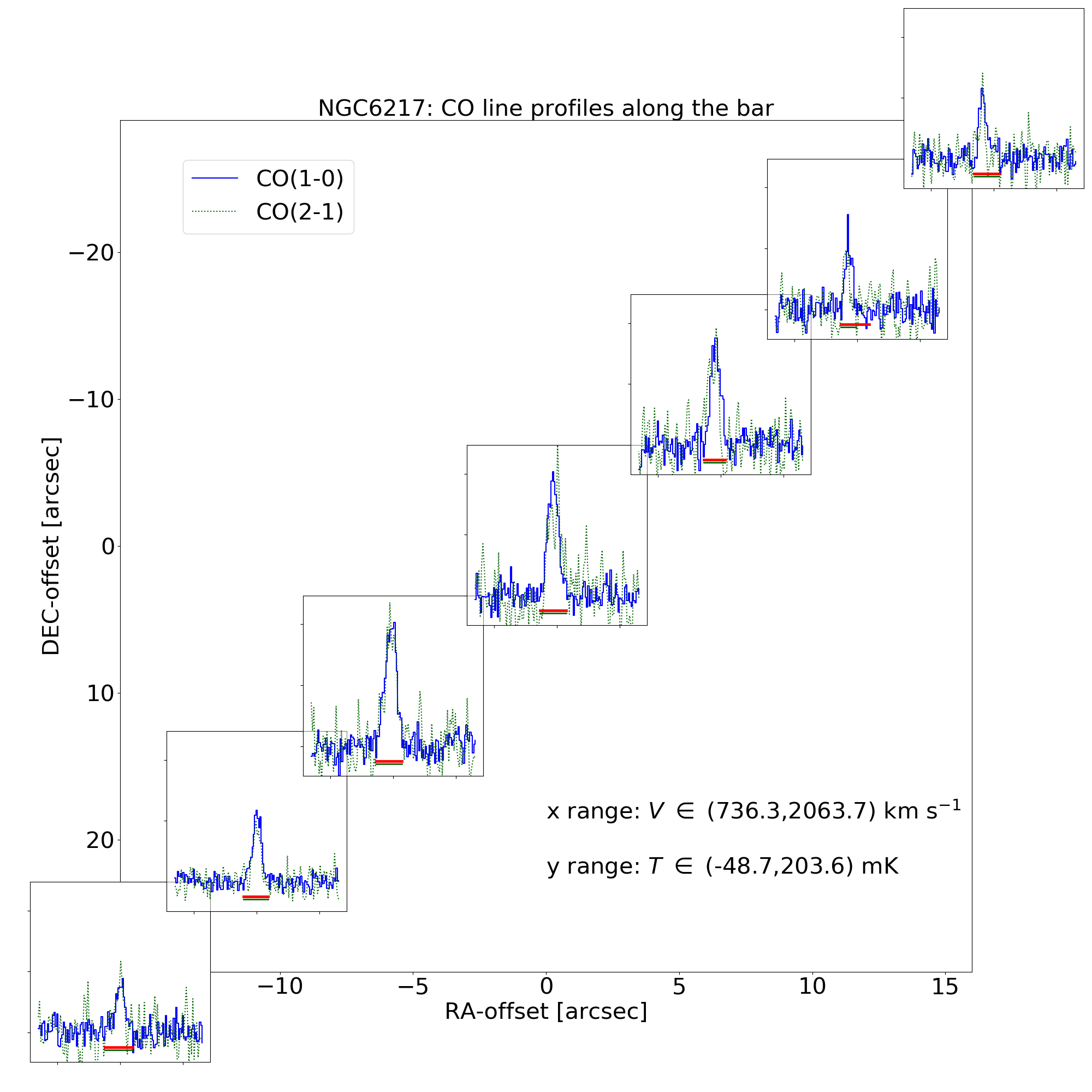}
\caption{
As in Fig.~\ref{plot_pointing_example}, but for NGC~6217.
}
\label{plots_spectra_app_10}
\end{figure}

\begin{figure}
\centering
\includegraphics[width=0.45\textwidth]{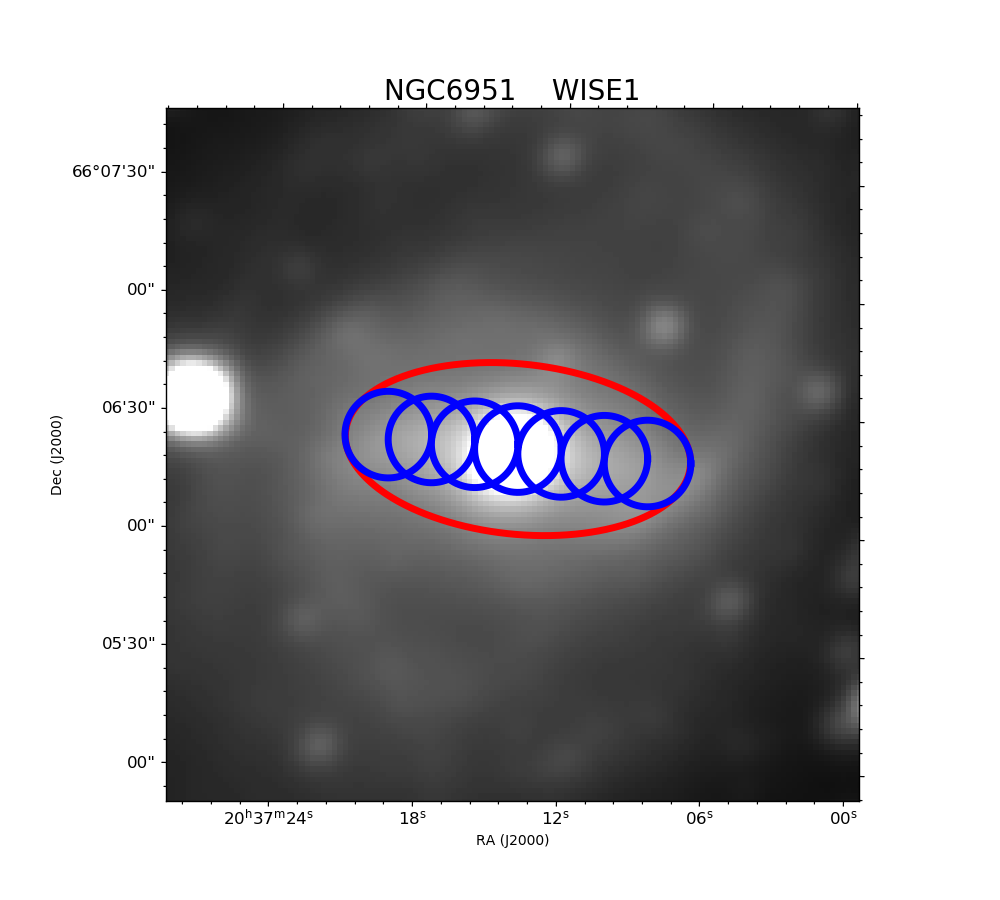}\\[-3ex]
\includegraphics[width=0.45\textwidth]{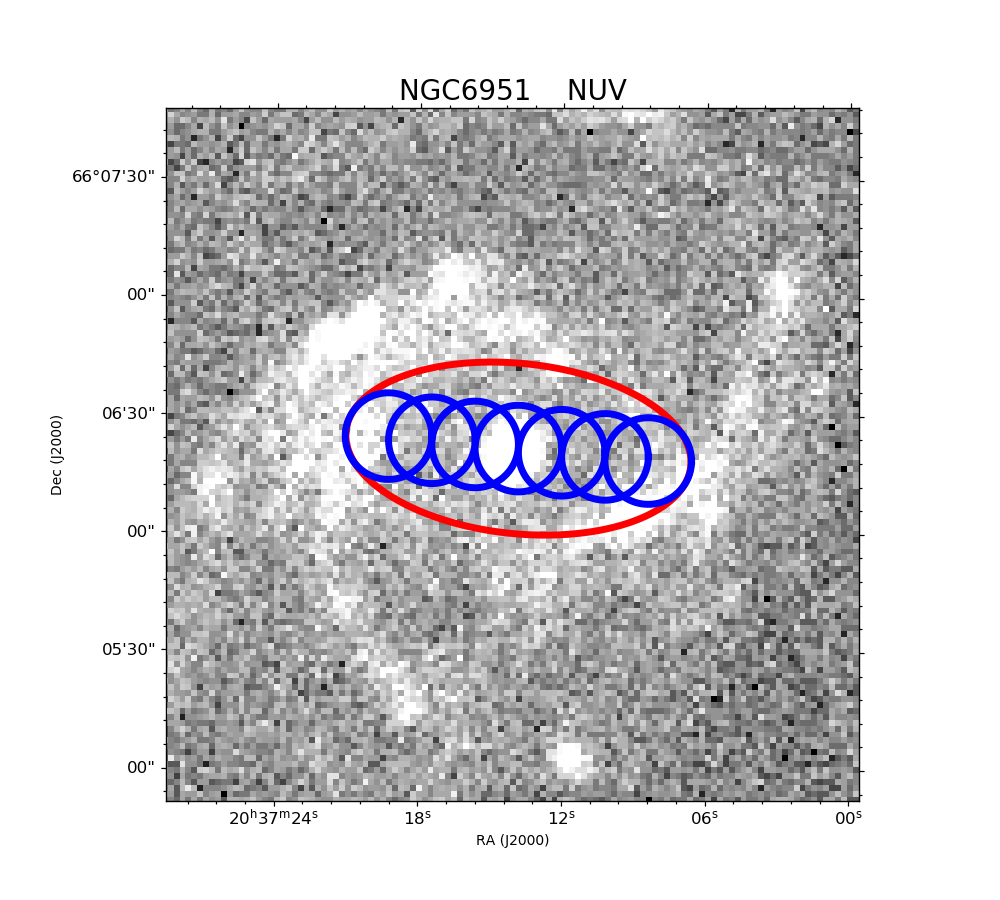}\\[-3ex]
\includegraphics[width=0.49\textwidth]{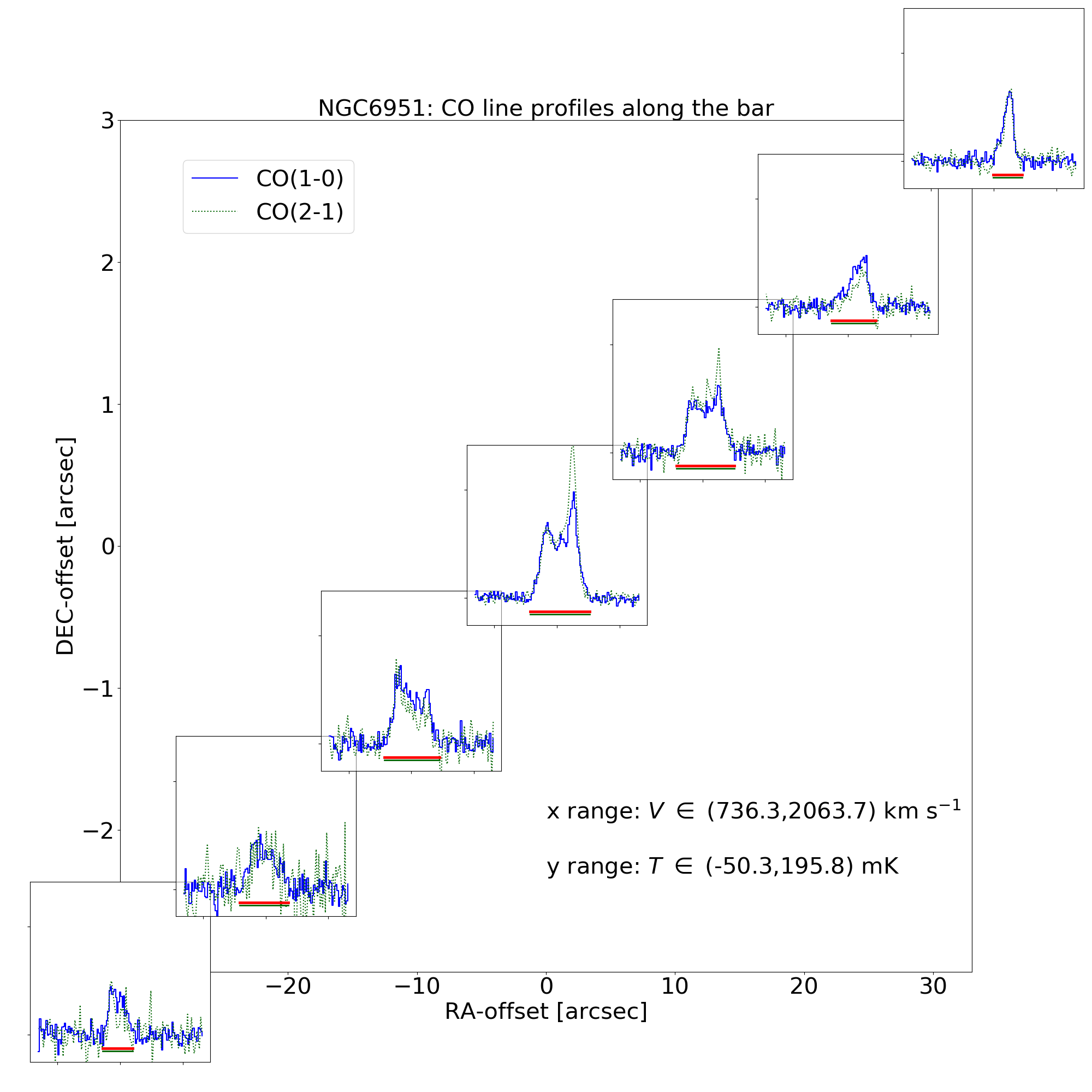}
\caption{
As in Fig.~\ref{plot_pointing_example}, but for NGC~6951.
}
\label{plots_spectra_app_11}
\end{figure}

\begin{figure}
\centering
\includegraphics[width=0.45\textwidth]{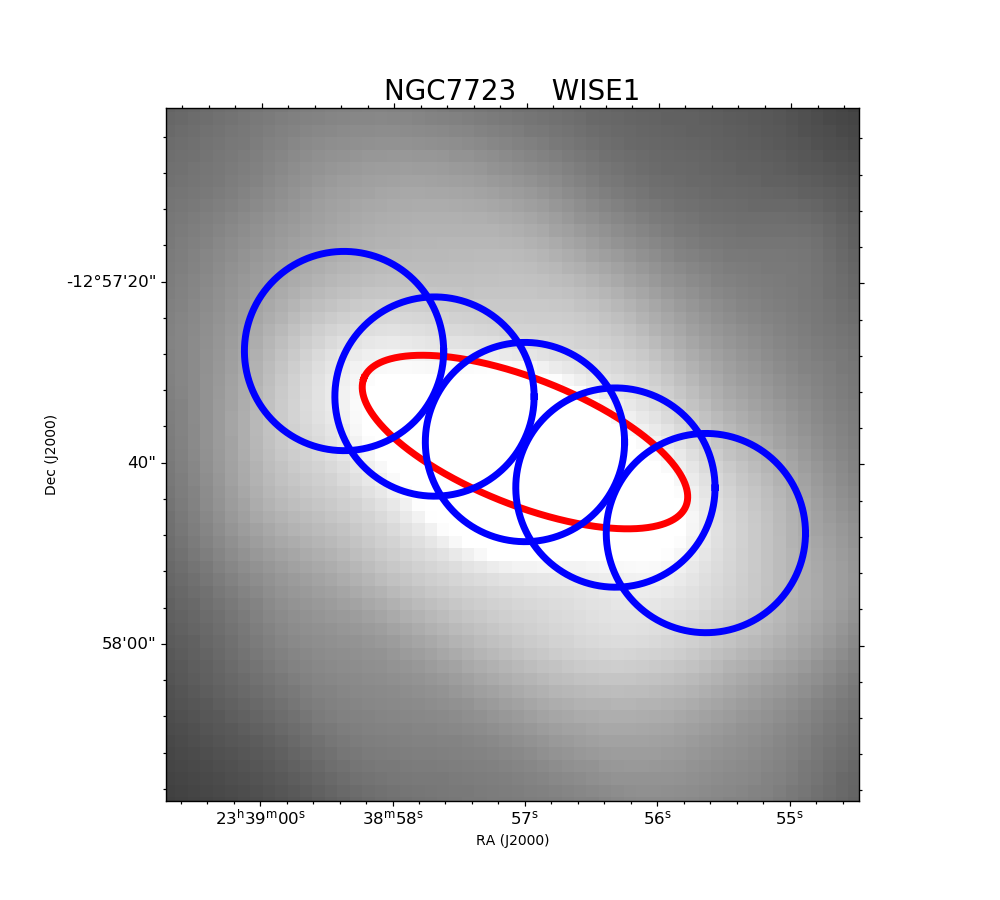}\\[-3ex]
\includegraphics[width=0.45\textwidth]{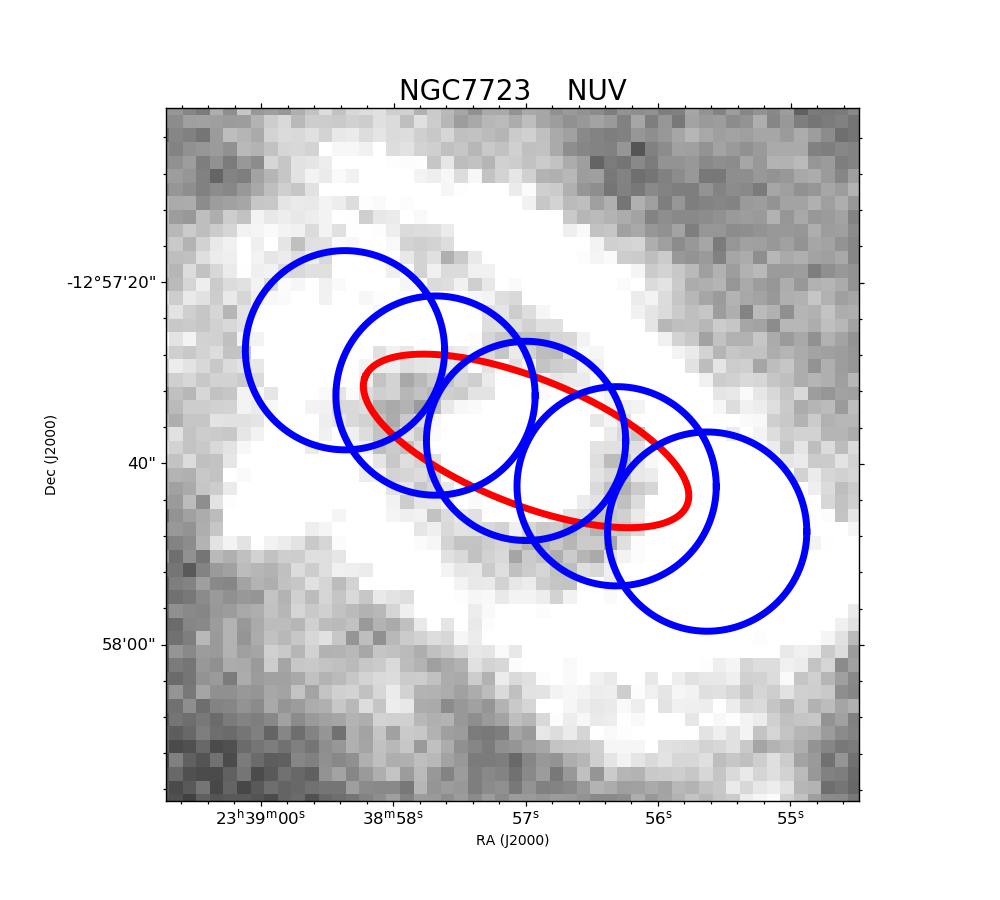}\\[-3ex]
\includegraphics[width=0.49\textwidth]{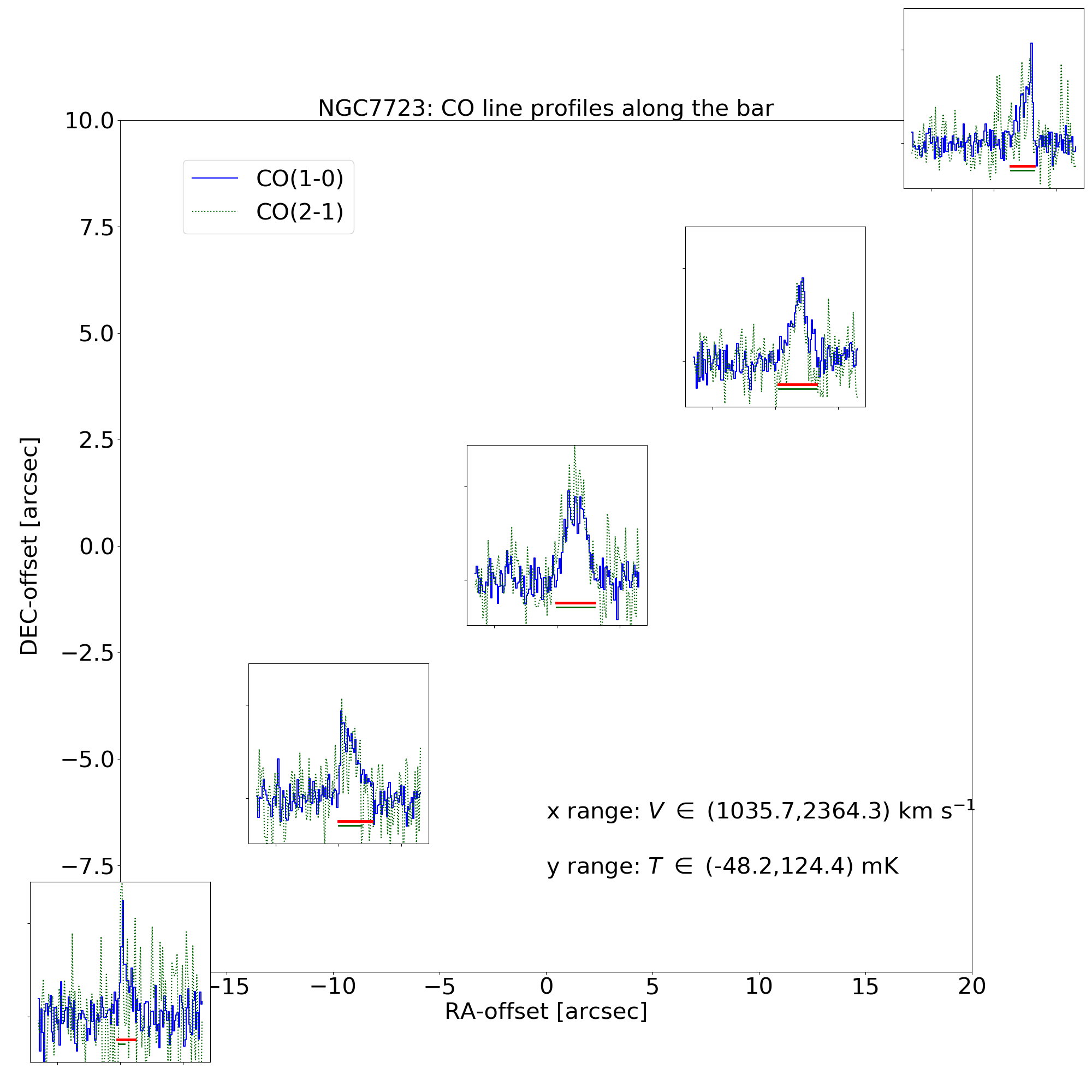}
\caption{
As in Fig.~\ref{plot_pointing_example}, but for NGC~7723.
}
\label{plots_spectra_app_12}
\end{figure}

%
%
We present the CO(1-0) and CO(2-1) spectra used in this work for the 12 galaxies in our sample 
(Figs. \ref{plots_spectra_app_0}-\ref{plots_spectra_app_12}). 
The upper and central panels correspond to the \emph{WISE} 1 and \emph{GALEX} NUV images. 
The bottom panel shows the spectra in every beam along bars, 
with every subplot located according to the RA and DEC offsets of the pointings. 
Velocity and intensity ranges are indicated in the lower right corner of the plots. 
In the images we show in blue the regions where the CO spectra were taken, 
which were chosen based on the visual estimate of the bar length and position angle. 
These are in good agreement with the values reported by \citet[][]{2016A&A...587A.160D}. 

We also show, with a red ellipse, the bar isophote measured by \citet[][]{2015A&A...582A..86H} via 
ellipse fitting (see Sect.~\ref{sample_data}), 
which is known to slightly underestimate the true size of 
bars \citep[][]{1991A&AS...88..325W,2002MNRAS.337.1118L,2003ApJS..146..299E,2006A&A...452...97M,2016A&A...587A.160D}. 
Visual analysis and ellipse fitting yield similar measurements, the only exception being NGC~5383 (Fig \ref{plots_spectra_app_7}) 
where the presence of H{\sc\,ii} regions in the spiral arms at the bar end make the ellipse fitting output less reliable. 
Uncertainties on bar sizes are not 
a big issue for the current analysis, given fairly large FWHM (21.5\arcsec) of the apertures where spectra were taken. 
%
%
%
%
%
%
%
%
%
%
%
%
%
%
%
%
%
%
%
%
%
%
%
%
%
%
\clearpage
\newpage
%
%
\onecolumn
\section{Velocity-integrated intensities}\label{H2_tables}
The following data are listed in Table \ref{CO_H2_int}:
\begin{itemize}
 \item Column 1: Galaxy name;
 \item Column 2: Right ascension of centre of the IRAM-30m pointing (in degrees);
 \item Column 3: Declination of centre of the IRAM-30m pointing (in degrees);
 \item Column 4: Velocity-integrated intensity $I_{\rm CO(1-0)}$ (in K km s$^{-1}$) and uncertainty on spectrum integration;
 \item Column 5: Root mean square (rms) noise (in mK) associated with each CO(1-0) spectrum at a velocity resolution of $\delta V = 10.5$ km s$^{-1}$;
 \item Column 6: Zero-level velocity width of the CO(1-0) line (in km s$^{-1}$) of each spectrum ($\Delta V$);
 \item Column 7: Velocity-integrated intensity $I_{\rm CO(2-1)}$ (in K km s$^{-1}$) and its error;
 \item Column 8: Root mean square (rms) noise (in mK) associated with each CO(2-1) spectrum at a velocity resolution of $\delta V = 10.5$ km s$^{-1}$;
 \item Column 9: Mass of molecular gas (in $M_{\odot}$) and its error (Eq. \ref{eq_mh2});
 \item Column 10: Surface density of molecular gas mass (in $M_{\odot}$ pc$^{-2}$) and its error (Eq. \ref{eq_sigma_h2}).
\end{itemize}
\input{./table_H2.txt}
\twocolumn
%
%
\clearpage
\newpage
%
%
\section{SFEs with different direct and hybrid recipes of SFRs}\label{SF_hybrid_non_hybrid}
\begin{figure}
\centering
\includegraphics[width=0.47\textwidth]{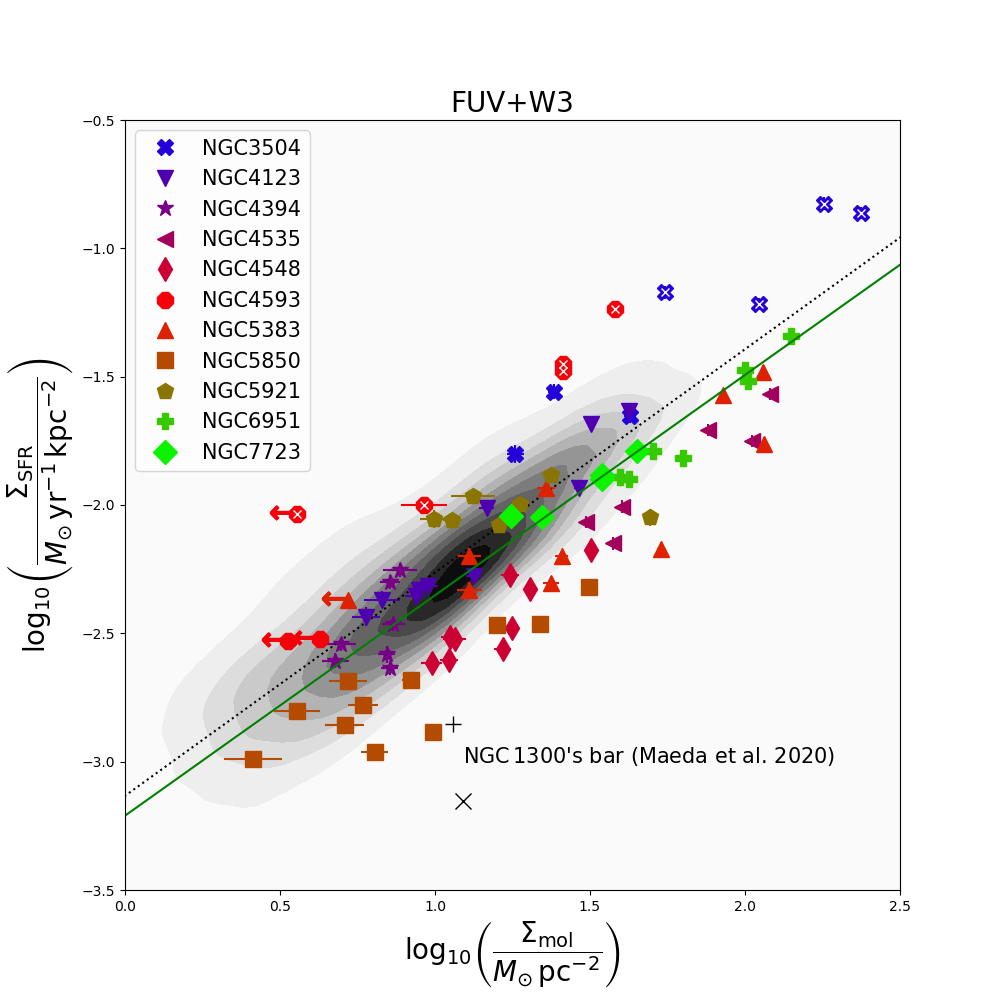}
\includegraphics[width=0.47\textwidth]{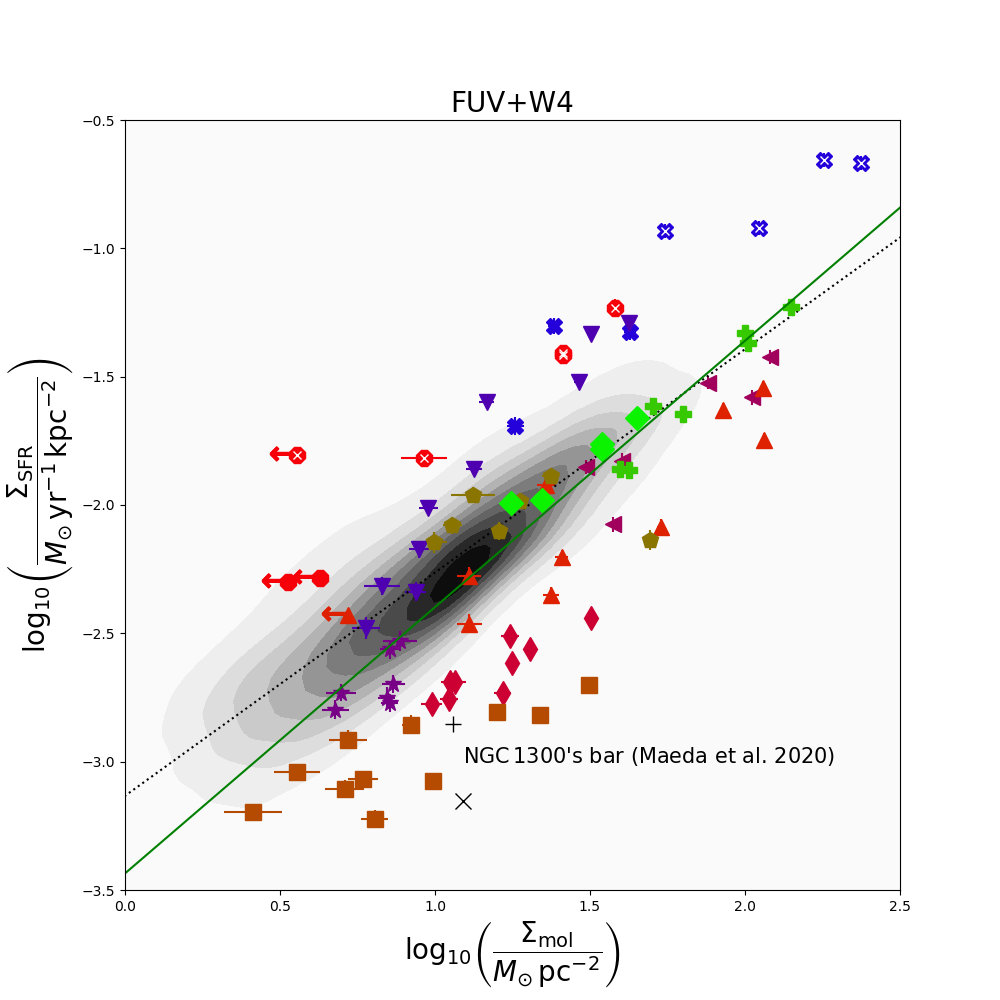}
\caption{
As in the left panels of Fig.~\ref{PLOTS_ALLTOGETHER_B}, 
but using hybrid SFR recipes corresponding to 
Eq. \ref{Eq_FUV_WISE3} (FUV+\emph{WISE} 3) and 
Eq. \ref{Eq_FUV_WISE4} (FUV+\emph{WISE} 4), 
as indicated on top of the panels.
}
\label{COMPARISON_TRACERS_GALS_ALL}
\end{figure}

\begin{figure*}
\centering
\includegraphics[width=0.49\textwidth]{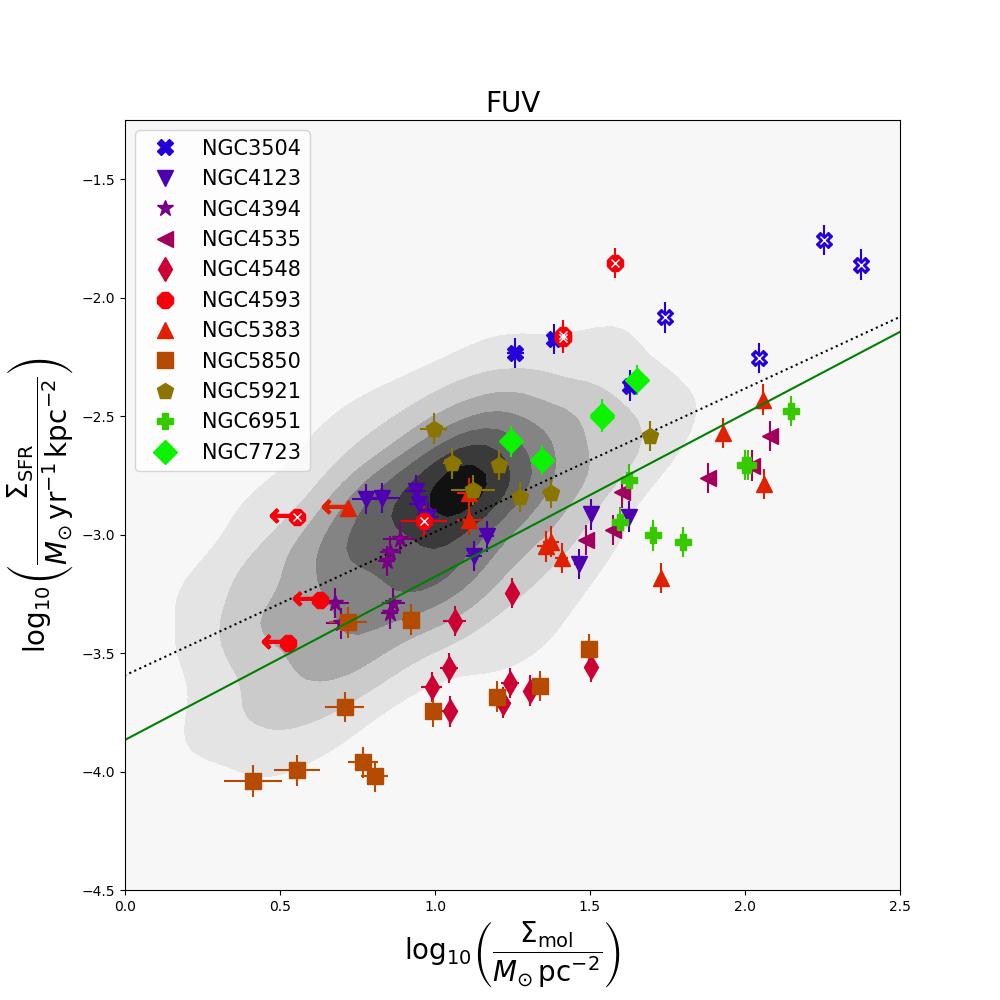}
\includegraphics[width=0.49\textwidth]{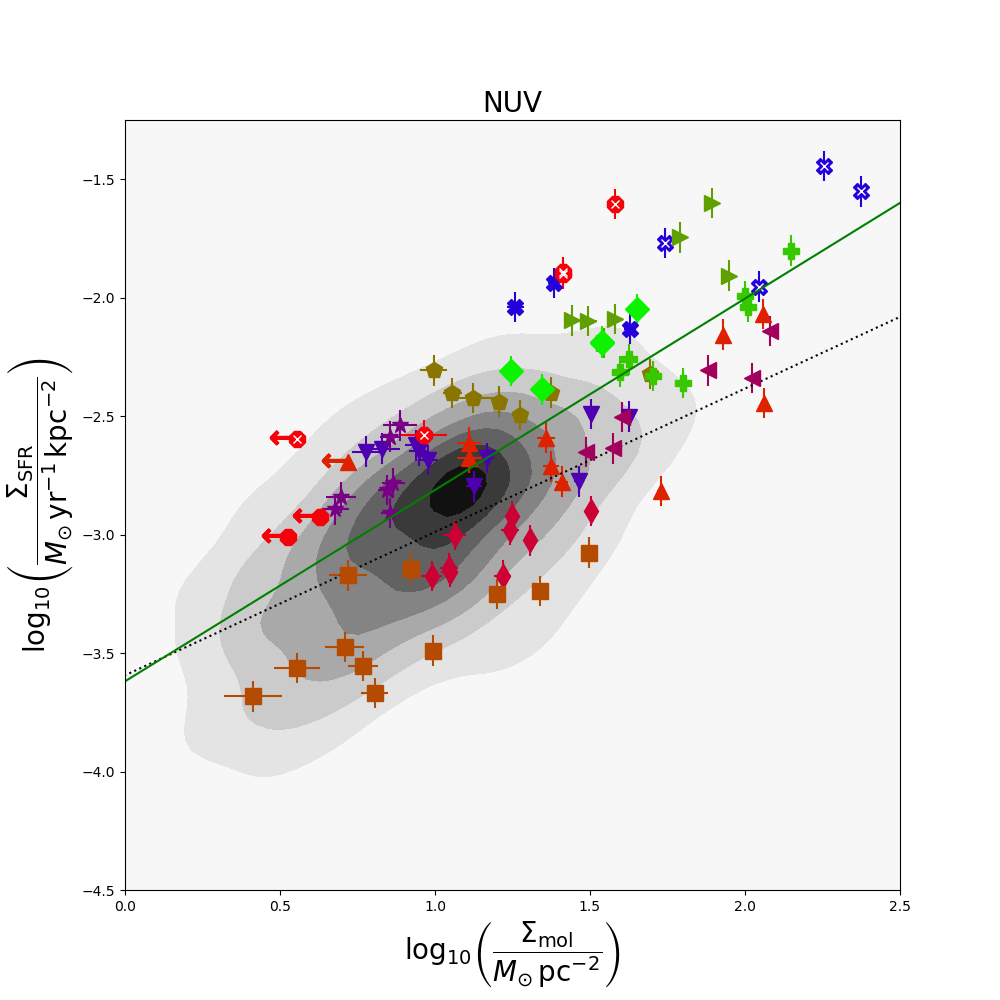}
\includegraphics[width=0.49\textwidth]{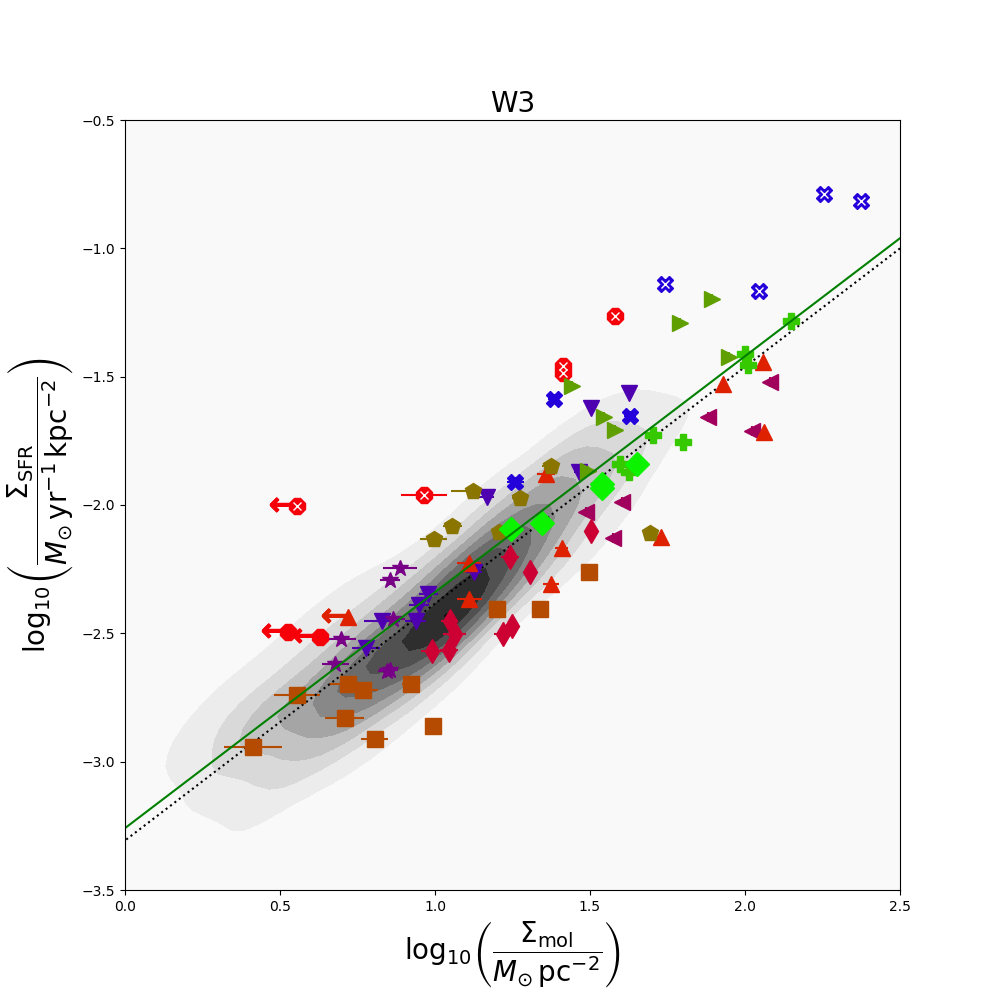}
\includegraphics[width=0.49\textwidth]{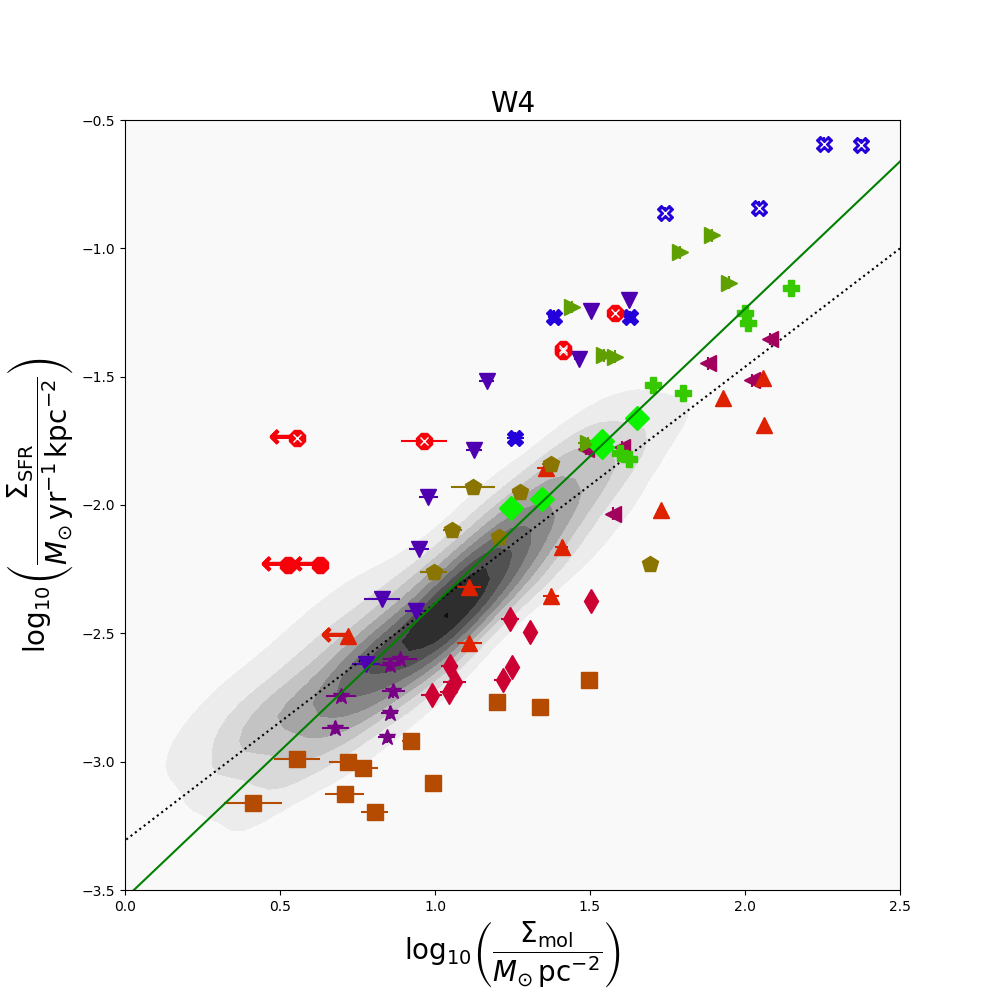}
\caption{
As in the left panels of Fig.~\ref{PLOTS_ALLTOGETHER_B}, but using direct SFR recipes corresponding to 
Eq. \ref{Eq_FUV} (FUV), 
Eq. \ref{Eq_NUV} (NUV), Eq. \ref{Eq_WISE3} (\emph{WISE} 3), and Eq. \ref{Eq_WISE4} (\emph{WISE} 4), as indicated on top of the panels.  
The grey contours trace the measurements from \citet[][]{2011ApJ...730L..13B} 
using FUV (upper panels) and \emph{Spitzer} 22 $\mu$m \citep[Eq. 1 in][]{2008AJ....136.2846B}.
}
\label{COMPARISON_TRACERS_GALS_ALL_2}
\end{figure*}

%
%
Here we show the relationship between star formation rate and molecular gas 
surface densities (Sects. \ref{ap_photometry} and \ref{SFE_sect}) using a variety of 
hybrid (Fig.~\ref{COMPARISON_TRACERS_GALS_ALL}) and direct (Fig.~\ref{COMPARISON_TRACERS_GALS_ALL_2}) 
recipes for the calculation of SF calibrated by \citet[][]{2019ApJS..244...24L}:

%
%
\begin{equation}\label{Eq_FUV}
{\rm SFR}_{FUV}[M_{\odot}\, {\rm yr^{-1}}] = \frac{L_{FUV}{\rm [erg\,s^{-1}]}}{10^{43.42}},
\end{equation}
\begin{equation}\label{Eq_NUV}
{\rm SFR}_{NUV}[M_{\odot}\, {\rm yr^{-1}}] = \frac{L_{NUV}{\rm [erg\,s^{-1}]}}{10^{43.24}},
\end{equation}
\begin{equation}\label{Eq_WISE3}
{\rm SFR}_{W3}[M_{\odot}\, {\rm yr^{-1}}] = \frac{L_{W3}{\rm [erg\,s^{-1}]}}{10^{42.70}},
\end{equation}
\begin{equation}\label{Eq_WISE4}
{\rm SFR}_{W4}[M_{\odot}\, {\rm yr^{-1}}] = \frac{L_{W4}{\rm [erg\,s^{-1}]}}{10^{42.63}}.
\end{equation}
%
%
We take advantage of the combination of UV and infrared (IR) data, which allows us to obtain dust-corrected SFRs:
%
%
\begin{equation}\label{Eq_FUV_WISE3}
{\rm SFR}_{FUV,W3}[M_{\odot}\, {\rm yr^{-1}}] = \frac{L_{FUV}{\rm [erg\,s^{-1}]}}{10^{43.42}}+\frac{L_{W3}{\rm [erg\,s^{-1}]}}{10^{42.79}},
\end{equation}
\begin{equation}\label{Eq_NUV_WISE3}
{\rm SFR}_{NUV,W3}[M_{\odot}\, {\rm yr^{-1}}] = \frac{L_{NUV}{\rm [erg\,s^{-1}]}}{10^{43.24}}+\frac{L_{W3}{\rm [erg\,s^{-1}]}}{10^{42.86}},
\end{equation}
\begin{equation}\label{Eq_FUV_WISE4}
{\rm SFR}_{FUV,W4}[M_{\odot}\, {\rm yr^{-1}}] = \frac{L_{FUV}{\rm [erg\,s^{-1}]}}{10^{43.42}}+\frac{L_{W4}{\rm [erg\,s^{-1}]}}{10^{42.73}},
\end{equation}
\begin{equation}\label{Eq_NUV_WISE4}
{\rm SFR}_{NUV,W4}[M_{\odot}\, {\rm yr^{-1}}] = \frac{L_{NUV}{\rm [erg\,s^{-1}]}}{10^{43.24}}+\frac{L_{W4}{\rm [erg\,s^{-1}]}}{10^{42.79}}.
\end{equation}
%
%

In Table \ref{table_slopes_scatter} we indicate the slope, $y$-intercept, and 1$\sigma$ forward scatter of the 
residuals associated with the linear fit (log space) of the KS law.
For a discussion of the differences of the KS law for the different recipes, 
see Sects. \ref{ap_photometry} and \ref{SFE_sect} in this paper.

\begin{figure*}
\centering
\includegraphics[width=0.49\textwidth]{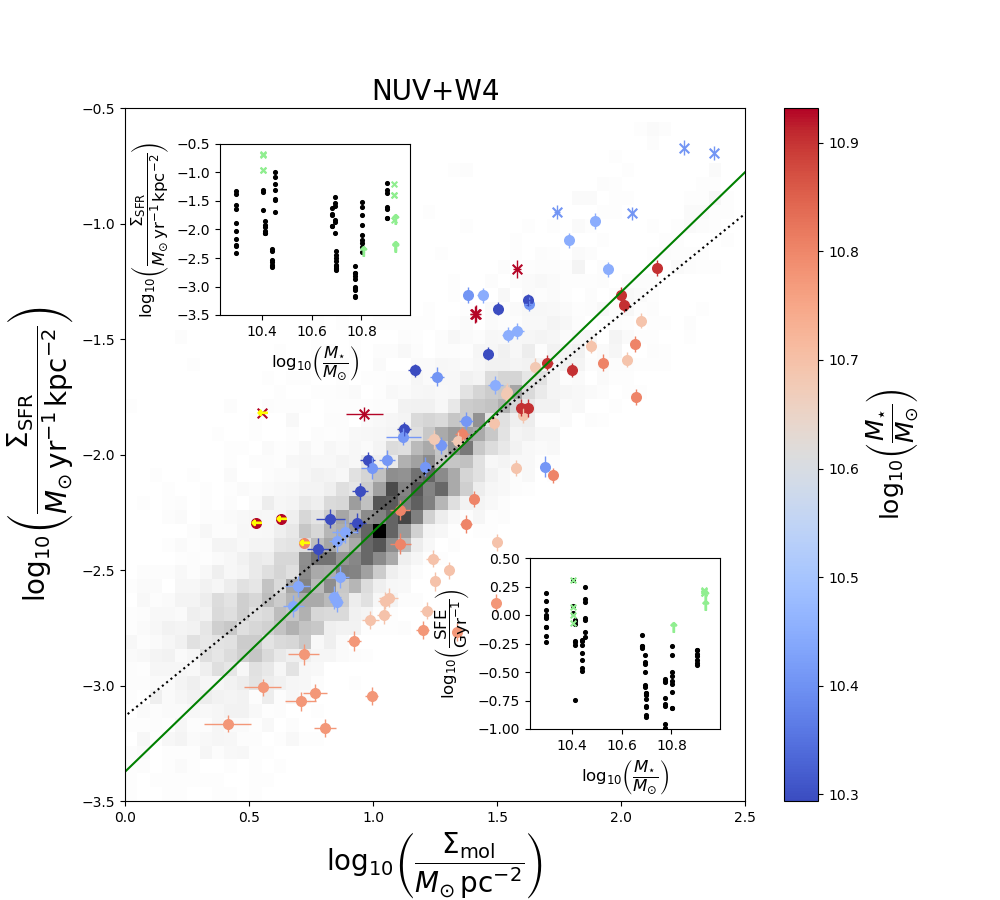}
\includegraphics[width=0.49\textwidth]{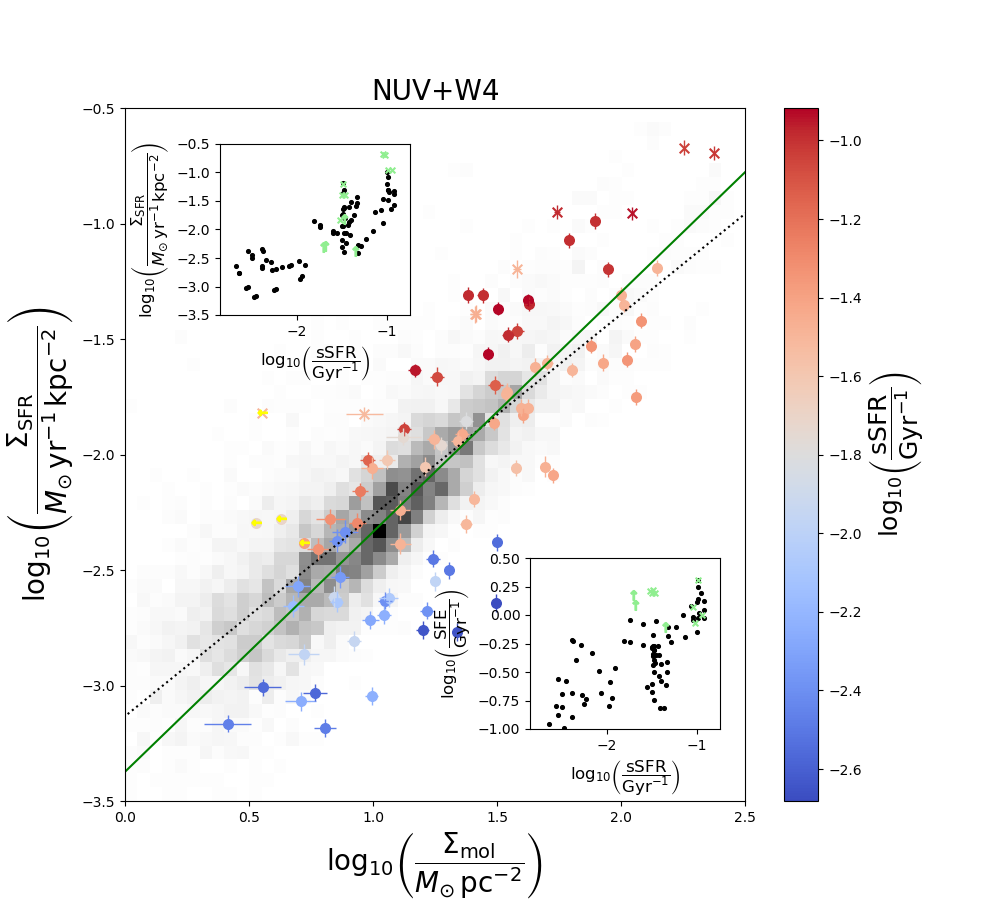}
\includegraphics[width=0.49\textwidth]{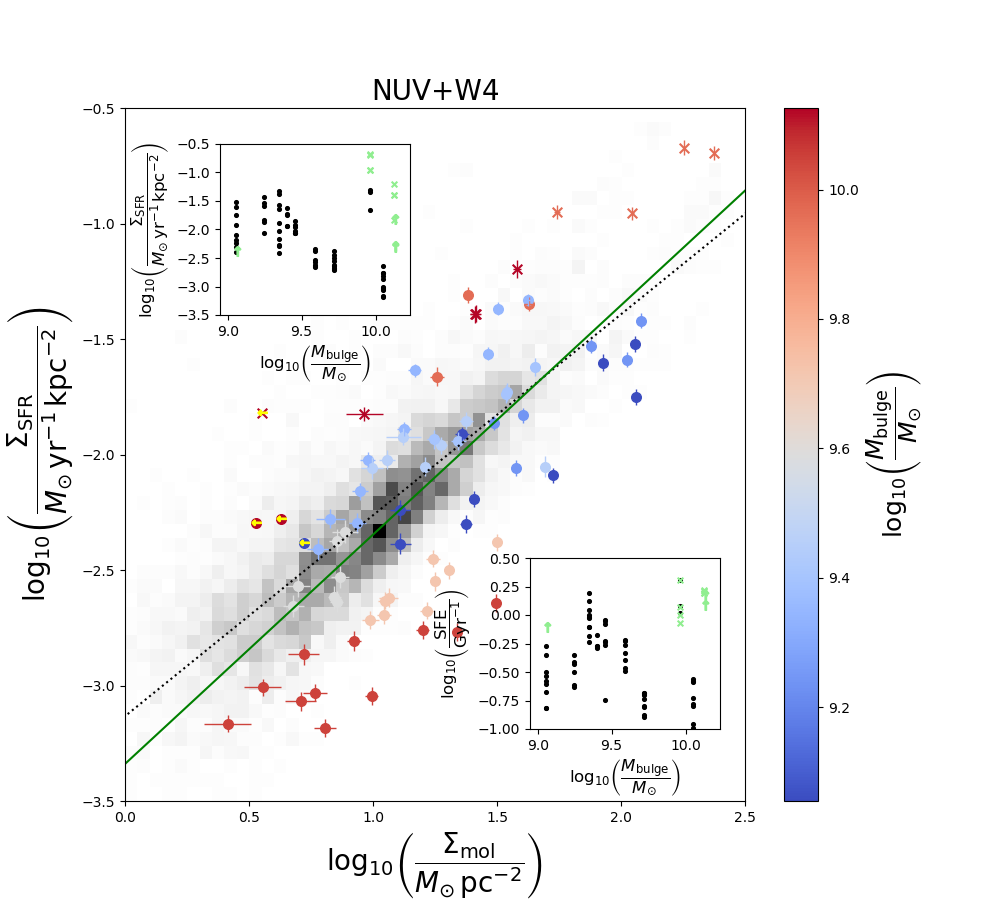}
\includegraphics[width=0.49\textwidth]{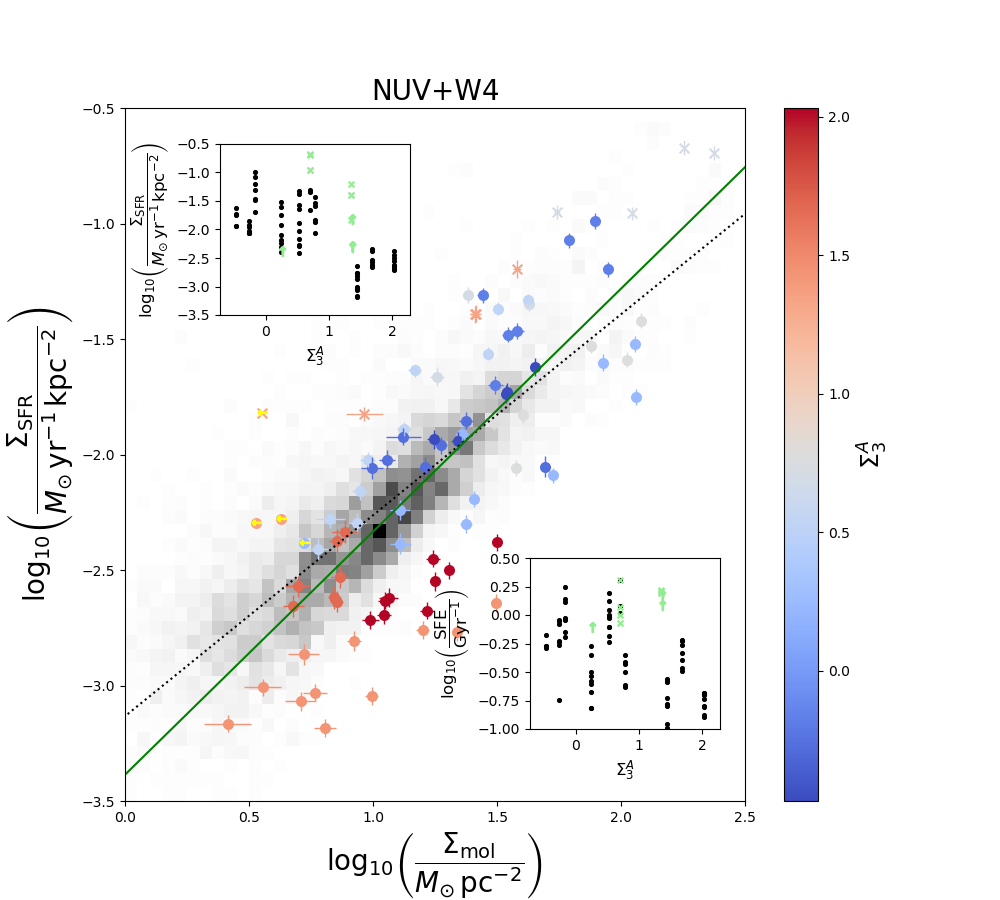}
\caption{
As in Fig.~\ref{BAR_VS_GLOBAL_2}, but calculating SFRs from the combination of NUV and WISE 4.
}
\label{BAR_VS_GLOBAL_2_W4}
\end{figure*}

We also estimate $\Sigma_{\rm SFR}$ from NUV and WISE 4 combined (Eq.~\ref{Eq_NUV_WISE4}) 
to study the KS relation (Fig.~\ref{BAR_VS_GLOBAL_2_W4}) as a function of total stellar mass, specific star formation rate, 
bulge stellar mass, and projected surface density to the third-nearest neighbour galaxy. 
The trends are qualitatively the same as reported in Sect.~\ref{SFE_sect}, 
where $\Sigma_{\rm SFR}$ was estimated from NUV and WISE 3 (Fig.~\ref{BAR_VS_GLOBAL_2}).
%
%
\input{table_KS.txt}
%
%
%
%
%
%
%
%
\clearpage      
\newpage
%
%
\onecolumn
\section{Aperture photometry output}\label{Aperture_photometry_fluxes}
The following data are listed in Table \ref{ap_phot}:
\begin{itemize}
 \item Column 1: Galaxy name;
 \item Column 2: Right ascension of centre of the IRAM-30m pointing (in degrees);
 \item Column 3: Declination of centre of the IRAM-30m pointing (in degrees);
 \item Column 4: Decimal logarithm of the flux within beam in the NUV images (in janskys) and its uncertainty;
 \item Column 5: Decimal logarithm of the flux within beam in the FUV images (in janskys) and its uncertainty;
 \item Column 6: Decimal logarithm of the flux within beam in the \emph{WISE} 1 images (in janskys) and its uncertainty;
 \item Column 7: Decimal logarithm of the flux within beam in the \emph{WISE} 2 images (in janskys) and its uncertainty;
 \item Column 8: Decimal logarithm of the flux within beam in the \emph{WISE} 3 images (in janskys) and its uncertainty;
 \item Column 9: Decimal logarithm of the flux within beam in the \emph{WISE} 4 images (in janskys) and its uncertainty;
 \item Column 10: $\Sigma_{\rm SFR}$ - Star formation surface density (in $M_\odot$ yr$^{-1}$ kpc$^{-2}$) obtained from 
 NUV and \emph{WISE} 3 images using Eqs. \ref{Eq_NUV_WISE3_paper} and \ref{Eq_NUV_WISE3_paper_2}, and its error;
 \item Column 11: $\Sigma_{\star}$ - Stellar mass surface density (in $M_\odot$ pc$^{-2}$) obtained from  \emph{WISE} 1 images using Eq. \ref{eq_sigma_mass}, 
 and its error.
\end{itemize}
\input{./table_SFR.txt}
\twocolumn
%
%
\clearpage
\newpage
%
%
\section{Use of CO(2-1) to estimate molecular gas masses and star formation efficiencies}\label{CO_2_1_masses}
\begin{figure}[hbt!]
\centering
\includegraphics[width=0.37\textwidth]{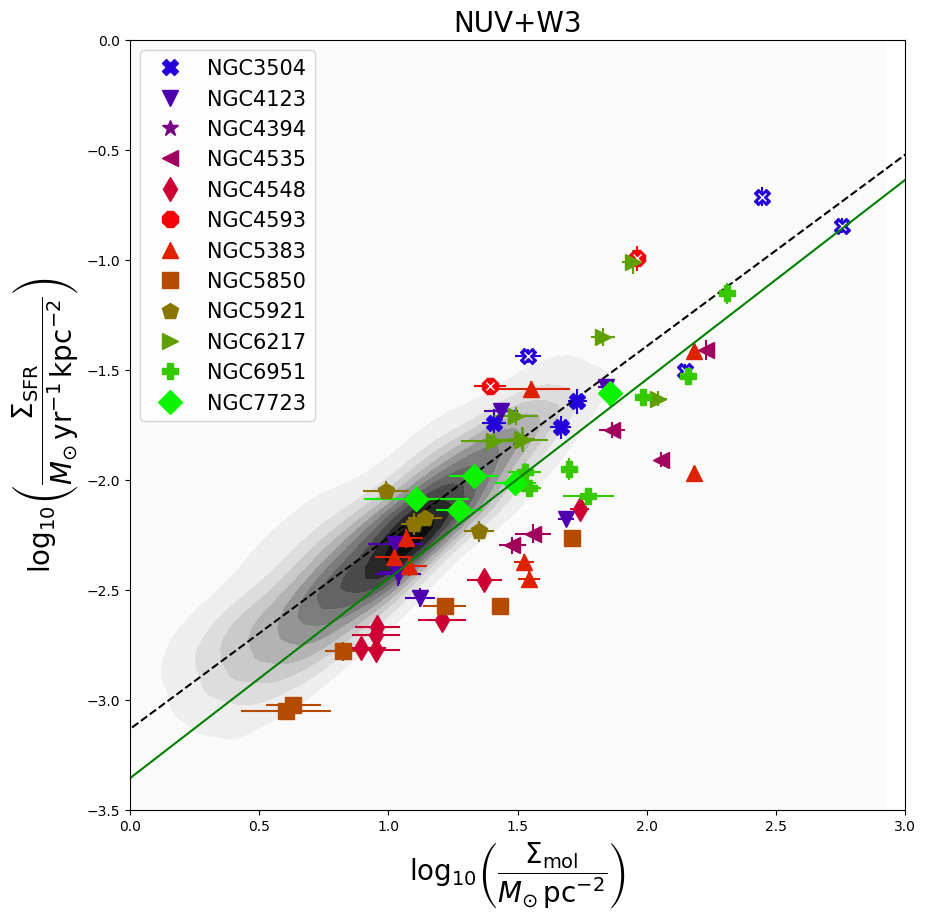}\\
\includegraphics[width=0.37\textwidth]{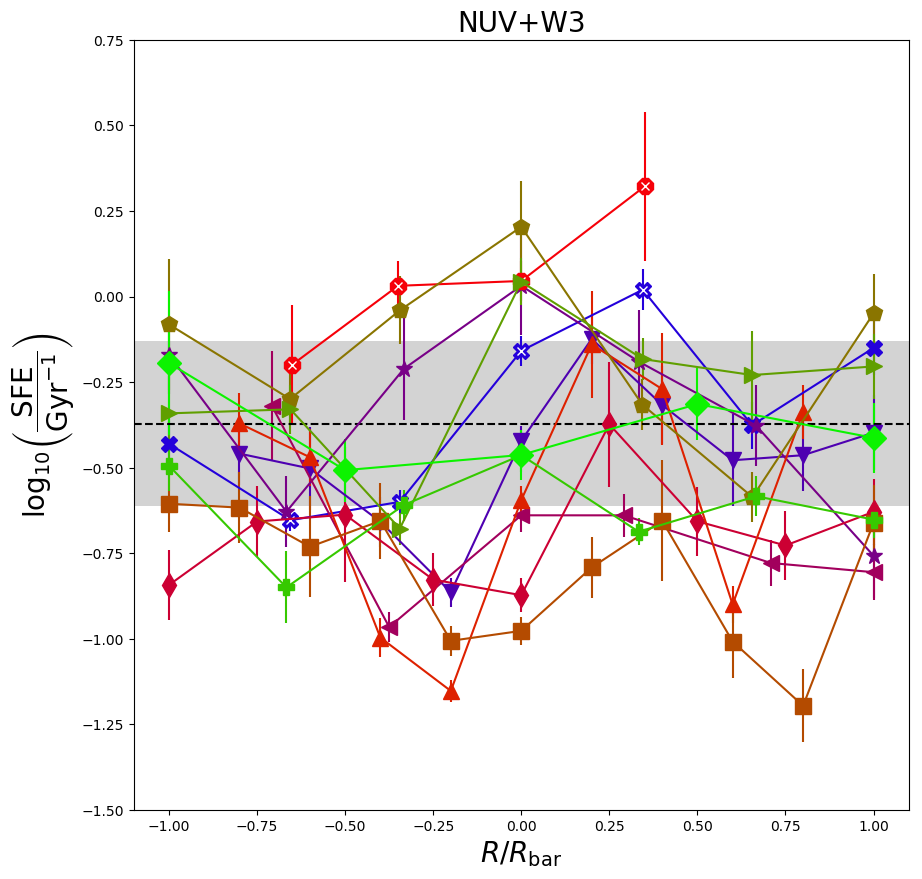}
\caption{
As in Fig.~\ref{PLOTS_ALLTOGETHER_B}, but using CO(2-1) spectra 
and estimating SFEs within $10.75 \arcsec$ apertures (based on the profiles shown in Fig.~\ref{CO_2_1_profiles}). 
Only  the pointings where CO(2-1) was detected are shown.
}
\label{CO_2_1_KS}
\end{figure}

\begin{figure}[hbt!]
\centering
\includegraphics[width=0.4\textwidth]{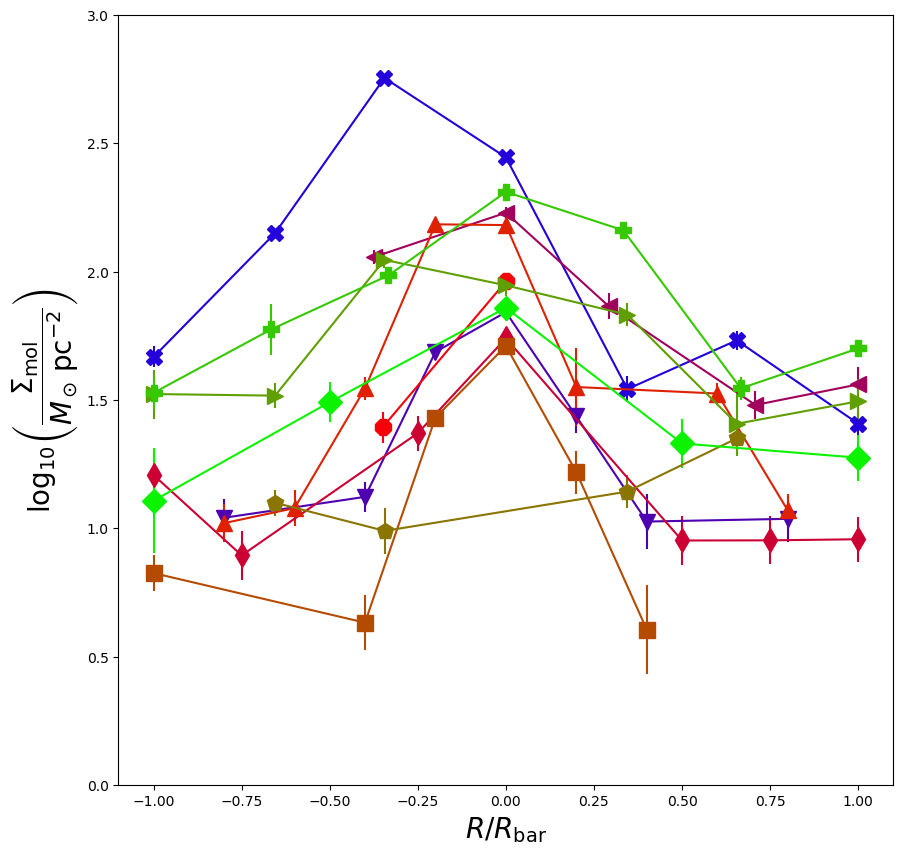}\\
\includegraphics[width=0.4\textwidth]{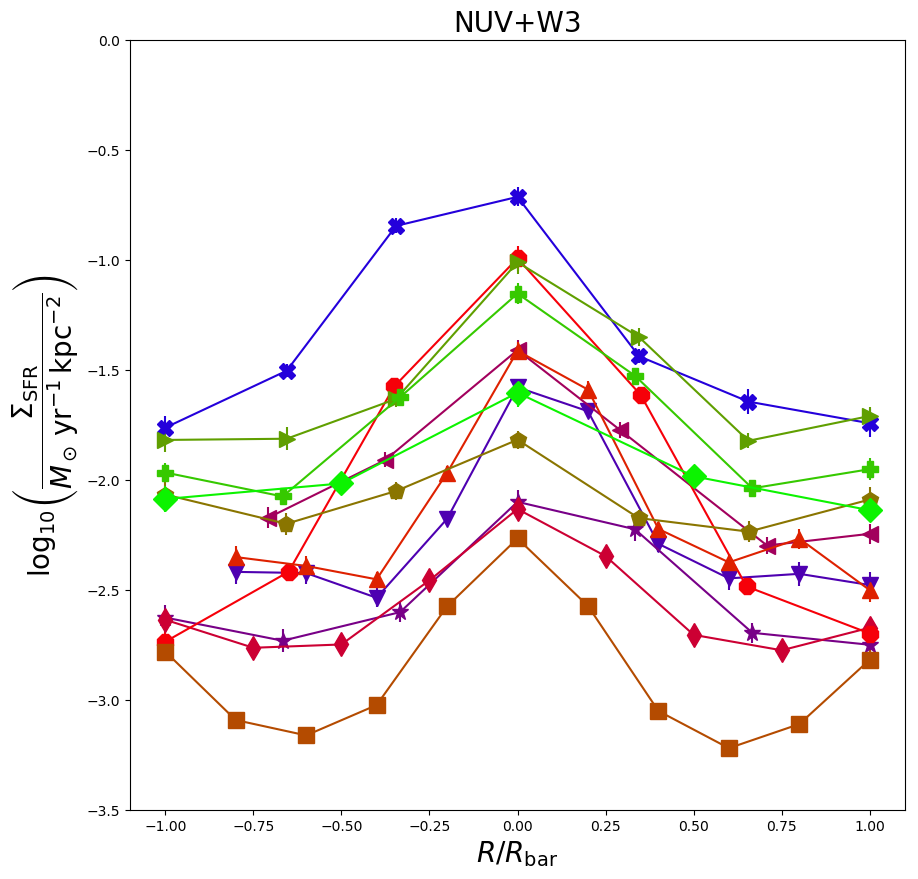}\\
 \includegraphics[width=0.4\textwidth]{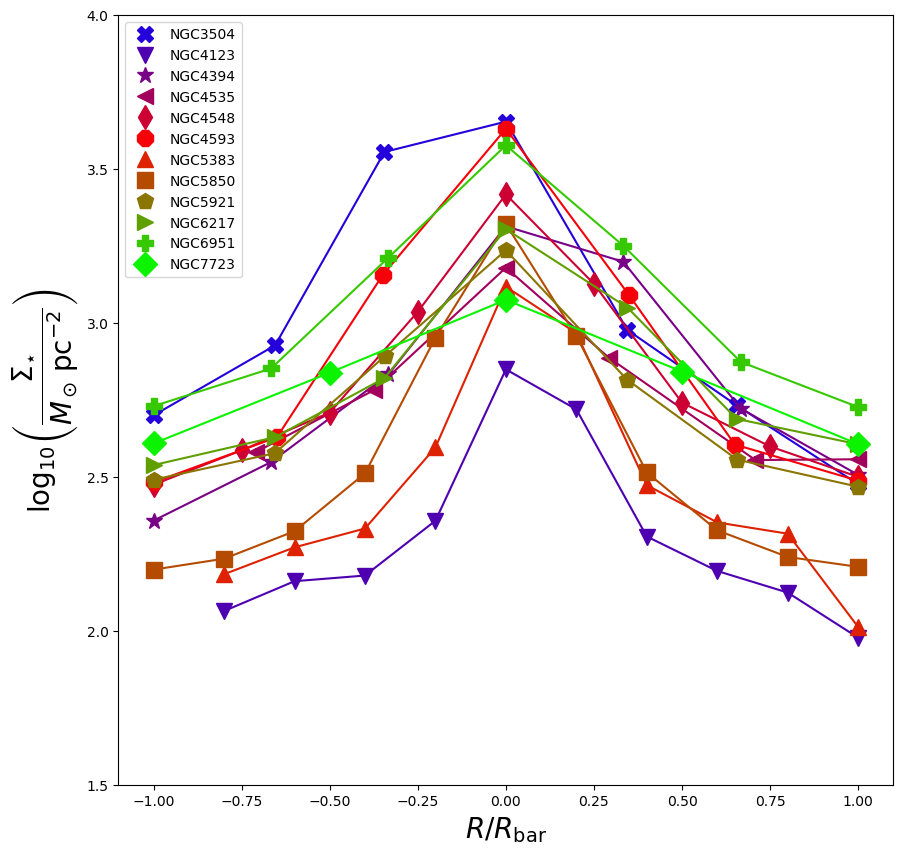}
\caption{
As in Fig.~\ref{PLOTS_ALLTOGETHER_A}, but computing molecular gas surface densities from CO(2-1) spectra (\emph{upper panel}) 
and calculating star formation rate (\emph{central panel}, from the combination of NUV and W3) and stellar mass (\emph{lower panel}) 
surfaces densities within the CO(2-1) pointings (${\rm FWHM}=10.75 \arcsec$). 
In the upper panel  only  the pointings where CO(2-1) was detected are shown.
}
\label{CO_2_1_profiles}
\end{figure}

%
%
In order to test the sensitivity to resolution of the KS law within bars, we use the CO(2-1) spectra instead of CO(1-0). 
The latter, exploited in the previous sections, have better quality and two times lower calibration errors, but also two times worse resolution. 
We calculated $M_{\rm mol}$ and $\Sigma_{\rm mol}$ from $I_{\rm CO(2-1)}$, 
assuming an intrinsic ratio $R_{21}=I_{\rm CO(2-1)}/I_{\rm CO(1-0)}= 0.8$ \citep[][]{2009AJ....137.4670L}, 
and then applied Eqs.~\ref{eq:lco}, \ref{eq_mh2}, and \ref{eq_sigma_h2}. 
We recalculate $\Sigma_{\rm SFR}$ and $\Sigma_{\rm \ast}$ following the steps described in Sect.~\ref{ap_photometry} 
for the smaller CO(2-1) beam (i.e. FWHM=10.75\arcsec, corresponding to $\sim 1-2$ kpc) if CO(2-1) was detected. The resulting profiles are  
shown in Fig.~\ref{CO_2_1_profiles}. These are more centrally peaked, as expected from the higher resolution, 
and the SF dips in quiescent bars are more clearly resolved. The $\Sigma_{\rm SFR}$-$\Sigma_{\rm mol}$ relation and SFE profiles 
(Fig.~\ref{CO_2_1_KS}) are similar to the ones reported in Sect.~\ref{SFE_sect}. 
The mean SFEs are 0.12 dex lower when using CO(2-1) to estimate $\Sigma_{\rm mol}$ as compared to CO(1-0); 
however, the differences in the CO(2-1)-based SFE amplitudes are  dependent on the adopted $R_{21}$, 
which is not necessarily constant in and across galaxies (see next Sect.~\ref{CO_2_1}). 
Despite the poorer data quality, the analysis of the KS law with CO(2-1) spectra reinforces the conclusions of this work: 
 low SFE in the massive galaxies NGC~4548 and NGC~5850, 
which are known to lie in denser environments and host massive bulges (Sect.~\ref{SFE_sect}).
%
%
%
%
%
%
\clearpage
\newpage
%
%
\section{Gas excitation traced from CO(2-1) and CO(1-0) spectra}\label{CO_2_1}
\begin{figure}
\centering
\includegraphics[width=0.49\textwidth]{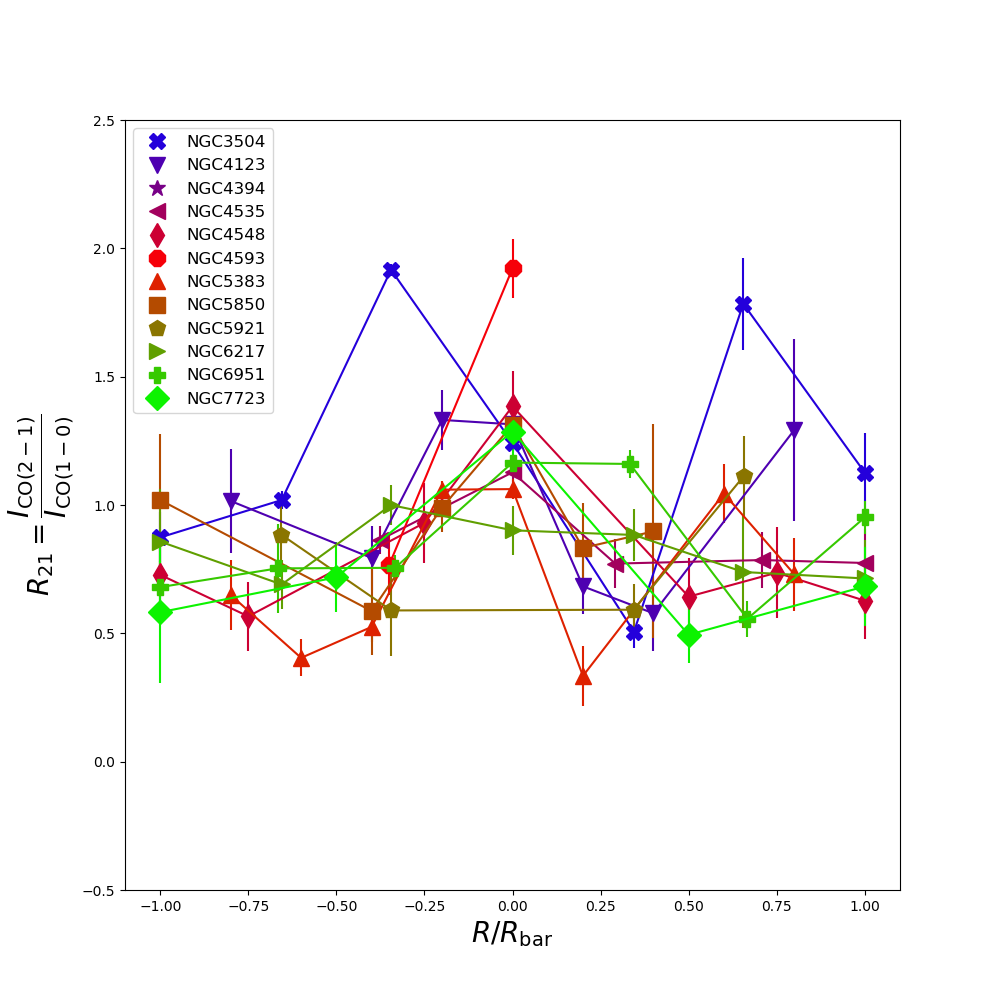}
\includegraphics[width=0.49\textwidth]{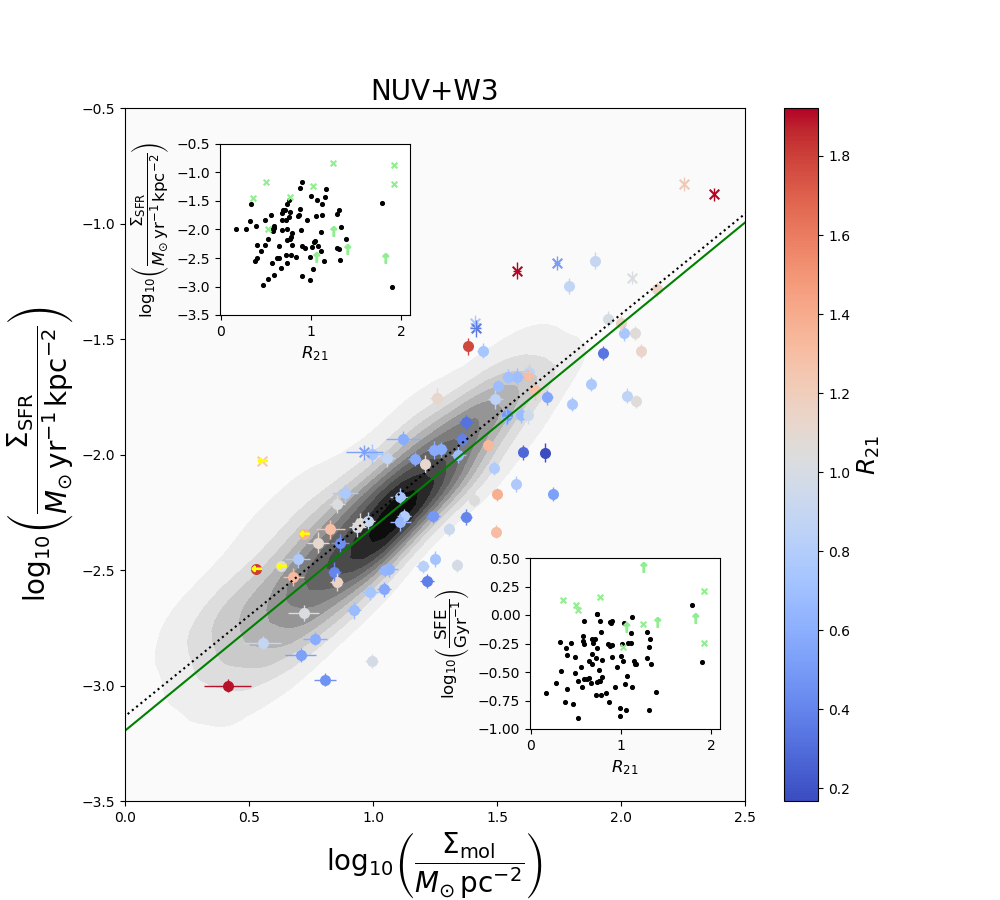}
\caption{
Gas excitation across bars. 
\emph{Upper panel}: Ratio of the velocity-integrated line intensity of CO(2-1) to CO(1-0) spectra along bars 
of each of the 12 galaxies in our sample (see legend). 
\emph{Lower panel}: As in Fig.~\ref{BAR_VS_GLOBAL_2}, but colour-coded by $R_{21}$. 
The insets show $\Sigma_{\rm SFR}$ (upper left corner) and SFE (lower right corner) 
vs $R_{21}$; the green points indicate non-reliable measurements. 
The grey contours trace the measurements from \citet[][]{2011ApJ...730L..13B}.
}
\label{plots_R21}
\end{figure}

%
%
In this appendix~we analyse the gas excitation across the bars in our sample. 
We calculate the ratio of the velocity-integrated line intensity 
of the IRAM-30m CO(1-0) and CO(2-1) spectra: $R_{21}=I_{\rm CO(2-1)}/I_{\rm CO(1-0)}$. 
The mean value is $R_{21}= 0.89 \pm 0.03$ $(\sigma=0.33)$. 
In the upper panel of Fig.~\ref{plots_R21} we show $R_{21}$ as a function of radius along the individual bars. 
There is no clear trend of $R_{21}$ along the bars, and no noticeable differences between the different galaxies. 
In the lower panel of Fig.~\ref{plots_R21} we show the KS relation, colour-coded by $R_{21}$. 
There is no correlation between $R_{21}$ and $\Sigma_{\rm SFR}$ or SFE (see inset plots of Fig.~\ref{plots_R21}).

In order to interpret $R_{21}$ we have to take into account several effects. 
$R_{21}$ depends on the excitation temperature and the opacity of the molecular gas. 
In addition, the observed $R_{21}$ depends on the distribution of the molecular gas relative to the beam size 
since in our measurements the beam size of $I_{\rm CO(2-1)}$ is half the beam size of $I_{\rm CO(1-0)}$. 
The intrinsic value, $R_{\rm 21, intrinsic}$, that would be measured with matched aperture observations 
is $R_{\rm 21, intrinsic} > 1$ for optically thin gas, and (the more realistic case) $R_{\rm 21, intrinsic} \sim 0.6 -1$ 
for optically thick gas in thermal equilibrium, where $R_{\rm 21, intrinsic}$ depends on the temperature of the gas. 
Values below $\sim 0.5$ are an indication of subthermal excitation due to a gas density well below the critical density 
so that the excitation temperature is below the kinetic temperature. 

If the source is more extended than the beams, we observe $R_{21} = R_{\rm 21, intrinsic}$. 
If the source is less extended than the beams, the observed ratio is higher than the 
intrinsic ratio with the limiting case of a point-like source, 
for which we would observe $R_{21}  = R_{\rm 21, intrinsic} \cdot \frac{{\rm FWHM(CO(1-0)})^2}{{\rm FWHM(CO(2-1)})^2} = 4 \cdot R_{\rm 21, intrinsic}$ 
\citep[see][]{1997ApJ...478..144S,2019A&A...627A.107L}.

Matched aperture observations of  CO(1-0) and CO(2-1) give  line ratios 
of $R_{21}=0.89\pm0.6$ \citep[][for a small sample of nearby spiral galaxies]{1993A&AS...97..887B}, 
$R_{21} \sim 0.8$ \citep[][for the SINGS sample]{2009AJ....137.4670L}, 
and $R_{21}\sim 0.6-0.9$ \citep[][for four low-luminosity AGNs from the NUclei of GAlaxies, NUGA, survey]{2015A&A...577A.135C}. 
$R_{21}\sim 0.8$ is consistent with optically thick gas with an excitation temperature of $\sim$10 K \citep[][]{2009AJ....137.4670L}. 
It is therefore very likely that a similar situation holds in our sample, meaning that $R_{21}$ $>1$ can be 
interpreted as optically thick thermalised gas with a spatial extension smaller than (at least) the CO(1-0) beam, 
and the values between $R_{21} \sim 0.5 -1$ as optically thick extended gas in thermal and kinetic equilibrium. 
There are a few pointings with a lower $R_{21}$ which indicate subthermal excitation due to a low gas density.

\citet[][]{2020MNRAS.495.3840M} found a tight correlation between the SFE and $R_{21}$ 
(observed with matched beams, so that $R_{21} = R_{\rm 21, intrinsic}$), in the sense that areas with a low SFE show a lower $R_{21}$. 
We cannot confirm this trend. Even though we took into account the different beam sizes of CO(1-0) and CO(2-1), 
a possibly small source size can only mean that our measured $R_{21}$ overestimates $R_{\rm 21, intrinsic}$. 
Thus, the low values of $R_{21}$ indicate that $R_{\rm 21, intrinsic}$ is even lower. 
From Fig.~\ref{plots_R21} (lower panel) it is hard to see how we could produce a trend between $R_{\rm 21, intrinsic}$ and the SFE, 
since the low values ($R_{\rm 21} < 0.5$) are rather evenly distributed in pixels of high and low SFEs. 
%
%
%
%
\end{appendix}
%
%
\end{document}

%% file: sample_table.dat
\begin{table*}
\begin{small}
\caption{
Sample and properties. Column 1: Galaxy identification.
Column 2: Morphological classification.
Column 3: Hubble type.
Column 4: Distance.
Column 5: Disk inclination.
Column 6: Total stellar mass.
Column 7: Total H{\sc\,i} mass.
Column 8: Projected surface density to the third nearest neighbour galaxy.
Column 9: Projected major axis at the isophotal level 25 mag arcsec$^{-2}$ in the $B-$band.
Column 10: Bar radius (calculated via ellipse fitting).
Column 11: Bar normalised $m=2$ Fourier amplitude.
Column 12: Bar deprojected ellipticity.
Column 13: Number of pointings done with IRAM-30m.
Column 14: CO(1-0) IRAM-30m FWHM in kpc. For CO(2-1) the FWHM is one half of the listed value.
}
\label{sample_12_props}
\centering
\begin{tabular}{| c| c| c| c| c| c| c| c| c| c| c| c| c| c|}
\hline\hline
Galaxy & Morphological & $T$ & $D$ & $i$ & log$_{10}$ $M_{\star}$ & log$_{10}$ $M_{\rm HI}$ & $\Sigma_{3}^{A}$ & $D_{25}$ & $R_{\rm bar}$ & $A_{2}$ & $\epsilon_{\rm bar}$ & $N_{\rm CO}$ & $\theta_{\rm beam}$ \\
       &  classification   &   & (Mpc) & ($^{\circ}$) & ($M_{\odot}$)         & ($M_{\odot}$) &             & (")        & (")           &         &      &   &  (kpc) \\
\tiny (1)     &\tiny (2) &\tiny (3) &\tiny (4) & \tiny(5)         & \tiny(6)           &\tiny (7)       & \tiny(8)           &\tiny    (9)     &  \tiny(10)    &  \tiny(11) &  \tiny(12) & \tiny (13) &\tiny (14) \\
\hline
NGC$\,$3504 & (R$\underline{1}$')SA$\underline{\rm B}$($\underline{\rm r}$s,nl)a & 1 & 20.0 & 12.8 & 10.40 &  8.68 &  0.70 & 147.3 & 33.97 & 1.09 & 0.63 & 7 & 2.1 \\
NGC$\,$4123 & SB$\underline{\rm x}$(rs)a$\underline{\rm b}$ & 2 & 21.9 & 46.9 & 10.29 &  9.64 &  0.54 & 189.7 & 52.51 & 0.62 & 0.68 & 10 & 2.3 \\
NGC$\,$4394 & ($\underline{\rm R}$L)SB(rs,bl,nl)0/a & 0 & 16.8 & 30.4 & 10.44 &  8.73 &  1.68 & 208.0 & 45.71 & 0.55 & 0.55 & 7 & 1.8 \\
NGC$\,$4535 & SAB(s)c & 5 & 17.0 & 25.6 & 10.69 &  9.63 &  0.78 & 487.7 & 43.08 & 0.52 & 0.65 & 6 & 1.8 \\
NGC$\,$4548 & SB(rs,bl)$\underline{\rm a}$b & 2 & 16.2 & 39.0 & 10.70 &  8.84 &  2.03 & 329.7 & 57.98 & 0.68 & 0.64 & 9 & 1.7 \\
NGC$\,$4593 & (R')SB(rs,bl)a & 1 & 33.9 & 33.4 & 10.93 &  9.31 &  1.34 & 143.9 & 52.51 & 0.81 & 0.62 & 7 & 3.6 \\
NGC$\,$5383 & SB(r$\underline{\rm s}$,nr)$\underline{\rm a}$b & 1 & 37.6 & 34.7 & 10.80 &  9.88 &  0.24 & 147.3 & 50.16 & 0.81 & 0.57 & 10 & 3.9 \\
NGC$\,$5850 & (R')SB(r,bl,nr,nb)$\underline{\rm a}$b & 2 & 23.1 & 36.0 & 10.77 &  9.27 &  1.44 & 198.7 & 62.76 & 0.74 & 0.65 & 11 & 2.4 \\
NGC$\,$5921 & SB($\underline{\rm r}$s)b & 3 & 19.9 & 33.5 & 10.41 &  9.36 & -0.26 & 181.2 & 45.55 & 0.40 & 0.73 & 7 & 2.1 \\
NGC$\,$6217 & (R'L)SB(rs)b & 3 & 23.9 & 23.5 & 10.45 &  9.75 & -0.17 & 134.3 & 36.76 & 0.56 & 0.74 & 7 & 2.5 \\
NGC$\,$6951 & SAB(rs)bc & 4 & 23.5 & 39.4 & 10.90 &  9.51 & - & 189.7 & 44.01 & - & 0.54 & 7 & 2.5 \\
NGC$\,$7723 & (R')SB(rs)a$\underline{\rm b}$ & 3 & 27.4 & 47.8 & 10.68 &  8.95 & -0.47 & 189.7 & 19.13 & 0.38 & 0.58 & 5 & 2.9 \\
\hline
\end{tabular}
\end{small}
\end{table*}

%% file: table_SFE_bars.txt
\begin{table*}
\begin{small}
\begin{center}
\caption{
Statistics of the star formation efficiency (SFE) along bars.
For each radial interval (2$^{\rm nd}$-5$^{\rm th}$ columns) and hybrid star formation tracer (1$^{\rm st}$ column, see text for details) we indicate the mean SFE.
In parenthesis we show standard deviation ($\sigma$). In brackets we indicate the number of elements (i.e. IRAM-30m pointings with CO(1-0) detection) within the radial bin.
}
\label{table_SFE_bars}
\centering
\begin{tabular}{| c| c| c| c| c|}
\hline\hline
log$_{10}\Bigg(\dfrac{SFE}{{\rm Gyr}^{-1}}\Bigg)$ & |$R/R_{\rm bar}$|<1/4 & 1/4<|$R/R_{\rm bar}$|<1/2 & 1/2<|$R/R_{\rm bar}$|<3/4 & 3/4<|$R/R_{\rm bar}$| \\
\hline
NUV+W3 & -0.41 ($\sigma$=0.30) [$N$=18] & -0.29 ($\sigma$=0.28) [$N$=20] & -0.33 ($\sigma$=0.25) [$N$=19] & -0.39 ($\sigma$=0.23) [$N$=24] \\ 
NUV+W4 & -0.39 ($\sigma$=0.45) [$N$=18] & -0.21 ($\sigma$=0.33) [$N$=20] & -0.25 ($\sigma$=0.34) [$N$=19] & -0.41 ($\sigma$=0.30) [$N$=24] \\ 
FUV+W3 & -0.44 ($\sigma$=0.29) [$N$=17] & -0.32 ($\sigma$=0.29) [$N$=18] & -0.38 ($\sigma$=0.24) [$N$=17] & -0.44 ($\sigma$=0.22) [$N$=22] \\ 
FUV+W4 & -0.45 ($\sigma$=0.46) [$N$=17] & -0.26 ($\sigma$=0.36) [$N$=18] & -0.32 ($\sigma$=0.34) [$N$=17] & -0.49 ($\sigma$=0.31) [$N$=22] \\ 
\hline
\end{tabular}
\end{center}
\end{small}
\end{table*}

%% file: table_H2.txt
\begin{center}
\begin{small}
\begin{longtable}{| c| c| c| c| c| c| c| c| c| c|}
\hline
\multicolumn{1}{|c}{\centering Galaxy} & \multicolumn{1}{|c}{\centering RA} & \multicolumn{1}{|c|}{\centering DEC} & 
\multicolumn{1}{|c}{\centering $I_{\rm CO(1-0)}$} & \multicolumn{1}{|c}{\centering rms$_{\rm CO(1-0)}$} & \multicolumn{1}{|c|}{\centering $\Delta V$} & 
\multicolumn{1}{|c}{\centering $I_{\rm CO(2-1)}$} & \multicolumn{1}{|c}{\centering rms$_{\rm CO(2-1)}$} & \multicolumn{1}{|c}{\centering $M_{\rm mol}$} & \multicolumn{1}{|c|}{\centering $\Sigma_{\rm mol}$} \tabularnewline
\multicolumn{1}{|c}{\centering } & \multicolumn{1}{|c}{\centering ($^{\circ}$)} & \multicolumn{1}{|c|}{\centering ($^{\circ}$)} & 
\multicolumn{1}{|c}{\centering (K km s$^{-1}$)} & \multicolumn{1}{|c}{\centering (mK)} & \multicolumn{1}{|c|}{\centering (km s$^{-1}$)} & 
\multicolumn{1}{|c}{\centering (K km s$^{-1}$)} & \multicolumn{1}{|c}{\centering (mK)} & \multicolumn{1}{|c}{\centering ($M_{\odot}$)} & \multicolumn{1}{|c|}{\centering ($M_{\odot}$ pc$^{-2}$)} \tabularnewline
\multicolumn{1}{|c}{\centering \tiny (1)} & \multicolumn{1}{|c}{\centering \tiny (2)} & \multicolumn{1}{|c|}{\centering \tiny (3)} & 
\multicolumn{1}{|c}{\centering \tiny (4)} & \multicolumn{1}{|c}{\centering \tiny (5)} & \multicolumn{1}{|c|}{\centering \tiny (6)} & 
\multicolumn{1}{|c}{\centering \tiny (7)} & \multicolumn{1}{|c}{\centering \tiny (8)} & \multicolumn{1}{|c}{\centering \tiny (9)} & \multicolumn{1}{|c|}{\centering \tiny (10)} \tabularnewline
\hline
NGC3504 & 165.79059 &  27.98028 & 10.03$\pm$0.59 &  11.34 & 260 &   8.77$\pm$0.83 &  15.92 &  8.32$\pm$1.25 &  42.58$\pm$1.84 \\
  & 165.79248 &  27.97750 & 26.11$\pm$0.49 &   9.06 & 283 &  26.62$\pm$0.77 &  14.10 &  8.74$\pm$1.73 & 110.81$\pm$1.54 \\
  & 165.79436 &  27.97500 & 55.88$\pm$0.58 &  10.43 & 293 & 107.10$\pm$0.68 &  12.32 &  9.07$\pm$1.99 & 237.18$\pm$1.80 \\
  & 165.79625 &  27.97222 & 42.30$\pm$0.53 &   9.41 & 300 &  52.53$\pm$0.68 &  12.22 &  8.95$\pm$1.91 & 179.51$\pm$1.64 \\
  & 165.79814 &  27.96944 & 12.97$\pm$0.52 &   9.54 & 281 &   6.57$\pm$0.76 &  14.02 &  8.43$\pm$1.42 &  55.06$\pm$1.61 \\
  & 165.80002 &  27.96694 &  5.70$\pm$0.32 &   9.54 & 105 &  10.17$\pm$0.85 &  25.66 &  8.08$\pm$1.28 &  24.20$\pm$0.99 \\
  & 165.80191 &  27.96417 &  4.28$\pm$0.37 &   9.23 & 155 &   4.82$\pm$0.52 &  12.92 &  7.95$\pm$1.10 &  18.15$\pm$1.16 \\
\hline
NGC4123 & 182.03407 &   2.88211 &  2.91$\pm$0.32 &   9.86 &  95 &   2.96$\pm$0.49 &  15.46 &  7.71$\pm$1.00 &   8.65$\pm$0.70 \\
  & 182.03685 &   2.88128 &  2.98$\pm$0.30 &  11.85 & 114 & <  3.23 &  23.52 &  7.72$\pm$1.04 &   8.87$\pm$0.66 \\
  & 182.03990 &   2.88044 &  4.49$\pm$0.38 &   7.95 & 144 &   3.57$\pm$0.48 &  12.34 &  7.90$\pm$1.10 &  13.36$\pm$0.84 \\
  & 182.04269 &   2.87961 &  9.83$\pm$0.50 &   8.83 & 226 &  13.08$\pm$0.92 &  18.91 &  8.24$\pm$1.32 &  29.22$\pm$1.09 \\
  & 182.04575 &   2.87878 & 14.23$\pm$0.43 &  10.21 & 230 &  18.70$\pm$1.35 &  27.63 &  8.40$\pm$1.54 &  42.31$\pm$0.93 \\
  & 182.04881 &   2.87794 & 10.74$\pm$0.48 &   8.77 & 241 &   7.34$\pm$1.12 &  22.23 &  8.27$\pm$1.37 &  31.94$\pm$1.04 \\
  & 182.05159 &   2.87711 &  4.94$\pm$0.38 &   9.87 & 175 &   2.86$\pm$0.70 &  16.41 &  7.94$\pm$1.15 &  14.70$\pm$0.83 \\
  & 182.05466 &   2.87628 &  3.20$\pm$0.31 &  10.27 & 146 & <  2.89 &  21.05 &  7.75$\pm$1.05 &   9.53$\pm$0.68 \\
  & 182.05743 &   2.87544 &  2.26$\pm$0.41 &   8.71 & 115 &   2.93$\pm$0.60 &  17.42 &  7.60$\pm$0.81 &   6.73$\pm$0.90 \\
  & 182.06049 &   2.87461 &  2.01$\pm$0.29 &   9.51 &  80 & <  2.24 &  16.31 &  7.55$\pm$0.91 &   5.99$\pm$0.62 \\
\hline
NGC4394 & 186.47528 &  18.22185 &  1.86$\pm$0.16 &   4.54 &  92 & <  0.75 &   5.45 &  7.38$\pm$1.11 &   6.99$\pm$0.44 \\
  & 186.47733 &  18.21935 &  1.27$\pm$0.17 &   5.90 &  79 & <  1.68 &  12.27 &  7.22$\pm$0.93 &   4.77$\pm$0.47 \\
  & 186.47937 &  18.21685 &  1.96$\pm$0.22 &   4.33 & 191 & <  0.87 &   6.33 &  7.40$\pm$1.00 &   7.35$\pm$0.60 \\
  & 186.48141 &  18.21435 &  2.05$\pm$0.34 &   5.09 & 174 & <  1.57 &  11.48 &  7.43$\pm$0.84 &   7.71$\pm$0.95 \\
  & 186.48347 &  18.21185 &  1.91$\pm$0.19 &   5.96 & 189 & <  1.97 &  14.35 &  7.39$\pm$1.04 &   7.17$\pm$0.53 \\
  & 186.48552 &  18.20935 &  1.32$\pm$0.20 &   4.88 & 114 & <  1.02 &   7.47 &  7.23$\pm$0.87 &   4.97$\pm$0.56 \\
  & 186.48756 &  18.20685 &  1.90$\pm$0.16 &   8.09 & 120 & <  2.17 &  15.82 &  7.39$\pm$1.11 &   7.14$\pm$0.44 \\
\hline
NGC4535 & 188.58041 &   8.19330 & 10.21$\pm$0.58 &   9.68 & 224 & <  2.85 &  20.81 &  8.15$\pm$1.27 &  40.08$\pm$1.68 \\
  & 188.58211 &   8.19553 & 26.85$\pm$0.60 &  10.99 & 281 &  23.22$\pm$1.38 &  25.44 &  8.57$\pm$1.66 & 105.35$\pm$1.72 \\
  & 188.58463 &   8.19803 & 30.66$\pm$0.51 &  12.04 & 262 &  34.57$\pm$1.83 &  34.91 &  8.63$\pm$1.79 & 120.34$\pm$1.46 \\
  & 188.58659 &   8.19997 & 19.38$\pm$0.60 &  11.73 & 249 &  14.95$\pm$1.77 &  34.75 &  8.43$\pm$1.52 &  76.04$\pm$1.73 \\
  & 188.58884 &   8.20275 &  7.84$\pm$0.48 &  11.36 & 169 &   6.16$\pm$0.76 &  18.11 &  8.03$\pm$1.24 &  30.78$\pm$1.38 \\
  & 188.59052 &   8.20469 &  9.57$\pm$0.56 &  12.22 & 204 &   7.41$\pm$1.18 &  25.59 &  8.12$\pm$1.25 &  37.56$\pm$1.63 \\
\hline
NGC4548 & 188.84856 &  14.49091 &  5.24$\pm$0.31 &   9.89 & 154 &   3.81$\pm$0.83 &  20.63 &  7.75$\pm$1.26 &  17.70$\pm$0.76 \\
  & 188.85143 &  14.49230 &  3.28$\pm$0.30 &   7.53 & 146 &   1.85$\pm$0.41 &  10.41 &  7.55$\pm$1.08 &  11.08$\pm$0.74 \\
  & 188.85429 &  14.49369 &  4.89$\pm$0.41 &   6.43 & 356 & <  1.83 &  13.37 &  7.72$\pm$1.11 &  16.52$\pm$1.03 \\
  & 188.85716 &  14.49508 &  5.97$\pm$0.35 &   8.14 & 323 &   5.56$\pm$0.87 &  14.98 &  7.81$\pm$1.25 &  20.19$\pm$0.88 \\
  & 188.86003 &  14.49647 &  9.40$\pm$0.42 &   7.64 & 360 &  13.01$\pm$1.14 &  18.66 &  8.01$\pm$1.37 &  31.77$\pm$1.05 \\
  & 188.86290 &  14.49786 &  5.17$\pm$0.46 &   7.59 & 308 & <  2.52 &  18.35 &  7.75$\pm$1.09 &  17.49$\pm$1.15 \\
  & 188.86577 &  14.49925 &  3.30$\pm$0.29 &   6.78 & 190 &   2.12$\pm$0.46 &  10.30 &  7.55$\pm$1.10 &  11.15$\pm$0.71 \\
  & 188.86864 &  14.50064 &  2.88$\pm$0.31 &   6.06 & 161 &   2.12$\pm$0.46 &  11.17 &  7.50$\pm$1.01 &   9.75$\pm$0.77 \\
  & 188.87151 &  14.50202 &  3.42$\pm$0.41 &   6.87 & 166 &   2.14$\pm$0.43 &  10.40 &  7.57$\pm$0.97 &  11.56$\pm$1.02 \\
\hline
NGC4593 & 189.90707 &  -5.34977 & < 0.93 &   8.52 & 600 & <  1.68 &  12.23 & < 7.68$\pm$0.44 & <  3.36$\pm$1.43 \\
  & 189.90958 &  -5.34783 &  2.53$\pm$0.59 &   7.19 & 283 & <  1.33 &   9.70 &  8.11$\pm$0.72 &   9.20$\pm$1.58 \\
  & 189.91209 &  -5.34616 &  7.11$\pm$0.47 &   7.87 & 424 &   5.45$\pm$0.76 &  11.46 &  8.56$\pm$1.21 &  25.81$\pm$1.26 \\
  & 189.91460 &  -5.34421 & 10.49$\pm$0.53 &   6.75 & 405 &  20.15$\pm$0.67 &  10.23 &  8.73$\pm$1.32 &  38.10$\pm$1.42 \\
  & 189.91711 &  -5.34227 &  7.14$\pm$0.53 &  10.88 & 433 & <  2.55 &  18.60 &  8.56$\pm$1.16 &  25.92$\pm$1.41 \\
  & 189.91962 &  -5.34060 & < 0.99 &   7.08 & 183 & <  1.22 &   8.87 & < 7.70$\pm$0.62 & <  3.58$\pm$0.84 \\
  & 189.92213 &  -5.33866 & < 1.17 &   8.17 & 118 & <  1.22 &   8.91 & < 7.78$\pm$0.69 & <  4.25$\pm$0.80 \\
\hline
NGC5383 & 209.25691 &  41.85528 &  3.61$\pm$0.46 &  10.69 & 147 &   2.34$\pm$0.39 &  10.01 &  8.35$\pm$0.94 &  12.90$\pm$1.22 \\
  & 209.25989 &  41.85305 &  6.62$\pm$0.51 &  10.40 & 192 &   2.68$\pm$0.43 &   9.64 &  8.61$\pm$1.14 &  23.70$\pm$1.35 \\
  & 209.26324 &  41.85083 & 14.92$\pm$0.60 &   9.51 & 322 &   7.83$\pm$0.80 &  13.77 &  8.97$\pm$1.41 &  53.39$\pm$1.57 \\
  & 209.26622 &  41.84861 & 32.24$\pm$0.70 &  11.48 & 347 &  34.17$\pm$0.73 &  12.19 &  9.30$\pm$1.67 & 115.36$\pm$1.83 \\
  & 209.26958 &  41.84639 & 31.96$\pm$0.65 &  11.86 & 369 &  33.93$\pm$1.04 &  16.79 &  9.30$\pm$1.70 & 114.34$\pm$1.72 \\
  & 209.27293 &  41.84417 & 23.75$\pm$0.67 &  11.46 & 373 &   7.94$\pm$2.76 &  44.22 &  9.17$\pm$1.56 &  84.99$\pm$1.78 \\
  & 209.27592 &  41.84195 &  6.38$\pm$0.58 &  10.31 & 247 & <  2.49 &  18.18 &  8.60$\pm$1.08 &  22.81$\pm$1.53 \\
  & 209.27928 &  41.83972 &  7.17$\pm$0.46 &  11.58 & 223 &   7.49$\pm$0.68 &  14.08 &  8.65$\pm$1.22 &  25.64$\pm$1.21 \\
  & 209.28226 &  41.83750 &  3.60$\pm$0.45 &  10.52 & 180 &   2.63$\pm$0.39 &   8.94 &  8.35$\pm$0.95 &  12.87$\pm$1.19 \\
  & 209.28561 &  41.83528 & < 1.47 &  10.81 & 188 & <  2.03 &  14.78 & < 7.96$\pm$0.61 & <  5.25$\pm$1.25 \\
\hline
NGC5850 & 226.76862 &   1.55108 &  1.49$\pm$0.29 &   7.45 & 102 &   1.52$\pm$0.25 &   7.52 &  7.54$\pm$0.79 &   5.25$\pm$0.74 \\
  & 226.77139 &   1.54969 &  1.45$\pm$0.29 &   7.88 &  93 & <  0.76 &   5.55 &  7.53$\pm$0.78 &   5.11$\pm$0.74 \\
  & 226.77417 &   1.54830 &  1.81$\pm$0.25 &   6.95 & 194 & <  0.84 &   6.15 &  7.62$\pm$0.92 &   6.39$\pm$0.64 \\
  & 226.77696 &   1.54691 &  1.66$\pm$0.25 &   5.51 & 125 &   0.97$\pm$0.24 &   6.63 &  7.58$\pm$0.88 &   5.84$\pm$0.65 \\
  & 226.77974 &   1.54552 &  6.19$\pm$0.40 &   9.18 & 247 &   6.13$\pm$0.44 &   8.69 &  8.16$\pm$1.22 &  21.81$\pm$1.04 \\
  & 226.78252 &   1.54413 &  8.89$\pm$0.41 &   8.78 & 294 &  11.66$\pm$0.52 &   9.44 &  8.31$\pm$1.35 &  31.30$\pm$1.07 \\
  & 226.78529 &   1.54274 &  4.51$\pm$0.37 &   7.96 & 212 &   3.76$\pm$0.73 &  15.43 &  8.02$\pm$1.12 &  15.89$\pm$0.97 \\
  & 226.78807 &   1.54136 &  1.02$\pm$0.24 &   6.44 & 133 &   0.91$\pm$0.37 &   9.82 &  7.37$\pm$0.72 &   3.59$\pm$0.62 \\
  & 226.79085 &   1.53997 &  0.74$\pm$0.22 &   6.73 &  98 & <  1.40 &  10.18 &  7.23$\pm$0.65 &   2.59$\pm$0.56 \\
  & 226.79362 &   1.53858 &  2.80$\pm$0.22 &   6.87 &  98 & <  2.77 &  20.18 &  7.81$\pm$1.14 &   9.88$\pm$0.57 \\
  & 226.79640 &   1.53719 &  2.37$\pm$0.20 &   6.39 &  98 & <  1.59 &  11.57 &  7.74$\pm$1.10 &   8.36$\pm$0.53 \\
\hline
NGC5921 & 230.48247 &   5.06178 & 13.62$\pm$0.90 &  12.00 & 321 & <  2.28 &  16.63 &  8.38$\pm$1.21 &  49.46$\pm$2.41 \\
  & 230.48331 &   5.06483 &  3.14$\pm$0.31 &   7.90 & 111 &   2.77$\pm$0.33 &   9.55 &  7.74$\pm$1.05 &  11.38$\pm$0.82 \\
  & 230.48415 &   5.06761 &  3.65$\pm$0.82 &   9.51 & 147 &   2.15$\pm$0.44 &  11.23 &  7.81$\pm$0.74 &  13.26$\pm$2.19 \\
  & 230.48526 &   5.07067 &  6.51$\pm$0.56 &  15.59 & 155 & <  2.10 &  15.30 &  8.06$\pm$1.10 &  23.65$\pm$1.50 \\
  & 230.48637 &   5.07372 &  5.17$\pm$0.41 &   9.02 & 175 &   3.06$\pm$0.46 &  10.78 &  7.96$\pm$1.14 &  18.77$\pm$1.09 \\
  & 230.48721 &   5.07650 &  4.44$\pm$0.24 &  20.96 &  89 &   4.94$\pm$0.64 &  21.03 &  7.89$\pm$1.29 &  16.13$\pm$0.64 \\
  & 230.48805 &   5.07956 &  2.73$\pm$0.37 &  13.97 &  93 & <  2.01 &  14.65 &  7.68$\pm$0.92 &   9.91$\pm$1.00 \\
\hline
NGC6217 & 248.14055 &  78.20620 &  7.79$\pm$0.76 &  15.09 & 222 &   6.68$\pm$1.47 &  30.50 &  8.34$\pm$1.05 &  31.10$\pm$2.23 \\
  & 248.14734 &  78.20342 &  9.54$\pm$0.79 &   9.14 & 199 &   6.58$\pm$0.75 &  16.44 &  8.43$\pm$1.12 &  38.07$\pm$2.31 \\
  & 248.15413 &  78.20092 & 22.27$\pm$0.76 &  16.06 & 201 &  22.26$\pm$1.60 &  34.89 &  8.79$\pm$1.48 &  88.90$\pm$2.24 \\
  & 248.16228 &  78.19814 & 19.65$\pm$0.65 &  15.74 & 208 &  17.71$\pm$1.77 &  37.89 &  8.74$\pm$1.49 &  78.45$\pm$1.91 \\
  & 248.17043 &  78.19537 & 15.40$\pm$0.68 &  17.27 & 170 &  13.60$\pm$1.43 &  34.00 &  8.63$\pm$1.38 &  61.49$\pm$1.99 \\
  & 248.17722 &  78.19286 &  6.92$\pm$0.45 &  16.65 & 227 &   5.12$\pm$1.48 &  30.33 &  8.29$\pm$1.22 &  27.64$\pm$1.31 \\
  & 248.18401 &  78.19009 &  8.77$\pm$0.68 &  13.93 & 197 &   6.26$\pm$1.23 &  27.05 &  8.39$\pm$1.14 &  35.02$\pm$2.01 \\
\hline
NGC6951 & 309.28613 &  66.10472 & 11.79$\pm$0.31 &  12.95 & 242 &   8.02$\pm$1.16 &  23.13 &  8.43$\pm$1.59 &  39.65$\pm$0.77 \\
  & 309.29367 &  66.10500 & 18.80$\pm$0.63 &  17.12 & 391 &  14.16$\pm$3.23 &  50.51 &  8.63$\pm$1.49 &  63.21$\pm$1.56 \\
  & 309.30121 &  66.10528 & 30.44$\pm$0.55 &  12.34 & 439 &  22.99$\pm$1.53 &  22.52 &  8.84$\pm$1.75 & 102.38$\pm$1.36 \\
  & 309.30875 &  66.10555 & 41.73$\pm$0.54 &   6.17 & 474 &  48.63$\pm$0.48 &   6.82 &  8.98$\pm$1.89 & 140.33$\pm$1.34 \\
  & 309.31628 &  66.10583 & 29.69$\pm$0.86 &   9.88 & 464 &  34.44$\pm$1.23 &  17.64 &  8.83$\pm$1.55 &  99.86$\pm$2.12 \\
  & 309.32382 &  66.10611 & 15.00$\pm$1.04 &   8.09 & 355 &   8.35$\pm$0.87 &  14.28 &  8.53$\pm$1.19 &  50.44$\pm$2.58 \\
  & 309.33139 &  66.10639 & 12.54$\pm$0.63 &   7.70 & 230 &  11.94$\pm$0.55 &  11.23 &  8.45$\pm$1.32 &  42.18$\pm$1.57 \\
\hline
NGC7723 & 354.73221 & -12.96361 &  6.02$\pm$0.49 &  16.66 & 140 &   3.50$\pm$1.64 &  42.96 &  8.21$\pm$1.12 &  17.60$\pm$1.06 \\
  & 354.73508 & -12.96222 & 11.80$\pm$0.62 &  12.15 & 268 &   8.48$\pm$1.54 &  29.13 &  8.50$\pm$1.30 &  34.52$\pm$1.33 \\
  & 354.73792 & -12.96083 & 15.37$\pm$0.54 &  12.87 & 305 &  19.77$\pm$1.58 &  27.96 &  8.62$\pm$1.47 &  44.97$\pm$1.17 \\
  & 354.74075 & -12.95944 & 11.84$\pm$0.68 &  11.70 & 299 &   5.85$\pm$1.28 &  22.82 &  8.50$\pm$1.27 &  34.64$\pm$1.46 \\
  & 354.74362 & -12.95806 &  7.54$\pm$0.73 &   9.63 & 184 &   5.15$\pm$1.05 &  24.04 &  8.31$\pm$1.05 &  22.07$\pm$1.57 \\
\hline
\caption{
CO integrated spectra and molecular gas.
}
\label{CO_H2_int}
\end{longtable}
\end{small}
\end{center}

%% file: table_KS.txt
\begin{table}[b]
\begin{center}
\begin{small}
\caption{
Slope, $y-$intercept, and 1-$\sigma$ forward scatter of residuals in
linear fit ${\rm log_{10}}\Bigg(\dfrac{\Sigma_{\rm SFR}}{M_{\odot} \, {\rm yr}^{-1} \, {\rm kpc}^{-2}}\Bigg)=m \cdot {\rm log}_{10}\Bigg(\dfrac{\Sigma_{\rm mol}}{M_{\odot} \, {\rm pc}^{-2}}\Bigg)+b$ 
for the different SF tracers and recipes (1$^{\rm st}$ column; see text).
}
\label{table_slopes_scatter}
\centering
\begin{tabular}{| c| c| c| c|}
\hline\hline
 KS & slope (m) & $y-$intercept (b) & scatter (1-$\sigma$) \\
\hline
NUV & 0.81$\pm$0.08 & -3.62$\pm$0.11 & 0.35 \\
FUV & 0.69$\pm$0.10 & -3.87$\pm$0.13 & 0.41 \\
W3 & 0.92$\pm$0.06 & -3.26$\pm$0.08 & 0.25 \\
W4 & 1.15$\pm$0.09 & -3.53$\pm$0.13 & 0.40 \\
NUV+W3 & 0.88$\pm$0.06 & -3.20$\pm$0.09 & 0.27 \\
NUV+W4 & 1.04$\pm$0.09 & -3.37$\pm$0.12 & 0.37 \\
FUV+W3 & 0.86$\pm$0.06 & -3.21$\pm$0.09 & 0.26 \\
FUV+W4 & 1.04$\pm$0.09 & -3.44$\pm$0.13 & 0.39 \\
\hline
\end{tabular}
\end{small}
\end{center}
\end{table}

%% file: table_SFR.txt
\begin{center}
\begin{tiny}
\begin{longtable}{| c| c| c| c| c| c| c| c| c| c| c|}
\hline
\multicolumn{1}{|c}{\centering Galaxy} & \multicolumn{1}{|c}{\centering RA} & \multicolumn{1}{|c|}{\centering DEC} & 
\multicolumn{1}{|c}{\centering log$_{10}$$F_{\rm NUV}$} & \multicolumn{1}{|c}{\centering log$_{10}$$F_{\rm FUV}$} & \multicolumn{1}{|c|}{\centering log$_{10}$$F_{\rm W1}$} & 
\multicolumn{1}{|c}{\centering log$_{10}$$F_{\rm W2}$} & \multicolumn{1}{|c}{\centering log$_{10}$$F_{\rm W3}$} & \multicolumn{1}{|c}{\centering log$_{10}$$F_{\rm W4}$} &
\multicolumn{1}{|c}{\centering log$_{10}$$\Sigma_{\rm SFR}$} & \multicolumn{1}{|c|}{\centering log$_{10}$$\Sigma_{\star}$} \tabularnewline
\multicolumn{1}{|c}{\centering } & \multicolumn{1}{|c}{\centering ($^{\circ}$)} & \multicolumn{1}{|c|}{\centering ($^{\circ}$)} & 
\multicolumn{1}{|c}{\centering (Jy)} & \multicolumn{1}{|c}{\centering (Jy)} & \multicolumn{1}{|c|}{\centering (Jy)} & 
\multicolumn{1}{|c}{\centering (Jy)} & \multicolumn{1}{|c}{\centering (Jy)} & \multicolumn{1}{|c}{\centering (Jy)} & 
\multicolumn{1}{|c}{\centering ($M_{\ast}$ yr$^{-1}$ kpc$^{-2}$)} & \multicolumn{1}{|c|}{\centering ($M_{\ast}$ pc$^{-2}$)} \tabularnewline
\multicolumn{1}{|c}{\centering \tiny (1)} & \multicolumn{1}{|c}{\centering \tiny (2)} & \multicolumn{1}{|c|}{\centering \tiny (3)} & 
\multicolumn{1}{|c}{\centering \tiny (4)} & \multicolumn{1}{|c}{\centering \tiny (5)} & \multicolumn{1}{|c|}{\centering \tiny (6)} & 
\multicolumn{1}{|c}{\centering \tiny (7)} & \multicolumn{1}{|c}{\centering \tiny (8)} & \multicolumn{1}{|c}{\centering \tiny (9)} &
\multicolumn{1}{|c}{\centering \tiny (10)} & \multicolumn{1}{|c|}{\centering \tiny (11)} \tabularnewline
\hline
NGC3504 & 165.79059 &  27.98028 & -2.99$\pm$0.06 & -3.23$\pm$0.06 & -1.88$\pm$0.01 & -2.08$\pm$0.01 & -1.33$\pm$0.02 & -0.75$\pm$0.02 & -1.64$\pm$0.03 &  2.62$\pm$0.01 \\
  & 165.79248 &  27.97750 & -2.81$\pm$0.06 & -3.11$\pm$0.06 & -1.52$\pm$0.01 & -1.64$\pm$0.01 & -0.84$\pm$0.02 & -0.32$\pm$0.02 & -1.24$\pm$0.03 &  2.99$\pm$0.01 \\
  & 165.79436 &  27.97500 & -2.41$\pm$0.06 & -2.71$\pm$0.06 & -1.18$\pm$0.01 & -1.26$\pm$0.01 & -0.49$\pm$0.02 & -0.08$\pm$0.02 & -0.87$\pm$0.03 &  3.33$\pm$0.01 \\
  & 165.79625 &  27.97222 & -2.30$\pm$0.06 & -2.61$\pm$0.06 & -1.14$\pm$0.01 & -1.22$\pm$0.01 & -0.46$\pm$0.02 & -0.07$\pm$0.02 & -0.83$\pm$0.03 &  3.36$\pm$0.01 \\
  & 165.79814 &  27.96944 & -2.63$\pm$0.06 & -2.94$\pm$0.06 & -1.47$\pm$0.01 & -1.59$\pm$0.01 & -0.81$\pm$0.02 & -0.34$\pm$0.02 & -1.17$\pm$0.03 &  3.03$\pm$0.01 \\
  & 165.80002 &  27.96694 & -2.80$\pm$0.06 & -3.03$\pm$0.06 & -1.83$\pm$0.01 & -2.02$\pm$0.01 & -1.26$\pm$0.02 & -0.74$\pm$0.02 & -1.53$\pm$0.04 &  2.68$\pm$0.01 \\
  & 165.80191 &  27.96417 & -2.90$\pm$0.06 & -3.09$\pm$0.06 & -2.11$\pm$0.01 & -2.31$\pm$0.01 & -1.58$\pm$0.02 & -1.22$\pm$0.02 & -1.75$\pm$0.04 &  2.40$\pm$0.01 \\
\hline
NGC4123 & 182.03407 &   2.88211 & -3.33$\pm$0.06 & -3.51$\pm$0.06 & -2.36$\pm$0.01 & -2.60$\pm$0.01 & -1.97$\pm$0.02 & -1.74$\pm$0.02 & -2.32$\pm$0.04 &  1.99$\pm$0.01 \\
  & 182.03685 &   2.88128 & -3.35$\pm$0.06 & -3.57$\pm$0.06 & -2.27$\pm$0.01 & -2.51$\pm$0.01 & -1.91$\pm$0.02 & -1.49$\pm$0.02 & -2.29$\pm$0.04 &  2.08$\pm$0.01 \\
  & 182.03990 &   2.88044 & -3.50$\pm$0.06 & -3.79$\pm$0.06 & -2.20$\pm$0.01 & -2.43$\pm$0.01 & -1.78$\pm$0.02 & -1.11$\pm$0.02 & -2.27$\pm$0.03 &  2.15$\pm$0.01 \\
  & 182.04269 &   2.87961 & -3.48$\pm$0.06 & -3.82$\pm$0.06 & -1.99$\pm$0.01 & -2.20$\pm$0.01 & -1.39$\pm$0.02 & -0.75$\pm$0.02 & -1.96$\pm$0.03 &  2.36$\pm$0.01 \\
  & 182.04575 &   2.87878 & -3.21$\pm$0.06 & -3.62$\pm$0.06 & -1.76$\pm$0.01 & -1.94$\pm$0.01 & -1.08$\pm$0.02 & -0.52$\pm$0.02 & -1.66$\pm$0.03 &  2.59$\pm$0.01 \\
  & 182.04881 &   2.87794 & -3.20$\pm$0.06 & -3.61$\pm$0.06 & -1.80$\pm$0.01 & -1.99$\pm$0.01 & -1.14$\pm$0.02 & -0.57$\pm$0.02 & -1.70$\pm$0.03 &  2.55$\pm$0.01 \\
  & 182.05159 &   2.87711 & -3.38$\pm$0.06 & -3.71$\pm$0.06 & -2.04$\pm$0.01 & -2.25$\pm$0.01 & -1.49$\pm$0.02 & -0.84$\pm$0.02 & -2.02$\pm$0.03 &  2.31$\pm$0.01 \\
  & 182.05466 &   2.87628 & -3.39$\pm$0.06 & -3.62$\pm$0.06 & -2.22$\pm$0.01 & -2.46$\pm$0.01 & -1.87$\pm$0.02 & -1.29$\pm$0.02 & -2.29$\pm$0.04 &  2.13$\pm$0.01 \\
  & 182.05743 &   2.87544 & -3.34$\pm$0.06 & -3.55$\pm$0.06 & -2.31$\pm$0.01 & -2.54$\pm$0.01 & -1.97$\pm$0.02 & -1.69$\pm$0.02 & -2.32$\pm$0.04 &  2.04$\pm$0.01 \\
  & 182.06049 &   2.87461 & -3.35$\pm$0.06 & -3.55$\pm$0.06 & -2.44$\pm$0.01 & -2.65$\pm$0.01 & -2.07$\pm$0.02 & -1.94$\pm$0.02 & -2.38$\pm$0.04 &  1.92$\pm$0.01 \\
\hline
NGC4394 & 186.47528 &  18.22185 & -3.62$\pm$0.06 & -3.91$\pm$0.06 & -2.16$\pm$0.01 & -2.42$\pm$0.01 & -2.26$\pm$0.02 & -2.33$\pm$0.02 & -2.51$\pm$0.04 &  2.30$\pm$0.01 \\
  & 186.47733 &  18.21935 & -3.70$\pm$0.06 & -4.09$\pm$0.06 & -1.94$\pm$0.01 & -2.20$\pm$0.01 & -2.24$\pm$0.02 & -2.29$\pm$0.02 & -2.53$\pm$0.04 &  2.51$\pm$0.01 \\
  & 186.47937 &  18.21685 & -3.59$\pm$0.06 & -4.09$\pm$0.06 & -1.64$\pm$0.01 & -1.90$\pm$0.01 & -2.06$\pm$0.02 & -2.15$\pm$0.02 & -2.38$\pm$0.04 &  2.81$\pm$0.01 \\
  & 186.48141 &  18.21435 & -3.35$\pm$0.06 & -3.82$\pm$0.06 & -1.40$\pm$0.01 & -1.67$\pm$0.01 & -1.86$\pm$0.02 & -2.03$\pm$0.02 & -2.17$\pm$0.04 &  3.06$\pm$0.01 \\
  & 186.48347 &  18.21185 & -3.39$\pm$0.06 & -3.87$\pm$0.06 & -1.44$\pm$0.01 & -1.71$\pm$0.01 & -1.91$\pm$0.02 & -2.05$\pm$0.02 & -2.21$\pm$0.04 &  3.01$\pm$0.01 \\
  & 186.48552 &  18.20935 & -3.65$\pm$0.06 & -4.18$\pm$0.06 & -1.73$\pm$0.01 & -1.99$\pm$0.01 & -2.14$\pm$0.02 & -2.17$\pm$0.02 & -2.45$\pm$0.04 &  2.72$\pm$0.01 \\
  & 186.48756 &  18.20685 & -3.71$\pm$0.06 & -4.13$\pm$0.06 & -2.00$\pm$0.01 & -2.26$\pm$0.01 & -2.26$\pm$0.02 & -2.24$\pm$0.02 & -2.55$\pm$0.04 &  2.45$\pm$0.01 \\
\hline
NGC4535 & 188.58041 &   8.19330 & -3.33$\pm$0.06 & -3.64$\pm$0.06 & -1.91$\pm$0.01 & -2.13$\pm$0.01 & -1.63$\pm$0.02 & -1.22$\pm$0.02 & -1.99$\pm$0.03 &  2.56$\pm$0.01 \\
  & 188.58211 &   8.19553 & -3.16$\pm$0.06 & -3.53$\pm$0.06 & -1.72$\pm$0.01 & -1.93$\pm$0.01 & -1.35$\pm$0.02 & -0.96$\pm$0.02 & -1.74$\pm$0.03 &  2.75$\pm$0.01 \\
  & 188.58463 &   8.19803 & -2.97$\pm$0.06 & -3.40$\pm$0.06 & -1.56$\pm$0.01 & -1.76$\pm$0.01 & -1.16$\pm$0.02 & -0.80$\pm$0.02 & -1.55$\pm$0.03 &  2.91$\pm$0.01 \\
  & 188.58659 &   8.19997 & -3.13$\pm$0.06 & -3.58$\pm$0.06 & -1.67$\pm$0.01 & -1.88$\pm$0.01 & -1.30$\pm$0.02 & -0.89$\pm$0.02 & -1.69$\pm$0.03 &  2.80$\pm$0.01 \\
  & 188.58884 &   8.20275 & -3.48$\pm$0.06 & -3.84$\pm$0.06 & -1.91$\pm$0.01 & -2.15$\pm$0.01 & -1.66$\pm$0.02 & -1.23$\pm$0.02 & -2.06$\pm$0.03 &  2.56$\pm$0.01 \\
  & 188.59052 &   8.20469 & -3.46$\pm$0.06 & -3.80$\pm$0.06 & -1.98$\pm$0.01 & -2.22$\pm$0.01 & -1.77$\pm$0.02 & -1.48$\pm$0.02 & -2.13$\pm$0.03 &  2.49$\pm$0.01 \\
\hline
NGC4548 & 188.84856 &  14.49091 & -3.68$\pm$0.06 & -4.00$\pm$0.06 & -1.98$\pm$0.01 & -2.23$\pm$0.01 & -2.05$\pm$0.02 & -2.01$\pm$0.02 & -2.45$\pm$0.03 &  2.43$\pm$0.01 \\
  & 188.85143 &  14.49230 & -3.90$\pm$0.06 & -4.32$\pm$0.06 & -1.87$\pm$0.01 & -2.13$\pm$0.01 & -2.14$\pm$0.02 & -2.11$\pm$0.02 & -2.58$\pm$0.03 &  2.54$\pm$0.01 \\
  & 188.85429 &  14.49369 & -3.93$\pm$0.06 & -4.46$\pm$0.06 & -1.69$\pm$0.01 & -1.95$\pm$0.01 & -2.08$\pm$0.02 & -2.06$\pm$0.02 & -2.55$\pm$0.03 &  2.72$\pm$0.01 \\
  & 188.85716 &  14.49508 & -3.78$\pm$0.06 & -4.41$\pm$0.06 & -1.41$\pm$0.01 & -1.65$\pm$0.01 & -1.83$\pm$0.02 & -1.88$\pm$0.02 & -2.32$\pm$0.03 &  3.00$\pm$0.01 \\
  & 188.86003 &  14.49647 & -3.66$\pm$0.06 & -4.31$\pm$0.06 & -1.24$\pm$0.01 & -1.48$\pm$0.01 & -1.68$\pm$0.02 & -1.76$\pm$0.02 & -2.17$\pm$0.03 &  3.17$\pm$0.01 \\
  & 188.86290 &  14.49786 & -3.74$\pm$0.06 & -4.38$\pm$0.06 & -1.36$\pm$0.01 & -1.60$\pm$0.01 & -1.78$\pm$0.02 & -1.82$\pm$0.02 & -2.27$\pm$0.03 &  3.05$\pm$0.01 \\
  & 188.86577 &  14.49925 & -3.92$\pm$0.06 & -4.50$\pm$0.06 & -1.65$\pm$0.01 & -1.90$\pm$0.01 & -2.02$\pm$0.02 & -2.01$\pm$0.02 & -2.50$\pm$0.03 &  2.76$\pm$0.01 \\
  & 188.86864 &  14.50064 & -3.94$\pm$0.06 & -4.40$\pm$0.06 & -1.85$\pm$0.01 & -2.11$\pm$0.01 & -2.14$\pm$0.02 & -2.12$\pm$0.02 & -2.60$\pm$0.03 &  2.56$\pm$0.01 \\
  & 188.87151 &  14.50202 & -3.76$\pm$0.06 & -4.12$\pm$0.06 & -1.96$\pm$0.01 & -2.22$\pm$0.01 & -2.07$\pm$0.02 & -2.07$\pm$0.02 & -2.50$\pm$0.03 &  2.44$\pm$0.01 \\
\hline
NGC4593 & 189.90707 &  -5.34977 & -3.80$\pm$0.06 & -4.24$\pm$0.06 & -2.02$\pm$0.01 & -2.25$\pm$0.01 & -2.10$\pm$0.02 & -1.64$\pm$0.02 & -2.50$\pm$0.03 &  2.42$\pm$0.01 \\
  & 189.90958 &  -5.34783 & -3.37$\pm$0.06 & -3.73$\pm$0.06 & -1.74$\pm$0.01 & -1.87$\pm$0.01 & -1.57$\pm$0.02 & -1.16$\pm$0.02 & -1.99$\pm$0.03 &  2.70$\pm$0.01 \\
  & 189.91209 &  -5.34616 & -2.68$\pm$0.06 & -2.94$\pm$0.06 & -1.34$\pm$0.01 & -1.37$\pm$0.01 & -1.06$\pm$0.02 & -0.81$\pm$0.02 & -1.43$\pm$0.03 &  3.10$\pm$0.01 \\
  & 189.91460 &  -5.34421 & -2.40$\pm$0.06 & -2.64$\pm$0.06 & -1.15$\pm$0.01 & -1.16$\pm$0.01 & -0.87$\pm$0.02 & -0.66$\pm$0.02 & -1.20$\pm$0.04 &  3.29$\pm$0.01 \\
  & 189.91711 &  -5.34227 & -2.69$\pm$0.06 & -2.96$\pm$0.06 & -1.37$\pm$0.01 & -1.40$\pm$0.01 & -1.09$\pm$0.02 & -0.80$\pm$0.02 & -1.45$\pm$0.03 &  3.07$\pm$0.01 \\
  & 189.91962 &  -5.34060 & -3.39$\pm$0.06 & -3.71$\pm$0.06 & -1.77$\pm$0.01 & -1.90$\pm$0.01 & -1.61$\pm$0.02 & -1.15$\pm$0.02 & -2.03$\pm$0.03 &  2.67$\pm$0.01 \\
  & 189.92213 &  -5.33866 & -3.72$\pm$0.06 & -4.06$\pm$0.06 & -2.02$\pm$0.01 & -2.25$\pm$0.01 & -2.12$\pm$0.02 & -1.64$\pm$0.02 & -2.48$\pm$0.03 &  2.42$\pm$0.01 \\
\hline
NGC5383 & 209.25691 &  41.85528 & -3.46$\pm$0.06 & -3.72$\pm$0.06 & -2.34$\pm$0.01 & -2.56$\pm$0.01 & -1.96$\pm$0.02 & -1.94$\pm$0.02 & -2.29$\pm$0.04 &  2.09$\pm$0.01 \\
  & 209.25989 &  41.85305 & -3.50$\pm$0.06 & -3.81$\pm$0.06 & -2.23$\pm$0.01 & -2.46$\pm$0.01 & -1.90$\pm$0.02 & -1.76$\pm$0.02 & -2.27$\pm$0.03 &  2.20$\pm$0.01 \\
  & 209.26324 &  41.85083 & -3.60$\pm$0.06 & -3.96$\pm$0.06 & -2.11$\pm$0.01 & -2.33$\pm$0.01 & -1.72$\pm$0.02 & -1.42$\pm$0.02 & -2.17$\pm$0.03 &  2.32$\pm$0.01 \\
  & 209.26622 &  41.84861 & -3.23$\pm$0.06 & -3.57$\pm$0.06 & -1.82$\pm$0.01 & -2.01$\pm$0.01 & -1.31$\pm$0.02 & -1.09$\pm$0.02 & -1.77$\pm$0.03 &  2.62$\pm$0.01 \\
  & 209.26958 &  41.84639 & -2.85$\pm$0.06 & -3.21$\pm$0.06 & -1.55$\pm$0.01 & -1.74$\pm$0.01 & -1.04$\pm$0.02 & -0.91$\pm$0.02 & -1.47$\pm$0.03 &  2.88$\pm$0.01 \\
  & 209.27293 &  41.84417 & -2.94$\pm$0.06 & -3.35$\pm$0.06 & -1.61$\pm$0.01 & -1.81$\pm$0.01 & -1.13$\pm$0.02 & -0.99$\pm$0.02 & -1.56$\pm$0.03 &  2.82$\pm$0.01 \\
  & 209.27592 &  41.84195 & -3.38$\pm$0.06 & -3.83$\pm$0.06 & -1.91$\pm$0.01 & -2.12$\pm$0.01 & -1.48$\pm$0.02 & -1.26$\pm$0.02 & -1.93$\pm$0.03 &  2.52$\pm$0.01 \\
  & 209.27928 &  41.83972 & -3.56$\pm$0.06 & -3.88$\pm$0.06 & -2.12$\pm$0.01 & -2.34$\pm$0.01 & -1.77$\pm$0.02 & -1.57$\pm$0.02 & -2.20$\pm$0.03 &  2.31$\pm$0.01 \\
  & 209.28226 &  41.83750 & -3.40$\pm$0.06 & -3.61$\pm$0.06 & -2.21$\pm$0.01 & -2.43$\pm$0.01 & -1.82$\pm$0.02 & -1.73$\pm$0.02 & -2.18$\pm$0.03 &  2.22$\pm$0.01 \\
  & 209.28561 &  41.83528 & -3.48$\pm$0.06 & -3.67$\pm$0.06 & -2.44$\pm$0.01 & -2.65$\pm$0.01 & -2.04$\pm$0.02 & -1.92$\pm$0.02 & -2.34$\pm$0.04 &  1.99$\pm$0.01 \\
\hline
NGC5850 & 226.76862 &   1.55108 & -3.95$\pm$0.06 & -4.14$\pm$0.06 & -2.32$\pm$0.01 & -2.58$\pm$0.01 & -2.29$\pm$0.02 & -2.40$\pm$0.02 & -2.69$\pm$0.03 &  2.11$\pm$0.01 \\
  & 226.77139 &   1.54969 & -4.25$\pm$0.06 & -4.50$\pm$0.06 & -2.24$\pm$0.01 & -2.51$\pm$0.01 & -2.42$\pm$0.02 & -2.52$\pm$0.02 & -2.87$\pm$0.03 &  2.18$\pm$0.01 \\
  & 226.77417 &   1.54830 & -4.45$\pm$0.06 & -4.79$\pm$0.06 & -2.13$\pm$0.01 & -2.40$\pm$0.01 & -2.50$\pm$0.02 & -2.59$\pm$0.02 & -2.97$\pm$0.03 &  2.29$\pm$0.01 \\
  & 226.77696 &   1.54691 & -4.33$\pm$0.06 & -4.73$\pm$0.06 & -1.89$\pm$0.01 & -2.15$\pm$0.01 & -2.31$\pm$0.02 & -2.42$\pm$0.02 & -2.80$\pm$0.03 &  2.54$\pm$0.01 \\
  & 226.77974 &   1.54552 & -4.02$\pm$0.06 & -4.41$\pm$0.06 & -1.54$\pm$0.01 & -1.82$\pm$0.01 & -1.99$\pm$0.02 & -2.19$\pm$0.02 & -2.48$\pm$0.03 &  2.88$\pm$0.01 \\
  & 226.78252 &   1.54413 & -3.85$\pm$0.06 & -4.26$\pm$0.06 & -1.39$\pm$0.01 & -1.66$\pm$0.01 & -1.85$\pm$0.02 & -2.08$\pm$0.02 & -2.33$\pm$0.03 &  3.04$\pm$0.01 \\
  & 226.78529 &   1.54274 & -4.03$\pm$0.06 & -4.46$\pm$0.06 & -1.54$\pm$0.01 & -1.81$\pm$0.01 & -2.00$\pm$0.02 & -2.17$\pm$0.02 & -2.48$\pm$0.03 &  2.88$\pm$0.01 \\
  & 226.78807 &   1.54136 & -4.34$\pm$0.06 & -4.77$\pm$0.06 & -1.88$\pm$0.01 & -2.15$\pm$0.01 & -2.33$\pm$0.02 & -2.39$\pm$0.02 & -2.81$\pm$0.03 &  2.54$\pm$0.01 \\
  & 226.79085 &   1.53997 & -4.46$\pm$0.06 & -4.81$\pm$0.06 & -2.13$\pm$0.01 & -2.39$\pm$0.01 & -2.53$\pm$0.02 & -2.56$\pm$0.02 & -3.00$\pm$0.03 &  2.30$\pm$0.01 \\
  & 226.79362 &   1.53858 & -4.27$\pm$0.06 & -4.52$\pm$0.06 & -2.24$\pm$0.01 & -2.50$\pm$0.01 & -2.45$\pm$0.02 & -2.48$\pm$0.02 & -2.89$\pm$0.03 &  2.19$\pm$0.01 \\
  & 226.79640 &   1.53719 & -3.92$\pm$0.06 & -4.13$\pm$0.06 & -2.29$\pm$0.01 & -2.56$\pm$0.01 & -2.29$\pm$0.02 & -2.32$\pm$0.02 & -2.68$\pm$0.03 &  2.13$\pm$0.01 \\
\hline
NGC5921 & 230.48247 &   5.06178 & -3.11$\pm$0.06 & -3.37$\pm$0.06 & -2.02$\pm$0.01 & -2.24$\pm$0.01 & -1.71$\pm$0.02 & -1.64$\pm$0.02 & -1.99$\pm$0.04 &  2.42$\pm$0.01 \\
  & 230.48331 &   5.06483 & -3.19$\pm$0.06 & -3.49$\pm$0.06 & -1.86$\pm$0.01 & -2.09$\pm$0.01 & -1.69$\pm$0.02 & -1.51$\pm$0.02 & -2.01$\pm$0.04 &  2.58$\pm$0.01 \\
  & 230.48415 &   5.06761 & -3.22$\pm$0.06 & -3.60$\pm$0.06 & -1.62$\pm$0.01 & -1.85$\pm$0.01 & -1.55$\pm$0.02 & -1.34$\pm$0.02 & -1.93$\pm$0.03 &  2.82$\pm$0.01 \\
  & 230.48526 &   5.07067 & -3.19$\pm$0.06 & -3.61$\pm$0.06 & -1.48$\pm$0.01 & -1.72$\pm$0.01 & -1.45$\pm$0.02 & -1.25$\pm$0.02 & -1.86$\pm$0.03 &  2.96$\pm$0.01 \\
  & 230.48637 &   5.07372 & -3.29$\pm$0.06 & -3.63$\pm$0.06 & -1.66$\pm$0.01 & -1.89$\pm$0.01 & -1.58$\pm$0.02 & -1.36$\pm$0.02 & -1.98$\pm$0.03 &  2.78$\pm$0.01 \\
  & 230.48721 &   5.07650 & -3.23$\pm$0.06 & -3.49$\pm$0.06 & -1.89$\pm$0.01 & -2.13$\pm$0.01 & -1.71$\pm$0.02 & -1.53$\pm$0.02 & -2.04$\pm$0.04 &  2.54$\pm$0.01 \\
  & 230.48805 &   5.07956 & -3.10$\pm$0.06 & -3.34$\pm$0.06 & -2.04$\pm$0.01 & -2.27$\pm$0.01 & -1.74$\pm$0.02 & -1.67$\pm$0.02 & -2.00$\pm$0.04 &  2.40$\pm$0.01 \\
\hline
NGC6217 & 248.14055 &  78.20620 & -2.93$\pm$0.06 & - & -2.04$\pm$0.01 & -2.24$\pm$0.01 & -1.51$\pm$0.02 & -1.21$\pm$0.02 & -1.76$\pm$0.04 &  2.44$\pm$0.01 \\
  & 248.14734 &  78.20342 & -2.92$\pm$0.06 & - & -1.90$\pm$0.01 & -2.10$\pm$0.01 & -1.35$\pm$0.02 & -0.87$\pm$0.02 & -1.66$\pm$0.03 &  2.58$\pm$0.01 \\
  & 248.15413 &  78.20092 & -2.74$\pm$0.06 & - & -1.68$\pm$0.01 & -1.86$\pm$0.01 & -1.07$\pm$0.02 & -0.59$\pm$0.02 & -1.41$\pm$0.03 &  2.80$\pm$0.01 \\
  & 248.16228 &  78.19814 & -2.43$\pm$0.06 & - & -1.47$\pm$0.01 & -1.63$\pm$0.01 & -0.84$\pm$0.02 & -0.40$\pm$0.02 & -1.16$\pm$0.03 &  3.01$\pm$0.01 \\
  & 248.17043 &  78.19537 & -2.58$\pm$0.06 & - & -1.56$\pm$0.01 & -1.73$\pm$0.01 & -0.94$\pm$0.02 & -0.47$\pm$0.02 & -1.27$\pm$0.03 &  2.92$\pm$0.01 \\
  & 248.17722 &  78.19286 & -2.93$\pm$0.06 & - & -1.80$\pm$0.01 & -1.98$\pm$0.01 & -1.18$\pm$0.02 & -0.68$\pm$0.02 & -1.55$\pm$0.03 &  2.68$\pm$0.01 \\
  & 248.18401 &  78.19009 & -3.02$\pm$0.06 & - & -1.98$\pm$0.01 & -2.16$\pm$0.01 & -1.30$\pm$0.02 & -0.87$\pm$0.02 & -1.66$\pm$0.03 &  2.50$\pm$0.01 \\
\hline
NGC6951 & 309.28613 &  66.10472 & -3.07$\pm$0.06 & -3.70$\pm$0.06 & -1.73$\pm$0.01 & -1.95$\pm$0.01 & -1.41$\pm$0.02 & -1.17$\pm$0.02 & -1.83$\pm$0.03 &  2.67$\pm$0.01 \\
  & 309.29367 &  66.10500 & -3.12$\pm$0.06 & -3.79$\pm$0.06 & -1.55$\pm$0.01 & -1.77$\pm$0.01 & -1.33$\pm$0.02 & -0.94$\pm$0.02 & -1.78$\pm$0.03 &  2.86$\pm$0.01 \\
  & 309.30121 &  66.10528 & -2.80$\pm$0.06 & -3.46$\pm$0.06 & -1.26$\pm$0.01 & -1.48$\pm$0.01 & -1.03$\pm$0.02 & -0.67$\pm$0.02 & -1.48$\pm$0.03 &  3.14$\pm$0.01 \\
  & 309.30875 &  66.10555 & -2.56$\pm$0.06 & -3.23$\pm$0.06 & -1.11$\pm$0.01 & -1.33$\pm$0.01 & -0.85$\pm$0.02 & -0.53$\pm$0.02 & -1.28$\pm$0.03 &  3.30$\pm$0.01 \\
  & 309.31628 &  66.10583 & -2.75$\pm$0.06 & -3.46$\pm$0.06 & -1.24$\pm$0.01 & -1.46$\pm$0.01 & -0.98$\pm$0.02 & -0.63$\pm$0.02 & -1.43$\pm$0.03 &  3.16$\pm$0.01 \\
  & 309.32382 &  66.10611 & -3.09$\pm$0.06 & -3.76$\pm$0.06 & -1.53$\pm$0.01 & -1.76$\pm$0.01 & -1.30$\pm$0.02 & -0.91$\pm$0.02 & -1.75$\pm$0.03 &  2.87$\pm$0.01 \\
  & 309.33139 &  66.10639 & -3.02$\pm$0.06 & -3.52$\pm$0.06 & -1.73$\pm$0.01 & -1.95$\pm$0.01 & -1.44$\pm$0.02 & -1.20$\pm$0.02 & -1.83$\pm$0.03 &  2.67$\pm$0.01 \\
\hline
NGC7723 & 354.73221 & -12.96361 & -3.01$\pm$0.06 & -3.30$\pm$0.06 & -1.79$\pm$0.01 & -2.03$\pm$0.01 & -1.60$\pm$0.02 & -1.33$\pm$0.02 & -1.98$\pm$0.04 &  2.56$\pm$0.01 \\
  & 354.73508 & -12.96222 & -2.88$\pm$0.06 & -3.19$\pm$0.06 & -1.59$\pm$0.01 & -1.83$\pm$0.01 & -1.45$\pm$0.02 & -1.09$\pm$0.02 & -1.84$\pm$0.04 &  2.75$\pm$0.01 \\
  & 354.73792 & -12.96083 & -2.75$\pm$0.06 & -3.04$\pm$0.06 & -1.49$\pm$0.01 & -1.74$\pm$0.01 & -1.35$\pm$0.02 & -0.98$\pm$0.02 & -1.72$\pm$0.04 &  2.85$\pm$0.01 \\
  & 354.74075 & -12.95944 & -2.89$\pm$0.06 & -3.19$\pm$0.06 & -1.59$\pm$0.01 & -1.83$\pm$0.01 & -1.43$\pm$0.02 & -1.07$\pm$0.02 & -1.83$\pm$0.04 &  2.76$\pm$0.01 \\
  & 354.74362 & -12.95806 & -3.08$\pm$0.06 & -3.38$\pm$0.06 & -1.79$\pm$0.01 & -2.03$\pm$0.01 & -1.58$\pm$0.02 & -1.29$\pm$0.02 & -2.00$\pm$0.04 &  2.55$\pm$0.01 \\
\hline
\caption{
Aperture protometry, star formation rate and stellar surface densities.
}
\label{ap_phot}
\end{longtable}
\end{tiny}
\end{center}